\newcommand{\mydriver}{pdftex} 

\documentclass[12pt,\mydriver]{thesis}  

\usepackage{styles/titlesec}
   \titleformat{\chapter}
      {\normalfont\large}{Chapter \thechapter:}{1em}{}

\usepackage{graphicx}
\usepackage{amsmath}
\usepackage{cite}
\usepackage{lscape}
\usepackage{styles/indentfirst}
\usepackage{styles/latexsym}
\usepackage{styles/multirow}
\usepackage{styles/tabls}
\usepackage{wrapfig}
\usepackage{styles/slashbox}
\usepackage{longtable}
\usepackage{styles/supertabular}
\usepackage{subfigure}
\usepackage[bookmarks, colorlinks=true, plainpages=false, citecolor=black, urlcolor=blue, filecolor=blue, linkcolor=black]{hyperref}
\usepackage{lineno}

\renewcommand{\baselinestretch}{2}
\begin{document}

\hbox{\ }

\renewcommand{\baselinestretch}{1}
\small \normalsize

\vspace{0.75in}

\begin{center}
\large{{ABSTRACT}}

\vspace{3em}

\end{center}
\hspace{-.15in}
\begin{tabular}{ll}
Title of dissertation:    & {\large  AN ALL-SKY SEARCH FOR}\\
&				      {\large  BURSTS OF VERY HIGH ENERGY} \\
&				      {\large  GAMMA RAYS WITH HAWC} \\
\ \\
&                          {\large  Joshua Wood, Doctor of Philosophy, 2016} \\
\ \\
Dissertation directed by: & {\large  Professor Jordan Goodman} \\
&  				{\large	 Department of Physics } \\
\end{tabular}

\vspace{3em}

\renewcommand{\baselinestretch}{2}
\large \normalsize
A new ground-based wide-field extensive air shower array known as the High-Altitude Water Cherenkov (HAWC) Observatory promises a new window to
monitoring the $\sim$100 GeV gamma-ray sky with the potential for detecting
a high energy spectral cutoff in gamma-ray bursts (GRBs). It represents a roughly 15 times sensitivity gain over
the previous generation of wide-field gamma-ray air shower instruments and
is able to detect the Crab Nebula at high significance ($>$5 $\sigma$) with each
daily transit. Its wide field-of-view ($\sim$2 sr) and $>$95\% uptime make it an
ideal instrument for detecting GRB emission at $\sim$100 GeV
with an expectation for observing $\sim$1 GRB per year based on existing
measurements of GRB emission.

An all-sky, self-triggered search for VHE emission produced by GRBs with
HAWC has been developed. We present the results of this search on three
characteristic GRB emission timescales, 0.2 seconds, 1 second, and 10
seconds, in the first year of the fully-populated HAWC detector which is
the most sensitive dataset to date. No significant detections were found,
allowing us to place upper limits on the rate of GRBs containing
appreciable emission in the $\sim$100 GeV band. These constraints exclude
previously unexamined parameter space.

\thispagestyle{empty}
\hbox{\ }
\vspace{1in}
\renewcommand{\baselinestretch}{1}
\small\normalsize
\begin{center}

\large{{AN ALL-SKY SEARCH FOR BURSTS OF\\
VERY HIGH ENERGY GAMMA RAYS WITH HAWC}}\\
\ \\
\ \\
\large{by} \\
\ \\
\large{Joshua Randall Wood}
\ \\
\ \\
\ \\
\ \\
\normalsize
Dissertation submitted to the Faculty of the Graduate School of the \\
University of Maryland, College Park in partial fulfillment \\
of the requirements for the degree of \\
Doctor of Philosophy \\
2016
\end{center}

\vspace{7.5em}

\noindent Advisory Committee: \\
Professor Jordan Goodman, Chair/Advisor \\
Professor Greg Sullivan \\
Professor Julie McEnery \\
Professor Peter Shawhan \\
Professor Chris Reynolds

\thispagestyle{empty}
\hbox{\ }

\vfill
\renewcommand{\baselinestretch}{1}
\small\normalsize

\vspace{-.65in}

\begin{center}
\large{\copyright \hbox{ }Copyright by\\
Joshua Randall Wood  
\\
2016}
\end{center}

\vfill

\pagestyle{plain}
\pagenumbering{roman}
\setcounter{page}{2}

\renewcommand{\baselinestretch}{1}
\small\normalsize
\tableofcontents 
\newpage
\listoffigures 
\newpage
\addcontentsline{toc}{chapter}{List of Abbreviations}

\renewcommand{\baselinestretch}{1}
\small\normalsize
\hbox{\ }

\vspace{-3em}

\begin{center}
\large{List of Abbreviations}
\end{center} 

\vspace{3pt}

\begin{tabular}{ll}
BATSE & Burst and Transient Source Experiment \\
CGRO & Compton Gamma Ray Observatory \\
DAQ & Data Acquisition System \\
IACT & Imaging Atmospheric Cherenkov Telescope \\
NTP & Network Time Protocol \\
GPS & Global Positioning System \\
GRB & Gamma-Ray Burst \\
HAWC & High-Altitude Water Cherenkov (Observatory) \\
pe & photoelectron \\
TDC & Time-to-Digital Converter \\
VHE & Very-High Energy ($>$ $\sim$100 GeV)\\
WCD & Water Cherenkov Detector \\
WCT & Water Cherenkov Telescope \\
\end{tabular}

\newpage
\setlength{\parskip}{0em}
\renewcommand{\baselinestretch}{2}
\small\normalsize

\setcounter{page}{1}
\pagenumbering{arabic}
\renewcommand{\thechapter}{1}
\chapter{Gamma-Ray Burst Science}\label{ch:intro}

\section{Introduction}

Gamma-ray bursts (GRBs) are the most luminous events in the known universe.
They consist of intense gamma-ray flashes coming from cosmological distances
with durations ranging from 10$^{-3}$ to 10$^3$ seconds. Their spectra show
non-thermal emission, predominantly at keV to MeV energies, that accounts for
a beaming-corrected energy release of $\sim$10$^{51}$ ergs, which is roughly
equivalent to the total energy output by the Sun over its entire lifetime.

The prompt gamma-ray flashes associated with GRBs are followed by long-lasting,
smoothly decaying afterglow signals at X-ray and optical frequencies that have lead to
the identification of extragalactic host galaxies. This resulted in the association
of long timescale GRBs, defined by timescales longer than $\sim$2 seconds, with
core-collapse supernovae in massive stars. Current population studies of host galaxies
in short duration bursts point to a progenitor class of compact-binary mergers for
durations less than $\sim$2 seconds \cite{Berger:2013jza}.

Observations support a model in which both progenitor classes form an accreting black hole powering
a highly relativistic jet with gravitational energy released during the infall of surrounding matter.
The jet interacts both with itself and surrounding material to form internal and external shocks where Fermi acceleration takes place.
Accelerated charged particles subsequently emit synchrotron radiation to generate the measured
non-thermal spectrum \cite{Meszaros:2006rc} \cite{Piran:1999kx}.

Yet despite the many advancements made in the field of GRB science in the nearly 50 years since
their discovery many open questions remain. In particular,
relatively little is known about the behavior of prompt GRB emission at the highest
energies which is a regime of interest both for its ability to probe the physical environment
of GRBs as well as the density of light in the high redshift Universe. The primary purpose of
this dissertation then is to provide a measurement of GRB emission in the very-high energy (VHE) regime.

We begin in this chapter with an overview of the major observational results that have
informed the theoretical model for GRB emission. In doing so we emphasize the 
observational difficulties associated with performing measurements of GRB emission
at VHE energies in current ground-based Imaging Atmospheric Cherenkov Telescopes and satellite-based experiments.
We use this to motivate the need for a new ground-based, wide-field gamma-ray experiment known
as the High-Altitude Water Cherenkov (HAWC) Observatory which expects to observe $\sim$1 GRB per year at
VHE photon energies. We also describe the current model behind GRB emission in Section \ref{sec:fireballmodel}
to inform the reader why we expect to see VHE emission from GRBs.

In Chapter \ref{ch:showers} we describe the physical processes behind the development of air showers measured in
ground-based observational techniques to inform our discussion of the experimental design of the HAWC
Observatory in Chapter \ref{ch:hawcobs} and its methodology for air shower reconstruction in Chapter \ref{ch:Reco}.
Chapter \ref{ch:method} focuses on the description of our all-sky, self-triggered search for VHE emission
from GRBs and \ref{ch:sens} describes its sensitivity. Chapter \ref{ch:results} presents the results
of our search from the first year of data available from the HAWC detector. No significant detections were found.

While we have yet to detect a GRB, we conclude that our analysis does have sensitivity
to known bursts GRB 090510 and GRB 130427A as well as a potential population of low fluence
bursts that do not trigger the Fermi Large Area Telescope. It may therefore be only another
year or two before we obtain our first detection of VHE emission from a GRB. Furthermore,
recent advances in the on-site reconstruction performed in real-time at the HAWC site now
allow us to run our algorithm in real-time with the same sensitivity as presented here.
This offers the tantalizing prospect using the HAWC Observatory to trigger
the Very Energetic Radiation Imaging Telescope Array System (VERITAS)
as both observe the same overhead sky. This, in principle, might lead to the first VHE
follow-up detection of a GRB by an Imaging Atmospheric Cherenkov Telescope as well.



\newpage

\section{Observations}\label{sec:obs}

\subsection{Discovery}

The first GRB detection occurred in 1967 when the Vela system of satellites
observed a brief flash of gamma-ray photons while monitoring for violations
of the Nuclear Test Ban Treaty \cite{Gehrels:2009qy}. A further 16 bursts were recorded
between 1969 and 1972 with durations ranging from 0.1 to 30 seconds and only one burst found to be associated with a solar flare \cite{Kleb:1973}.
Both the Sun and Earth were eliminated as sources for the remaining bursts,
leading researches to conclude they were observing phenomena of cosmic origin. 
The exact nature of the cosmic sources producing GRBs was unknown at the time
as the small data set of available bursts failed to correlate with known astrophysical transients, such as supernovae.

\newpage

\subsection{BATSE}\label{sec:BATSE}

While other satellites continued to contribute to the data set of known GRBs after their discovery by the Vela network,
the first major experiment specifically designed to study GRBs was launched on-board the Compton Gamma-ray Observatory (CGRO)
in 1991 \cite{CGRO:1991}. This experiment, known as the Burst and Transient Source Experiment (BATSE), was sensitive
to gamma-ray energies from 15 keV - 2 MeV with a 4$\pi$ sr field-of-view and an angular resolution of $\sim$2$^\circ$ \cite{BATSE:1999}.
BATSE observed nearly 3000 GRBs during its operational period from 1991-2000, revealing a rate of two to three visible bursts
occurring in the Universe each day after accounting for burst occultation by the Earth \cite{Gehrels:2009qy}.

\begin{figure}[ht!]
\begin{center}
\includegraphics[height=3in]{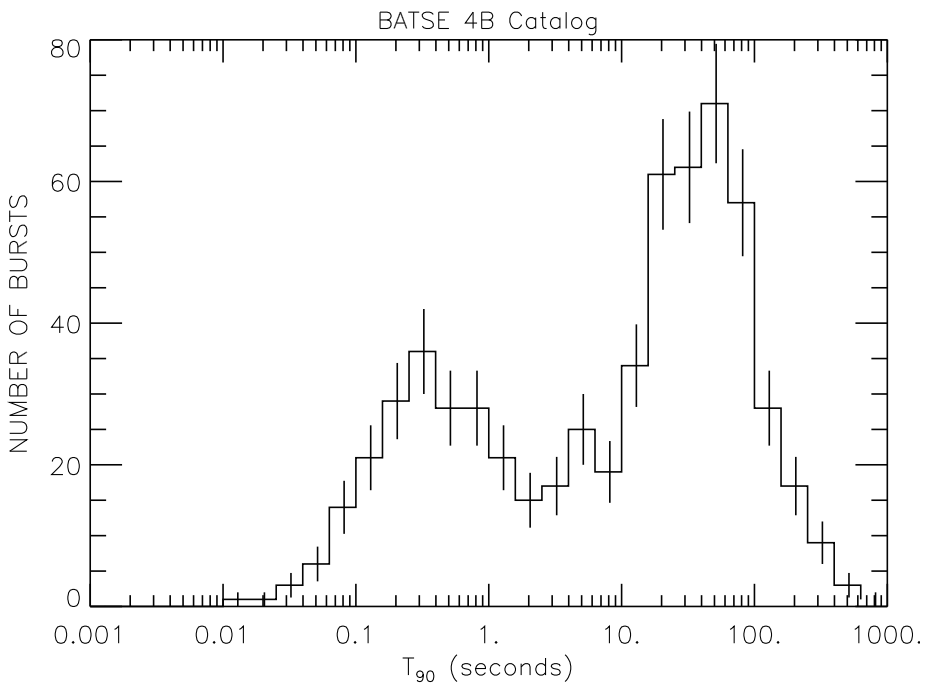}
\caption{Duration distribution ($T_{90}$) of BATSE GRBs \cite{Mallozzi:2014a}.
         $T_{90}$ is defined as the timescale over which 90\% of the measured GRB photons arrive.
         This distribution reveals two populations of bursts, short and long, divided by $T_{90} = 2$ s.}
\label{fig:t90distbatse}
\end{center}
\end{figure}

\begin{figure}[ht!]
\begin{center}
\includegraphics[height=5.5in]{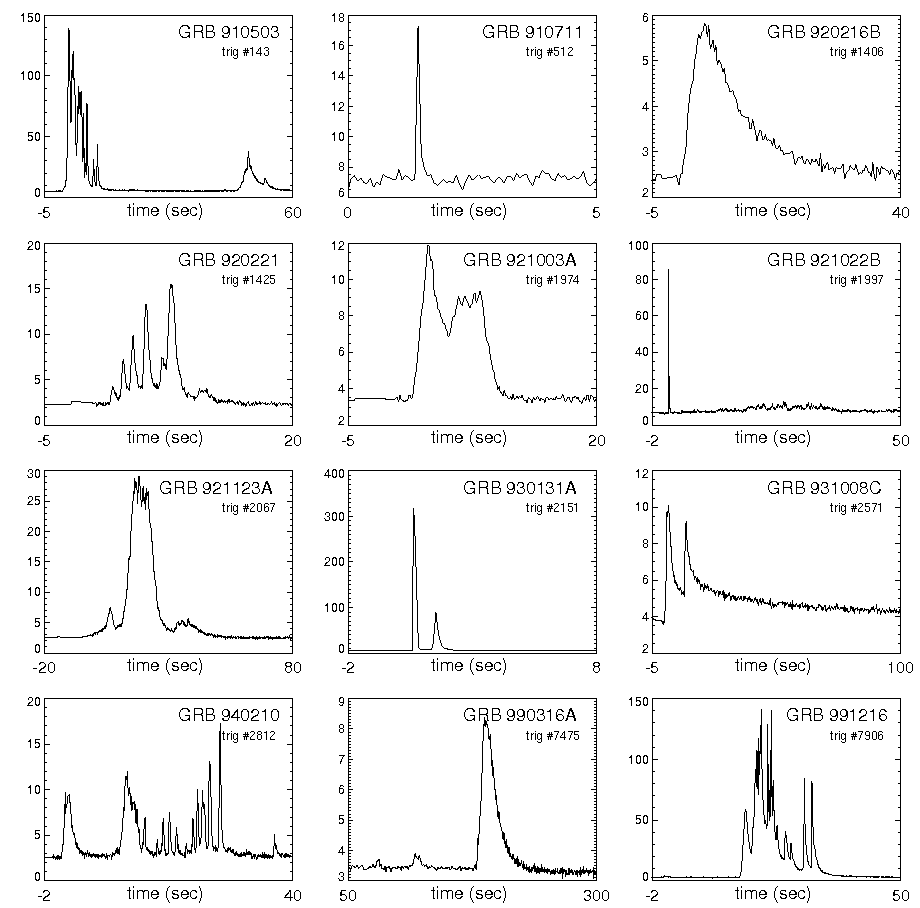}
\caption{Twelve light curves for measured BATSE GRBs \cite{Mallozzi:2014b}.
         The short timescales of the peaks observed over the duration of each light curve
         denotes a variability time much shorter than $T_{90}$ and suggests that emission
         is produced by a compact object.}
\label{fig:batsepanel}
\end{center}
\end{figure}

An important result of the BATSE data set is the fact that the distribution of burst durations, measured by the time
in which 90\% of the observed gamma-ray photons arrive ($T_{90}$), is bimodal suggesting two
different progenitor populations (Figure \ref{fig:t90distbatse}). $T_{90} = 2$ seconds marks the transition point
between the two halves of this distribution in BATSE leading to the general classification of bursts possessing durations less
than 2 seconds as short-duration bursts with longer bursts referred to as long-duration bursts, although some overlap
between the populations is known to occur. Both the short and long duration populations were seen to exhibit variability
on timescales much smaller than $T_{90}$ (Figure \ref{fig:batsepanel}) suggesting that emission is produced by a compact object in both cases.

\newpage

\newpage

\newpage

\begin{figure}[ht!]
\begin{center}
\includegraphics[height=4in]{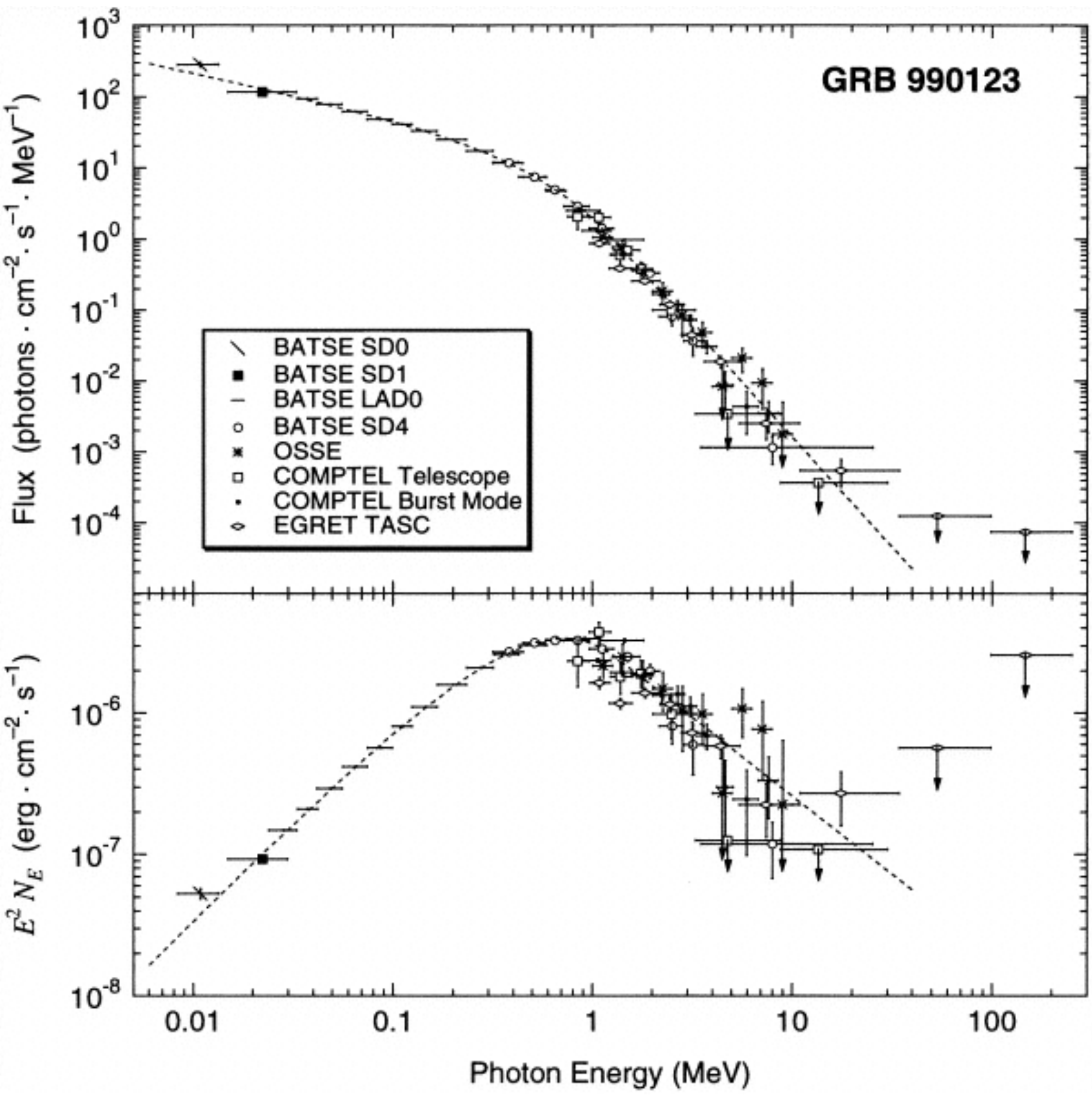}
\caption{Band function fit to the spectrum of GRB 990123 shown as both the number of photon flux $N_E$ and in $E^2 N_E = \nu f_\nu$ units \cite{Briggs:1999}.
         The crosses mark measurements made by BATSE and the Imaging Compton Telescope, two instruments on-board the Compton Gamma-Ray Observatory.
         The dashed line marks the Band function fit to the data.}
\label{fig:bandfit}
\end{center}
\end{figure}

Another development during the BATSE era was the success achieved by fitting time-integrated burst spectra with the phenomenological
Band function defined as
\begin{equation}
f_{BAND}(E) =  \Bigg\{
  \begin{array}{lr}
  A \, ( \frac{E}{100 \, keV} )^\alpha  \, e^{ - \frac{E (\alpha - \beta)}{E_c} }, \,\,\, E < E_c \\
  A \, ( \frac{E_c}{100 \, keV} )^\alpha \, e^{\, \beta -  \alpha} \, ( \frac{E}{E_c} )^\beta, \,\,\, E \ge E_c
  \end{array}
\end{equation}
where $E_c$ is related to the peak energy in plots of $\nu f(\nu)$ according to
\begin{equation}
E_c = (\alpha - \beta) \frac{ E_{peak}}{2 + \alpha}
\end{equation}
and $A$ is the spectrum normalization in photons / cm$^2$ s keV \cite{Kaneko:2006}.
Figure \ref{fig:bandfit} on the following page shows the shape of a Band function fit to GRB 990123.
Most GRBs have values of $\alpha \approx -1$ and $\beta \approx -2$ \cite{Gruber:2014}.
While this fit is empirically motivated, we will show in Section \ref{sec:fireballmodel} that the overall shape of a smoothly joined power law with a peak energy is
expected from synchrotron emission by a population of energetic electrons accelerated at collisionless shocks.

\newpage

\newpage

\begin{figure}[b!]
\begin{center}
\includegraphics[height=2.8in]{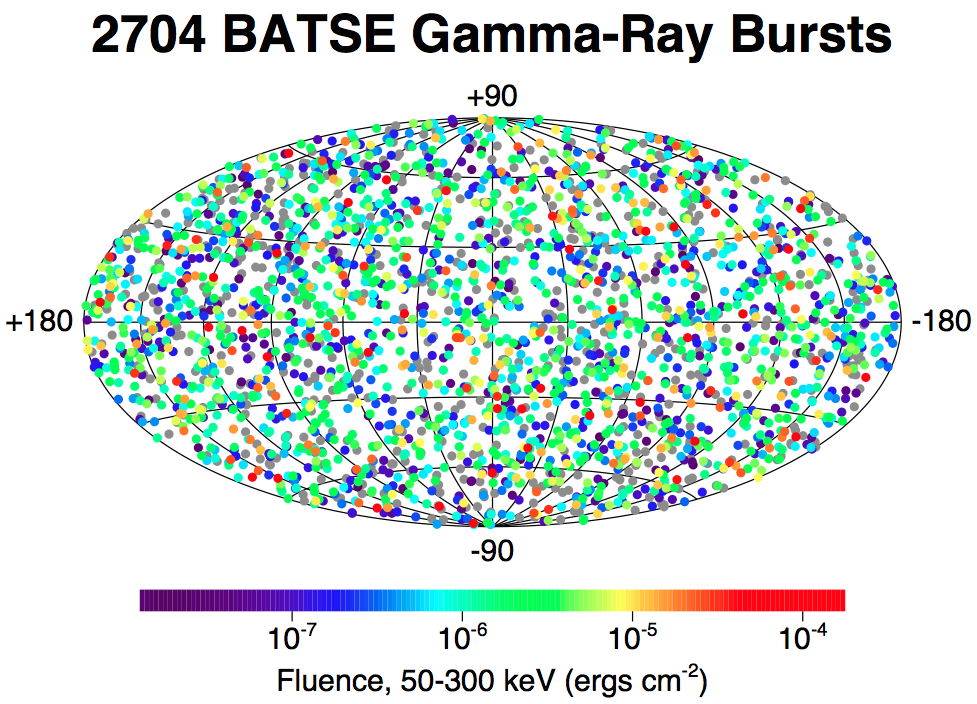}
\caption{Spatial distribution of BATSE GRBs \cite{Briggs:2014a}.}
\label{fig:batsespatial}
\end{center}
\end{figure}

The last major result that we will discuss from the BATSE data set is the isotropic distribution of burst locations throughout the sky
(Figure \ref{fig:batsespatial}). This suggests GRBs are extragalactic in origin as the distribution is expected
to be non-uniform for bursts occurring within the Milky Way \cite{Gehrels:2009qy}. However, optical observations
were unable to confirm this through redshift measurements of BATSE GRBs as the provided angular resolution
was much too large to locate the host galaxies of GRB events. Better measurements of the prompt GRB spectrum were needed
at X-ray energies where the incident photons can be reflected and focused to provide resolutions on the order of 1 arcmin.

\newpage

\subsection{Afterglow Follow-up}

A major breakthrough in the observational study of GRBs occurred when the BeppoSAX 
satellite detected X-ray emission from GRB 970228 \cite{Bloom:2001}. The X-ray measurements provided
a localization of $\sim$1 arcmin which allowed the first successful optical follow-up to be performed, confirming
an extragalactic origin in a host galaxy at redshift $z = 0.70$. This proved that GRBs originated outside the
Milky Way but also implied a very large isotropic energy release ($\sim$10$^{54}$ erg) given the high flux of
photons measured in the keV-MeV range in GRB spectra at Earth over the cosmological distance to the source.

\begin{figure}[ht!]
\begin{center}
\includegraphics[height=3in]{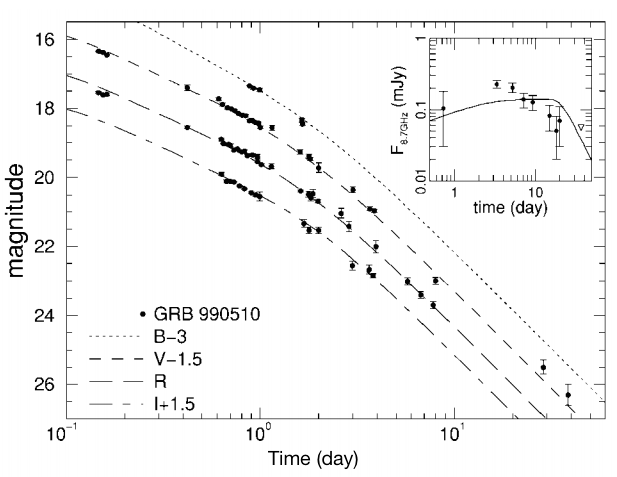}
\caption{Observed break in the afterglow emission at different wavelengths from GRB 990510 \cite{kumar:2000}.
         This break results when the jetted relativistic outflow from the GRB source collides
         with the external burst environment and slows to the level where relativistic
         beaming is on the order of the jet opening angle.}
\label{fig:jetbreak}
\end{center}
\end{figure}

The high photon flux combined with the compact distance scale required by the short variability times of GRB sources
also caused an issue known as the compactness problem where the expected photon density in the keV-MeV range at the source
was expected to be high enough to entirely absorb the observed non-thermal emission via photon-photon pair production \cite{Ruderman:1975}.
This problem was solved with the realization that photon emission occurring in the rest frame of a jetted relativistic outflow with $\Gamma >$ 100 from the GRB source would place the keV to MeV
photons observed at Earth below the pair production threshold at the source itself. This relativistic outflow was expected to produce a characteristic
steepening of afterglow emission at late times when relativistic beaming effects reached the order of the opening angle of the jet as it slowed in the external burst environment,
which was confirmed in the behavior of typical afterglow observations (Figure \ref{fig:jetbreak}).

\newpage

Jet opening angles measured from breaks in afterglow emission were subsequently
used to correct previously estimated values for the isotropic energy release to account
for beaming of the source photons, yielding measured energy releases on the order of typical supernovae ($\sim10^{51}$ erg) \cite{Frail:2001}.
The connection to supernovae was confirmed in the case long GRB 980425 when afterglow emission was observed to be coincident with the
type Ic supernova SN 1998bw \cite{Woosley:1999}.
Afterglow observations of other long GRBs have since provided a number of type Ic supernovae associations, overwhelmingly supporting the interpretation of
core collapse supernovae as the progenitors for long GRBs \cite{Woosley:2006}.

\begin{figure}[ht!]
\begin{center}
\includegraphics[height=2.8in]{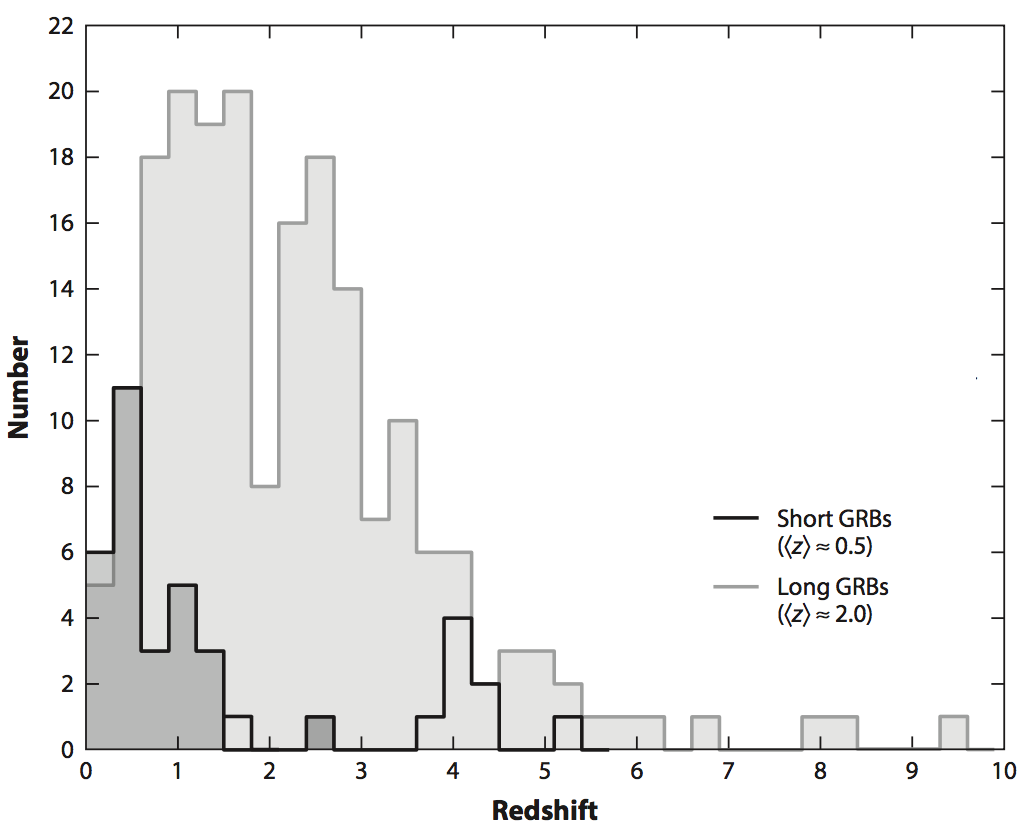}
\caption{Redshift distribution for short (BLACK) and long (GREY) GRBs for
         GRBs with measured redshift \cite{Berger:2013jza}.
         The open portions of the histogram for short GRBs indicates
         upper limits based the lack of spectral features in afterglow
         and/or host-galaxy optical detections.}
\label{fig:redshift}
\end{center}
\end{figure}

Despite the success of long GRB observations in the early period of afterglow
observations, follow-up of prompt emission in short GRB events remained elusive with the first generation of X-ray satellites
designed to detect GRBs as their shorter timescales stymied follow-up observations.
This changed in 2005 with the launch of the \textit{Swift} satellite which uses a suite of instruments to automatically detect and perform
rapid afterglow follow-ups of GRBs \cite{Gehrels:2004}. \textit{Swift} detected the first
short GRB afterglow in GRB 050509b \cite{Castro:2005} and ushered in a new era of successful optical follow-up
that significantly expanded measurements of the GRB redshift distribution in both the short and long GRBs. Figure \ref{fig:redshift}
shows the short and long GRB redshift distributions for the set of GRBs with measured redshifts available at the time of a recent GRB review.

The current set of known host-galaxies obtained with optical follow-ups
further confirms the association of long GRBs with core-collapse supernova as they consist
exclusively in star-formation galaxies where core-collapse supernovae are known to occur \cite{Perley:2016}.
Similar host-galaxy studies, in addition to the compact source size and energetics required by prompt emission,
in observations of short GRBs currently suggest progenitors of either a merger between a binary neutron star pair
or a neutron star in a binary system with a black hole \cite{Berger:2013jza}.

\newpage

\newpage

\subsection{Fermi Satellite}

While the \textit{Swift} satellite revolutionized optical follow-up
of GRBs at lower energies, the launch of the Fermi Gamma-Ray Space Telescope in 2008
opened the window to high energy observations of the prompt emission phase above 100 MeV
with its pair of instruments, the Gamma-Ray Burst Monitor (GBM) \cite{Meegan:2009} and the Large Area Telescope (LAT) \cite{Atwood:2009ez}.
Together these instruments offer the unique ability to trigger on the keV-MeV photons typical of spectra associated with GRB emission
and immediately follow with measurements at GeV energies. Triggers are provided by the GBM which covers the energy range from 8 keV-40 MeV
with full view of the unocculted sky and the LAT provides high energy measurements from 20 MeV to $>$300 GeV over a 2.4 sr field-of-view.

One of the first major results to come from the Fermi mission was the observation of high energy
emission during the prompt phase of GRB 080916C which was observed to both start later and last longer than emission
in the keV-MeV energy range \cite{Abdo:2009}. This was again confirmed in the short-hard gamma-ray burst GRB 090510
which also yielded the first LAT detection of significant spectral deviation from the empirical Band function
in the form of an additional high energy power law component \cite{Ackermann:2010}.
Soon a picture began emerging about the high energy emission which was characterized by
starting later than emission observed at lower energies, lasting longer, and decaying as a power law $t^{-\alpha}$
after the end of the low energy emission \cite{Ackermann:2013a}.

\newpage

\begin{figure}[ht!]
\begin{center}
\includegraphics[height=2.65in]{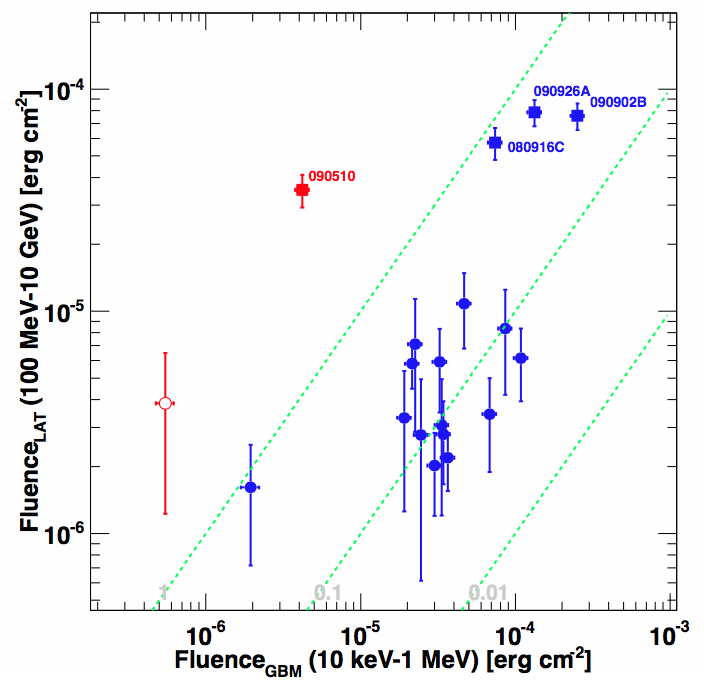}
\caption{Fluence measured at high energies in Fermi LAT versus fluence measured at low energies Fermi GBM \cite{Ackermann:2013a}.
         Red symbols indicate short GRBs and blue symbols indicate long GRBs.
         The lines mark fluence ratios of 0.01, 0.1, and 1 with the two short bursts having
         a ratio of $\sim$1 and the group of long bursts exhibiting a ratio of $\sim$0.1.}
\label{fig:fluratio}
\end{center}
\end{figure}

Additionally, two sets of high energy GRBs became apparent. 
First, there was a large group of long duration bursts with high energy fluences on the order of 10\%
the fluence measured in the GBM  (Figure \ref{fig:fluratio}) and $E^{-2}$ power laws at the highest energies in addition to the Band component.
And second, there was a set of short-hard bursts with high energy fluences on the order of 100\% the low energy fluence detected in the GBM
with hard, high energy power laws like the $E^{-1.6}$ component found in GRB 050910.

\vspace{2cm}

The distinctness of the picture presented in high energy Fermi LAT data
compared to the GBM data at lower energies implies a different emission mechanism
in addition to the synchrotron emission in the region producing the majority of prompt emission at low energies.
This is particularly true in the case of the extraordinary burst GRB 130427A
for which the highest energy measured photon of 95 GeV occurred 244 seconds after the
start of the burst. Such a high energy photon cannot occur in standard
interpretations of electron synchrotron models as the cooling time is much too short \cite{Zhu:2013ufa}.
Yet a key feature needed for distinguishing between possible high energy emission mechanisms 
is largely missing from Fermi data, namely a high-energy
cutoff related to the intrinsic environment of the high energy emission region.

\begin{figure}[ht!]
\begin{center}
\includegraphics[height=2.65in]{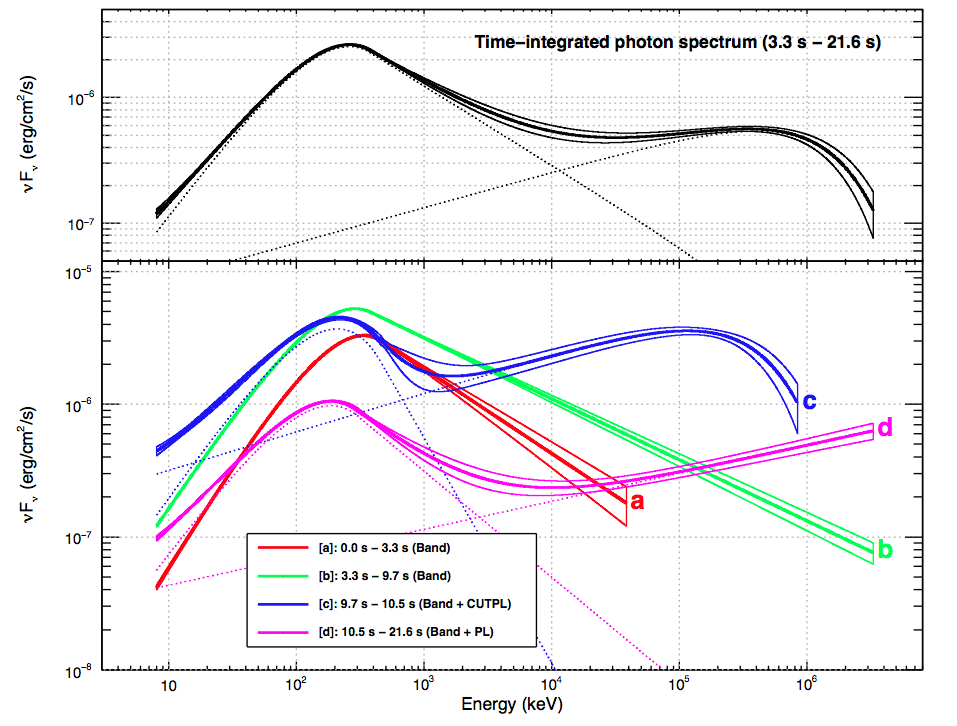}
\caption{Fits to the spectra observed in GRB 090926A by the Fermi satellite \cite{Bregeon:2011bu}.
         The presence of a spectral break at 1.4 GeV is detected with good significance ($\sim4\sigma$).
         To date, this is the best measurement of a cutoff at GeV energies.}
\label{fig:spectralbreak}
\end{center}
\end{figure}

Such a cutoff must occur at some point in GRB spectra as the finite
Lorentz boost of the relativistic jet powering GRB emission cannot prevent the highest
photon energies from pair producing off the observed flux of keV-MeV photons inside the GRB source.
To date, however, the strongest evidence for a cutoff remains the 4$\sigma$ detection of a cutoff at 1.4 GeV in GRB 090926C
(Figure \ref{fig:spectralbreak}) despite the $\sim$10 GRBs detected by the LAT each year. This implies
then that most spectral cutoffs occur well into the GeV energy range 
where the $\sim 1$ m$^2$ effective area provided by the Fermi LAT simply is not large enough to accumulate the
statistics needed to determine a cutoff given the steeply falling flux of typical GRB spectra.
This motivates the need for ground-based detections of GRB emission as the current generation
of ground-based gamma-ray observatories have effective areas to $\sim$100 GeV photons that are $\ge$100x
the size of the Fermi LAT.



\newpage

\newpage

\subsection{Ground-based Non-observations}\label{sec:gndbase}

Given the single 95 GeV photon seen from GRB130427A in Fermi alone,
one expects ground-based TeV gamma-ray detectors to be capable of observing on the order of $\sim$100 VHE
photons or more arriving from a similar GRB as the effective areas of current generation
detectors are greater than 100x the size of the Fermi satellite for photon energies above 50 GeV \cite{VERITAS:2013} \cite{HESS:2005} \cite{MAGIC:2016} \cite{Abeysekara:2013tza}. Yet all available experiments have thus far only reported non-detections. As we shall see below, this is largely
due to the design of ground-based experiments built prior to the HAWC Observatory.

\begin{figure}[ht!]
\begin{center}
\subfigure[][]{\includegraphics[height=2.5in]{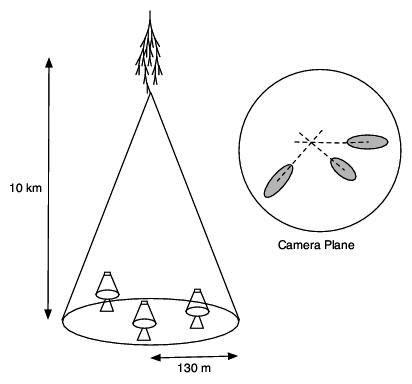}}
\quad\quad\quad\quad
\subfigure[][]{\includegraphics[height=2.2in]{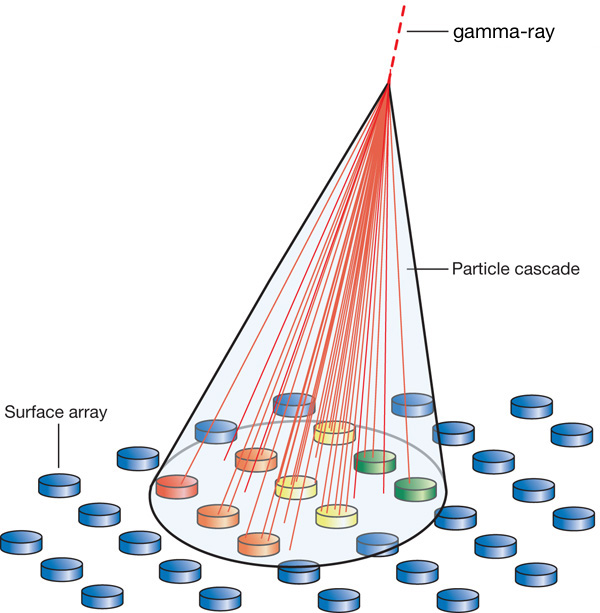}}
\end{center}
\caption{Diagram of (a) IACT technique and (b) WCT technique.
         The IACT method uses multiple telescopes to image air shower propagation through the atmosphere.
         The WCT method uses a ground array of water tanks to measure the energy deposited by electromagentic
         shower particles at ground level.}
\label{fig:technique}
\end{figure}

Ground-based gamma-ray experiments generally fall into two main classes, Imaging Atmospheric Cherenkov Telescopes (IACTs) and Water Cherenkov Telescopes (WCTs),
based on the different techniques used to measure air showers generated by gamma-rays in the upper atmosphere. IACTs employ mirrors to focus
Cherenkov light generated by secondary air shower particles as they move through the air onto a camera that allows them to track shower progression
through the atmosphere in the 2D plane of the camera. The resulting image from one telescope is then combined with a set of images from other IACTs
placed nearby and acting in unison to obtain a complete picture of the air shower trajectory (Figure \ref{fig:technique}). In contrast, WCTs measure
the energy of secondary air shower particles reaching ground level in a surface array of water tanks. The shower trajectory is reconstructed
from the arrival times of particles across the array.

\newpage

The benefit of the IACT technique is very good angular resolution ($\sim$0.1$^\circ$)
because the full shower progression is tracked but it comes at a cost of a very small field-of-view ($\sim4^\circ$).
IACTs are therefore pointed instruments and must be triggered to slew to a GRB transient. The latency associated
with receiving a trigger from another experiment combined with the time needed to slew to the position of a burst
means that IACT follow-ups of GRBs have occurred, at best, minutes after the end of the burst T90 \cite{MAGIC:2009} \cite{HESS:2009} \cite{VERITAS:2011}
where the high energy signal is already expected to have rapidly decayed.
Furthermore,  the IACT technique only works on clear, dark nights
resulting in duty cycles of $\sim$15\% that can inhibit follow-up studies until the following day as in the case of the
reported VERITAS observation of GRB 130427A \cite{VERITAS:2014}. We therefore conclude that the current set
of non-observations by IACT instruments is the result of their low duty cycles and small fields-of-view.

WCTs compensate for the short-comings of IACTs by being wide-field instruments
capable of detecting transients over the entire overhead sky without the need to point.
In addition, the sealed water tanks used to detect secondary air shower particles can be operated regardless of 
atmospheric conditions, such as daylight and cloud coverage, allowing them to operate continuously. This means
that a GRB event will be recorded, even prior to receiving an external trigger, as long as it is within the overhead sky.

\begin{figure}[ht!]
\begin{center}
\includegraphics[height=2.8in]{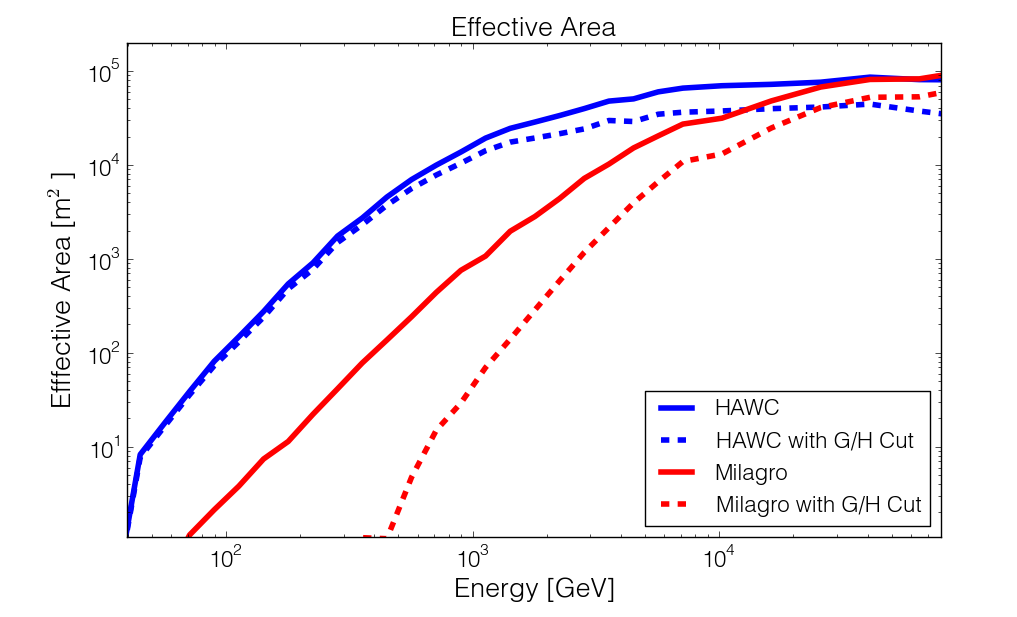}
\caption{Effective area of the Milagro experiment (RED) compared to the
         effective area of the HAWC experiment (BLUE) as a function of
         energy. The dashed curves represent the effective area in each
         experiment after applying typical point-source analysis cuts.
         The point-source analysis cuts were not used in the Milagro search for
         GRB emission as they eliminated the expected GRB signal at 100 GeV.
         The HAWC experiment maintains a much higher effective area even 
         after applying point source analysis cuts due to its higher
         altitude (4100 m in HAWC vs 2630 m in Milagro) which yields
         a lower attenuation of air shower signals.
        }
\label{fig:hawceffarea}
\end{center}
\end{figure}

However, WCTs naturally have lower effective areas for low energy photons
compared to an IACT experiment. This results from the large attenuation experienced by low energy showers as
they travel to ground level (See Section \ref{ch:showers}). Figure \ref{fig:hawceffarea} demonstrates this in
the effective area of the Milagro Gamma-Ray Observatory \cite{Atkins:2000}, the precursor experiment to the HAWC Observatory,
which has a large effective area at high energies but only provides $\sim$3 m$^2$ at 100 GeV after relaxing the analysis cuts typically used for point source analysis because they completely remove signals below $\sim$500 GeV.
While this is comparable to the Fermi satellite, the Milagro experiment was 
much less sensitive to low energy photons as the hadronic air shower background
for ground-based WCTs is much larger than backgrounds in the Fermi LAT.
We conclude then that Milagro's null-detection \cite{Vlasios:2008} after 7 years of 
operations is the result of a detector design which was not sensitive enough to 100 GeV photons to provide
an appreciable detection of VHE emission from a GRB.

\newpage

Figure \ref{fig:hawceffarea} also shows the effective area of the newly completed HAWC Observatory.
This experiment addresses the problem of low energy sensitivity in Milagro by moving the detector
plane to a much higher altitude, 4100 m a.s.l., compared to the 2630 m altitude of the Milagro experiment.
Doing so yields the $\sim$100 m$^2$ area needed to appreciably detect a VHE cutoff during the prompt emission
phase of a GRB. As we shall see in Section \ref{sec:outlook}, this results in an expectation for observing
$\sim$1 GRB per year and represents the most promising prospect of detecting prompt emission from ground-level.

\newpage

\newpage

\section{Theoretical Model}\label{sec:fireballmodel}

Since the Fermi satellite has measured GRB emission at $\sim$100 GeV, there is no question that this emission
exists. However, there is a great deal of uncertainty about how it is made. We will now demonstrate this with a discussion
of the underlying model for GRB emission as supported by the measurements reported in Section \ref{sec:obs}.
Our goal here is to reveal the difficulties associated with determining the mechanisms behind GeV emission
given the current set of measurements and thereby motivate the need for a new type of measurement,
namely a significant detection of a spectral cutoff at the highest energies.

\begin{figure}[ht!]
\begin{center}
\includegraphics[height=2.8in]{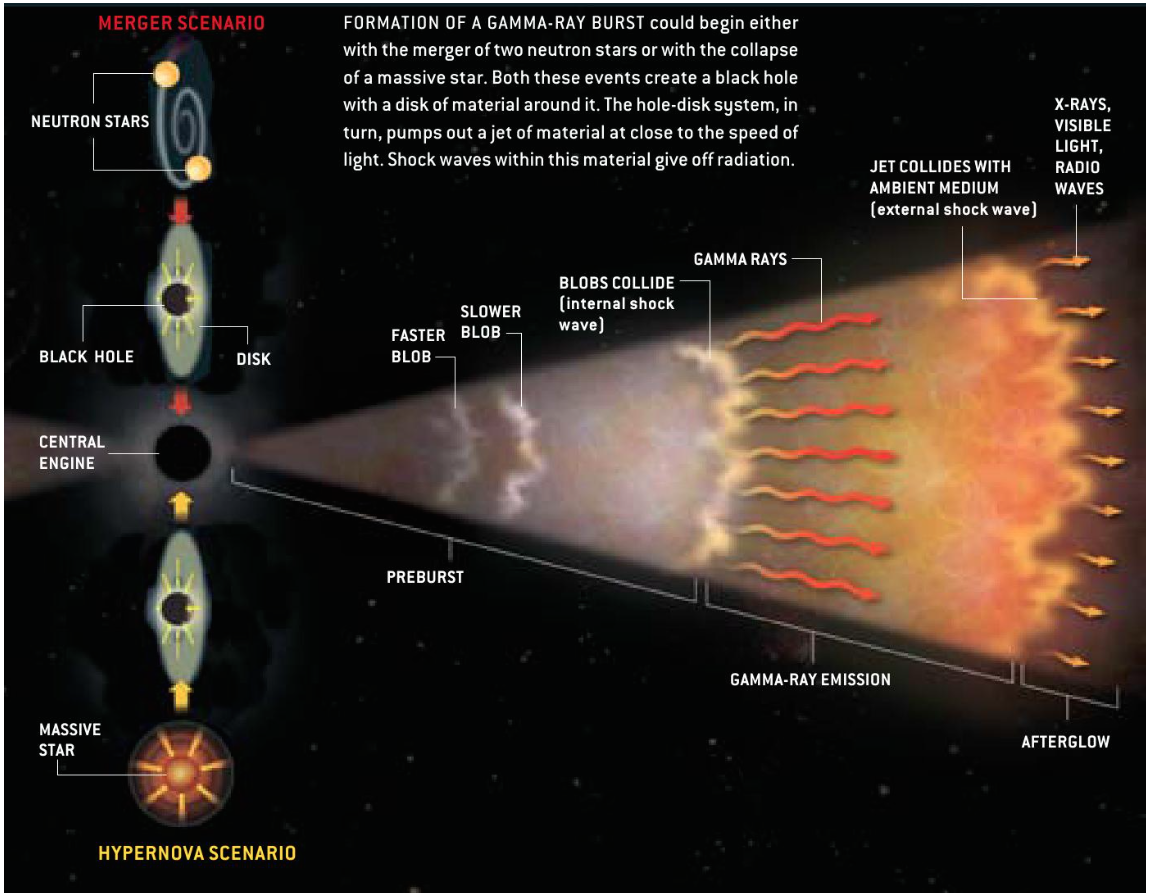}
\caption{Diagram of the fireball model for GRB emission \cite{Gehrels:2002}.
         It consists of both progenitor populations, core-collapsing massive stars and compact binary mergers,
         resulting in a black hole powering a highly relativistic jet. Clumps of matter within the jet are
         believed to collide, resulting in shocks that accelerate particles which subsequently radiate via the synchrotron process. }
\label{fig:fireball}
\end{center}
\end{figure}

The current theoretical model, referred to as the fireball model \cite{Meszaros:2006rc} \cite{Piran:1999kx}, that accounts for both the high temporal variability of GRB light curves
as well as the fluxes measured in the keV - MeV band at Earth is that of a newly formed black
hole powering a highly relativistic jet with gravitational energy released from the infall of surrounding matter
(Figure \ref{fig:fireball}). 
Clumps of matter within the jet travel at different
speeds and form collisionless shock boundaries where electrons are accelerated to high energies by Fermi acceleration \cite{Gehrels:2009qy}.
These electrons subsequently produce synchrotron radiation that yields a low energy spectral index between -3/2 and -2/3, depending
on whether they are in the fast or slow cooling regime, and a high energy
spectral index of about -2, which reflects the steepness in the underlying energy distribution of electrons undergoing Fermi acceleration \cite{Lloyd:2000}.
While this is not a perfect description of all GRB spectra \cite{Preece:1998}, it broadly matches
the overall Band-fit shape of most bursts and must therefore 
play the dominant role in prompt emission.

\newpage

GRB afterglow is also well described by electron synchrotron emission 
with the distinction being that it occurs when the expanding jet of relativistic material
collides with the external burst environment \cite{Sari:1998}. This
accounts for the characteristic temporal decay seen in afterglow light curves, which corresponds to the ejecta slowing
as it sweeps up more matter from the external medium \cite{Mezaros:2006}. It also accounts for the delayed onset
of the afterglow as the internal material producing prompt emission must first expand to the radius
where the density of swept up material is appreciable enough to begin slowing the ejecta and forming a shock boundary \cite{Gehrels:2009qy}.

High energy GRB emission, on the other hand, is in direct contradiction with
simple synchrotron emission models. This is largely because its decay timescale is much
longer than expected from the efficient energy loss associated with electron
synchrotron emission at the highest energies \cite{Zhu:2013ufa}. However, this does not mean that synchrotron emission
is not involved in the production of GeV photons. It could be that the low energy photons
from the prompt emission phase and the afterglow provide the seed photons for inverse Compton
scattering to occur in the external blast wave, which would explain why high energy emission is delayed
 \cite{Beloborodov:2014}. Other possible mechanisms include proton-synchrotron radiation, photo-hadronic interactions,
 and photon pair annihilation cascades \cite{Gehrels:2013}.

A number of different models are therefore being considered for the production
of high energy photons from GRBs, each with their own set of unique constrains.
Yet one commonality is the fact that a measurement of a spectral cutoff at the highest 
energies would provide a better understanding of the environment responsible
for producing high energy emission and allow more differentiate between models.
Specifically, we will show in Section \ref{sec:vhecutoff} that the observation
of a high energy cutoff can provide an estimate of the bulk Lorentz factor of
the material producing high energy GRB photons.

\newpage

\section{Absorption of VHE Emission}\label{sec:vhecutoff}

As mentioned in Section \ref{sec:fireballmodel}, the measurement
of a spectral cutoff at high energies would provide key insights
into understanding current models for high energy emission in GRBs.
We will now motivate this statement by discussing the photon pair production
process involved in creating a high energy cutoff as well as what
a cutoff measurement tells us about the environment where high energy emission occurs.
We will also describe how photon-photon pair production on extragalactic background light (EBL)
causes absorption of VHE photons as they travel to Earth from
the large redshifts of GRB sources. This is an important feature to account for when
studying photon energies $\ge$100 GeV where we expect to make a detection with HAWC.




We begin with the cross section for photon-photon pair production given by
\begin{equation}
\sigma_{\lambda \lambda}(y) = \sigma_T \, g(y), \,\,\,\,\, g(y) = \frac{3}{16} (1 - y^2) \Bigg[ (3 - y^4) ln \frac{1+y}{1-y} - 2 y (2 - y^2) \Bigg]
\label{eq:pair_cross}
\end{equation}
where $\sigma_T = 6.65\times10^{-25}$ cm$^2$ is the Thomson cross section and
\begin{equation}
y^2 \equiv 1 - \frac{ 2 m_e^2 c^4 }{ E_1 \, E_2 \, (1-cos\theta)}
\end{equation}
with $\theta$ being the collision angle and $E_1$ and $E_2$ being the photon energies. Noting that $y$ must be real results in the condition
\begin{equation}
\sqrt{E_1 \, E_2 \, (1-cos\theta)/2 } \ge m_e c^2
\end{equation}
for pair production to occur. This requires the target photon to have at least an energy of $E_2 \approx 0.2$ eV to absorb an incident photon with $E_1 =$ 1 TeV,
 which is easily satisfied inside typical GRB emission environments where we expect the keV-MeV photons measured in typical Band fits
 to provide a target population of photons.


\subsection{Intrinsic Cutoff}

As mentioned above, the flux of photons in the keV-MeV range measured at Earth during the prompt emission phase of GRBs
implies that pair production targets for VHE photons are also present in the source itself. Historically, this resulted
in a contradiction known as the compactness problem because the short variability timescales of the prompt emission phase
implied a compact emission region which should have been opaque to the observed MeV photon flux if keV and MeV photon production
occurred co-spatially \cite{Ruderman:1975}. This problem was resolved in the MeV regime with the understanding that the jets producing
photon emission are highly relativistic with bulk Lorentz factors of $\Gamma \ge$ $\sim100$, placing typical BAND-spectrum photons
below the pair production threshold \cite{Berger:2013jza}. Cutoffs are therefore not relevant to low-energy instruments like the Fermi GBM.

While large, the bulk Lorentz factor of the region producing prompt GRB emission must be finite and therefore requires the existence of an intrinsic
pair-production cutoff at VHE energies. One can derive the cutoff location in a simple one-zone model where all photons are created co-spatially by treating the target photons as coming from
the high energy component of the band fit with the measured fluence
\begin{equation}
F(E) = T_{90} \, A \, \Bigg( \frac{E_c}{100 \, keV} \Bigg)^\alpha \, e^{\beta - \alpha} \, \Bigg( \frac{E}{E_c} \Bigg)^\beta = F(E_c) \, \Bigg( \frac{E}{E_c} \Bigg)^\beta \,\,\,\,\, E \ge E_c
\end{equation}
Doing so results in the following expression for opacity due to pair production in the co-moving emission frame
\begin{equation}
\tau_{\gamma\gamma}(E) = \sigma_T \, \Bigg( \frac{d_L(z)}{c \Delta t} \Bigg)^2 \, E_c F(E_c) \, (1 + z)^{-2(\beta + 1)} \, \Gamma^{2(\beta-1)} \, \Bigg( \frac{E E_c}{m_e^2 c^4} \Bigg)^{-\beta-1} \, G(\beta)
\label{eq:opacity_intrinsic}
\end{equation}
where $E_0$ is the energy of a VHE photon measured at Earth, $z$ is the source redshift, $d_L(z)$ is the luminosity distance to the source, $\Gamma$ is the bulk Lorentz factor of the emission region, $\Delta t$ is the measured variability time of prompt emission, and $G(\beta)$ is a factor associated with the integral of the pair production cross section in Equation \ref{eq:pair_cross} over all possible interaction angles. See Appendix \ref{appendixC} for a full derivation of this result.

\begin{figure}[ht!]
\begin{center}
\includegraphics[height=3.5in]{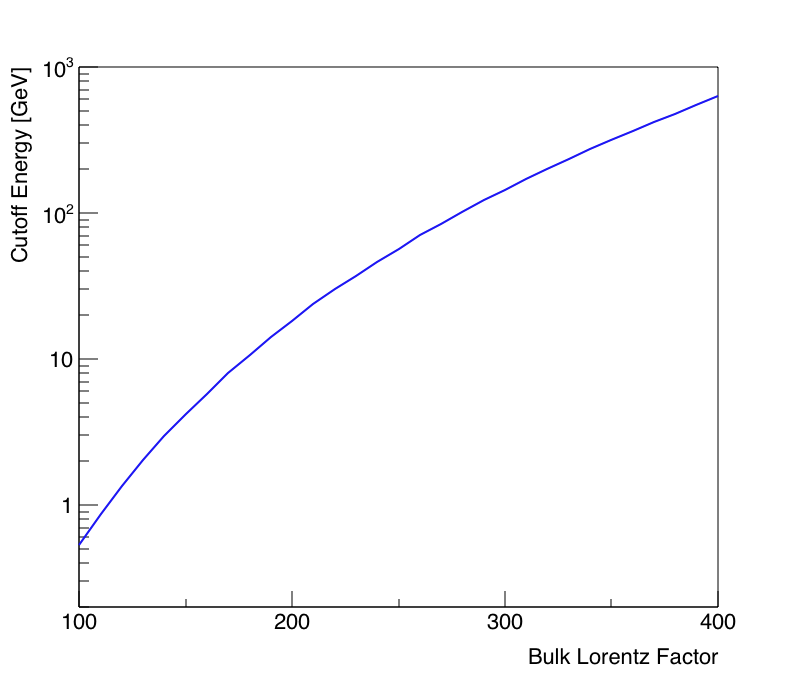}
\caption{Intrinsic cutoff energy as a function of the bulk Lorentz factor $\Gamma$
        for the one-zone model described by setting Equation \ref{eq:opacity_intrinsic} equal to unity and using the parameters in
        Table \ref{medianband} for a GRB at $z = 0.5$ with $T_{90} = \Delta t =$ 1 second.}
\label{onezonecutoff}
\end{center}
\end{figure}

The cutoff location is found by setting Equation \ref{eq:opacity_intrinsic} equal to unity and solving for $E_0$ in terms of estimated source properties.
Doing so for a GRB at redshift $z = 0.5$ with $T_{90} = \Delta t =$ 1 second and the median Band fit parameters from the second GBM catalog (Table \ref{medianband}) yields the curve shown
in Figure \ref{onezonecutoff}. This curve demonstrates that intrinsic cutoffs above several hundred GeV are not unreasonable in the one-zone model as most lower limits
on estimates of the bulk Lorentz factor span the range from 100-400 \cite{Racusin:2011}. This argument is strengthened by the observation of a 95 GeV photon from GRB 130427A \cite{Maselli:2013uza}. This indicates that we do expect to see VHE emission from a GRB in HAWC as the cutoff is above the $\sim$100 GeV threshold where the effective area of HAWC is $\sim$100x the size of the Fermi LAT.

And while the one-zone model can produce extremely large estimates for the bulk Lorentz factor in the highest energy GRBs, 
lower values of $\Gamma$ can be made consistent with observations simply by extending the model to account for high energy photon production at larger radii than low energy emission \cite{ZHao:2011}.
This is referred to as the two-zone model and it acts to reduce the density of low energy photons in the region where VHE photons are made, thereby reducing the required relativistic boosting by a factor of $\sim$2 for the same cutoff energy. Measurements of the spectral cutoff can therefore
be used to distinguish between the location of high energy emission relative to the location of low energy emission on the basis whether
the emission model results in a reasonable bulk Lorentz factor for the production of the observed cutoff.
This is crucial for determining the mechanism of high energy emission
as some models treat high energy emission as occurring co-spatially with the synchrotron emission producing keV-MeV energies whereas
others assume the regions are separate \cite{Gehrels:2013}

\vspace{6cm}
\renewcommand{\arraystretch}{0.75}
\begin{table}[ht!]
\begin{center}
\begin{tabular}{|c|c|c|c|} \hline
  $\alpha$ & $\beta$ & $E_{peak}$ [keV] & Flux [photons/cm$^2$/s] \\ \hline
  -0.86$^{+0.33}_{-0.25}$ & -2.29$^{+0.30}_{-0.39}$ & 174$^{+286}_{-73}$ & -3.16$^{+4.85}_{-1.55}$ \\ \hline
\end{tabular}
\caption{Median parameter values and the 68\% CL of the distributions for Band spectrum fits to data in Reference \cite{Gruber:2014}.}
\label{medianband}
\end{center}
\end{table}
\renewcommand{\arraystretch}{1.0}

\newpage

\newpage

\subsection{Extragalactic Background Light}




EBL emitted by stars and active galactic nuclei (AGN) over the entire age of the Universe
also acts as a source of low-energy target photons for VHE photons as they traverse the cosmological distances between GRB
sources and the Earth. Any cutoff measurement, as well as expectations for the ability to detect high energy photons from a given source,
must therefore account for pair production off the EBL. In this section, we will describe the features of EBL attenuation that will
be relevant to our modeling of GRB signals viewable by the HAWC Observatory in Section \ref{ch:sens}.

\begin{figure}[ht!]
\begin{center}
\includegraphics[height=2.65in]{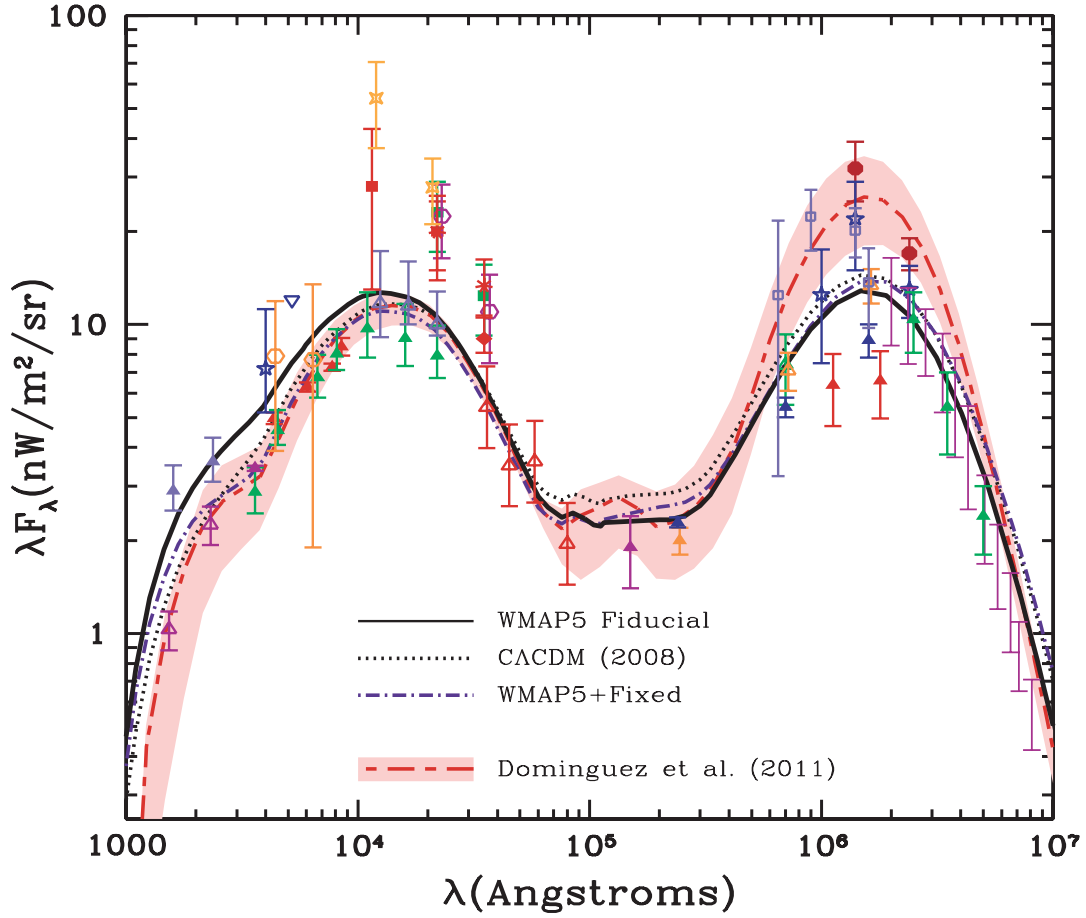}
\caption{Measured EBL intensity for redshift $z = 0$ from a number of different experiments. Data points with upward pointing triangles represent lower limits
         while the rest result from direct detection measurements. See \cite{GILMORE2012} for a full description of each data set. Also shown
         are the predicted curves from four different theoretical models, WMAP5 Fiducial, WMAP5+Fixed, Dom\'{i}nguez et al., and C$\Lambda$CDM (2008).}
\label{fig:ebl_intensity}
\end{center}
\end{figure}

Figure \ref{fig:ebl_intensity} presents measurements of EBL intensity over the range of wavelengths relevant to
VHE photon propagation. These measurements come from a combination of direct techniques, which often 
have large systematic errors at infrared wavelengths due to the subtraction of foreground light, and indirect techniques that provide upper and lower limits \cite{GILMORE2012}.
The relative lack of precise EBL measurements, particularly as a function of redshift, leaves room for a number of different theoretical models describing the available data.
These models generally fall into four main categories: (1) forward evolution models which start from measurements of initial cosmological parameters derived from experiments
like the Wilkinson Microwave Anisotropy Probe (WMAP) and evolve them forward to present day with a combination of analytic and numeric techniques, (2) backward evolution models
which begin with present-day measurements for galaxy emission and evolve them back in time, (3) inferred evolution models which use an empirical parameterization of the star formation rate
combined with theoretical models for stellar emission, and (4) more empirical approaches which attempt to derive both the initial model
parameters and evolution from data \cite{Dwek:2013} \cite{GILMORE2012} \cite{Dominguez:2011}.

\begin{figure}[ht!]
\begin{center}
\includegraphics[height=3in]{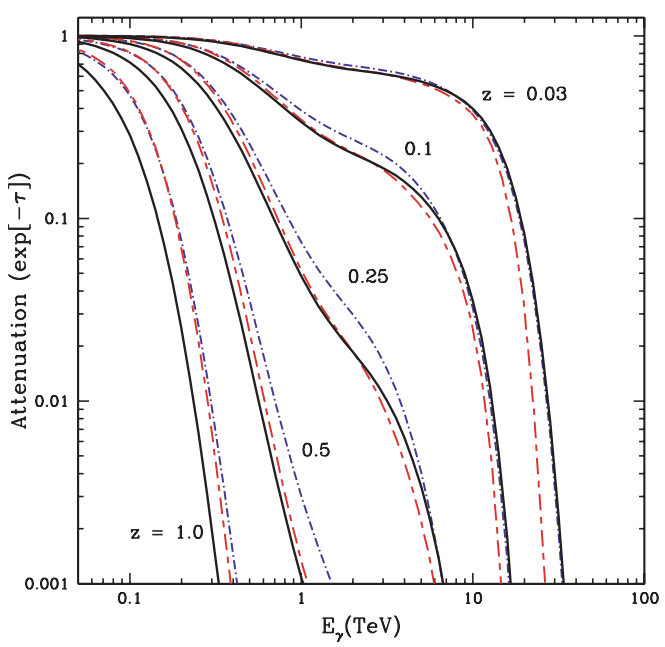}
\caption{The attenuation $e^{-\tau}$ of gamma-rays versus energy for redshifts $z = 0.03, 0.25, 0.5, 1.0$ for the WMAP5 Fiducial (solid black), WMAP5+Fixed (dash-dotted violet),
         and Dom\'{i}nguez et al. (dash-dotted red) models. A y-axis value of 1.0 indicates no attenuation. This figure is reproduced from \cite{GILMORE2012}.}
\label{fig:atten_v_z}
\end{center}
\end{figure}

Three forward-evolution models, WMAP5 Fiducial, WMAP5+Fixed, and \\ C$\Lambda$CDM (2008), are compared to the EBL intensity data in Figure \ref{fig:ebl_intensity}.
Also shown is the curve for the largely empirical model of Dom\'{i}nguez et al \cite{Dominguez:2011}. In general, the differences between each model in this figure
are indistinguishable as the models are tuned to reproduce current data for small redshifts. Larger differences appear between the models at higher redshifts, which
can be seen in gamma-ray attenuation curves in Figure \ref{fig:atten_v_z}. We choose to use the WMAP5 Fiducial model in the remainder of our work as it is specifically
developed with the intent of describing the attenuation of VHE photons coming from high redshift sources \cite{GILMORE2012}. This
results in attenuation of nearly all photons above 100 GeV for a redshift of 1, which defines the viewable volume of GRB bursts in HAWC given that the effective
area shown in Section \ref{sec:gndbase} falls as $\sim$$E^{2}$ between 1 TeV and 100 GeV but then drastically drops off even faster at energies below 100 GeV.


\vspace{2cm}

\newpage

\section{Outlook}\label{sec:outlook}

A new ground-based wide-field extensive air shower array known as the High-Altitude Water Cherenkov (HAWC) Observatory promises a new window to
monitoring the $\sim$100 GeV gamma-ray sky with the potential for detecting
spectral cutoffs in GRBs. It represents a roughly 15 times sensitivity gain over
the previous generation of wide-field gamma-ray air shower instruments for hard spectrum galactic sources and
is able to detect the Crab nebula at high significance ($>$5$\sigma$) with each
daily transit. The sensitivity gain is even greater at 100 GeV gamma-ray energies
where the effective area of HAWC is $\sim$100 m$^2$ which is much larger than the $\sim$3
m$^2$ achievable in Milagro, the precursor experiment to HAWC.
Its wide field-of-view ($\sim$2 sr), $>$95\% uptime, and $>$100x larger effective
area compared to the Fermi LAT instrument at energies above 100 GeV make it an
ideal instrument for discovering prompt gamma-ray burst (GRB) emission from the ground.

Combining existing GRB measurements made by the Fermi GBM instrument with the
sensitivity of the HAWC Observatory yields an expected rate of $\sim$1 observed GRB per year
from triggered observations of GBM-detected bursts alone \cite{Taboada:2013uza}. Performing
an all-sky, self-triggered search for GRB emission in HAWC relaxes the requirement of an overhead
GBM observation and raises the expected number of observed GRBs by about a
factor of two prior to accounting for trials. As will be shown in Section \ref{sec:senswtrials}, accounting
for trials in the all-sky search requires a 2x increase in the flux needed for discovery compared to the triggered search.
Given that measured GRB fluxes follow a power law distribution with index -3/2 \cite{Bhat:2016}, this approximately
balances the increase in the number of expected discoveries in the self-triggered search to yield an identical expectation of $\sim$1 observed
GRB per year from an all-sky, self-triggered search.

Both a triggered GRB search and an all-sky, self-triggered GRB search are currently being pursued in available HAWC data.
To date, no significant detections of VHE emission were found in either search after approximately 1 year of operating the full
HAWC detector. The focus of this dissertation is to present the methodology behind the all-sky, self-triggered search and
describe its null detection in the context of upper limits on the rate of VHE emission in a previously unconstrained parameter space.
These limits will become more sensitive in time, but our belief is that the search algorithm described here will provide a positive
detection of VHE emission during future years of HAWC operations.

\renewcommand{\thechapter}{2}

\chapter{Extensive Air Showers}\label{ch:showers}

The HAWC Observatory measures extensive air showers (EASs) produced when high-energy cosmic-ray primaries interact in the upper atmosphere. There are two main classes of primaries, gamma-ray primaries and hadronic primaries consisting of fully ionized nuclei and gamma-ray primaries.
The charged nature of hadronic primaries leads to their directional randomization in galactic magnetic fields \cite{Ptuskin:2012} for energies relevant
to HAWC and results in a highly isotropic arrival distribution at the Earth with levels of anisotropy measured to a relative intensity below 10$^{-3}$ \cite{Abeysekara:2014a}.
This lack of pointing means that information about individual cosmic-ray sources cannot be determined from measurements of the cosmic-ray particles themselves.

High-energy gamma-ray primaries on the other hand are electrically neutral, allowing them to point directly back to their origin at astrophysical particle acceleration sites.
The main science mission of the HAWC Observatory then is to provide measurements of high-energy gamma-ray photons through the air showers they create in the upper atmosphere to identify
and better understand sources of high-energy particle acceleration. The following sections in this chapter detail the physical processes
involved in air shower production and how they relate to measurements of gamma-ray primaries made with HAWC. 

We discuss gamma-ray air shower development in Section \ref{sec:photshower} followed by 
the development of hadronic air showers in Section \ref{sec:crshower}, which act as the main background for gamma-ray analyses in HAWC.
We then describe the processes related to the development of the shower plane measured
at ground level in HAWC for both gamma-ray and hadronic primaries in Section \ref{sec:showplan}.
We finish by noting the major observable differences between hadronic and gamma-ray showers at ground level in Section \ref{sec:diffthatthing},
which allow us to distinguish between the two types of showers as they appear within HAWC data in Chapter \ref{ch:Reco}.


\newpage

\section{Gamma-ray Air Showers}\label{sec:photshower} 

Gamma-ray induced air showers begin when the incident gamma-ray converts to an electron-positron pair in the coulomb field of an atom in the upper atmosphere
\begin{equation}
\gamma \rightarrow e^- + e^+
\end{equation}
The electron and positron pair subsequently generate more photons via Bremsstrahlung radiation and an electromagnetic cascade develops.
The full energy of the incident gamma-ray therefore immediately enters a single electromagnetic cascade,
in stark contrast to the hadronic air shower case discussed in Section \ref{sec:crshower} where both hadronic and electromagnetic cascades develop and only a fraction of the incident energy manifests
in the form of photons, electrons and positrons.

While it is possible for a high-energy photon to create muon and tau lepton pairs, their heavier masses compared to the electron mass suppress their production.
Modeling of gamma-ray air shower development can therefore be approximated as the interaction properties of electrons, positrons and photons alone. In particular, development is
largely described by the radiation length, $\chi_{0, brem}$, for electron bremsstrahlung to occur and the mean free path for pair
production by high-energy photon, $\chi_{0, pair}$. These quantities are related according to
\begin{equation}
\chi_{0, pair} = \frac{9}{7} \chi_{0, brem}
\end{equation}
despite having distinct physical interpretations with radiation length defined as the mean distance over which
an electron loses a factor of $e^{-1}$ in energy and mean free path defined as the mean distance between interactions.

\begin{figure}[ht!]
\begin{center}
\includegraphics[height=2.5in]{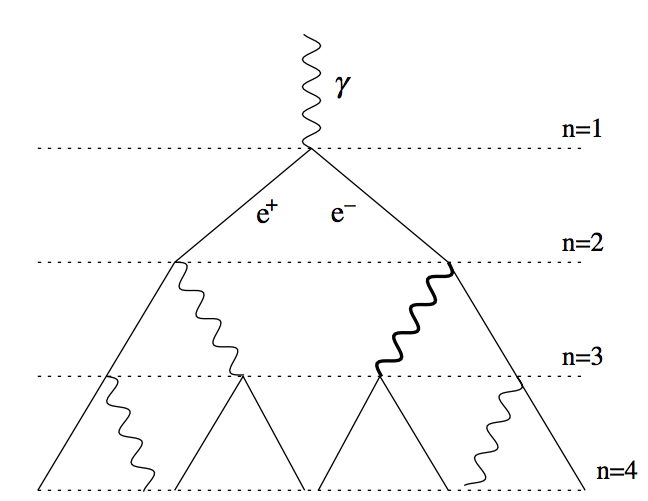}
\caption{Simplistic gamma-ray air shower model in the approximation where $\chi_{0, pair} = \chi_{0, brem}$.
         This figure is reproduced from Reference \cite{Matthews:2005}.}
\label{fig:gamma_prof_model}
\end{center}
\end{figure}

Figure \ref{fig:gamma_prof_model} presents a simplified gamma-ray air shower model under the approximation
$\chi_{0, pair} \approx \chi_{0, brem} = \chi_{0}$ originally developed by Heitler to demonstrate many features
of gamma-air showers \cite{Heitler:1954}. In this model, all particles undergo a splitting after traveling
a distance $d = \chi_{0} \, ln(2)$ which results in electrons and positrons producing Bremsstrahlung photons of exactly half their energy.
The photons split into equal energy electron-positron pairs after the same distance, resulting in the creation of
2$^n$ equal energy particles in the shower after a distance of $d \times n$. Particle multiplication continues until the depth when the particle energy
falls below the critical energy $E_c \approx$ 84 MeV and ionization energy losses dominate over Bremsstrahlung radiative losses. This depth is known as shower maximum
and can be calculated by setting the particle energy after $n$ foldings
\begin{equation}
E_p = \Big( \frac{1}{2} \Big)^n E_0
\end{equation}
equal to $E_c$ and solving for $n$ to find
\begin{equation}
X_{max} = d \times n = \chi_{0} \,  ln\Big( \frac{E_0}{E_c} \Big)
\label{eq:showmax}
\end{equation}

\newpage

\begin{figure}[ht!]
\begin{center}
\includegraphics[height=2.2in]{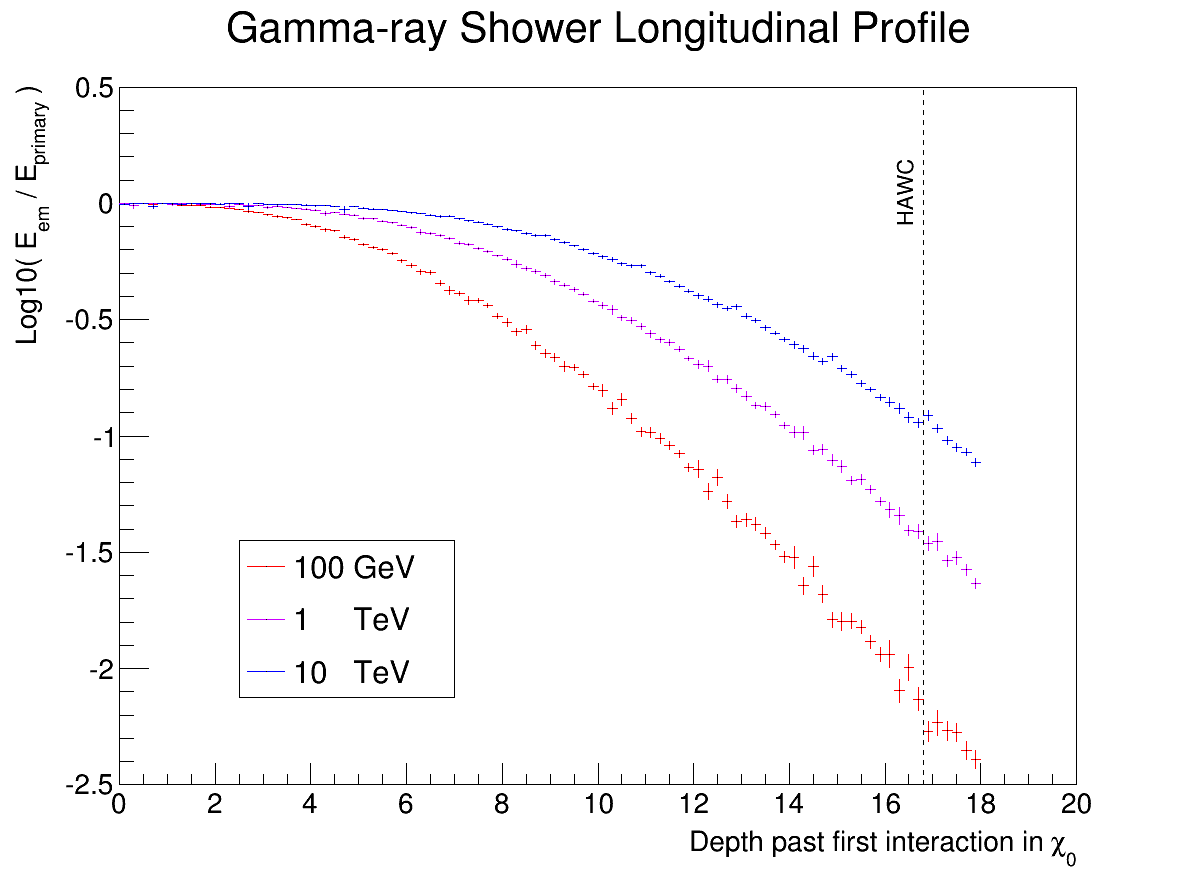}
\caption{Longitudinal shower profiles for the fraction of energy remaining
         in electromagnetic shower particles as a function of shower depth
         past the first interaction point for simulated gamma-ray primaries at 100 GeV, 1 TeV, and 10 TeV.
         Data points are the average from 1000 vertical primary particles at each energy modeled using CORSIKA.
         Shower depth is written in terms of the Bremsstrahlung radiation length in air, 37.15 g cm$^{-2}$ \cite{Grieder:2010}.
         The HAWC detector is designed to measure the electromagnetic shower energy remaining at a penetration depth
         of 16.8 radiation lengths for showers starting at the top of the atmosphere.}
\label{fig:gamma_prof}
\end{center}
\end{figure}

Equation \ref{eq:showmax} indicates that higher energy showers are able to penetrate deeper into the upper atmosphere
before energy loss due to ionization becomes significant. This is confirmed in the simulated gamma-ray shower profiles shown in Figure \ref{fig:gamma_prof}
which reveal that 10 TeV showers retain a larger fraction of the incident gamma-ray energy at a given depth when compared to 100 GeV showers.
Using the 1976 U.S. Standard Atmospheric Model, we find that the observation altitude of the HAWC observatory is at an atmospheric depth of approximately 16.8 radiation lengths for vertical showers.
This yields in an average of only $\sim$1\% the total incident energy reaching the observation level for a 100 GeV gamma-ray interacting at the top of the atmosphere, although
an additional factor of $\sim$1.5 is gained for every additional radiation length the primary gamma-ray travels before undergoing the first interaction.
The minimum detectable energy of a gamma-ray primary in HAWC is therefore determined both by the original photon energy as well as the first interaction depth.

\newpage

\newpage

\section{Hadronic Air Showers}\label{sec:crshower} 

Although gamma-ray shower detections are the main focus behind the HAWC Observatory's science mission, hadronic primaries still induce
the majority of extensive air showers and represent a formidable background that must be understood in order to perform any gamma-ray air shower analysis.
The cosmic-ray energy range relevant to producing backgrounds in HAWC is approximately 100 GeV - 100 TeV
with the overall number of primaries falling as a $E^{-2.7}$ power law \cite{Kampert:2012}. These primaries
are predominantly energetic protons \cite{PDG}. 

In the simple case of a cosmic-ray proton colliding with an atomic nucleus $A$ in the atmosphere we can represent the interaction as
\begin{equation}
p + A \rightarrow p + X + \pi^{\pm, 0} + K^{\pm,0} \, ...
\label{eq:proton_react}
\end{equation}
where $X$ is the fragmented nucleus, $\pi^{\pm, 0}$ are secondary pions, $K^{\pm, 0}$ are secondary kaons, and the ellipsis indicates other secondary particles.
Although secondary particles other than pions are created, their production cross sections are reduced by a factor of 10 compared to that of pions and can largely
be ignored \cite{Kampert:2012}. Shower development is therefore dominated by the subsequent interactions of the secondary pions and the fraction of incident energy
they carry with them.

On average, 1/3 of the shower energy at each generation of interactions goes into neutral pions which immediately decay to gamma-ray pairs \cite{Sommers:2004}.
The remaining energy enters charged pions which continue to produce additional pions via hadronic interactions until their energy falls below the pion
creation threshold \cite{Grieder:2010}. As a result, the majority of the original shower energy is transferred into gamma-ray pairs with only (2/3)$^n$ remaining in the
hadronic cascade after $n$ generations of interactions. Charged pions remaining at the end of the hadronic cascade will predominantly decay into muons
\begin{eqnarray}
\pi^+ &= \mu^+ + \nu_\mu \\
\pi^- &= \mu^- + \bar{\nu}_\mu
\end{eqnarray}
as the electron channel is suppressed by the muon-electron mass ratio due to helicity requirements.

\begin{figure}[ht!]
\begin{center}
\includegraphics[height=3.5in]{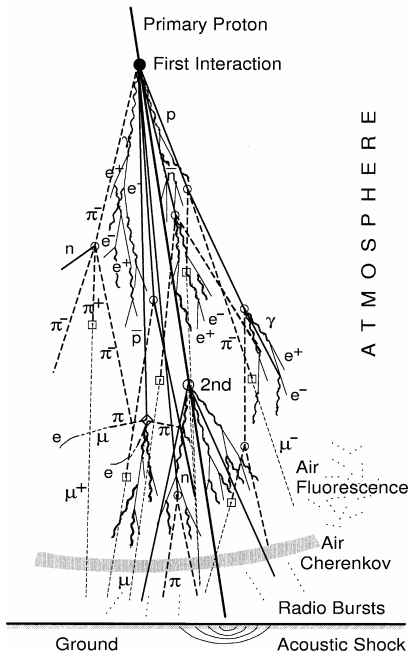}
\caption{Diagram of hadronic extensive air shower induced by a cosmic-ray proton interacting in the upper atmosphere. Reproduced from \cite{Grieder:2010}.}
\label{fig:cascade}
\end{center}
\end{figure}

Gamma-rays produced by the decay of neutral pions go on to create electron-positron pairs in the fields of nearby atoms in the atmosphere which subsequently generate additional photons through Bremsstrahlung radiation.
Again, a cascade develops as the Bremsstrahlung photons re-interact to produce electron-positron pairs until the resulting pairs enter the regime where the cross section for ionization
is comparable to the Bremsstrahlung cross section. This occurs at an electron energy of  $\sim$84 MeV in air \cite{Grieder:2010}. The net result is a series of electromagnetic cascades branching out from neutral pion nodes of the hadronic cascade, which are visible in the hadronic air shower diagram shown in Figure \ref{fig:cascade}.

\newpage

\begin{figure}[ht!]
\begin{center}
\includegraphics[height=3in]{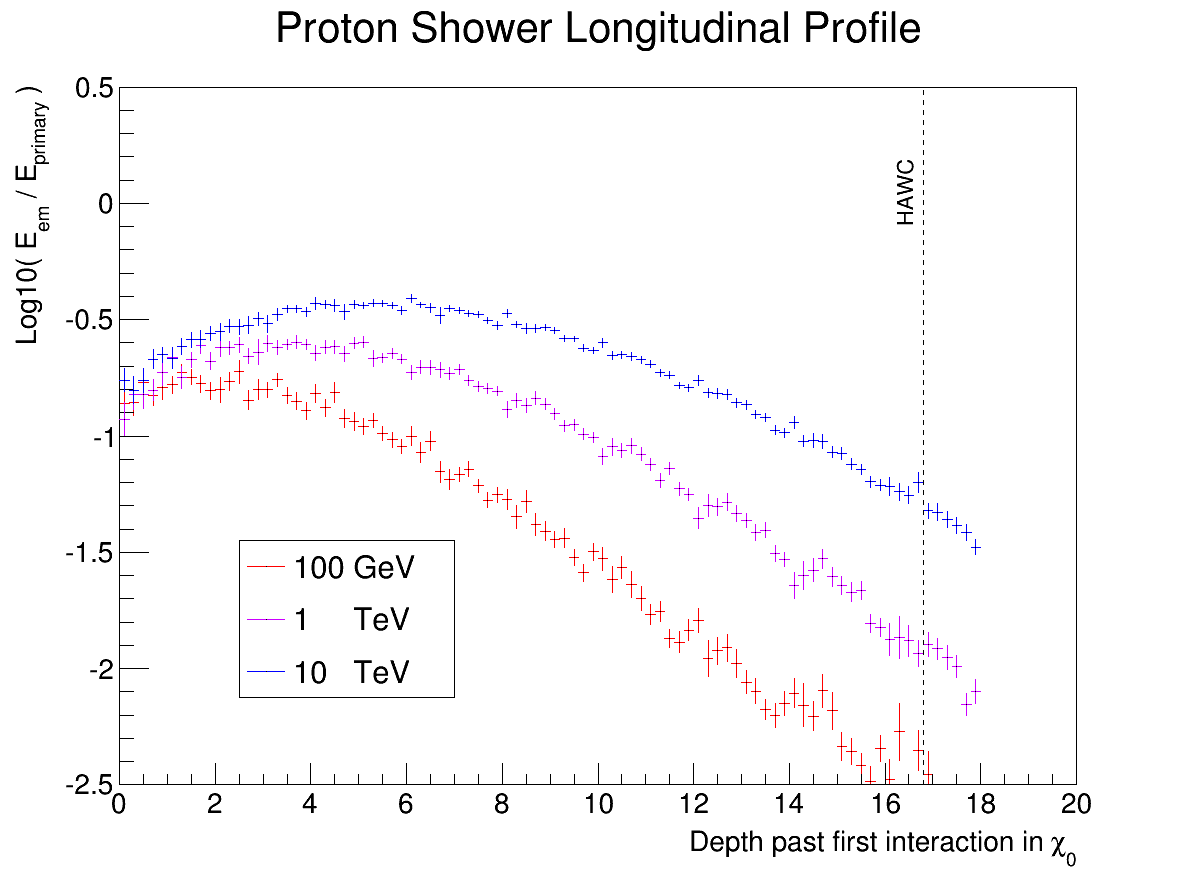}
\caption{Longitudinal shower profiles for the fraction of energy remaining
         in electromagnetic shower particles as a function of shower depth
         past the first interaction point for simulated proton primaries at 100 GeV, 1 TeV, and 10 TeV.
         Data points are the average from 1000 vertical primary particles at each energy modeled using CORSIKA.
         Shower depth is written in terms of the Bremsstrahlung radiation length in air, 37.15 g cm$^{-2}$ \cite{Grieder:2010}.
         The HAWC detector is designed to measure the electromagnetic shower energy remaining at a penetration depth
         of 16.8 radiation lengths for showers starting at the top of the atmosphere.}
\label{fig:proton_prof}
\end{center}
\end{figure}

\begin{figure}[ht!]
\begin{center}
\includegraphics[height=3in]{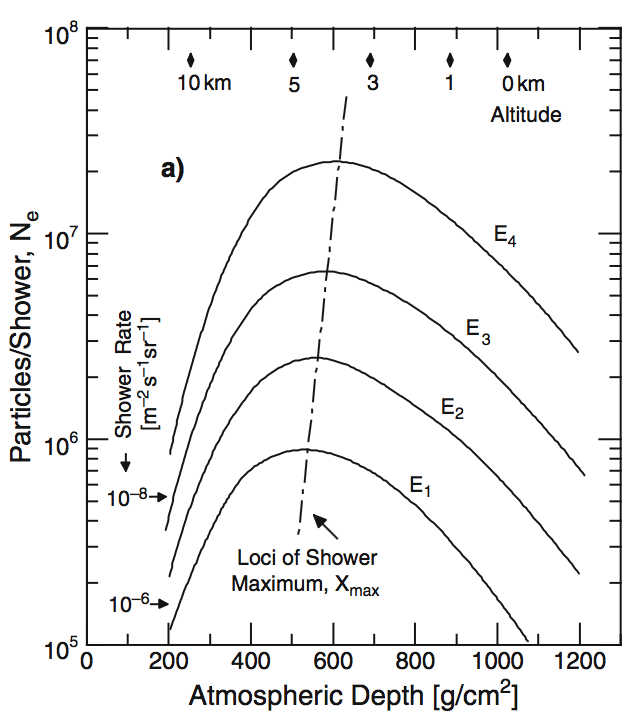}
\caption{Longitudinal shower profiles of the number of electromagnetic particles in
         a proton-induced EAS as a function of atmospheric depth for different energies \cite{Grieder:2010}.}
\label{fig:proton_number}
\end{center}
\end{figure}

The average energy in the electromagnetic portion of a proton shower, defined as the energy carried by photons, electrons and positrons, as a function of shower 
depth can be calculated using CORSIKA simulations of primary protons at different energies (Figure \ref{fig:proton_prof}).
These profiles begin with positive slope as the hadronic portion of the shower initially converts larger and larger fractions
of the hadronic shower energy into gamma-ray pairs through neutral pion decay. They reach a maximum at the end of pair production and begin
to fall as more and more particles in the electromagnetic cascade transition to ionization energy loss. This behavior is similar to 
the profile of total charged particle number for proton showers shown in Figure \ref{fig:proton_number}, however the two profiles are not exactly equivalent as individual particle energies are reduced
with each generation of interactions. The profile most relevant to HAWC is the electromagnetic shower energy profile because
the HAWC detector is designed to measure the electromagnetic shower energy remaining at an elevation of 4100 m, which
corresponds to penetration depth of 16.8 radiation lengths in Figure \ref{fig:proton_prof} for showers beginning at the top
of the atmosphere.

The longitudinal profiles in Figure \ref{fig:proton_prof} show that the maximum amount of energy available in the electromagnetic
portion of the proton shower is well below the energy of the primary particle. This is a direct result of the proton exiting
the initial interaction in Equation \ref{eq:proton_react}, which can carry away as much as 50\% of the initial proton energy \cite{Kampert:2012}.
The energy transferred to the secondary shower particles is referred to as the inelasticity of the original interaction.

\newpage

\newpage

\newpage

\section{Lateral Shower Development and Curvature}\label{sec:showplan} 

\begin{figure}[ht!]
\begin{center}
\includegraphics[height=4in]{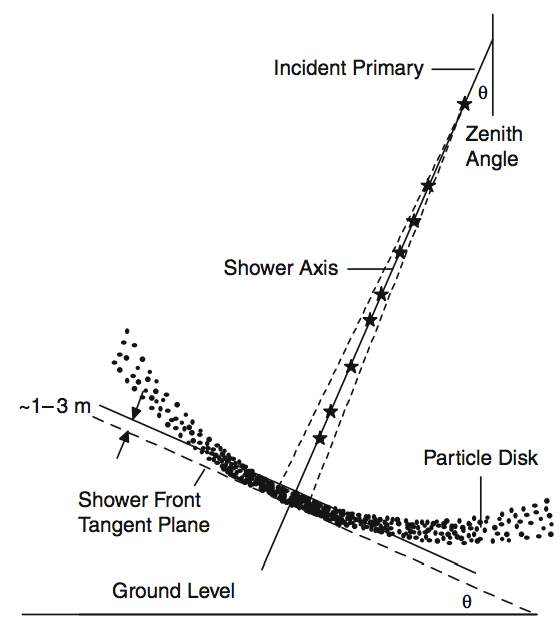}
\caption{Diagram of extensive air shower plane showing the width and curvature of the particle distribution with respect to the
        shower axis \cite{Grieder:2010}.}
\label{fig:shower_timing}
\end{center}
\end{figure}

The electromagnetic energy measured at ground level in HAWC arrives in 
the form of electrons, positrons, and photons traveling in a curved
shower front centered on the trajectory of the original primary particle (Figure \ref{fig:shower_timing}).
This results from the low density of the atmosphere, which allows shower particles to spread laterally away from the axis 
defined by the original direction of the primary particle over time as interactions in both gamma-ray and hadronic air showers introduce transverse momenta.
Conservation of momentum dictates that this happen symmetrically about the original trajectory thereby forming a
disk centered on the shower axis. This disk has a roughly spherical curvature with respect to the location of the
first interaction because shower particles are all moving at approximately the speed of light.

As will be shown in Chapter \ref{ch:Reco}, detections of the arrival time of lateral energy in the particle disk
are used to determine the original direction of the primary particle in HAWC. These are affected by the finite shower plane width shown
in Figure \ref{fig:shower_timing} that results from larger path length differences near the edges of the shower
where the average particle energy is lower causing greater scattering angles as well as local variations in the 
individual particle energies at each point along the disk \cite{Grieder:2010}. This is because the timing measurements
are determined by the arrival time of the first particle at a given location in the shower disk which will fluctuate within
the range dictated by the width of the disk depending how the air shower randomly develops in the atmosphere. Measurements
at locations of high particle density (high shower energy) increase the chance of the first particle arriving at the earliest
extent of the shower disk and will therefore be biased to earlier times \cite{Atkins:2000}. This effect must
be accounted for as a function of shower plane width, determined by the lateral distance from the shower axis, 
and the total measured energy.

\newpage

\section{Shower Differentiation}\label{sec:diffthatthing}

\begin{figure}[ht!]
\begin{center}
\includegraphics[height=4.5in]{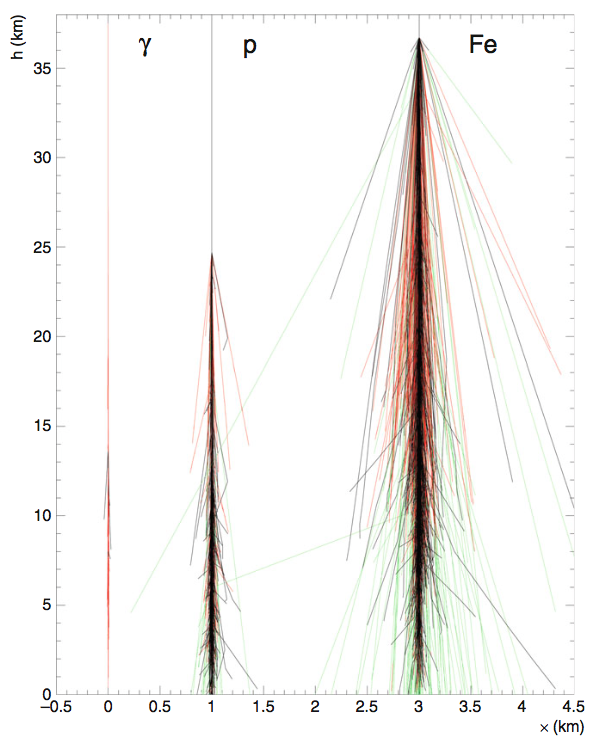}
\end{center}
\caption{Side view of simulated 100 TeV gamma-ray, proton, and Iron induced air showers.
         Trajectories are displayed for all secondary particles of energy $\ge$ 10 GeV with
         electromagnetic component shown in RED, hadrons in BLACK, and muons in GREEN.
         The gamma-ray air shower exhibits the fewest off-axis trajectories as the hadronic
         interactions present in proton and Iron showers are much more efficient at generating
         off-axis momenta compared to multiple Coulomb scattering.
         This figure is reproduced from Reference \cite{Grieder:2010}.}
\label{fig:spreadthatstuffout}
\end{figure}

A major distinction occurs between gamma-ray and hadronic air showers when
we examine the underlying interactions that drive the lateral shower development described in the previous section.
In the case of gamma-ray showers these interactions are multiple Coulomb scatterings which are much less efficient
at transporting shower energy off-axis compared to the hadronic interactions driving shower development in hadronic air showers.
This results in a large fraction of the electromagnetic energy in the gamma-ray shower remaining along the axis. Furthermore, the distribution
of energy about the shower axis is fairly smooth and uniform as the electrons and positrons that make up the gamma-ray air shower have uniform mass.
By comparison, the interactions of pions in typical hadronic air showers produce sub-showers that carry a significant amount of electromagnetic energy off-axis
as can be seen by the trajectories of $\ge$10 GeV particles in Figure \ref{fig:spreadthatstuffout}. These sub-showers result in non-uniformity of
the distribution of energy within the shower disk with large amounts of shower energy appearing in localized groups of secondary particles far from the shower axis.
Figure \ref{fig:spreadthatstuffout} also clearly demonstrates the large number of energetic muons produced
in hadronic air showers which are not present in gamma-ray air showers. These effects are shown in Chapter \ref{ch:Reco} to provide significant separation of gamma-ray primaries
from the hadronic air shower background in HAWC data.

\renewcommand{\thechapter}{3}

\chapter{The HAWC Observatory}\label{ch:hawcobs}

The High-Altitude Water-Cherenkov (HAWC) Observatory is a ground-based air shower
array comprised of 300 Water Cherenkov Detectors (WCDs) located at an elevation of 4100 meters
above sea level in central Mexico at a longitude of 97.3$^\circ$ West and a latitude of 19$^\circ$ North. It was completed in March 2015 and is sensitive to extensive air showers produced by
cosmic-ray primaries interacting in the upper atmosphere with energies between
50 GeV and 100 TeV \cite{ICRC2013}. It is currently the most sensitive WCT to gamma-ray primaries and, unlike IACTs, 
its wide field-of-view and near 100\% duty cycle yield an unbiased survey of the sky between -31$^{\circ}$ and 69$^\circ$ in declination each day
with 2.2 sr of overhead sky available at any given moment for the 50$^\circ$ zenith cut applied in our analysis. This makes the HAWC Observatory an ideal
instrument for searching for very-high energy (VHE) transients.

Each WCD is a 7.3 m diameter steel tank containing a light-tight plastic lining filled with 188,000 liters
of purified water. There are four photomuliplier tubes (PMTs) positioned on the tank floor: a centrally located high-quantum efficiency Hamamatsu 10" R7081 PMT surrounded
by three Hamamatsu 8" R5912 PMTs. The three 8" PMTs are a radial distance of 1.85 m away from the central PMT
with 120$^\circ$ spacings between them. All the PMTs face upward to observe Cherenkov light produced in the 4 m height of water
overburden by secondary air shower particles and convert it to electrical signals measurable by our data acquisition system.

Altogether, the HAWC Observatory's 300 WCDs account for an active area of 12,500 m$^2$ covering a total area of 22,000 m$^2$.
A single building exists in the center of the array to house the data acquisition (DAQ) system responsible for recording the signals produced
by all 1200 PMTs as well as the calibration system.
The following sections in this chapter describe the
components used to measure, calibrate, and record air shower signals.

\vspace{5cm}

\begin{figure}[ht!]
\begin{center}
\includegraphics[width=4in]{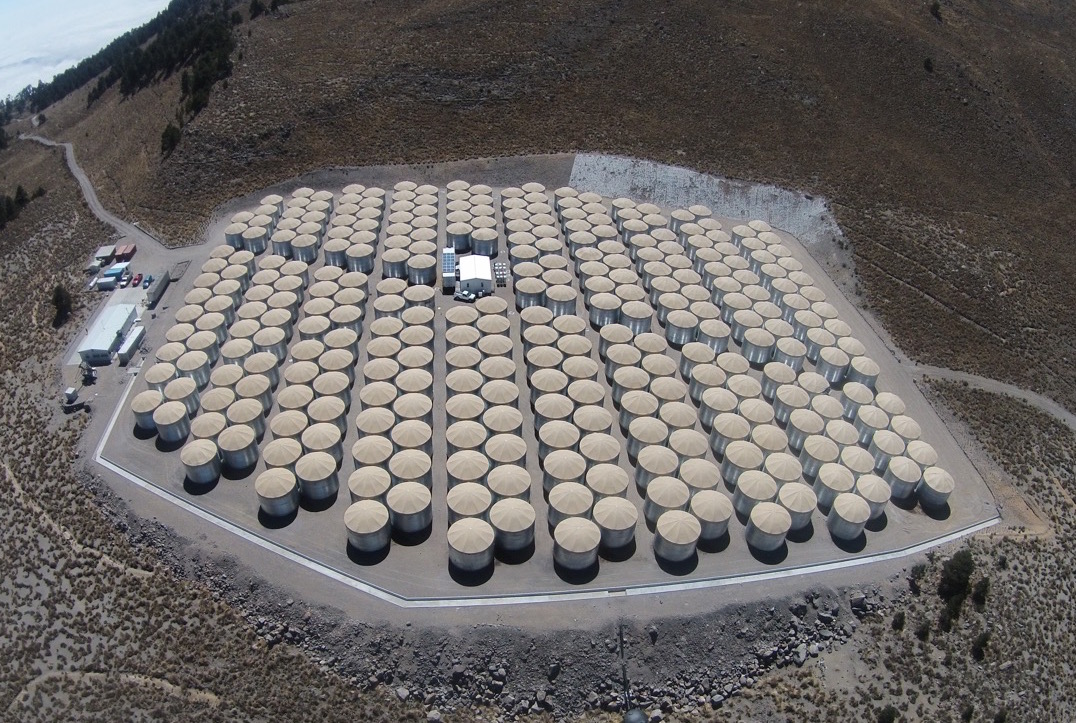}
\end{center}
\caption{Aerial photograph of the HAWC Observatory.}
\end{figure}

\newpage

\section{Water Cherenkov Detectors (WCDs)}

\begin{figure}[hb!]
\begin{center}
\includegraphics[height=2.5in]{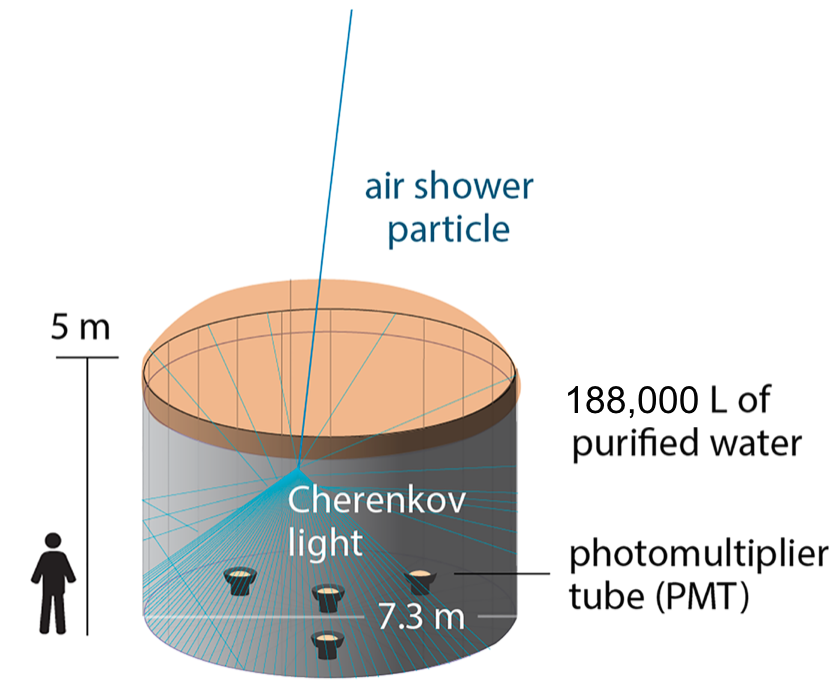}
\quad\quad\quad\quad
\includegraphics[height=1.5in]{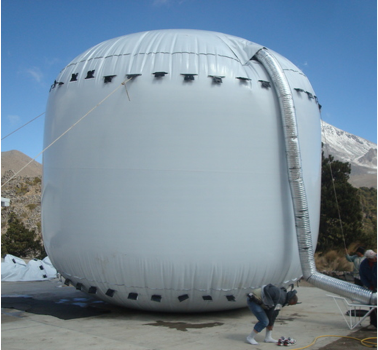}
\caption{Rendering of a secondary air shower particle producing Cherenkov radiation inside a WCD (Left)
         and a photograph of the light-tight plastic lining inflated during testing (Right).}
\label{tanky}
\end{center}
\end{figure}

As described above, WCDs are 7.3 m diameter steel tanks containing four upward facing photomultiplier tubes in a light-tight plastic liner (Figure \ref{tanky}).
They are filled with water because it has a large index of refraction that aids in the production of Cherenkov light in the tank
and is transparent to photons over the operating range of the PMTs.
The water is filtered to remove contaminants, producing an attenuation length of $\sim$10 meters for photon wavelengths detectable
by the PMTs, ensuring a large light yield even for photons traversing the full tank height.

The tank height is large enough that the electromagnetic particles in the air shower disk at ground level will range out in the water before reaching the tank bottom. This produces a direct
proportionality between the total light yield in the tank and the total electromagnetic energy in the shower at the tank's location as all particle energy is deposited inside the tank.
The PMTs therefore effectively measure the amount of energy reaching ground level in the electromagnetic portion of air showers. 

Each tank is optically isolated to aid in identification of local variations in the ground energy, which can be used for distinguishing between gamma-ray and hadronically initiated showers (Chapter \ref{ch:showers}). Furthermore, the water overburden of 4 meters is chosen to allow muons possessing the median muon energy produced in air showers to penetrate the full water height. This yields an additional level of discrimination between air shower progenitors for muons arriving far from the shower axis, as are expected in hadronically induced showers. Such muons produce an asymmetric response from the four PMTs when their final position on the tank floor is near one of the PMTs. This results in a large light yield in a single PMT far from the shower axis in hadronic air showers which is not expected in gamma-ray air showers where the lateral energy distribution is both highly peaked near the shower axis and relatively smooth.

\newpage

\section{Photomultiplier Tubes (PMTs)}

PMTs are a class of light-sensing vacuum tubes that operate on the basic principle
of the photoelectric effect. They are sensitive enough to detect single photons and
have extremely fast response speeds on the order of tens of nanoseconds. In addition,
they are available with large collection areas thereby reducing the total number of
devices needed to instrument an area the size of the HAWC observatory.

\begin{figure}[ht!]
\begin{center}
\subfigure[][]{\includegraphics[height=2.5in]{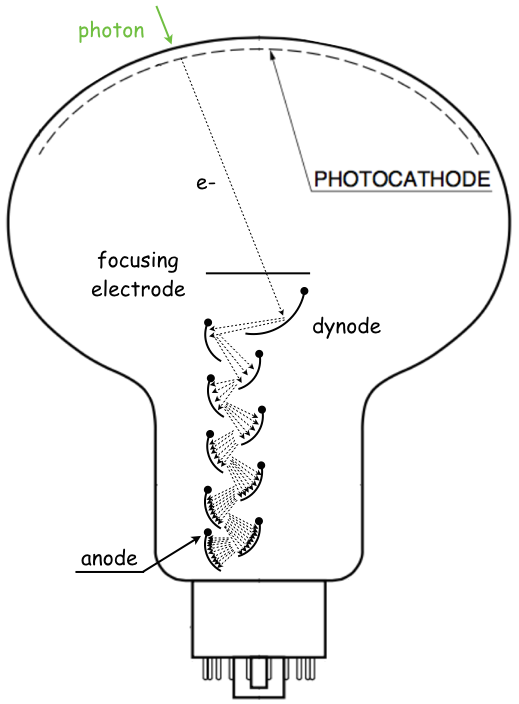}}
\quad\quad\quad\quad
\subfigure[][]{\includegraphics[height=2.3in]{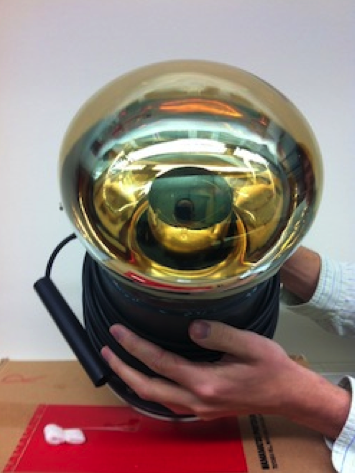}}
\end{center}
\caption{Diagram (a) and photograph (b) of an 8" R5912 Hamamatsu PMT.}
\label{fig:pmt_diagram}
\end{figure}

PMTs are typically comprised of an evacuated glass casing whose inner surface is lined
with a vapor-deposited semiconductor with a low work function, referred to as the
photocathode (Figure \ref{fig:pmt_diagram}). Photons reaching the photocathode
can liberate electrons from its surface via the photoelectric effect. These free electrons are
then accelerated towards a metal plate, called a dynode, located behind the 
photocathode and held at a significantly higher voltage. Each primary electron liberates a new
group of electrons when it collides with the first dynode. The new group of electrons is accelerated
towards the next dynode with each secondary electron now creating another group of electrons at the
second dynode. As a result, the number of electrons flowing through the dynode chain continues to
grow until they reach the final dynode.

Upon reaching the final dynode, all electrons are transferred to the anode where they are collected
and delivered to the PMT output for measurement. The ratio of the mean output charge for a solitary
photon signal producing a single primary electron to the fundamental electron charge
gives the gain, or amplification factor, of the PMT. This factor depends on the high voltage at
which the PMT is operated and the total number of dynodes. The PMTs used in HAWC consist of 8" Hamamatsu R5912 PMTs inherited
from the Milagro experiment supplemented by newer 10" Hamamatsu R7081 PMTs chosen during the initial design phase of the HAWC Observatory
to provide additional low-energy sensitivity. Both populations are operated with a positive high voltage of $\sim$1700 V and have a 10 stage dynode chain.
The exact value of HV applied in each channel is tuned to gain-match all PMTs thereby producing uniform electronics response. The average gain of 1.6$\times10^7$
is designed to give very good charge resolution for single photoelectron signals.

\newpage

\begin{figure}[ht!]
\begin{center}
\includegraphics[width=4.5in]{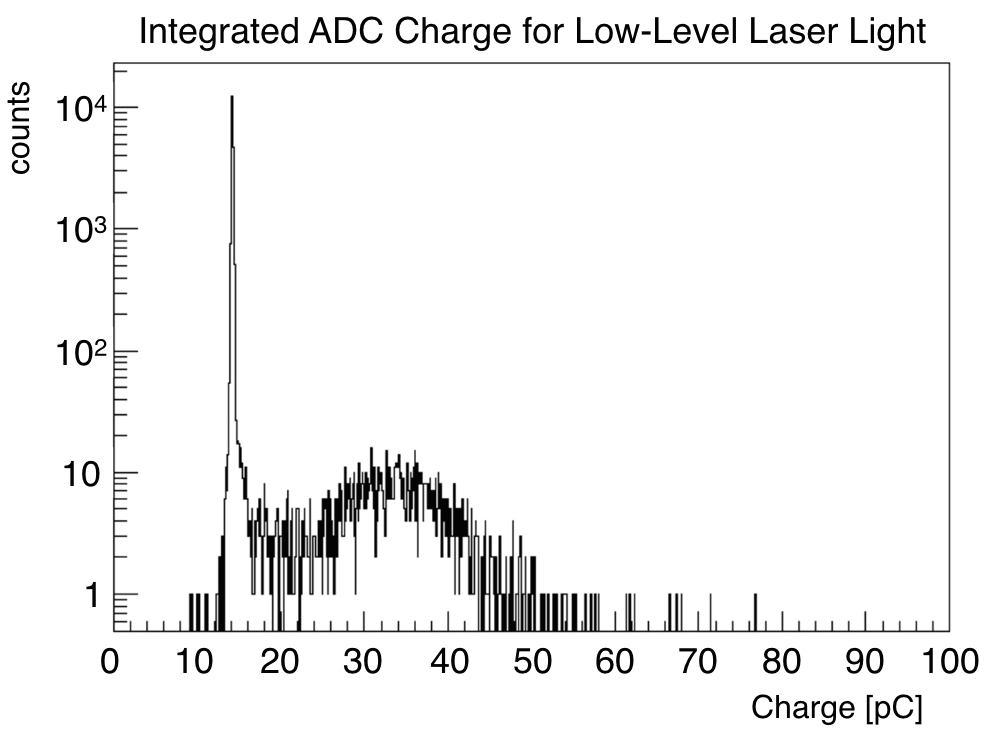}
\end{center}
\caption{Histogram of integrated ADC charge for an 8" HAWC PMT running at 2100 V and exposed to low-level laser light during initial testing.
        The sharp peak at 14 pC represents the pedestal of the electronics setup. The broader
        gaussian-like feature centered at 34 pC is the charge distribution associated with single
        photon measurements. The mean value of 34 pC indicates an average gain of $\sim$1.2$\times10^8$ after pedestal subtraction.
        The width of this peak is the result of differences in electron trajectories from different locations on the PMT
        surface to the first dynode. Typical HAWC PMTs are operated at a high voltage of 1700 V with a gain of $\sim10^7$ in the full detector configuration.
        }
\label{fig:single_pe_charge}
\end{figure}

PMT gain also depends on where the initial photon was absorbed
as asymmetries in the geometry of the PMT, particularly for locations far from the photocathode center,
produce different final electron velocities at the first dynode and therefore different numbers of secondary electrons \cite{DCHOOZ2011}.
This results in the broad spread of output charge for single photon measurements shown in Figure \ref{fig:single_pe_charge},
which acts as an uncertainty of about 35\% to any calibration relating total charge to photon number.
 
\newpage

\newpage 

\begin{figure}[ht!]
\begin{center}
\subfigure[][]{\includegraphics[height=2.5in]{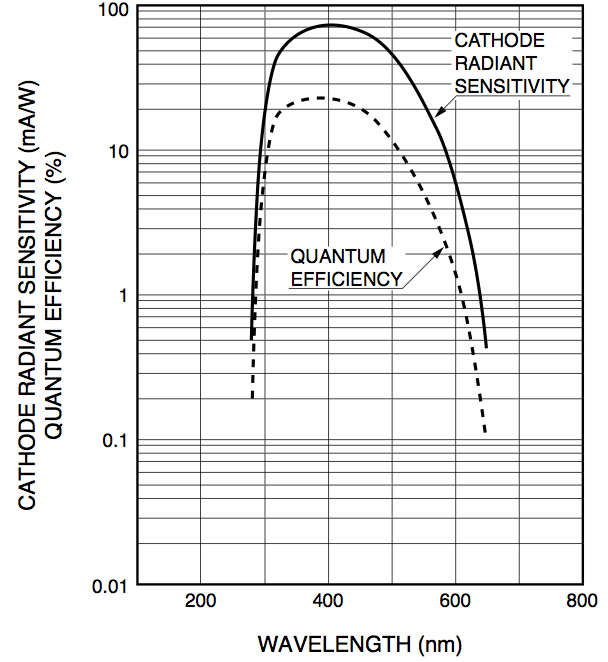}}
\quad\quad\quad\quad
\subfigure[][]{\includegraphics[height=2.5in]{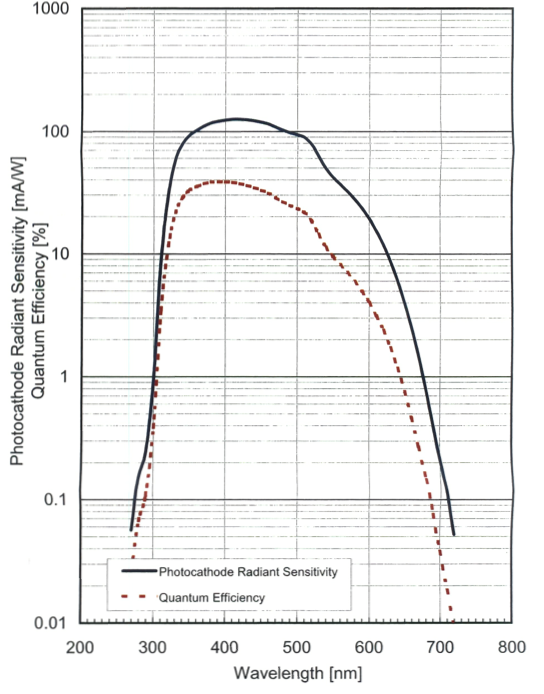}}
\end{center}
\caption{Hamamatsu reported quantum efficiencies for 8" R5912 (a) and 10" R7081 (b) PMTs.
         Note that these values are actually convolutions of the quantum and collection efficiencies
         as the measurement procedure consists of uniformly illuminating the photocathode with single
         photons and dividing the output current by a precisely calibrated reference sensor to determine
         the fraction of detected photons. There is no distinction between photons that fail to produce
         free electrons at the photocathode and photons that do produce an electron which subsequently fails
         to reach the first dynode.}
\label{fig:QE_CE}
\end{figure}

Although PMTs are highly sensitive, not every photon incident on the photocathode produces
a free electron as the photoelectric effect is determined by a probabilistic quantum process.
Quantum efficiency is the per-photon probability for creating a free electron and it depends
on the exact photocathode material and the wavelength of incident light. It is difficult to measure in practice
because of its convolution with collection efficiency, the probability for a free electron to land on the first dynode,
during photon measurements. Manufacturers like Hamamatsu therefore report quantum efficiency numbers
which are interpreted as the product of quantum and collection efficiencies (Figure \ref{fig:QE_CE}). The 10" HAWC PMTs 
have a $\sim$2x larger total collection efficiency for photons compared to the 8" HAWC PMTS after accounting for the convolution
of the quantum efficiencies in Figure \ref{fig:QE_CE} with the spectrum of Cherenkov light and the larger size of the 10" PMTs.

\newpage

\newpage

\section{Signal Processing}

\begin{figure}[ht!]
\begin{center}
\includegraphics[width=5in]{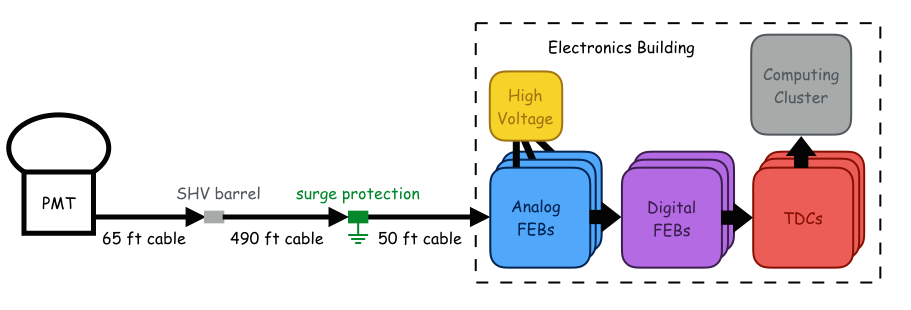}
\end{center}
\caption{Overview of signal processing. Arrows indicate the path taken by PMT signals
         as they travel towards the computing cluster responsible for performing air
         shower reconstruction.
        }
\label{fig:DAQ_block_diagram}
\end{figure}

The positive voltage applied to the HAWC PMTs allows them to be serviced by a single RG-59 coaxial cable
supplying high voltage to the anode while also transmitting measurements back to a centrally located electronics building (\ref{fig:DAQ_block_diagram}).
The central electronics building contains the data acquisition (DAQ) system which receives PMT signals with
a set of analog front-end electronics boards (FEBs). These separate PMT signals from the HV baseline via a blocking capacitor and then amplify each signal and apply two thresholds, a low
threshold and a high threshold. They are also responsible for distributing high-voltage provided by
an external high-voltage power supply to each PMT. The next DAQ component is a set of digital
FEBs which apply basic emitter-coupled logic (ECL) to reduce the low and high threshold outputs from the analog FEBs to a single
digital waveform. This waveform is then recorded by a group of CAEN time-to-digital converters (TDCs) which
transmit the results to an on-site computing cluster for air shower reconstruction and analysis. The following subsections describe each
part of Figure \ref{fig:DAQ_block_diagram} in more detail.


\newpage

\newpage

\subsection{Cable Propagation}

Each PMT converts Cherenkov light produced within a WCD to an electrical current which is transmitted
back to a centrally located electronics building using three RG-59 coaxial cables (Figure \ref{fig:DAQ_block_diagram}).
The first coaxial cable is 65 ft in length and attaches directly to the waterproofed encapsulation at the base of the PMT.
It is long enough to rise over the wall of the WCD and return to ground level where it terminates
at an SHV connector for easy inspection and maintenance of PMTs at the tank. From there, an SHV barrel
connector joins it to a 490 ft cable running underground to the electronics building. This long cable terminates at
a surge protection module, commonly referred to as a spark gap, just outside the electronics building.
A final 50 ft cable runs between the output of the surge protection module and the data DAQ
system located inside the electronics building. The total cable run for each PMT is therefore an identical 605 ft.

The reason for using identical cable lengths, as well as burying cable underground to minimize temperature variations,
is that the propagation delay and signal dispersion within a cable depend on both these quantities. Keeping them the same for
all PMTs therefore helps ensure uniform signal propagation throughout the array. To demonstrate this, we'll now discuss
some of the basic theories behind signal transmission along coaxial cables.

\begin{figure}[ht!]
\begin{center}
\includegraphics[width=2.in]{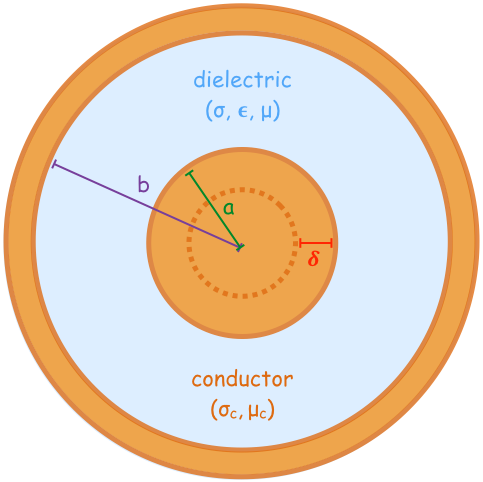}
\end{center}
\caption{Cross section of a coaxial cable. $\sigma_c$ and $\mu_c$ denotes the conductivity and permeability of the copper that comprises the central conductor and shielding. $\sigma$, $\epsilon$, and $\mu$
        denote the conductivity, permittivity, and permeability of the insulator. $\delta$ represents the skin depth within the conductor for a given
        frequency.}
\label{fig:coax_diagram}
\end{figure}

Coaxial cables consist of a central copper conductor surrounded by polyethylene insulator and
an outer copper braid (Figure \ref{fig:coax_diagram}). The central conductor supplies positive high-voltage to the PMT
anode and transmits signals back to the counting house. The outer braid is grounded inside the counting house to 
shield PMT signals from external noise as they travel along the central conductor. It is also connected directly to the
photocathode to provide a negative reference for the positive high-voltage anode. The insulator acts as protection 
against high-voltage breakdown between the center conductor and the grounded shielding.

Signal attenuation due to the finite resistance per unit length of the conductor becomes important
over long cable runs. Additionally, the structure of the coaxial cable introduces inductance,
capacitance, and conductance per unit length behavior that must be considered on the short timescales
of typical PMT signals \cite{HAYT}. Table \ref{tab:coaxcab_hf} presents the expressions for resistance, inductance,
capacitance, and conductance per unit length in terms of the properties of the conductor and insulator shown in Figure \ref{fig:coax_diagram}. 

\newpage

\begin{table}[ht!]
\begin{center}
\begin{tabular}{|c|c|c|} \hline
  Value & Expression Per Unit Length & Units \\ \hline
  Resistance & $r = \frac{1}{2 \pi \delta \sigma_c} \Big(\frac{1}{a} + \frac{1}{b} \Big)$ & $\Omega /$m \\ \hline
  Inductance & $l = \frac{\mu}{2 \pi} ln(b/a)$ & H$/$m \\ \hline
  Capacitance & $c = \frac{2 \pi \epsilon}{ln(b/a)}$ & f$/$m\\ \hline
  Conductance & $g = \frac{2 \pi \sigma}{ln(b/a)}$ & S$/$m \\ \hline
\end{tabular}
\end{center}
\caption{Expressions for resistance, inductance, capacitance, and conductance per unit length of the coaxial
         cable depicted in Figure \ref{fig:coax_diagram} in the high-frequency signal limit. $\delta$ is the
         skin depth for the conductor defined as $1 / \sqrt{\pi \nu \mu_c \sigma_c}$. See \cite{HAYT}
         for a full derivation of these quantities.}
\label{tab:coaxcab_hf}
\end{table}

\vspace{0.75in}

\begin{figure}[ht!]
\begin{center}
\includegraphics[width=4in]{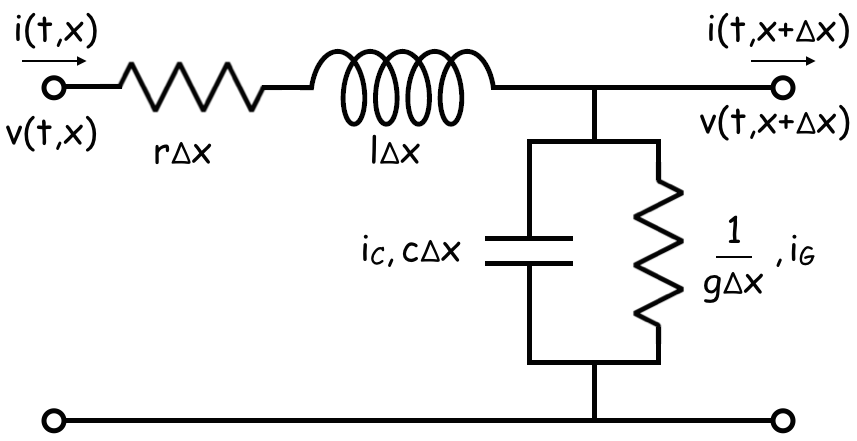}
\end{center}
\caption{Circuit diagram for high-frequency response of coaxial cable in Figure \ref{fig:coax_diagram} to a
         time-dependent input signal V(t, x) over a small length $\Delta x$. }
\label{fig:coax_circuit}
\end{figure}

\newpage

Considering the parameters in Table \ref{tab:coaxcab_hf} over a small length of cable $\Delta x$ results in the circuit
diagram shown in Figure \ref{fig:coax_circuit} on the previous page. Applying Kirchhoff's
voltage and current loop laws to this circuit in the limit $\Delta x \rightarrow \infty$ gives the following differential
equation
\begin{equation}
\frac{d^2v}{dx^2}(x,t) = rc \,\, \frac{dv}{dt} (x,t)  + lc \,\, \frac{d^2v}{dt^2} (x,t)
\label{wave_eq}
\end{equation}
which can be solved for wave solutions of the form $v(x,t) = V e^{i(\omega t - kx)}$ where $\omega$
is the angular frequency of the wave and
\begin{equation}
k = \sqrt{ lc \, \omega^2 + \frac{K c \, \omega^{3/2}}{\sqrt{2}} - i \frac{ K c \, \omega^{3/2}}{\sqrt{2}} }, \,\,\, K \equiv \frac{1}{2 \pi a} \sqrt{ \frac{\mu_c}{\sigma_c} }
\label{eq:k_w_K_intext}
\end{equation}
Appendix \ref{appendixB} presents the complete derivation of this result.

Noting that $k$ is a complex number, the signal amplitude after traveling distance $x$ along the cable will be
\begin{equation}
|v(x,t)| = V e^{\, Im(k) x}
\end{equation}
In the high frequency limit, $\frac{K^2}{2 \omega l^2} \ll 1$ resulting in
\begin{equation}
Im(k) \approx - \frac{K}{2 Z_0} \sqrt{ \frac{\omega}{2} }
\end{equation}
where $Z_0 \equiv \sqrt{l / c}$ is the intrinsic impedance of the cable.
Note that $Im(k)$ is negative and therefore causes signal attenuation in long cables.
The Belden 8241 cables used in HAWC have an attenuation of 3.4 dB for every 100 feet of cable for 100 MHz signals.
While significant, cable attenuation in the Belden 8241 cables is not problematic in HAWC as the DAQ electronics are sensitive enough to measure low signal levels
and the central location of the electronics building minimizes the total length of cable needed to reach the furthest tanks.

Although PMT signals are more complex than the wave solution used to derive the results so far, we can use a fast Fourier transform analysis to numerically reduce
the PMT signal at the beginning of the cable to a sum over a discrete set of frequencies
\begin{equation}
v(x = 0, t) = \sum_{i=0}^N V_i e^{i\omega_i t}
\end{equation}
where $V_i$ are the complex coefficients determined from the fast Fourier method. Each coefficient can then be separately propagated according to the wave solution dictated by Equation \ref{eq:k_w_K_intext} for frequency $\omega_i$. Performing the inverse Fourier transform on the propagated result yields the shape of the PMT signal after traversing the cable. 

\begin{figure}[ht!]
\begin{center}
\includegraphics[width=4.5in]{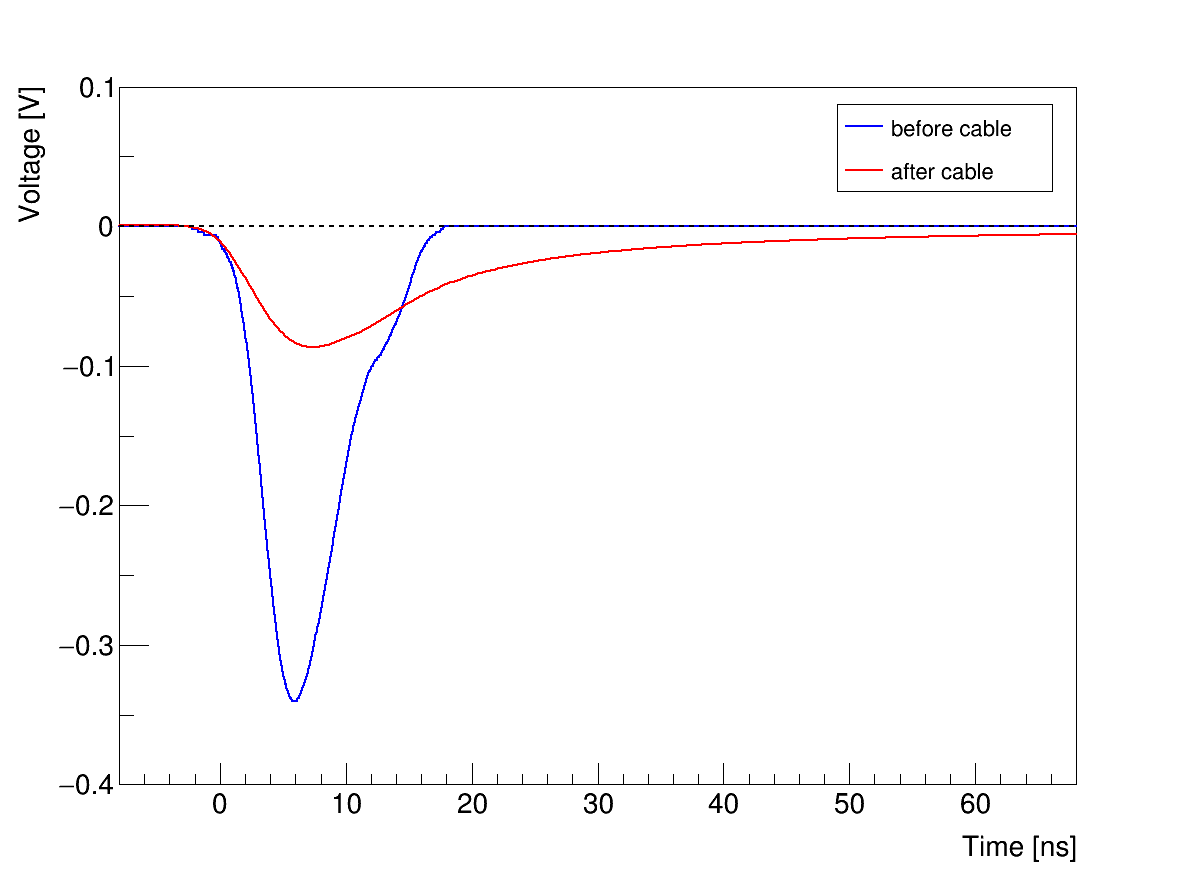}
\end{center}
\caption{A 10 photoelectron waveform before and after cable propagation for an 8" PMT operating at 1700V. The initial waveform was measured with an oscilloscope attached directly to the PMT base with a short RG-59 cable and 75 $\Omega$ terminating resistor. The cable propagation is performed by reducing the initial waveform to Fourier coefficients with a fast Fourier transform and propagating them using Equation \ref{eq:k_w_K_intext} and the parameters for Belden 8241 cable.}
\label{fig:cable_prop}
\end{figure}

Figure \ref{fig:cable_prop} presents the waveform of a 10 photoelectron PMT signal propagated along 605 ft of Belden 8241 cable using the fast Fourier method. As expected, the signal is attenuated following propagation through the cable. In addition, the output signal from the cable is noticeably elongated in time compared to the original PMT signal. This occurs because the wave speed through the cable is frequency dependent
\begin{equation}
v_i = \frac{\omega_i}{Re(k_i)} = \frac{1}{\sqrt{lc} + K / 2Z_0 \, \sqrt{2\omega_i}}
\label{eq:cable_prop_speed}
\end{equation}
(See Appendix \ref{appendixB}) and therefore results in dispersion of the original signal as its individual wave components propagate at different speeds.

Signal dispersion is not necessarily detrimental because the analog front end board electronics integrate signals over a relatively long timescale. However, there is an implicit temperature dependence
of the capacitance, inductance, and resistance per length of cable not shown in Equation \ref{eq:cable_prop_speed} which can result in unequal propagation times between different cables if they are not held at the same temperature. As a result, all HAWC cables are buried underground to minimize cable temperature variations across the array.

\newpage

\subsection{Analog Front End Boards}

The analog front end boards (FEBs) attach directly to the coaxial cables leading to PMTs.
Each board services a set of 16 PMTs and is responsible for distributing high voltage to the center conductor of each cable.
In addition, the analog FEBs perform several signal processing functions on the waveforms coming from each PMT.
First, they separate PMT signals from the DC high voltage baseline of the coaxial cable
with a blocking capacitor. They also terminate the transmission line with a resistance
equal to the intrinsic impedance of the coaxial cable to minimize signal reflections. Finally, they amplify
and apply two thresholds, a low threshold and a high threshold, to the PMT pulse. We will now present
a simplified model of the circuitry in a single channel on the analog FEB to gain an understanding of
how it affects waveforms coming from the PMT that it services.

\begin{figure}[ht!]
\begin{center}
\includegraphics[width=4in]{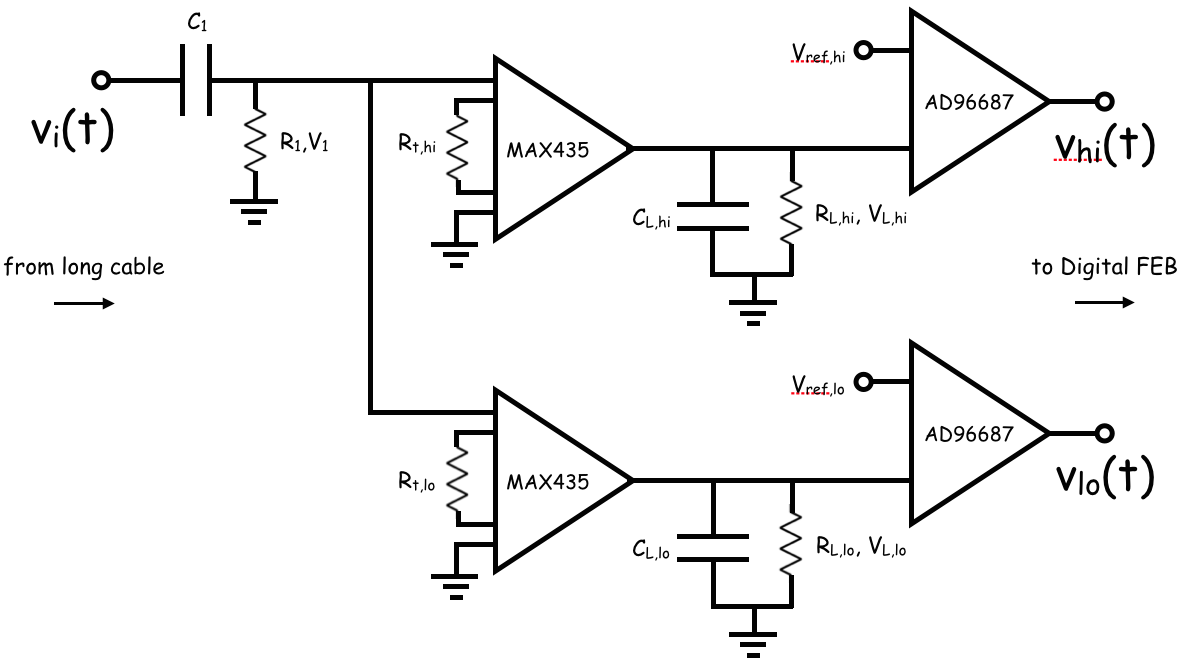}
\end{center}
\caption{Simplified circuit diagram for a single channel on the analog FEB. The actual electronics channel uses differential signal processing, but the circuit response is functionally the same as the response discussed in this section.}
\label{fig:analog_circuit}
\end{figure}

Figure \ref{fig:analog_circuit} presents a simplified circuit diagram for a single channel on the analog FEB. The input circuit consists of
the blocking capacitor, $C_1 = 0.66 \, \mu$F, and the termination resistor, $R_1 = 75 \, \Omega$. Using Kirchhoff's voltage loop law,
we can write down the voltage response $V_1(t)$ output from $C_1$ for an arbitrary input signal under the condition $v_i(t=0) = 0$ as an integral
\begin{equation}
V_1(t) = e^{ -t / R_1 C_1 } \int_0^t dt' \, \frac{dv_i}{dt'}(t') \, e^{ t' / R_1 C_1 }
\end{equation}
In the case of a steady-state sinusoidal input, we can express the resulting signal amplitude as
\begin{equation}
|V_1| = \frac{ \omega \, R_1 C_1 }{ \sqrt{ 1  + \omega^2 \, R_1^2 C_1^2 } } | v_i |
\end{equation}
which effectively shows this is a high-pass filter as DC signals with $\omega = 0$ produce zero response
behind the blocking capacitor. The characteristic frequency of this filter is $1 / 2 \pi R_1 C_1 \approx 30$ kHz which is much lower
than frequencies associated with typical PMT response times, allowing PMT signals to pass through the blocking capacitor with nothing 
more than a phase shift.

After passing through $C_1$, signals are received by two MAX435 transconductance amplifiers which mark the beginning of the high and low threshold discriminator circuits. Each amplifier
converts the input voltage to an output current according to
\begin{equation}
I_{out} = \frac{K}{R_t} V_{in}
\end{equation}
where $K = 4$ and $R_t$ is $680 \, \Omega$ in the high threshold circuit and $169 \, \Omega$ in the low threshold circuit. The output current flows across a capacitor $C_L$ and resistor $R_L$ placed in series. This creates a load voltage of
\begin{equation}
V_L(t) = \frac{K}{C_L R_t} \, e^{ -t / R_L C_L } \, \int_0^t \, dt' \, V_1(t') \, e^{t' / R_L C_L}
\label{eq:amp_response}
\end{equation}
in terms of the time-dependent signal $V_1(t)$ from the input circuit. This reduces to
\begin{equation}
|V_L| = K \frac{R_L}{R_t} \frac{|V_1|}{\sqrt{1 + (\omega R_L C_L)^2}}
\end{equation}
in the case of a steady-state sinusoidal input, giving a voltage amplification of 1 for the low threshold circuit and 7 for the high threshold circuit at frequency $\nu = 11$ MHz.
Refer to Table \ref{tab:analog_param}
for the values of $R_L$ and $C_L$ in each threshold circuit.

\begin{table}[h]
\begin{center}
\begin{tabular}{c|c|c|c|c|c}
Threshold & $R_t \, [\Omega]$ & $R_L \, [\Omega]$ & $C_L$ [pF] & $R_L C_L$ [ns] & Charge [pe]\\
\hline
low (-30 mV) & 169 & 1210 & 47 & 57 & $\sim$0.25 \\
\hline
high (-50 mV) & 680 & 680 & 100 & 68 & $\sim$5 \\
\end{tabular}
\end{center}
\caption{Component values for the low and high threshold circuits.
         The final column represents the approximate integrated
         charge needed prior to amplification to cross the threshold levels after amplification.
         These are given in units of the mean charge for single photoelectron signals generated by the PMT.}
\label{tab:analog_param}
\end{table}

\begin{figure}[ht!]
\begin{center}
\includegraphics[width=4in]{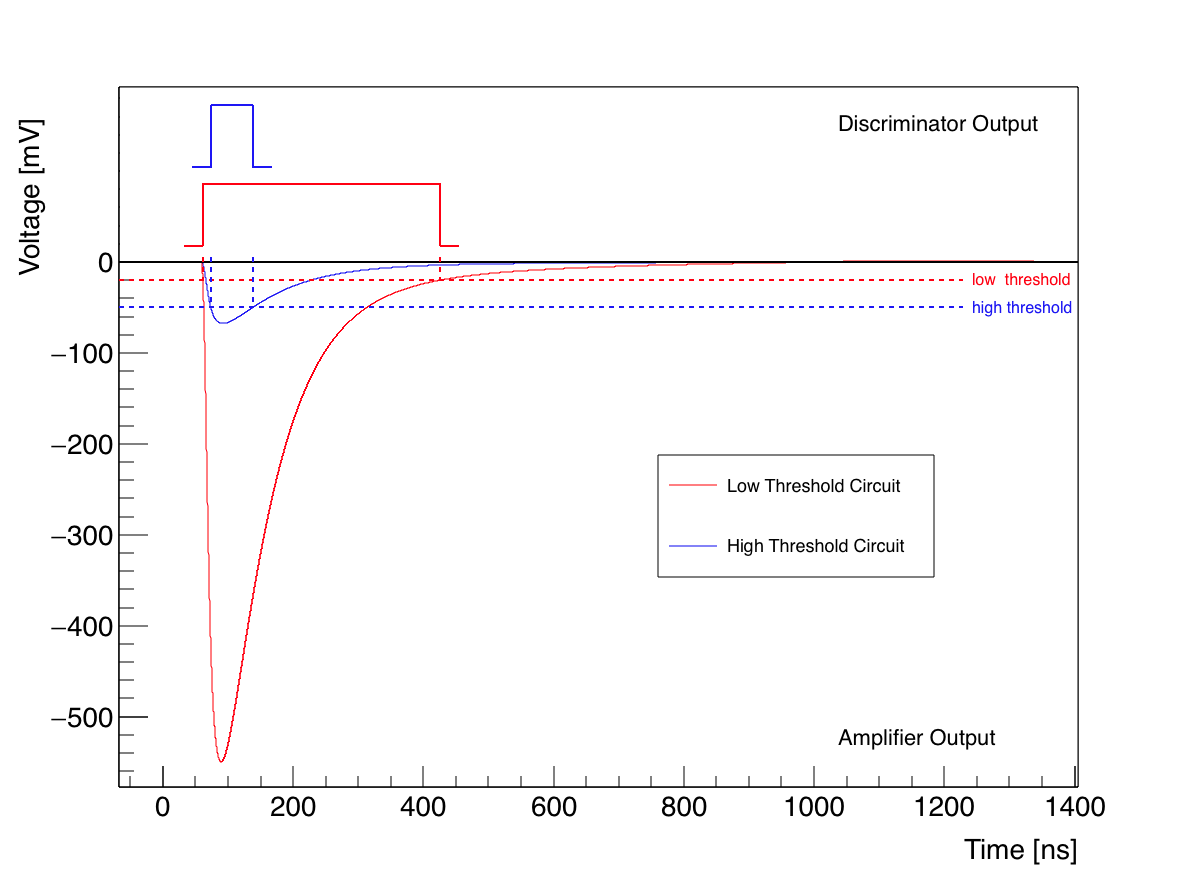}
\end{center}
\caption{Voltage response of the low (RED) and high (BLUE) threshold circuits in a single analog FEB channel to the 10 photoelectron waveform after cable propagation shown in Figure \ref{fig:cable_prop}.
         The smooth curves at negative voltages represent $V_L(t)$ calculated according to Equation \ref{eq:amp_response} with numerical integration using a time step of 0.1 ns. The square pulses at
         positive voltages represent the ECL logic signals output from the low and high threshold discriminators after applying -30 mV and -50 mV thresholds to $V_L(t)$, respectively. The logic signals are shown
         here with arbitrary units.}
\label{fig:4edge_analog}
\end{figure}

The time constant $R_L C_L$ is long compared to the duration of the waveforms shown in Figure \ref{fig:cable_prop} for both the low and high threshold circuits.
As a result, the load voltage $V_L(t)$ will exhibit a sharp rise for typical PMT pulses as the capacitor $C_L$ accumulates charge faster than it can discharge to ground through $R_L$. This behavior
smoothes out the response to individual multi-photoelectron signals, which can vary depending on the exact photoelectron arrival times, in favor of consistently integrating the total waveform charge.
When the input pulse has finished, the load voltage $V_L(t)$ drains back to ground level with an exponential folding time of $R_L C_L$. This is shown in Figure \ref{fig:4edge_analog}, which presents the numerically
calculated behavior of $V_L(t)$ for the 10 photoelectron waveform after cable propagation shown in Figure \ref{fig:cable_prop}.

\begin{figure}[ht!]
\begin{center}
\includegraphics[width=4in]{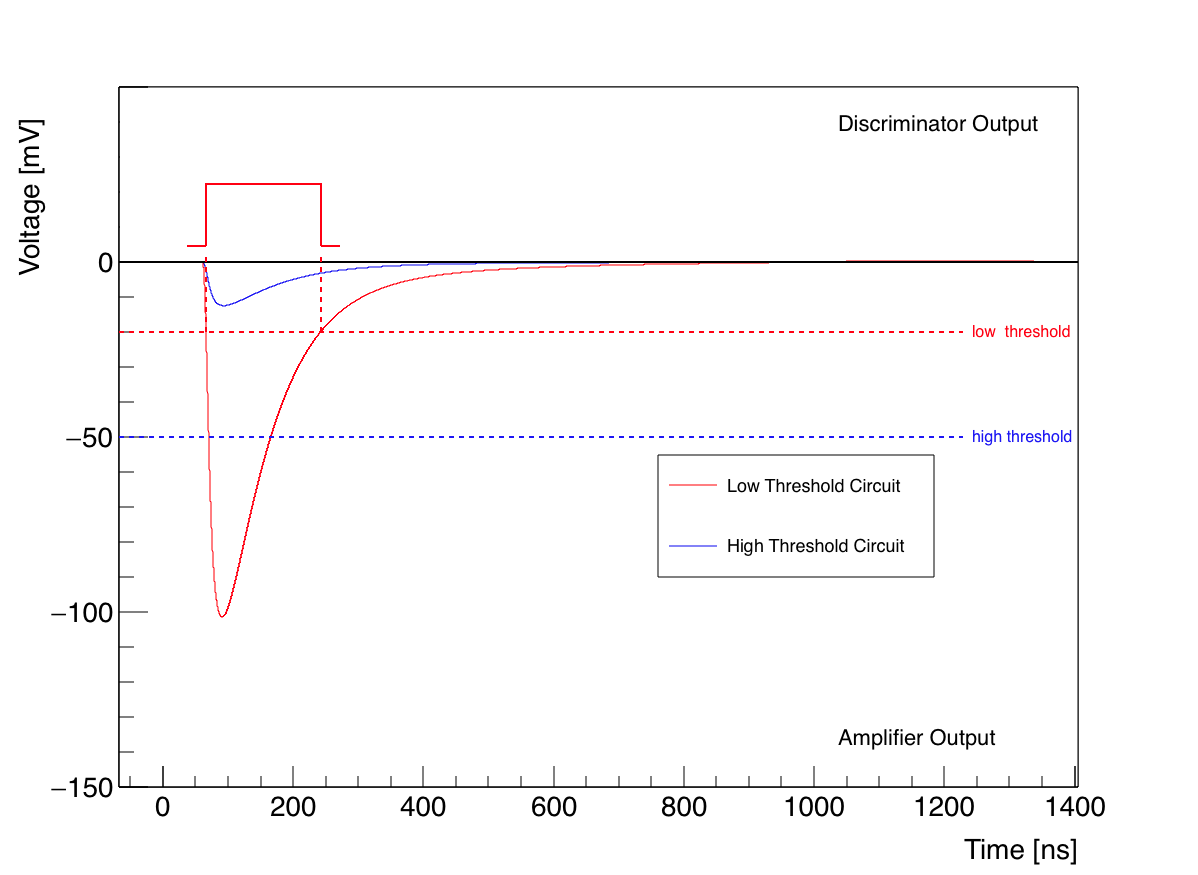}
\end{center}
\caption{Voltage response of the low (RED) and high (BLUE) threshold circuits in a single analog FEB channel to a 1 photoelectron signal in an 8" PMT operating at 1700V after cable propagation.
         The smooth curves at negative voltages represent $V_L(t)$ calculated according to Equation \ref{eq:amp_response} with numerical integration using a time step of 0.1 ns. The square pulse at
         positive voltage represents the ECL logic signal output from the low threshold discriminator after applying a -30 mV threshold. There is no response from the high threshold discriminator as the
         amplified signal does not exceed the -50 mV threshold. The logic signals are shown here with arbitrary units.}
\label{fig:2edge_analog}
\end{figure}

Following amplification, the signals in each threshold circuit are passed to AD96687 ultra-fast comparator chips
where they are compared to two different reference voltages, -30 mV in the low threshold circuit and -50 mV in the
high threshold circuit. This implies a pre-amplification threshold level of about -4 mV to PMT signals
in the low threshold circuit given the 7x amplification for 11 MHz signals. The 1x amplification in the high threshold
circuit means we can interpret the -50 mV high threshold as applying directly to the original PMT waveform.
These correspond approximately to the amplitudes of 0.25 photoelectron and 5 photoelectron signals prior to amplification.
Each comparator outputs an ECL logic pulse that begins when the signal voltage drops below the reference voltage and ends when the signal rises above the reference voltage again, thereby creating a square pulse whose width equals the time the original pulse was below the reference threshold. This width is called the time-over-threshold (TOT) and is typically shortened to LoTOT and HiTOT when referring to TOT from the low and high threshold circuits, respectively. Large PMT signals like the 10 photoelectron signal shown in Figure \ref{fig:4edge_analog} cross both thresholds and therefore have a LoTOT and HiTOT. However, small PMT signals only have a LoTOT as they are not large enough to cross the high threshold (Figure \ref{fig:2edge_analog}).

\newpage

\newpage

\newpage

\subsection{Digital Front End Boards}\label{sec:DFEB}

Each digital front end board (FEB) services a single analog FEB and therefore provides additional signal processing 
to 16 PMT channels. In each channel, it accepts the low and high threshold discriminator outputs generated in the analog FEB
and applies additional digital ECL logic to combine the low and high threshold signals into a single waveform. This results in a large
cost savings as the waveform can then be processed with a single time-to-digital converter (TDC) channel instead of two channels,
one for the low threshold and one for the high threshold. The digital front end boards also apply a number of checks on each digital waveform to ensure
the final result is measurable by the TDCs.

\begin{figure}[ht!]
\begin{center}
\includegraphics[width=4in]{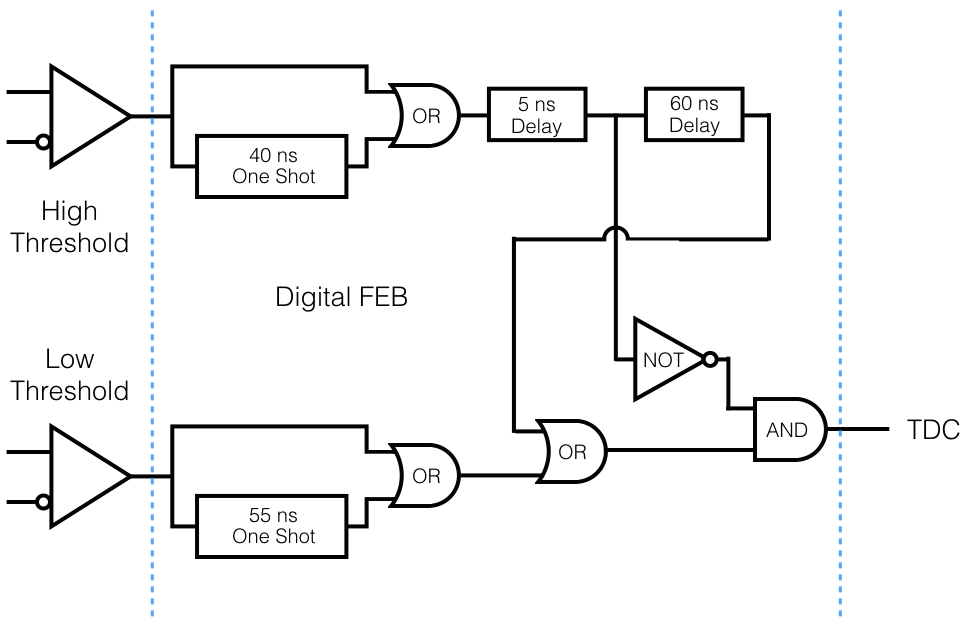}
\end{center}
\caption{Overview of digital FEB logic in a single PMT channel. The digital FEB accepts the analog FEB discriminator outputs from the far left, processes it through the diagram marked between the two dashed lines, and then transmits the final result to a single TDC channel.}
\label{fig:digital_logic}
\end{figure}

Figure \ref{fig:digital_logic} presents an overview of the digital logic applied to a single PMT channel inside the digital FEB.
Signals in this diagram begin on the left side, where they are received from the low and high threshold discriminator outputs, and travel to the right where they are output to a single TDC channel.
The first operation applied to both thresholds is a comparison between the incoming discriminator pulse and a fixed-width logic pulse implemented with a 1-shot circuit triggered by the incoming signal.
This is done to ensure the widths of both HiTOT and LoTOT are greater than the 5 ns edge pair resolution of the TDCs.

\begin{figure}[ht!]
\begin{center}
\includegraphics[width=3.5in]{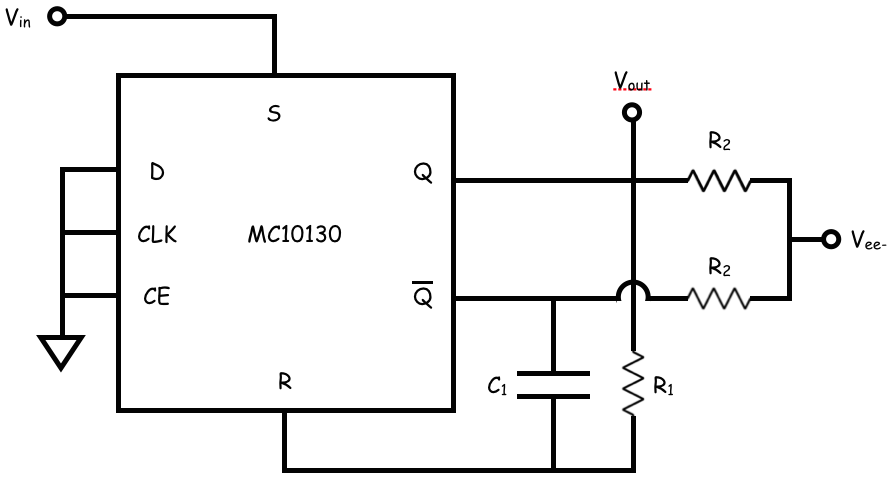}
\end{center}
\caption{One-shot latch circuit diagram}
\label{fig:latch_circuit}
\end{figure}

\begin{figure}[ht!]
\begin{center}
\subfigure[][]{\includegraphics[height=1.5in]{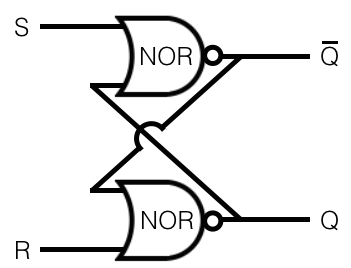}}
\quad
\subfigure[][]{\includegraphics[height=1.5in]{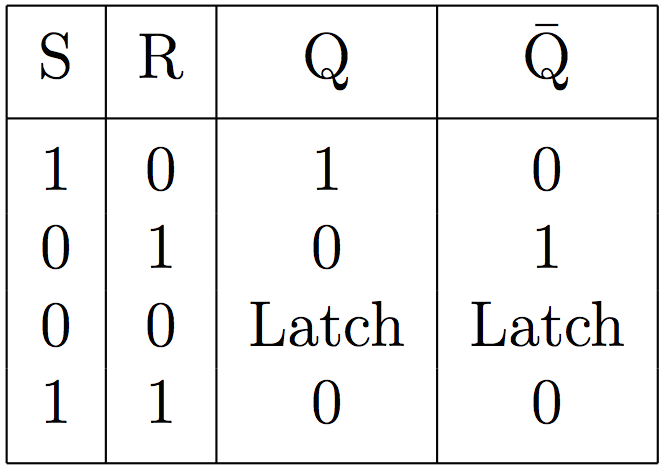}}
\end{center}
\caption{Flip-flop logic diagram (a) and logic table (b) for the MC10130 D-type latch with D, CE, and CLK inputs tied to ground.}
\label{fig:flip_flop}
\end{figure}

The 1-shot circuit design is shown in Figure \ref{fig:latch_circuit}. It employs a MC10130 D-type latch where
the data and clock inputs are tied to ground, forcing them to remain in a high state and allowing the set (S) and reset (R) inputs
to modify the outputs Q and \={Q} at all times. As a result, the latch operates as a simple SR flip-flop circuit according to the logic diagram
and truth table presented in Figure \ref{fig:flip_flop} where logic "1" corresponds to an output voltage $V_{high} = -0.8$ V and logic "0"
corresponds to an output voltage of $V_{low} = -1.6$ V. 

\newpage

\begin{figure}[ht!]
\begin{center}
\includegraphics[width=3.5in]{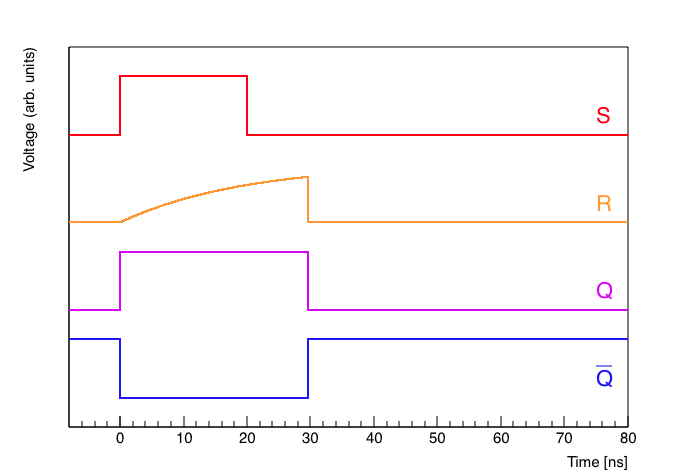}
\end{center}
\caption{Latch response to a 20 ns TOT pulse applied to the S input of an ideal one-shot circuit where $R_1C_1 \approx$ 20 ns and there are no delays associated with changing Q and \={Q}.}
\label{fig:latch_resp}
\end{figure}

Under normal operation, the circuit remains latched with S, R, and Q in a low state and \={Q} in a high state.
The arrival of a TOT signal at $V_{in}$ switches S to the high state and transforms the output such that Q is now in a low state
and \={Q} is in a high state. This allows current to flow from Q to \={Q}, charging the capacitor $C_1$ with a characteristic exponential folding time
of $R_1C_1$. The voltage difference across the charging capacitor reaches the threshold for registering input R as a high state after
$\sim1.5 R_1C_1$, which resets the circuit to its original configuration if the TOT signal is no longer active. This truncates the output at Q
and results in a square pulse with an approximate width of 1.5$R_1C_1$ (Figure \ref{fig:latch_resp}).

\newpage

In the case where the original TOT pulse is still active when the input R attempts to reset the circuit, both Q and \={Q} are set
to 0 rather than the original configuration. This causes the capacitor $C_1$ to discharge until the input at R falls below the threshold
for registering as a high state, briefly allowing the circuit to return to the set state to charge $C_1$ again and trigger another
reset. The average voltage output at Q during this cycle remains less than is required to register as a high state at the next electronics component
so the output at Q is effectively the same as the case where the input TOT is shorter than the circuit reset time.

\newpage

\begin{figure}[ht!]
\begin{center}
\includegraphics[width=4in]{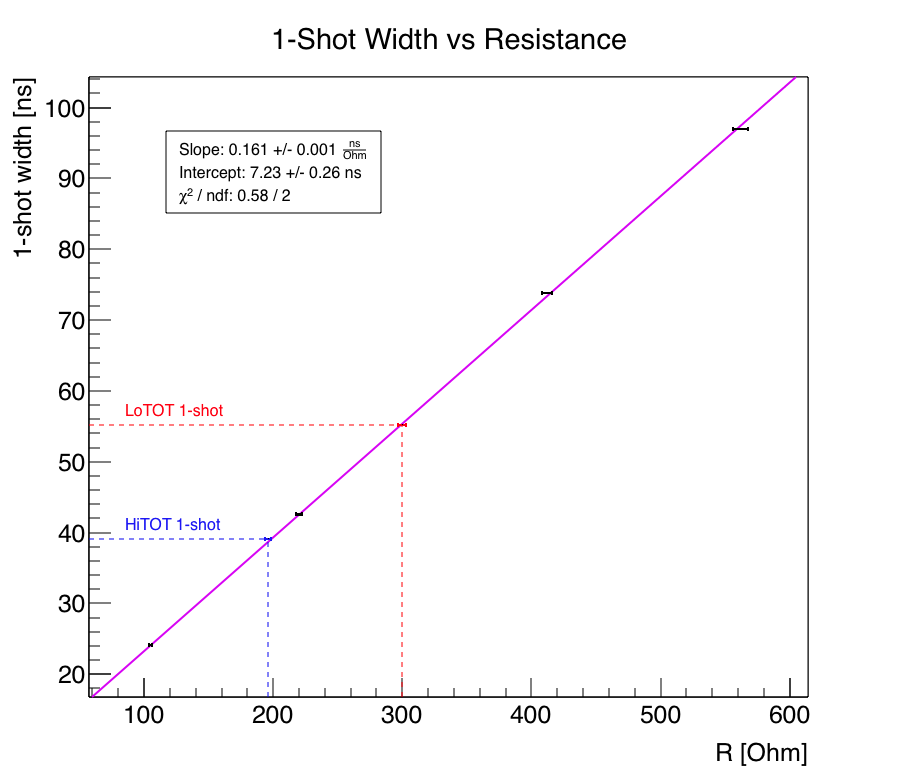}
\end{center}
\caption{Measured one-shot width vs resistance $R_1$ for $C_1 = 100$ pF. The blue dashed line marks the 197 $\Omega$ resistance value chosen to set
         the minimum HiTOT one-shot width of $\sim40$ ns and the red dashed line marks the 300 $\Omega$ resistance value chosen to set the minimum LoTOT one-shot width of $\sim55$ ns.}
\label{fig:one_shot_v_r}
\end{figure}

Figure \ref{fig:one_shot_v_r} presents the measured widths of one-shot pulses produced by different values of $R_1$ for $C_1 = 100$ pF.
These data follow a linear fit as the one-shot pulse width scales directly with the circuit reset time determined by $R_1C_1$. From
the slope we deduce the exponential folding factor in the one-shot circuit is 1.61. The non-zero intercept is the result of the summed propagation delay, setup time, 
and hold time associated with changing the Q and \={Q} outputs on the MC10130 chip.

Values of $R_1 = 300 \Omega$ and $C_1 = 100$ pF were chosen for the one-shot circuit associated with LoTOT to enforce a minimum LoTOT of 55 ns which is much larger than typical PMT rise times.
This aids discrimination between small and large PMT pulses after combination of the LoTOT and HiTOT waveforms. The 55 ns minimum is still a factor of 3 smaller than most
LoTOT associated with single photoelectron waveforms so the majority of PMT pulses remain unaffected by this requirement. 

Values of $R_1 = 197 \, \Omega$ and $C_1 = 100$ pF were chosen during the Milagro experiment for the one-shot circuit associated with HiTOT to enforce a minimum HiTOT of 40 ns.
This ensured HiTOT was always greater than the minimum edge pair resolution of the TDCs used in Milagro. In principle, this means the minimum HiTOT duration could be set lower in
HAWC because the minimum edge pair resolution of the newer HAWC TDCs is 5 ns but there is no evidence to suggest doing so would significantly affect the overall sensitivity of the experiment.
The minimum HiTOT setting of 40 ns is therefore kept for the HAWC electronics setup.

\begin{figure}[ht!]
\begin{center}
\includegraphics[width=4in]{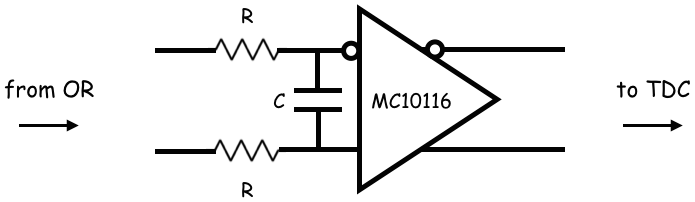}
\end{center}
\caption{Delay circuit implemented with an MC10116 receiver on a differential input.
         The total circuit delay equals the propagation delay of the MC10116 chip plus the exponential folding time for input signals
         to reach the high state given the time constant set by $RC$.}
\label{fig:delay_circuit}
\end{figure}

Two delays are applied to the HiTOT discriminator pulse after it passes the OR gate with the minimum HiTOT one-shot pulse.
These are implemented by placing in-line resistors bridged with a capacitor on the differential inputs of MC10116 chips
receiving the differential output from the OR gate (Figure \ref{fig:delay_circuit}). This setup introduces an
exponential response to the input pulse with a characteristic timescale of $RC$. As with the one-shot circuit, there is then
a linear relationship between $RC$ and the total circuit delay where the slope indicates the number of exponential foldings
required for the input to reach a high state and the intercept is the sum of propagation delays inside the MC10116 chip.

\newpage

\begin{table}[ht!]
\begin{center}
\begin{tabular}{|c|c|c|c|}
\hline
R [$\Omega$] & C [pF] & RC [ns] & Delay [ns] \\
\hline
105 $\pm$ 1 & 100 $\pm$ 5 & 10.50 $\pm$ 0.54 & 25.01 $\pm$ 0.01 \\
105 $\pm$ 1 & 15.00 $\pm$ 0.75 & 1.57 $\pm$ 0.08 & 8.38 $\pm$ 0.01 \\
\hline
\end{tabular}
\end{center}
\caption{Measured time delays from the first HiTOT delay circuit for two different values of capacitance $C$ at fixed $R = 105 \, \Omega$.}
\label{tab:risetimes}
\end{table}

Table \ref{tab:risetimes} presents the measured delay times for two different capacitance values $C$ at fixed resistance $R = 105 \, \Omega$ in the first HiTOT delay.
These are obtained from the time offset between the beginning of LoTOT and the beginning of HiTOT at the digital FEB output for PMT pulses with 225 ns $<$ HiTOT $<$ 230 ns.
This range of HiTOT values was chosen because it minimizes the intrinsic rise time of the original PMT waveform between the low and high thresholds and therefore yields
a time offset approximately equal to the first HiTOT delay. The value $C = 15$ pF was chosen for use in HAWC to ensure the time offset between the start of LoTOT and HiTOT
is both greater than the 5 ns edge pair resolution of the TDCs and smaller than the minimum LoTOT one-shot.

The second HiTOT delay is set to 60 ns using values of $R = 400 \, \Omega$ and $C = 100$ pF which are retained from the Milagro electronics.
It is applied to produce a signal which is guaranteed to return to the low state after the end of the HiTOT signal
seen by the AND gate in Figure \ref{fig:digital_logic}. An OR gate then combines the twice delayed HiTOT signal with LoTOT to extend the end of the LoTOT
pulse to at least 60 ns after the end of HiTOT. This ensures the time offset between the end of LoTOT and HiTOT is greater than the 5 ns minimum edge pair resolution of the TDCs.

\begin{figure}[ht!]
\begin{center}
\includegraphics[width=6.5in]{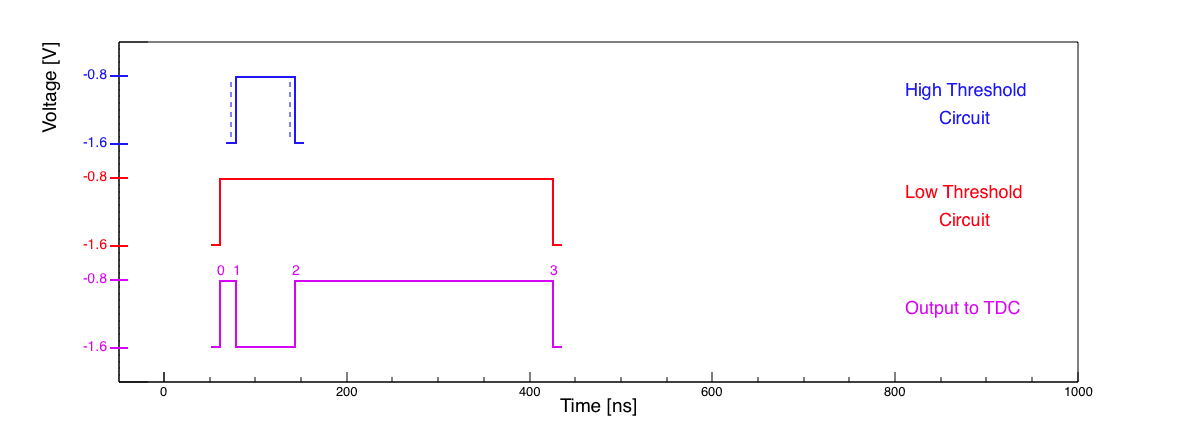}
\end{center}
\caption{Digital FEB circuit response to the low (RED) and high (BLUE) threshold discriminator signals from
         the analog FEB for the 10 photoelectron waveform in Figure \ref{fig:4edge_analog}.
         The BLUE dashed lines denote the position of the high threshold discriminator signal
         prior to the first HiTOT delay. The MAGENTA waveform represents the final output from the digital FEB
         after combining the low and high threshold signals at the final AND gate in Figure \ref{fig:digital_logic}.}
\label{fig:fouredge}
\end{figure}

The LoTOT signal produced by the OR gate with the twice delayed HiTOT signal passes to the AND gate in Figure \ref{fig:digital_logic}
where it is compared to the complement of the HiTOT signal after the first HiTOT delay.
This results in the waveform shown in Figure \ref{fig:fouredge} for the analog output of the 10 photoelectron pulse from Figure \ref{fig:4edge_analog}.
It is comprised of two square pulses delimited by four edges labeled from 0 to 3.

The four edges define three independent timing parameters. First, the time difference between edges 0 and 1 (T01) represents the rise time
of the input waveform from the low to high threshold plus the first HiTOT delay. Second, the time difference between
edges 1 and 2 (T12) marks the duration of HiTOT. And lastly, the time difference between
edges 2 and 3 (T23) marks the fall time of the input waveform. The sum of these three parameters equals LoTOT and is denoted by the difference between edges 0 and 3.
Refer to Table \ref{tab:timingpar} for a summary of the minimum values for each timing parameter based on the discussion in this section.

\newpage

\newpage

\renewcommand{\arraystretch}{0.5}
\begin{table}[ht!]
\begin{center}
\begin{tabular}{|c|c|c|c|}\hline
Timing & Description & Minimum & Enforcing \\
Parameter & & Value [ns] & Component \\
\hline
T01 & Rise Time & $\ge$ 5 & 1st HiTOT delay\\
T12 & HiTOT & $\ge$ 40 & HiTOT one-shot\\
T23 & Fall Time & $\ge$ 60 & 2nd HiTOT delay\\
T03 & LoTOT & T01+T12+T23 $\ge$ 105 & OR between LoTOT/HiTOT \\
\hline
\end{tabular}
\caption{Summary of timing parameters for the 4 edge waveform shown in Figure \ref{fig:fouredge}. Each parameter is
         labeled as T(i)(j) where i is the beginning edge and j is the final edge. Also shown are the
         minimum values enforced by the digital FEB. Note that T12 is equivalent to HiTOT and
         T03 is equivalent to LoTOT from Figure \ref{fig:4edge_analog} for waveforms that satisfy all minimum values.}
\label{tab:timingpar}
\end{center}
\end{table}

\newpage 

\begin{figure}[ht!]
\begin{center}
\includegraphics[width=6.5in]{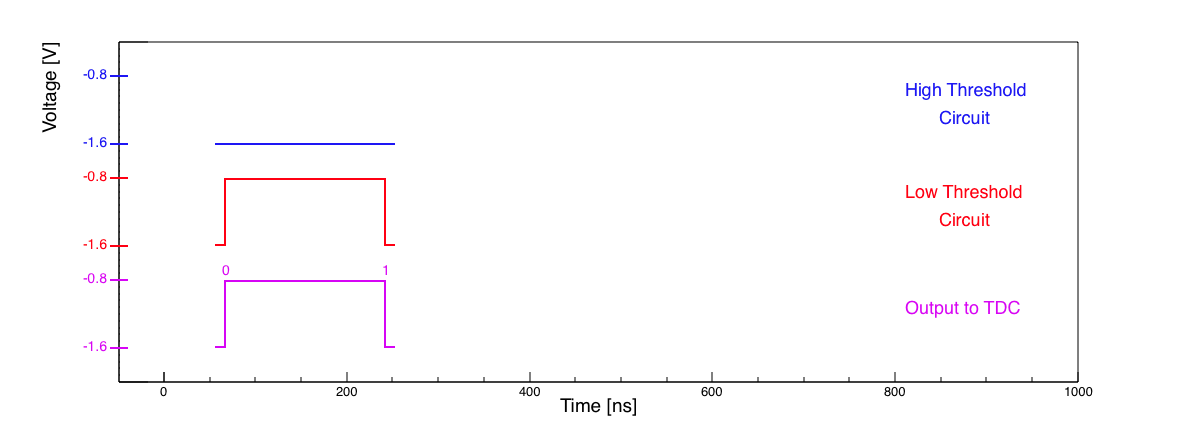}
\end{center}
\caption{Digital FEB circuit response to the low (RED) and high (BLUE) threshold discriminator signals from
         the analog FEB for the 1 photoelectron waveform in Figure \ref{fig:2edge_analog}. Note that there is
         no signal in the high threshold circuit because the 1 photoelectron waveform never crosses the high threshold.
         There is therefore a single timing parameter, T01, which equals LoTOT.}
\label{fig:twoedge_dig}
\end{figure}

A simpler waveform is produced by the digital FEB for the 1 photoelectron signal in Figure \ref{fig:2edge_analog}
because the high threshold is never crossed. In this case, the high threshold signal remains 0 and the low threshold
signal passes through the digital FEB with only a comparison to the LoTOT one-shot pulse width to ensure it has a value
larger than 55 ns. The digital FEB output is therefore a single square pulse with two edges, 0 and 1,
that give a single timing parameter, T01, equal to LoTOT (Figure \ref{fig:twoedge_dig}).

\newpage

\subsection{Time-to-Digital Converters}\label{sec:TDCs}

The HAWC Observatory uses 10 CAEN VX1190A TDCs, each with 128 channels, to record the waveforms output by the digital FEBs
which are shown in Figures \ref{fig:fouredge} and \ref{fig:twoedge_dig}. Each TDC channel records the rising and falling edges of
the digital waveforms with an absolute time precision of 100 picoseconds and a minimum edge pair resolution of 5 ns. This represents
a significant improvement over the TDCs used in Milagro which had an absolute time precision of 0.5 ns and a minimum edge pair resolution
of 15 ns.

As discussed in Section \ref{sec:DFEB}, the digital FEBs are designed to ensure that all waveform timing parameters are greater than the minimum edge pair resolution.
This is because the second edge in a pair of edges separated by less than 5 ns will be discarded by the TDC. The loss of an edge renders the measurement unusable
as one or more of the waveform's timing parameters will be incalculable. However, rare cases do occur when separate signals in the same channel arrive closely in time
and produce a single edge without a corresponding rising or falling edge companion. These single edges are flagged in the data stream to prevent their use in the
triggering and reconstruction of  air shower events.

\begin{figure}[ht!]
\begin{center}
\includegraphics[width=5in]{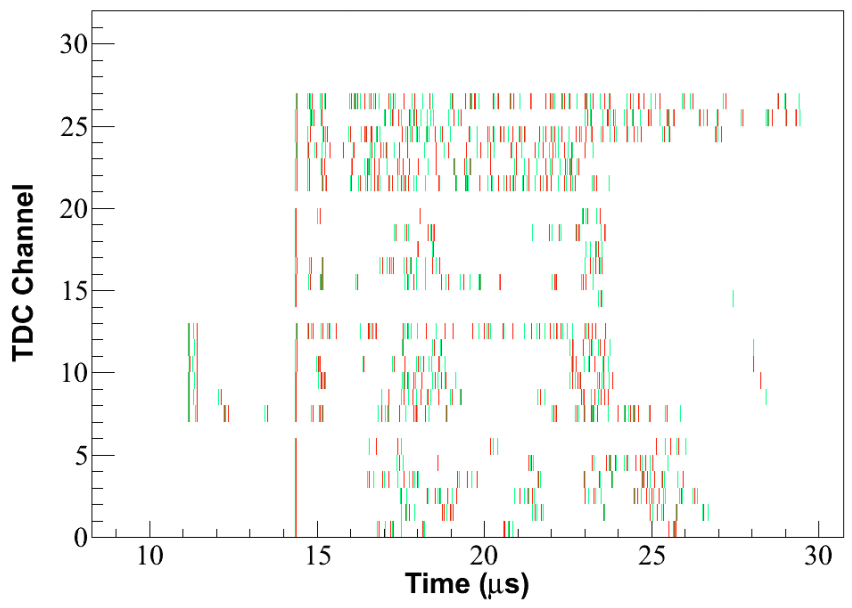}
\end{center}
\caption{Edge times for a TDC event recorded during early testing of the HAWC Observatory DAQ in the VAMOS array.
         There were four tanks operating at this time, each containing seven 8" PMTs
         rather than the standard HAWC configuration with three 8" PMTs and a central 10" PMT.
         GREEN lines mark rising edges and RED lines mark falling edges of square pulses
         measured by the TDCs. A set of 4 edge waveforms can be seen near 11 $\mu$s in channels 7-13 which are
         all in the same tank. This most likely marks a single muon event. An air shower event producing
         simultaneous hits in all 4 tanks can be seen at $\sim$14.2 $\mu$s.}
\label{fig:edgeevent}
\end{figure}

The TDCs are continuously operated using a 40 kHz clock to trigger buffered edge data into 25 microsecond long blocks known as TDC events (Figure \ref{fig:edgeevent}).
Each TDC event receives a GPS timestamp derived from the NTP time inside the first TDC resulting in millisecond
timing accuracy which is precise enough for analyzing the search timescales described in Chapters \ref{ch:method}-\ref{ch:results}.
A true GPS timing system is currently in development and will provide $\sim$1$\mu$s accuracy when complete.
Every set of 1000 sequential TDC events are further grouped into a data block known as a timeslice which is passed to the online server clients
that perform air shower triggering and reconstruction. The total data load to the servers is $\sim$450 MB/s with each TDC contributing 45 MB/s to the data stream. 

Since it would be prohibitively expensive to save the total data load of all waveforms, an air shower trigger criterion of observing 28 waveforms inside a 150 ns window
is applied to record waveform data. This criterion is discussed in detail in Section \ref{sec:trigger}. Additionally, the first 10 timeslices in every sequential group of 5000 timeslices
are saved to disk to provide a minimum bias dataset for low-level data studies such as trigger development. The total data load recorded to disk from both the triggered and minimum bias
sample is $\sim$20 MB/s.

\newpage

\newpage

\section{Air Shower Triggering}\label{sec:trigger}

As discussed in Section \ref{sec:TDCs}, we apply an air shower trigger to reduce
the total data rate within HAWC from $\sim$450 MB/s to $\sim$20 MB/s. The trigger criterion
requires 28 waveforms to arrive inside a 150 ns window. It results
in a $\sim$24 kHz rate of triggered events recorded to disk. This rate fluctuates by $\sim$10\%
over the course of each day as atmospheric pressure variations change the amount of atmospheric
overburden above HAWC.

\begin{figure}[ht!]
\begin{center}
\includegraphics[width=5in]{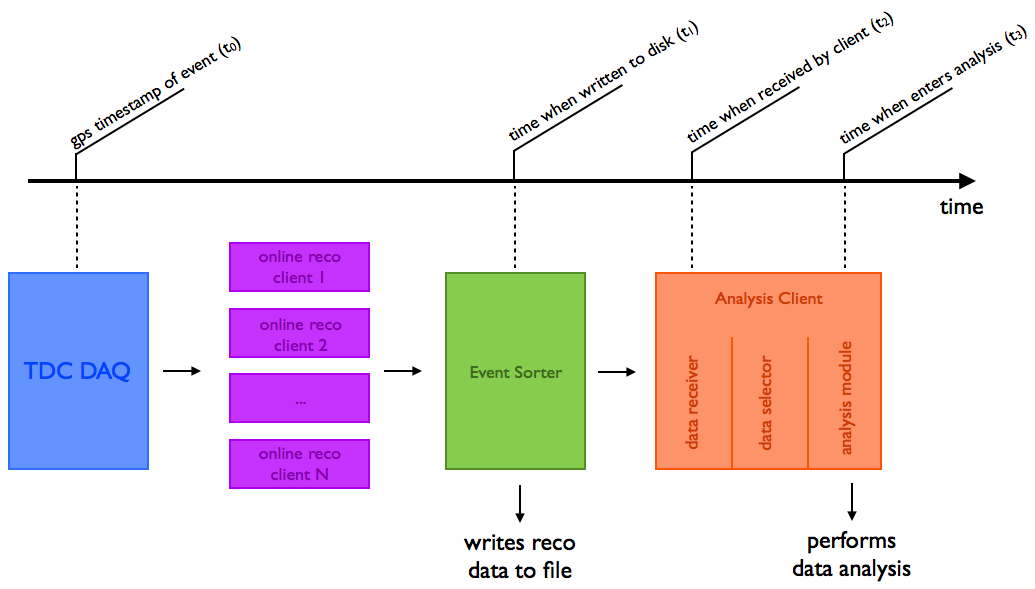}
\caption{Overview of air shower processing processing. TDC events containing waveform data are passed
         to reconstruction nodes in the on-site server farm where we
         apply a trigger criteria of observing 28 waveforms inside a 150 ns window.
         Events passing this trigger are referred to as air shower events and receive charge and timing calibrations
         followed by application of the reconstruction algorithms described in the remaining sections of this chapter.
         An event sorter receives both the original triggered events prior to calibration and reconstructed air showers
         from the online reconstruction nodes. The triggered events are time sorted according to their original trigger times
         and written to disk to allow retroactive reconstruction. The reconstructed events are also time ordered and written
         to disk but have the additional benefit of being directly accessible over socket connection, eliminating the need
         to wait for write completion of reconstructed data files while performing real-time analysis.
         }
\label{fig:onlinechain}
\end{center}
\end{figure}

\begin{figure}[ht!]
\begin{center}
\includegraphics[width=5in]{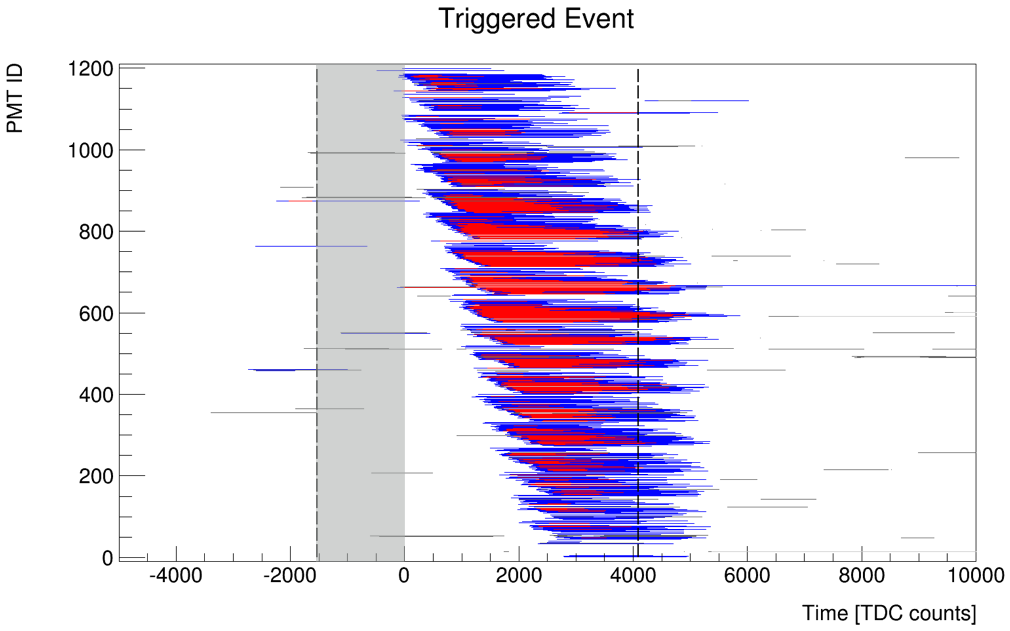}
\caption{Measured HiTOT (RED) and LoTOT (BLUE) for all PMTs in a triggered air shower event in HAWC data.
         The time axis is in units of TDC counts where 1 count is approximately equal to 0.1 ns.
         The GREY region marks the 150 ns trigger window in which 28 waveforms were observed. The dashed vertical
         lines mark the time selection of hits used in the reconstruction of this event. GREY horizontal lines
         denote hits that are excluded from the reconstruction either because they are outside the window
         for reconstruction or they fail the requirements discussed in Section \ref{sec:edges}.
         }
\label{fig:airshowtrig}
\end{center}
\end{figure}

Figure \ref{fig:onlinechain} gives an overview of the process involved in triggering air shower events within HAWC.
In it the TDC events described in Section \ref{sec:TDCs} are passed to online reconstruction nodes in the
on-site server farm where we apply the trigger criterion.
When this threshold is met, a window containing all measured waveforms from -0.5 $\mu$s to 1 $\mu$s around
the trigger is saved to form a triggered air shower event (Figure \ref{fig:airshowtrig}). The online reconstruction
nodes then apply the air shower reconstruction discussed in Chapter \ref{ch:Reco} in real-time to produce a data stream of
reconstructed events in addition to the triggered event data set.


\newpage

\newpage

\newpage

\begin{figure}[ht!]
\begin{center}
\includegraphics[width=4in]{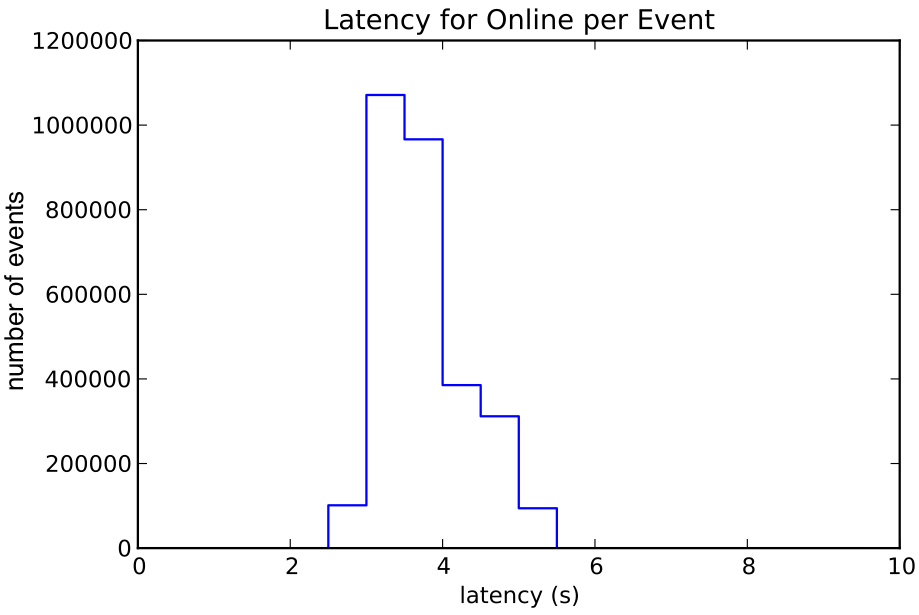}
\caption{Measured latency of the on-site air shower reconstruction.
         Latency is defined as the time difference between the GPS timestamp
         of a reconstructed air shower event and the time when it received
         by an analysis client (See Figure \ref{fig:onlinechain}). Latencies
         associated with recording events inside the TDC DAQ are on the
         order of milliseconds and can be ignored compared to the  
         than the $\sim$4 second latency shown in this plot.}
\label{fig:latency}
\end{center}
\end{figure}

A program known as the event sorter receives both the triggered and reconstructed air shower events from all online reconstruction
nodes and time orders them according to the original GPS timestamps applied by the TDCs. The sorter then
writes these events to disk in what are known as triggered and reconstructed data files, respectively.
Additionally, the sorter offers direct access to reconstructed events over a socket connection, eliminating
the need to wait for write completion of reconstructed data files while performing real-time analysis.
This yields a total system latency of $\sim$4 seconds from when the GPS timestamp is applied to
when reconstructed showers are ready for analysis (Figure \ref{fig:latency}).

\newpage

\newpage

\section{Calibration System}\label{calmeup}

\begin{figure}[ht!]
\begin{center}
\includegraphics[width=5in]{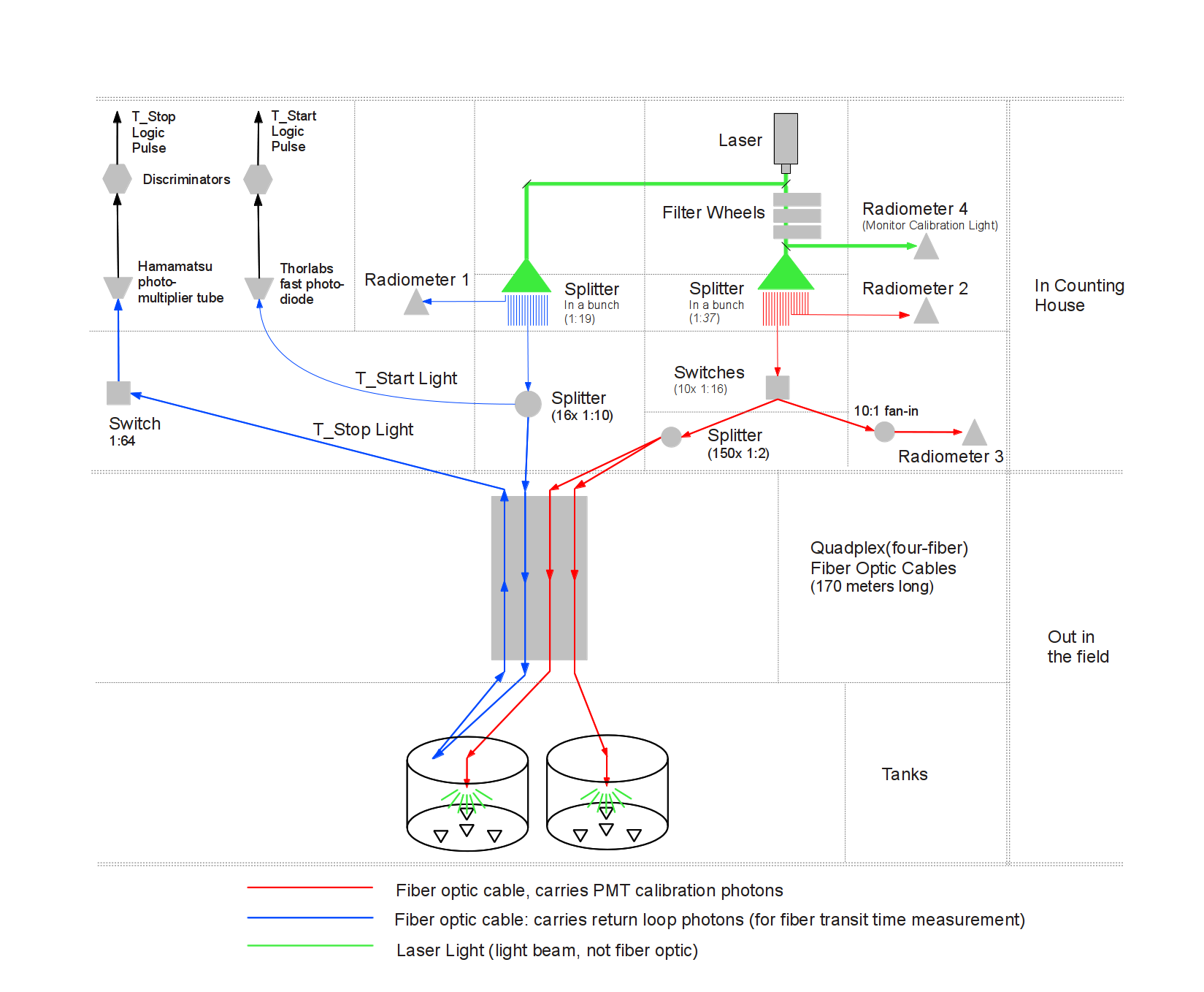}
\caption{Simplified overview of the laser calibration system responsible for sending light to tanks.}
\label{caloverview}
\end{center}
\end{figure}

\begin{figure}[ht!]
\begin{center}
\includegraphics[width=5in]{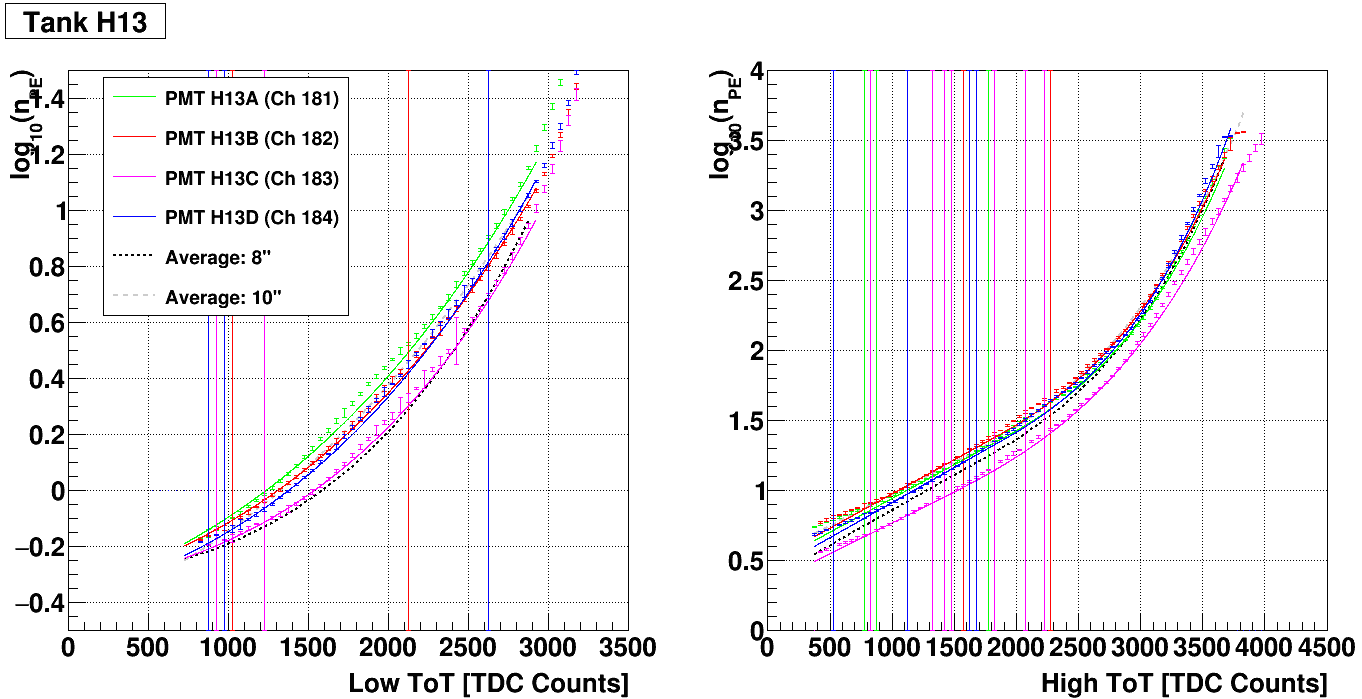}
\caption{Charge calibration curves for LoTOT (Left) and HiTOT (Right) in tank H13 in calibration run 5213. The ordinate represents the mean TOT value 
         associated with the number of measured photoelectrons at a given laser light intensity shown on the abscissa.
         The data points mark measurements from the calibration run whereas the solid lines represent fits to the data.
         The different colors mark the four PMTs within this tank. In practice, the LoTOT curves are used to calculate
         the charge of waveforms less than $\sim$5 pe because the high threshold is not crossed. The HiTOT curves are used
         when HiTOT is present. HiTOT is approximately linear in log-space at small values of TOT
         until saturation effects cause an upturn in the calibration curve. 
        }
\label{chargecurve}
\end{center}
\end{figure}

The purpose of the calibration system of the HAWC Observatory is
to convert the amplitude of PMT waveforms as measured by TOT to 
the corresponding number of photoelectrons generating the original waveform
as well as to correct for amplitude dependent timing effects, referred to as slewing.
To do this, a pulsed laser with a 1 ns pulse width is used to send light 
through optical fibers to a diffuser located
at the top of each HAWC tank (Figure \ref{caloverview}). The light level is varied to produce curves of total waveform charge versus TOT
(Figure \ref{chargecurve}). These curves are reported in units of the mean single photoelectron charge which corresponds to the total
number of detected photons. 

The calibration curve for LoTOT is used to calculate the charge of waveforms less than $\sim$5 pe
during air shower reconstruction because the high threshold is not crossed. The calibration curve for HiTOT is used when HiTOT is present
and extends up to signals with several thousand photoelectrons. This is possible because the conversion to TOT in the analog FEB acts
as a logarithmic amplifier which provides good charge resolution over a wide dynamic range.

\begin{figure}[ht!]
\begin{center}
\includegraphics[height=2.2in]{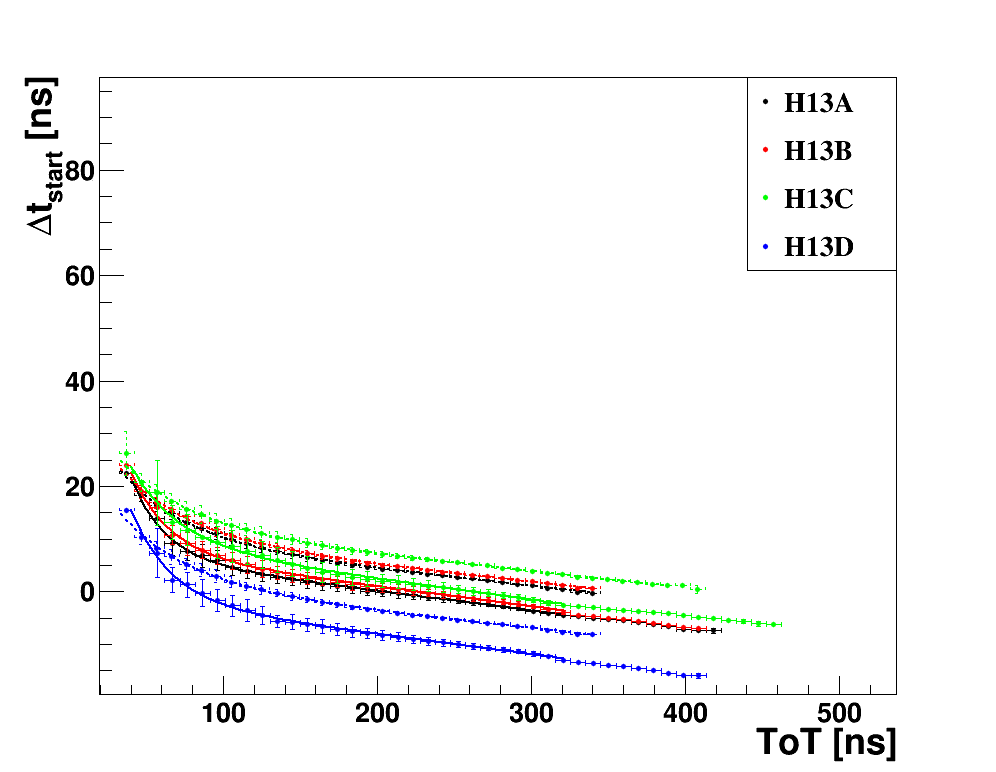}
\caption{Timing calibration curves for LoTOT (solid-lines/errors) and HiTOT (dashed-lines/errors) in tank H13 in calibration run 5213 after accounting for the length of optical fiber to the tank.
         The ordinate represents the mean TOT value at a given laser light intensity and the abscissa represents the mean threshold crossing time relative to the laser trigger.
         Data points mark measurements from the calibration run whereas the lines represent fits to the data.
         The different colors mark the four PMTs within this tank. The curves for HiTOT in a given PMT are typically above
         the curves for LoTOT as the time required to rise to the level of the high threshold
         is longer than the time needed to reach the low threshold. This is not strictly true at small values of TOT because
         the HiTOT curves should be compared to the LoTOT curves at higher values of TOT given that LoTOT is the sum of T01, HiTOT and T23.
         The LoTOT curves are used to calculate timing either until the threshold of measurable prepulsing
         or until HiTOT start has a smaller RMS than the LoTOT start distribution. See Section \ref{sec:prepulse} for a full discussion of when
         these transitions occur.
        }
\label{slewingcurve}
\end{center}
\end{figure}

The start time of LoTOT and HiTOT relative to the time of the laser trigger is used to correct for the overall electronics
delay in the channel as a function of TOT. This delay depends on the value of TOT as larger pulses have faster rise times, causing them
to cross the fixed low and high threshold levels faster than smaller amplitude pulses. This effect is visible in Figure \ref{slewingcurve} and is known as slewing.

\newpage

\newpage

\newpage

\newpage

\section{Prepulsing}\label{sec:prepulse}

In some cases, a photon passes through the photocathode and interacts at the first dynode to
produce a photoelectron \cite{Lubsand:2000}. The photoelectron is then accelerated through the remaining dynode chain and yields a lower gain
signal compared to photoelectrons initiated at the photocathode because the amplification obtained from collision with the first dynode is lost. This effect is
undesirable because it produces a signal, known as a prepulse, that precedes the arrival of the main pulse in multi-photoelectron signals and artificially changes the calibrated pulse timing.
However, we will show in this section that the pre-pulsing effect is negligible for small amplitude waveforms and can be avoiding for large amplitude waveforms by using the start of HiTOT for timing rather
than the start of LoTOT.

One can estimate the time difference between the prepulse and main pulse by assuming a linear electric potential between the photocathode and first dynode
\begin{equation}
V(s) = V_0 \, s / L
\end{equation}
where $L$ is approximately half the PMT diameter, $V_0$ is the potential between the photocathode and first dynode, and $s$ measures
the distance to the photocathode. $V_0$ is 545 V in the HAWC PMT design (Appendix \ref{appendixbase}) resulting in $q V_0 \ll m_e c^2$ so we can apply non-relativistic mechanics
for an electron starting from rest at the photocathode to find the following expression for transit time to the first dynode
\begin{equation}
dt = \int_0^L \sqrt{ \frac{m_e}{2 q V(s)} } ds = \sqrt{ \frac{2 m_e}{q V_0} }
\end{equation}
where $q$ is the magnitude of the electron charge. This yields a time of 15 ns for the electron to travel from the photocathode to the first dynode, which is much larger than the light crossing time in both classes of PMTs used in HAWC. The typical timescale for prepulsing is therefore around 15 ns.

\begin{figure}[ht!]
\begin{center}
\subfigure[][]{\includegraphics[width=2.5in]{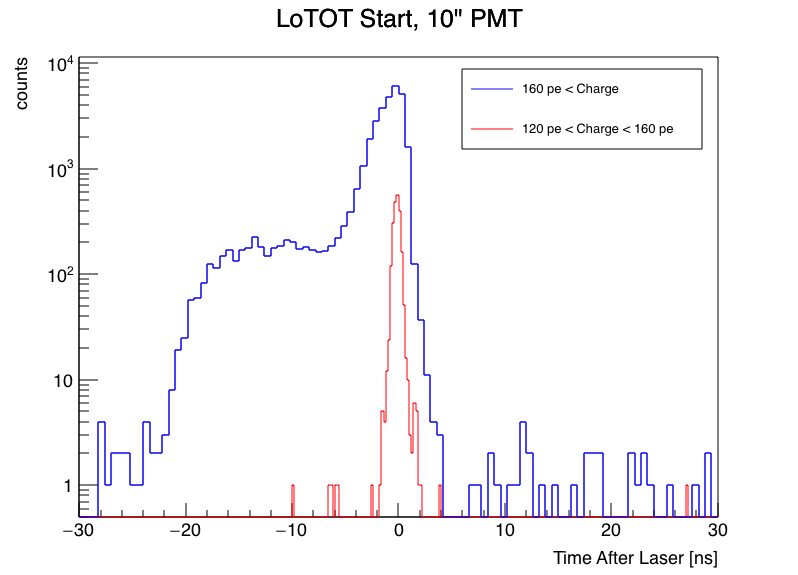}}
\quad\quad\quad\quad
\subfigure[][]{\includegraphics[width=2.5in]{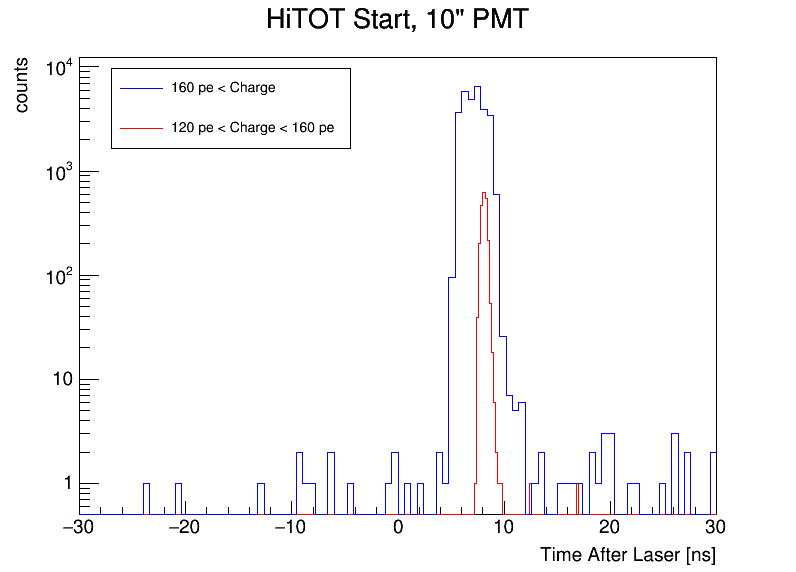}}
\caption{Start of (a) LoTOT and (b) HiTOT relative to the laser trigger time during calibration run 4505
         for a characteristic 10" PMT. The RED curve indicates 4 edge waveforms produced by laser light with calibrated charge values
         between 120 and 160 photoelectrons (pe).
         Its width distribution is narrow, indicating consistent crossing of both thresholds relative to the laser time.
         The BLUE curve indicates 4 edge wave waveforms produced by laser light with calibrated charge values $>$ 160 pe.
         It shows a large number of events with LoTOT starting at early times, but the peak of HiTOT start times
         is roughly consistent with the 4 edge waveform selection between 120 and 160 pe. This indicates the presence of measurable prepulsing
         effects in the 10" PMT population above 160 pe large enough to cross the low threshold at early times but do not cross the high
         threshold. The minimum extent of LoTOT start times for $>$160 pe signals is approximately consistent with the 15 ns time associated with electron travel to
         the first dynode. 
        }
\label{fig:preTENpulse}
\end{center}
\end{figure}

Prepulsing is a noticeable effect in calibration data for the 10" PMTs in HAWC. It can clearly be seen in the start time of LoTOT for 4 edge waveforms
produced by laser light with a calibrated charge level greater than 160 photoelectrons (Figure \ref{fig:preTENpulse}) which reveals a significant distribution of waveforms
arriving at early times compared to the main arrival time peak near 0 ns. They account for about 15\% of the total number of waveforms at this charge level. The minimum extent of this distribution is roughly consistent with the 15 ns expectation calculated from the electron transit time to the first dynode. Deviations from the expected value of 15 ns are explained by the non-linearity of the actual electric potential inside the PMT and a secondary form of prepulsing
that results from electrons generated at the photocathode whose initial trajectories cause them to miss the first dynode and travel directly to the second dynode \cite{Fleming:2002}. This secondary form of prepulsing occurs on a smaller timescale because the distance between the first and second dynodes is small than the distance between the photocathode and first dynode.

The start time of HiTOT relative to the laser trigger time for 4 edge waveforms with greater than 160 photoelectron in Figure \ref{fig:preTENpulse} is not effected by prepulsing because
the high threshold setting in HAWC corresponds to about 5 photoelectrons. This is much greater than the amplitude of typical prepulsing signals, which are smaller than the response to single photoelectrons initiated at the photocathode. The events at times prior to the main peak in Figure \ref{fig:preTENpulse} (b) are entirely consistent with the $\sim$40 kHz operating hit rate of the PMT. Note, however, that this peak occurs later than the start time of LoTOT because PMT pulses take longer to cross the high threshold compared to the low threshold. This rise time also explains the smaller width of HiTOT start distribution for the 120 to 160 photoelectron selection because the tighter range of pulse amplitudes yields a smaller selection of rise times. This dependence is accounted for during air shower reconstruction with the timing corrections discussed in Section \ref{calmeup}.

The 4 edge waveforms with calibrated light levels between 120 and 160 photoelectrons in Figure \ref{fig:preTENpulse} do not
show the prepulsing effects demonstrated in the $>$160 photoelectron sample. This dependance on the incident light level indicates
that prepulsing effects are only detectable above the low threshold when multiple prepulses are present. As a result, the start of LoTOT
will remain unaffected below 160 pe and can be used to provide good timing measurements. Above this value, HiTOT is used to determine the
timing of waveforms for 10" PMTs.

\begin{figure}[ht!]
\begin{center}
\subfigure[][]{\includegraphics[width=2.5in]{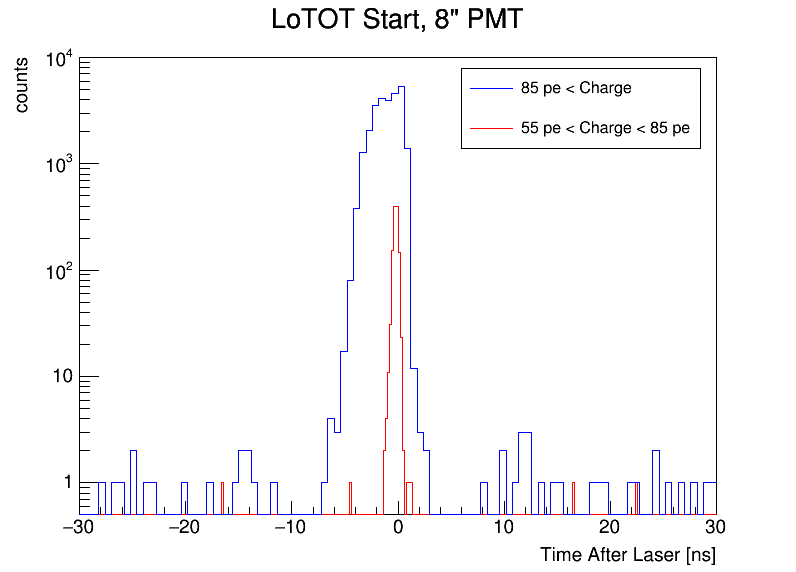}}
\quad\quad\quad\quad
\subfigure[][]{\includegraphics[width=2.5in]{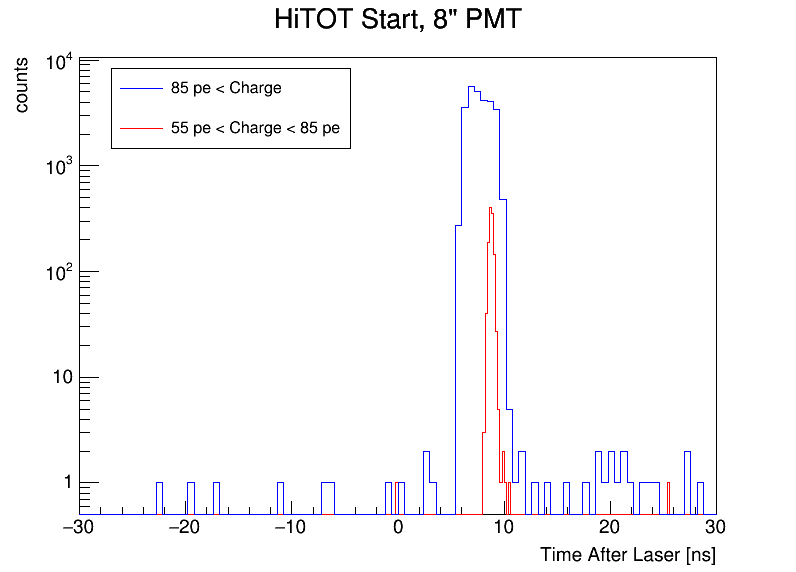}}
\caption{Start of (a) LoTOT and (b) HiTOT relative to the laser trigger time during calibration run 4505
         for a characteristic 8" PMT. The RED curve indicates 4 edge waveforms produced by laser light with calibrated charge values
         between 55 and 85 photoelectrons (pe).
         Its width distribution is narrow, indicating consistent crossing of both thresholds relative to the laser time.
         The BLUE curve indicates 4 edge wave waveforms produced by laser light with calibrated charge values $>$ 85 pe.
         Its width is larger than the RED curve in both LoTOT and HiTOT start because of the broader selection of rise times
         associated with the $>$ 85 pe cut, but there are no significant prepulsing effects.
        }
\label{fig:preEIGHTpulse}
\end{center}
\end{figure}

Applying the same type of analysis to 8" PMTs in HAWC at a 2x lower detected light level to account for differences
in the quantum efficiency reveals the 8" PMT population is much less susceptible to pre-pulsing
effects (Figure \ref{fig:preEIGHTpulse}). This agrees with initial testing of the HAWC PMTs \cite{Karn:2013}.
Our hypothesis is that the high quantum efficiency design in the 10" PMT involves a thinner photocathode coating, allowing
more photons to pass through the photocathode and interact with the first dynode, but this is speculation as the exact internal construction of the PMT
is not disclosed by the manufacturer. The absence of prepulsing effects means there is no explicit need to transition to HiTOT timing, however we choose to do so
above 85 pe on the basis that the width of the HiTOT start distribution is smaller compared to the width of the LoTOT start distribution in Figure \ref{fig:preEIGHTpulse}.
This minimizes the necessary timing correction discussed in Section \ref{calmeup}.

\newpage

\newpage

\newpage
\section{Afterpulsing}

While an ideal PMT would contain complete vacuum, real PMTs contain low quantities of the same molecules and atoms present in air.
This allows for electrons traversing the distance between the photocathode and first dynode to strike neutral atoms
and ionize them \cite{Ma:2009}. The electrons will continue to travel to the first dynode after this interaction and initiate an electronic signal while
the resulting ion subsequently drifts back to the photocathode where it can collide to liberate another electron. As with the original photoelectron, this electron
will accelerate towards the first dynode and initiate a second electronic signal, known as an afterpulse. Afterpulsing is especially prevalent in older populations of PMTs, like the 8" PMTs used in HAWC, as the vacuum inside a PMT slowly degrades over time.

\begin{figure}[ht!]
\begin{center}
\includegraphics[height=2.2in]{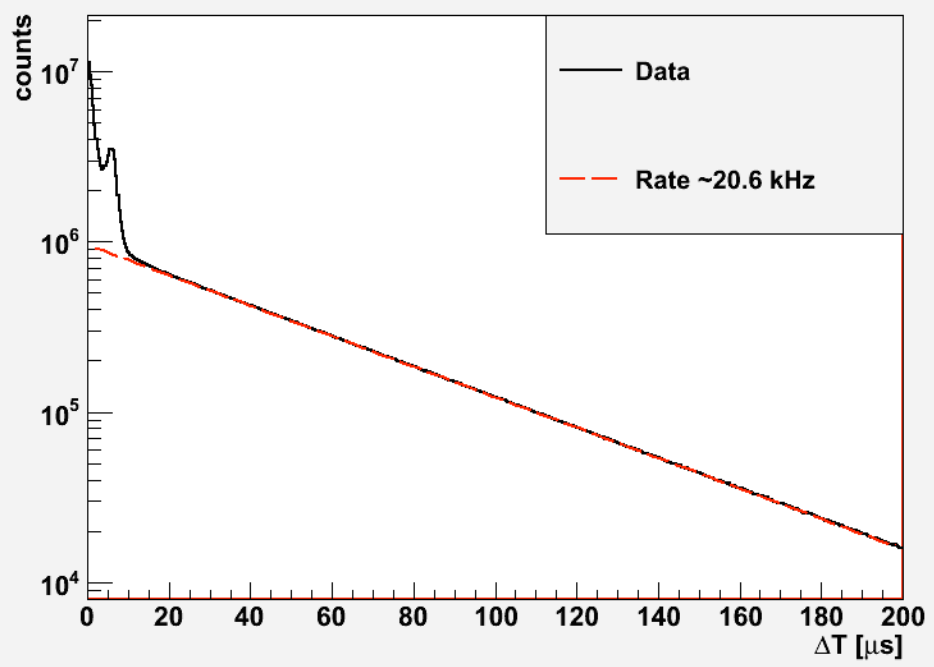}
\caption{Time between subsequent hits in the same 8" PMT channel.
         The distribution follows a pure exponential fit (red dashed line) at long timescales, which corresponds to a random
         rate of $\sim$20 kHz produced by real photon signals from air showers, single muons, and random noise. The deviations from the exponential fit to dT $<$ $\sim$10 $\mu$s represent a heightened hit
         rate produced from afterpulse events that are correlated to the prior hit. These events account for the remaining $\sim$9 kHz rate
         in the channel.
        }
\label{fig:rateofhits}
\end{center}
\end{figure}

Afterpulses can adversely effect the total charge and timing of calibrated pulses because their correlation to the original pulse yields a much higher noise rate immediately following real signals (Figure \ref{fig:rateofhits}) thereby increasing the chance for waveform overlap. As a result, veto windows are applied in each channel to flag waveforms falling inside the typical afterpulsing time ranges described in the remainder of this section. 
Flagged waveforms are excluded from both the air shower triggering and reconstruction algorithms. 

\newpage

\newpage

The time difference between the original signal and afterpulse can be calculated in a similar manner to the prepulse timescale from Section \ref{sec:prepulse}.
In this case, we now consider an ion with mass $m$ traversing the distance from the first dynode to the photocathode, $L$, assuming a linear
electric potential. This results in a transit time of
\begin{equation}
dt = \sqrt{ \frac{2 m_e}{q V_0} \frac{(L - s)}{L} }
\end{equation}
for ionization occurring a distance $s$ from the photocathode. The afterpulse timescale will therefore differ depending on the ion involved
as the transit time depends on the ion mass. Afterpulsing timescales are around 2 $\mu$s for He$^+$ and range from 5-8 $\mu$s for O$^+$, O$_2^+$, N$_2^+$, and CO$_2^+$
in a typical 10" PMT \cite{Ma:2009}.

\begin{figure}[ht!]
\begin{center}
\subfigure[][]{\includegraphics[width=2.5in]{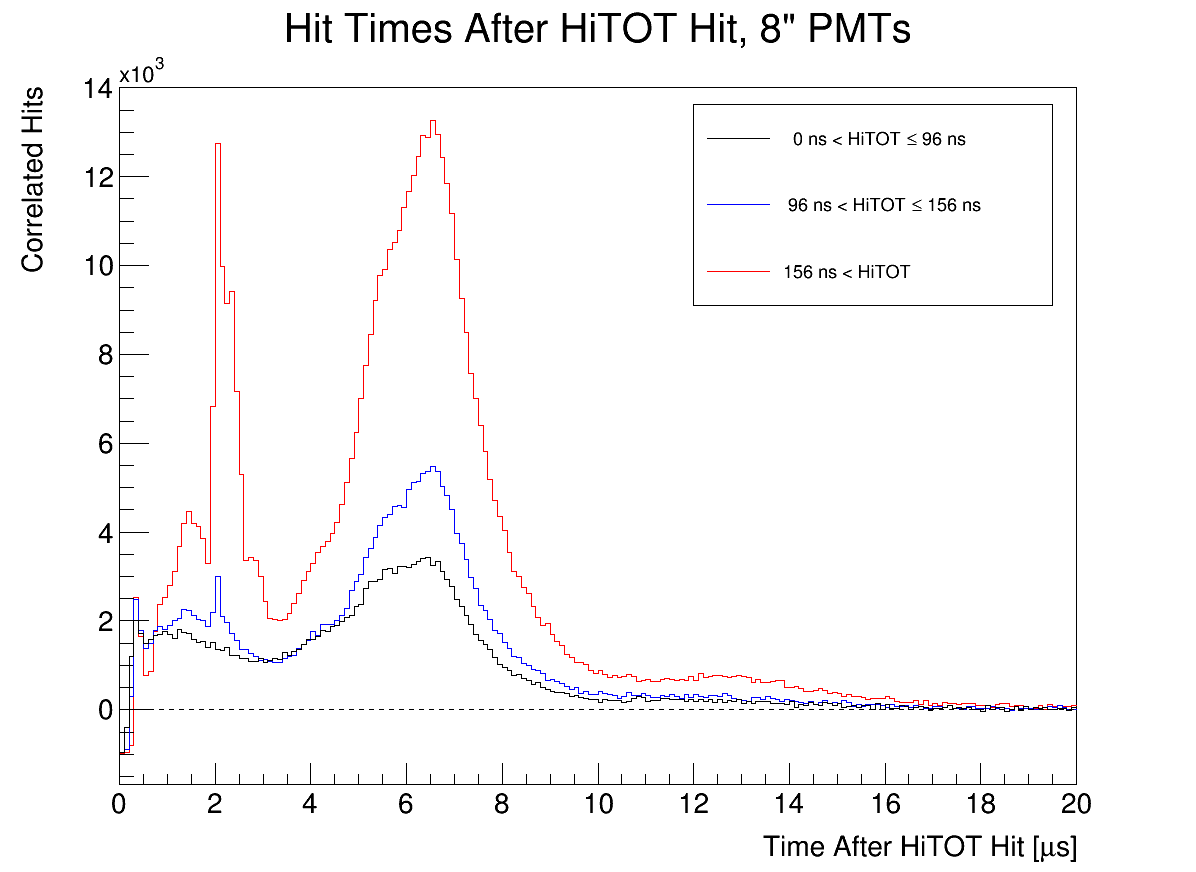}}
\quad\quad\quad\quad
\subfigure[][]{\includegraphics[width=2.5in]{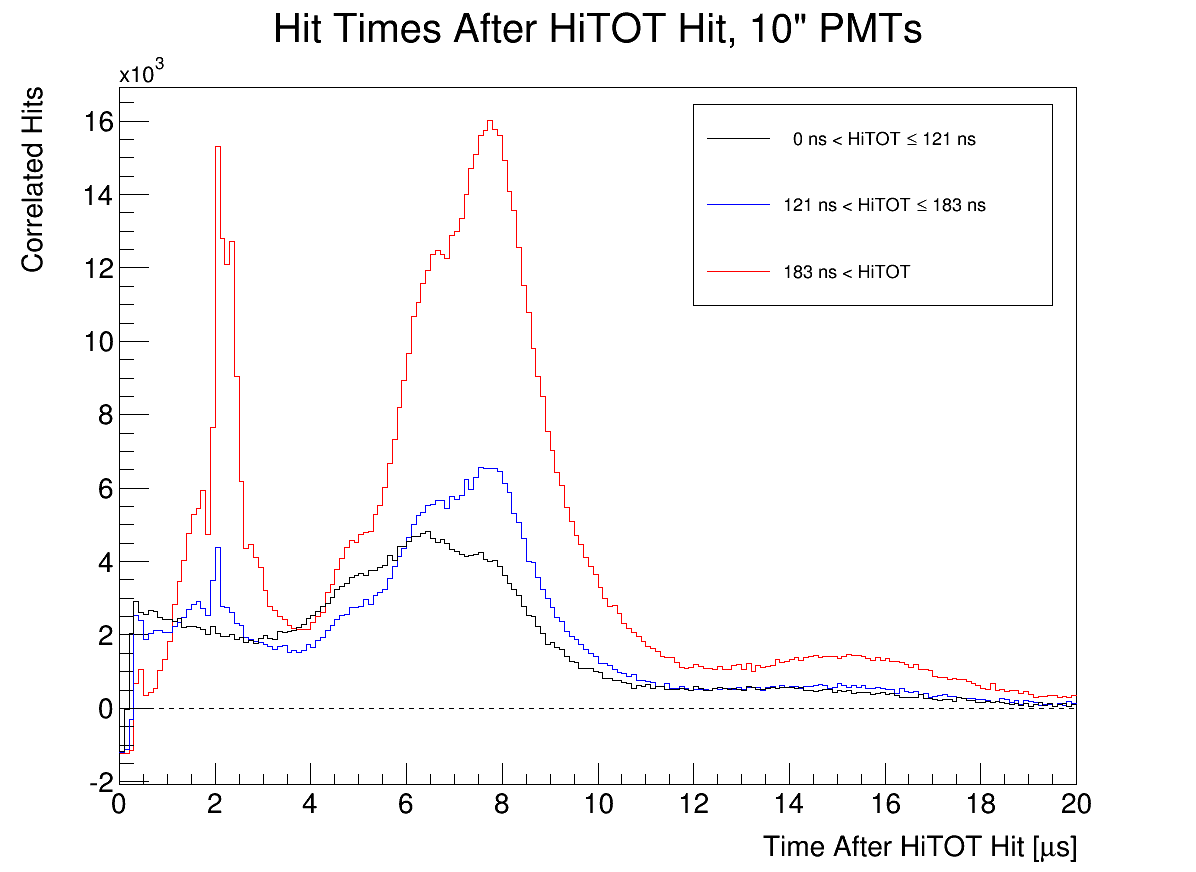}}
\caption{Correlated signal times following the measurement of a HiTOT waveform in the same channel for (a) 8" PMTs and (b) 10" PMTs in 125 seconds of TDC data.
         All times are relative to the start time of the original HiTOT event.
         This figure is produced by recording the times of all signals following within 20 $\mu$s of the HiTOT event and then subtracting
         the PMT rate at long timescales (30 - 50 $\mu$s) to remove non-correlated hits that follow the exponential rate in Figure \ref{fig:rateofhits}.
         The peaks near 2 $\mu$s and 5-8 $\mu$s represent afterpulsing populations. The small feature extending out to $\sim$15 ns corresponds to
         a population of secondary afterpulses initiated by the primary afterpulses that compose the peaks at 2 $\mu$s and 5-8 $\mu$s. The different color
         curves denote equal quantiles of the full HiTOT distribution, each containing 33\% of the total number of waveforms where HiTOT is present.
         The increasing amplitude of the afterpulse peaks with HiTOT results from the probability for observing an afterpulse from a single photoelectron signal
         compounding with every addition photoelectron present in the progenitor pulse, yielding a larger total probability for afterpulsing.
         }
\label{fig:aftthatpulse}
\end{center}
\end{figure}

Histogramming the arrival times of waveforms relative to the start time of a prior HiTOT waveform in the same channel confirms
that two dominant populations of afterpulses at 2 $\mu$s and 5-8 $\mu$s exist in both classes of HAWC PMTs (Figure \ref{fig:aftthatpulse}).
The HiTOT waveform trigger is used to define a consistent start time, as the threshold crossing effects described in Section \ref{calmeup}¬
are negligible compared to afterpulsing timescales, as well as to ensure low-level, single photoelectron electronics noise cannot contribute to the
triggering waveforms. The low-amplitude feature extending to $\sim$15 $\mu$s corresponds to a population of secondary afterpulses initiated by the
primary afterpulses that compose the peaks at 2 $\mu$s and 5-8 $\mu$s. The veto window applied after waveforms in each channel is 15 $\mu$s wide
in order to capture the majority of primary and secondary afterpulses.

\newpage

The 15 $\mu$s veto window is not applied after all waveforms because this would result
in a large dead time per channel given the $\sim$30 kHz and $\sim$52 kHz total rates for the 8" and 10" PMTs, respectively.
Instead, we apply it only after waveforms with HiTOT $>$ 200 ns which is the regime where afterpulsing effects are strongest (Figure \ref{fig:aftthatpulse}).
The value of 200 ns was chosen during the initial design phase of the experiment to yield a 1\% dead time for a 33\% reduction in the
number of afterpulses in 8" PMT channels. It currently gives a 4\% dead time for a 50\% reduction in the number of afterpulses in 10" PMT channels.
Further optimization studies are underway, but have yet to be completed.

\newpage

\newpage

\renewcommand{\thechapter}{4}

\chapter{Air Shower Reconstruction}\label{ch:Reco}

The DAQ system described in Chapter \ref{ch:hawcobs} connects to a server farm
located at the HAWC site which performs real-time ($\sim$4 second latency) triggering
and reconstruction of air shower events.
Air shower triggers are saved to portable disks and transferred to off-site server farms for
retroactive reconstruction as new calibrations and reconstruction algorithms become available.
Real-time reconstructed events use preliminary calibrations with a lower sensitivity
compared to off-site reconstructions but provide the ability to promptly
follow-up of external triggers, such as a satellite-detected GRB, as well as
disseminate internal triggers found at the HAWC site by the all-sky search method discussed in Chapter \ref{ch:method}.

Recent improvements in both the methodology of calibrations and reconstruction have resulted in an off-site
reconstruction known as the Pass 4 data set that contains a 2x increase in sensitivity compared to earlier
HAWC data \cite{Braun:2016}. This represents the most sensitive data set to date from any wide-field, ground-based
gamma-ray observatory and forms the input used to generate the results in Chapter \ref{ch:results} from
the search method described in Chapter \ref{ch:method}. Sections \ref{sec:edges} - \ref{sec:gammahadron} 
of this chapter discuss the algorithms used to perform air shower reconstruction and Section \ref{sec:performance} presents
their overall performance as verified with the Crab Nebula.


\newpage

\section{Edge Finding}\label{sec:edges}

\begin{figure}[ht!]
\begin{center}
\includegraphics[width=4in]{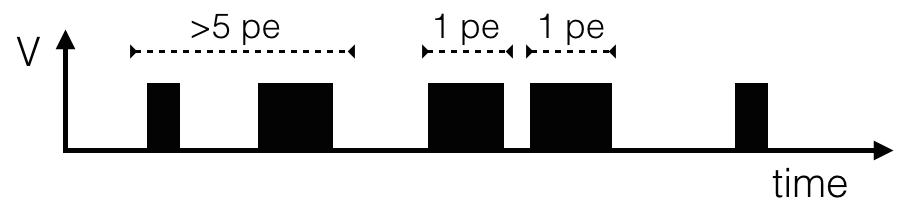}
\caption{Overview of single edge stream. Edge finding
         is applied to determine if pairs of pulses correspond
         to a single 4 edge waveform ($>$5 photoelectrons) 
         or pairs of 2 edge waveforms ($\sim$1 photoelectron).
         }
\label{fig:edgestream}
\end{center}
\end{figure}

The first algorithm applied during reconstruction
searches the continuous stream of data from each PMT
channel to identify the two and four edge waveforms discussed
in Chapter \ref{ch:hawcobs} (Figure \ref{fig:edgestream}).
This needs to be done to separate the pairs of square pulses
produced by four edge waveforms from pairs of distinct two edge waveforms
produced by lower charge PMT signals. It is essential to reconstruction
because the misindentification of two low charge waveforms as a 
single high charge waveform in a gamma-ray air shower
will cause the gamma-ray air shower to be erroneously identified
as a hadronic shower by the shower separation techniques 
discussed in Section \ref{sec:gammahadron}.

\vspace{5cm}

This algorithm works by assessing groups of two sequential square pulses. It determines
the presence of a four edge waveform based on the time width
of the leading square pulse. This pulse corresponds to T01 in
four edge waveforms and is, by design, much smaller than the minimum
allowed width for a two edge waveform (Chapter \ref{ch:hawcobs}).
The selection criterion for identifying a four edge waveform is then
given by T01 $<$ minimum two edge width, which is equal to 55 ns.

In addition to identifying four edge waveforms, the edge finding algorithm
provides quality checks to ensure that each waveform satisfies
the timing requirements imposed by the digital FEBs. These criteria are
shown in Table \ref{tab:edgecuts}. Waveforms failing these criteria are marked as bad and excluded from
the reconstruction.

\begin{figure}[ht!]
\begin{center}
\subfigure[][]{\includegraphics[width=2.5in]{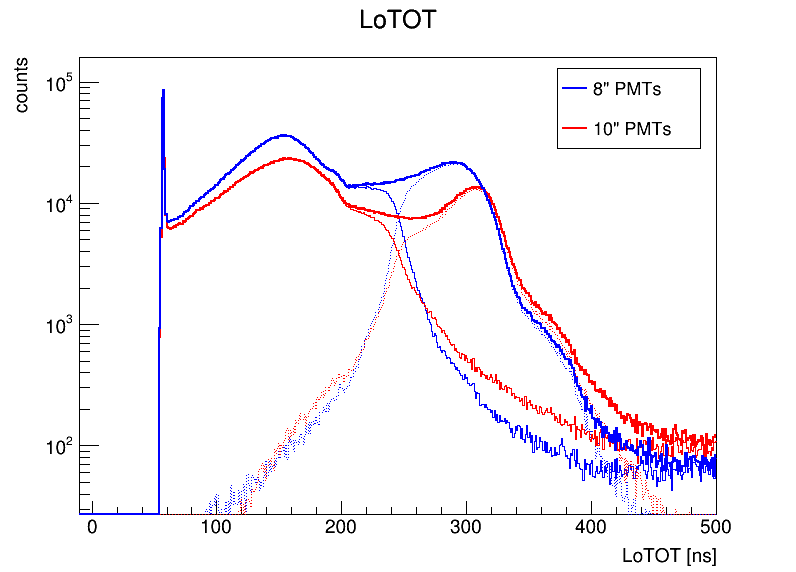}}
\quad\quad\quad\quad
\subfigure[][]{\includegraphics[width=2.5in]{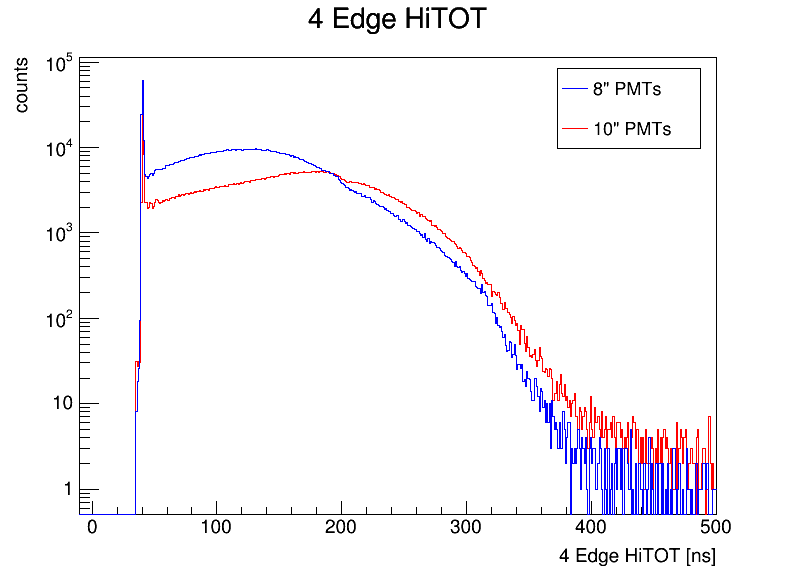}}
\subfigure[][]{\includegraphics[width=2.5in]{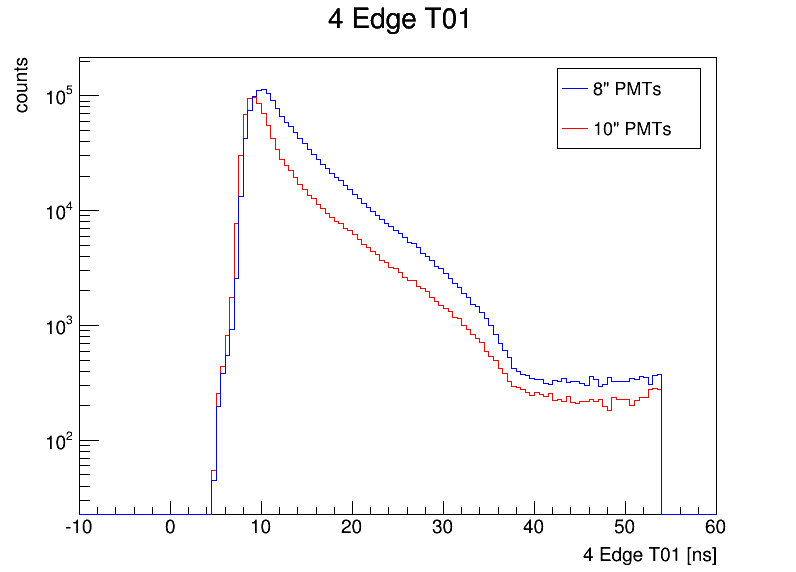}}
\quad\quad\quad\quad
\subfigure[][]{\includegraphics[width=2.5in]{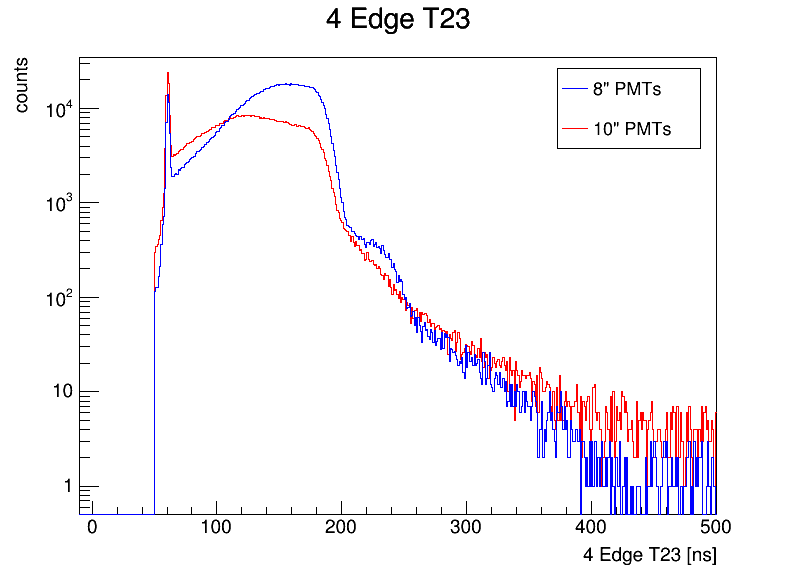}}
\caption{Measured T01, HiTOT, T23, and LoTOT for two and
         four edge waveforms identified by the edge finding algorithm
         in the raw TDC data stream. The data quality cuts in Table \ref{tab:edgecuts}
         are applied. The LoTOT distribution is the sum of LoTOT 
         from 2 edge waveforms at small values and the LoTOT from 4 edge
         waveforms at large values, marked by the dashed curves. A LoTOT
         of 150 ns approximately corresponds to single photoelectron signals.}
\label{fig:measured_tots}
\end{center}
\end{figure}

Figure \ref{fig:measured_tots} shows the timing parameters after quality selection
of two and four edge waveforms identified by the edge finder in the raw TDC data stream. The sharp
features that begin the LoTOT, HiTOT, and T23 plots represent the minimum
values enforced by the digital FEBs. They appear as sharp peaks because
the OR gate enforcing each minimum assigns TOT value at the peak
to the integral of PMT waveforms
that would appear to the left of the peak. The bump near a LoTOT of 150 ns
corresponds to single photoelectron signals, which make up the bulk of
the signals in the raw TDC data stream.

The distribution of T01 from four edge waveforms
continues out to the cut at $\sim$50 ns used to identify four edge waveforms.
This indicates that some four edge waveforms are misidentified as pairs of
two edge events. However, there are two reasons why this is not significant.
First, the distribution of four edge T01 about 36 ns represents only 1\% of the
total distribution, indicating we correctly identify the majority of four edge waveforms.
Second, long T01 correspond to PMT pulses at the threshold level of HiTOT.
This level is only $\sim$5 photoelectrons.
Breaking such a pulse into a pair of two edge events, which are typically around 1 photoelectron,
results in a error that is on the order of the fluctuations in the shower plane and is not significant
at the rate of 1\% of four edge hits.

\renewcommand{\arraystretch}{0.5}
\begin{table}[ht!]
\begin{center}
\begin{tabular}{|c|c|r|} \hline
  Waveform Type & Quality Selection [TDC Counts] \\ \hline
   2 edge & 540 $\ge$ LoTOT $<$ 5000 \\ \hline
          & T01 $<$ 540 \\
   4 edge & 350 $\ge$ HiTOT $<$ 5000 \\
          & 500 $<$ T23 \\ \hline
\end{tabular}
\end{center}
\caption{Quality selections applied to two and four edge waveforms during 
         edge finding in units of TDC Counts. These enforce the minimum
         timing parameters discussed in Chapter \ref{ch:hawcobs}. Maximum
         values represent the largest possible TOT producible by a PMT.
         The selection applied to T01 in 4 edge waveforms is used to identify them
         within the continuous TDC data stream.}
\label{tab:edgecuts}
\end{table}
\renewcommand{\arraystretch}{1.0}

\newpage

\newpage

\section{Core Fit}\label{sec:corefit}

\begin{figure}[ht!]
\begin{center}
\includegraphics[width=3.5in]{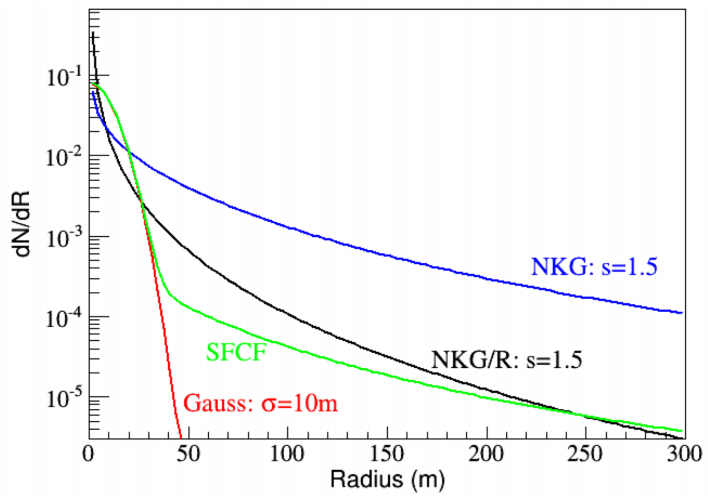}
\caption{Lateral distribution functions for the expected mean number of particles (NKG) and electromagnetic shower energy (NKG/R) in a gamma-ray air shower.
         as a function of radius to the shower axis measured in the shower plane.
         Both use a shower age parameter of 1.5.
         Also shown is the SFCF lateral distribution function used to successfully fit shower core positions in HAWC.
         It approximates the expected electromagnetic shower energy from the NKG/R distribution in the limit of large distances from the shower axis
         and matches a a two-dimensional Gaussian with a width of 10 m at small distances from the shower axis. Figure reproduced from \cite{Braun:2016}.}
\label{fig:coredist}
\end{center}
\end{figure}

As discussed in Chapter \ref{ch:showers}, the majority of air shower energy propagates
along the axis of the original primary particle despite the presence of interactions
which cause the shower to spread outwards in a disk perpendicular to this axis.
This is particularly true in the case of gamma-ray air showers where the multiple Coulomb
scattering of electrons in the air shower are less effective than hadronic interactions
in cosmic-ray showers at distributing momentum in the transverse direction and results in
the steeply shaped Nishimura-Kamata-Greisen (NKG) profiles \cite{Greisen:1960} for the mean lateral distribution of electromagnetic particles (NKG) and energy (NKG/R) 
shown in Figure \ref{fig:coredist}. 
The location of the shower axis at zero radius is referred to as the shower core and corresponds to the location of
maximal energy deposition. Measuring its location at ground level is essential in determining the expected
shower curvature and sampling corrections needed to accurately fit the shower timing plane and determine the original direction of a gamma-ray primary.

\newpage

\newpage

\begin{figure}[ht!]
\subfigure[][]{\includegraphics[height=1.8in]{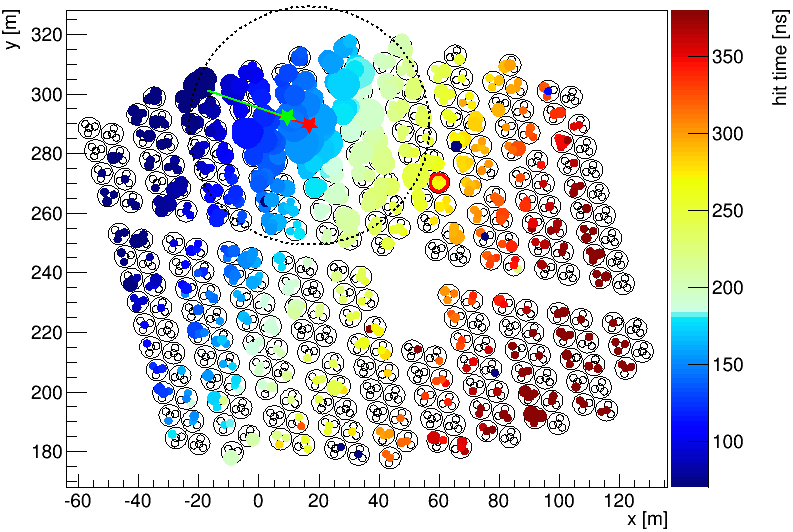}}
\quad\quad
\subfigure[][]{\includegraphics[height=1.8in]{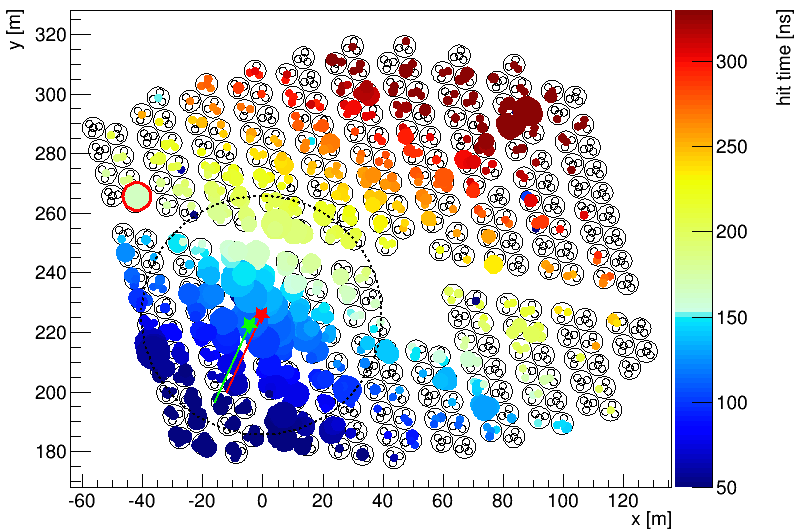}}
\caption{Diagram of PMT measurements for a (a) 47 TeV simulated gamma-ray shower and (b) 7 TeV simulated proton shower.
         Both register in 75\% of PMTs which are marked by colored circles. The size of each circle represents the
         total number of photoelectrons measured at a PMT, which is proportional to electromagnetic shower energy deposited in the tank.
         The largest charges appear near the true core location, marked in a GREEN star, where most of the shower energy arrives in the HAWC detector plane.
         The GREEN line pointing away from the core location denotes the shower axis of the simulated primary.
         Color indicates the start time of waveforms measured in each PMT. The RED line marks the reconstructed shower
         direction with a RED star marking the core position determined during reconstruction. The single circle
         outlined in RED represents the value of $Q_{max}(R>40\textrm{m})$ used for the compactness variable described
         in Section \ref{sec:compactness}. The dashed circle centered on the reconstructed core marks R=40 m.}
\label{fig:simshower}
\end{figure}

\begin{figure}[ht!]
\subfigure[][]{\includegraphics[height=1.8in]{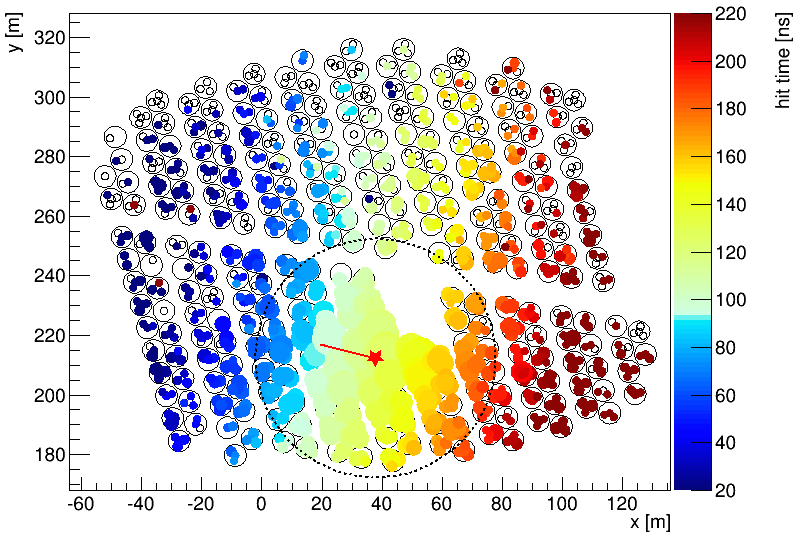}}
\quad\quad\quad\quad
\subfigure[][]{\includegraphics[height=1.8in]{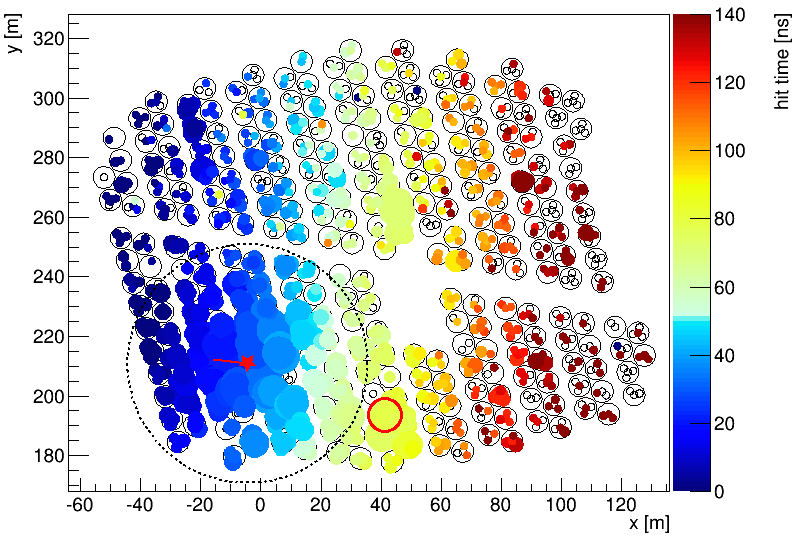}}
\caption{Diagram (a) gamma-ray like shower and (b) hadron-like shower in HAWC data.
         Both register in 75\% of PMTs which are marked by colored circles. The size of each circle represents the
         total number of photoelectrons measured at a PMT, which is proportional to electromagnetic shower energy deposited in the tank.
         The largest charges appear near the reconstructed location of the shower core, marked in a RED star, where most of the shower energy arrives in the HAWC detector plane.
         The RED line pointing away from the core location denotes the reconstructed direction of the original primary.
         Color indicates the start time of waveforms measured in each PMT. The single circle
         outlined in RED represents the value of $Q_{max}(R>40\textrm{m})$ used in the compactness variable described
         in Section \ref{sec:compactness}. The dashed circle centered on the reconstructed core marks R=40 m.}
\label{fig:realshower}
\end{figure}

Individual PMT measurements in HAWC (Figures \ref{fig:simshower} and \ref{fig:realshower}) record the total electromagnetic shower energy at ground level
in the form of photoelectron charge and fall according to the curve for NKG/R as a function of radial distance from the shower axis. 
The exact amount of charge at any given radius $\vec{x}$ from the shower core is
distributed about the mean expected charge $Q(\vec{x})$ of the NKG/R form due to the underlying Poisson distribution
of photoelectrons produced in the water of WCDs by fluctuations in shower development and the finite charge resolution of the PMTs.
The likelihood of observing a set of $N$ charge measurements $Z_i$ from the mean expected charges $Q(\vec{x})$ determined by the
shower direction and core location in the HAWC plane is
\begin{equation}
-2 log \mathcal{L} = \sum_{i=1}^{N} \frac{ \big(Z_i - Q(\vec{x})_i \big) } { Q(\vec{x})_i + \sigma_i^2 }
\label{eq:corelikely}
\end{equation}
This quantity is maximized for all PMTs, including null measurements, in each air shower trigger to determine the location of the shower
core prior to fitting the timing profile of the shower plane.

Two approximations are applied during the maximization process to greatly reduce the computational
load associated with maximizing Equation \ref{eq:corelikely}. First, the shower is assumed to be vertical
and, second, the NKG/R shape is approximated with a Gaussian core that smoothly transitions to the 1/R$^3$
behavior of the NKG/R function at large radii given by
\begin{equation}
S_i = S(A, \vec{x}, \vec{x_i}) = A \Big( \frac{1}{2 \pi \sigma^2} e^{-\frac{|\vec{x_i} - \vec{x}|^2}{2 \sigma^2}} + \frac{N}{(0.5 + \frac{|\vec{x_i} - \vec{x}|}{R_m})^3} \Big)
\end{equation}
This form yields a median core resolution of $\sim$5 meters for shower cores landing inside th HAWC array. This is equivalent to what can be attained from
using the full form of the NKG/R function and is 10x faster \cite{Braun:2016}.


\vspace{2cm}

\newpage

\newpage

\section{Angle Fit}\label{sec:anglefit}

\begin{figure}[ht!]
\begin{center}
\includegraphics[width=3.5in]{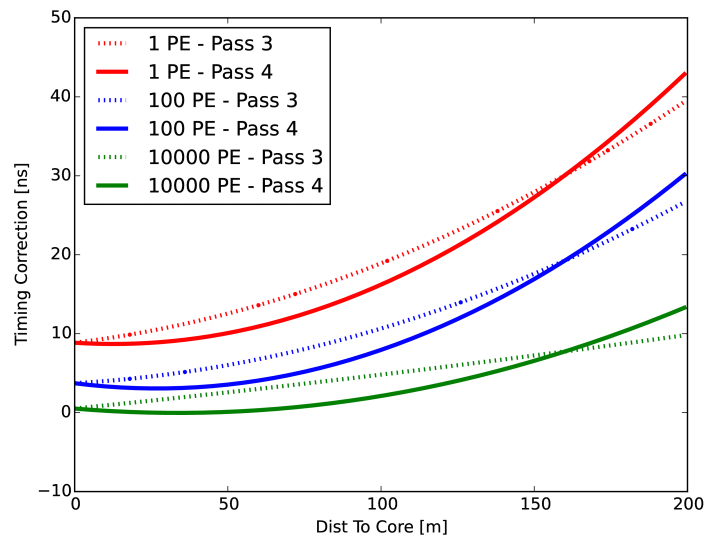}
\caption{Shower front timing correction applied during angle fitting for three different charge levels measured in units
         of the mean photoelectron charge (pe).
         Solid lines indicate the correction applied in the Pass 4 data set, which were determined from
         a pure sample of gamma-rays coming from the Crab Nebula that register in $>$75\% of PMTs.
         These corrections yield a $\sim$2x improvement in the angular resolution produced by the timing corrections
         applied prior to Pass 4 (dashed-lines) which were broadly based on the timing corrections used in the Milagro experiment \cite{Braun:2016}.
         The sampling effect described in Chapter \ref{ch:showers} causes the timing correction to be smallest at the highest charge level.}
\label{fig:curvecorrdata}
\end{center}
\end{figure}

Once the core location is determined, air shower reconstruction
proceeds with a fit to the direction of the incident particle. 
This is done by fitting the start time of all PMT waveforms to the expectation of a flat timing plane corrected
for the shower curvature and sampling effects described in Chapter \ref{ch:showers}
as a function of total measured charge and radius to the shower core. The timing correction is shown
for three different charge levels as a function of distance to the shower core in Figure \ref{fig:curvecorrdata}.
It is determined from a pure simple of reconstructed gamma-ray showers coming from a 0.25$^\circ$ region centered on the Crab Nebula.
The pure sample is selected by applying strict compactness (See Section \ref{sec:compactness}) and PINCness (See Section \ref{sec:pincness}) cuts
for showers that register in $>$75\% of PMTs.

\begin{figure}[ht!]
\begin{center}
\includegraphics[width=3.5in]{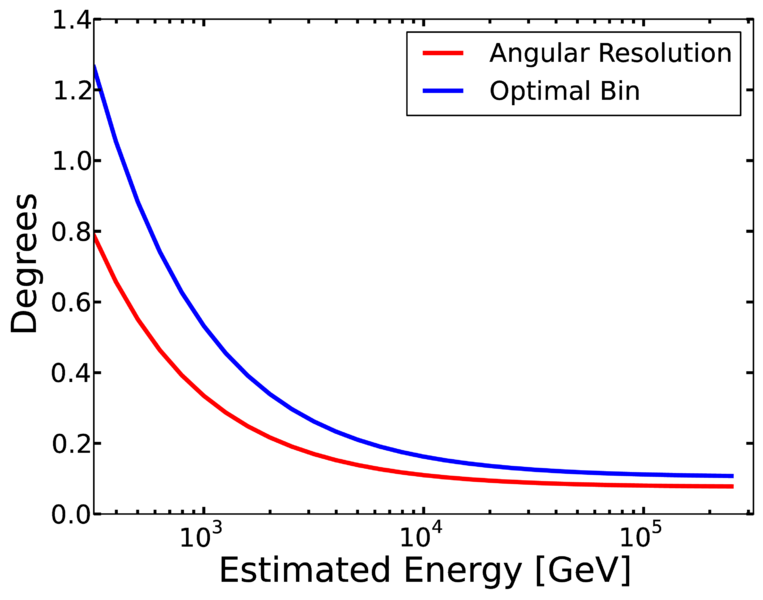}
\caption{Angular resolution of the HAWC Observatory as a function of energy.
         The angular resolution (RED) is the standard deviation of a 2D Gaussian fit to simulated air showers
         and matches the angular resolution measured with gamma-rays from the Crab Nebula using Pass 4 reconstruction.
         The optimal bin size (BLUE) corresponds to 70\% containment of gamma-rays from a point source and is used in standard point source analyses within HAWC. }
\label{fig:angleres}
\end{center}
\end{figure}

The timing correction is less than 0.15 nanoseconds per meter of distance to the shower core
but plays a large role in the overall angular resolution of the experiment shown in Figure \ref{fig:angleres}. The current correction
accounts for a $\sim$2x improvement in the angular resolution produced by the timing corrections applied
prior to Pass 4 that were based on early simulations of the HAWC detector rather than gamma-ray air shower data \cite{Braun:2016}.
This allows for a 2x smaller optimal spatial bin size in point source analyses, reducing the cosmic ray background,
which scales as bin area, by a factor of 4. It is the main reason the Pass 4 reconstruction is the most sensitive HAWC reconstruction to date.

\newpage

\newpage

\newpage

\section{Gamma/Hadron Separation}\label{sec:gammahadron}

While the angular resolution improvements described in the previous section help reduce
the hadronic air shower backgrounds for gamma-ray point source analyses, a good angular
resolution alone is not enough to provide good sensitivity to typical gamma-ray point
sources given the overwhelming rate of hadronic air shower events. Further criteria, referred to as gamma-hadron separation cuts, are
needed to distinguish between the different types of air showers. In HAWC these criteria are quantified in the form
of two variables, compactness and PINCness, which are described below.

Both of these parameters operate on the principle that hadronic showers, as described in Chapter \ref{ch:showers},
 contain interactions that are much more efficient compared to multiple Coulomb scattering in gamma-ray showers at carrying large amounts of energy far from the shower core via sub-showers
that lead to large asymmetries in the lateral energy distribution of the shower disk.
They also take advantage of the fact that hadronic showers support the generation of energetic muons which spread widely from the shower axis and travel to ground level with enough
energy to penetrate through HAWC tanks and create large, asymmetric signals when passing close to the location
of an individual PMT at the bottom of the tank. These features manifest themselves in the different shower types in Figures \ref{fig:simshower} and \ref{fig:realshower}
as large asymmetries in the total number of photoelectrons seen at PMTs located far from the shower core in hadronic showers which are not present in the relatively smooth and quickly
decaying distrubition of PMT signals found in gamma-ray air showers.

\newpage

\subsection{Compactness}\label{sec:compactness}

\begin{figure}[ht!]
\begin{center}
\includegraphics[width=4in]{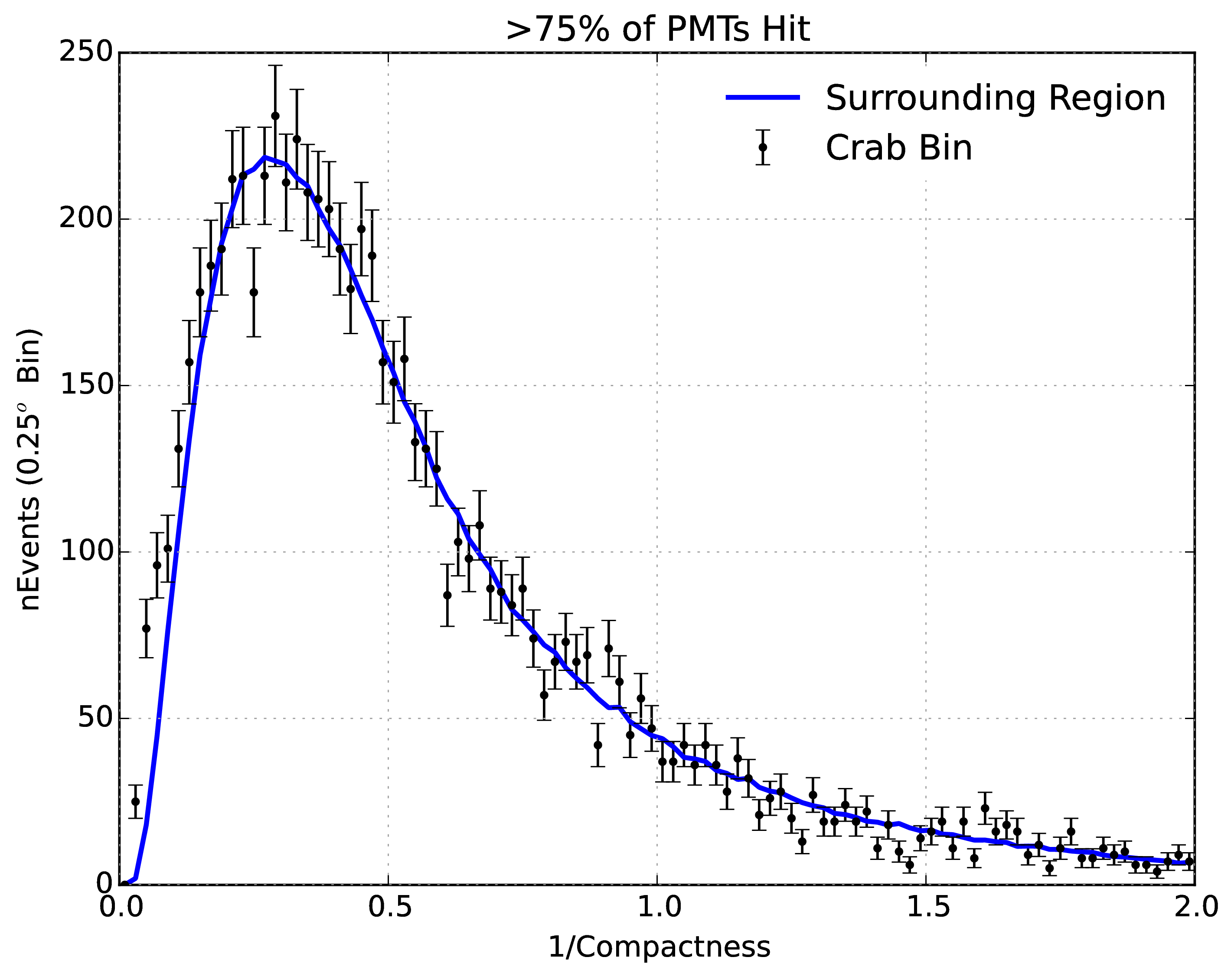}
\caption{Inverse compactness (1/$\mathcal{C}$) distribution for air shower events registering in $>$75\%
         of the HAWC PMTs. The BLUE curve represents a selection of cosmic-ray showers
         arriving in a 1$^\circ$-3$^\circ$ annulus surrounding the location of the Crab Nebula
         where there are no known high energy photon sources, only background events.
         The data points represent a selection
         of showers arriving within 0.25$^\circ$ degrees of the Crab Nebula. The bulk of this
         distribution is comprised of cosmic-ray air showers that match the BLUE curve. The deviation
         of data points above the BLUE curve between 1/$\mathcal{C}$ values near zero represent the population
         of high energy gamma-ray photons coming from the Crab Nebula which exhibit large values of compactness
         compared to background showers of the same size. }
\label{fig:invcompactness}
\end{center}
\end{figure}

Compactness is a gamma-hadron separation variable that describes the largest local deposition of energy far from the
shower axis relative to overall shower size. It is formulated according to
\begin{equation}
\mathcal{C} = \frac{N_{20}}{Q_{max}(R>40\textrm{m})}
\end{equation}
where $N_{20}$ is the number of PMTs signals measured within 20 ns of the reconstructed shower front
and $Q_{max}(R>40\textrm{m})$ is the maximum single PMT amplitude measured outside a distance of 40 meters from the reconstructed core location.
It is typically small for hadronic showers as muons and off-axis sub-cascades will generate large $Q_{max}(R>40\textrm{m})$.
A gamma-ray shower of the same footprint in HAWC will tend to have a larger value of compactness due to the sharp lateral distribution
describing gamma-ray air shower energy, as shown in Section \ref{sec:corefit}, which yields small values of $Q_{max}(R>40\textrm{m})$.
This is shown in Figure \ref{fig:invcompactness} for large showers reconstructing within 0.25$^\circ$ of the Crab Nebula.


The ratio of $Q_{max}(R>40\textrm{m})$ and $N_{20}$ is taken because the shower energy measured by the
maximum detected charge loosely scales with the shower size, allowing a single compactness value to
effectively discriminate between gamma-ray and hadronic air showers over a range of shower footprints.
This allows our analysis described in Chapters \ref{ch:method} and \ref{ch:sens} to obtain appreciable
sensitivity using a single compactness cut over the range of energies expected from typical GRB signals.
Overall, the compactness cut we apply reduces the background rate in simulations in of the HAWC detector 
by a factor of 10 while retaining $\sim75$\% of simulated gamma-ray air showers arriving within the optimal
bin size used in our analysis.


\vspace{2cm}

\newpage

\subsection{PINCness}\label{sec:pincness}

\begin{figure}[ht!]
\begin{center}
\includegraphics[width=3.5in]{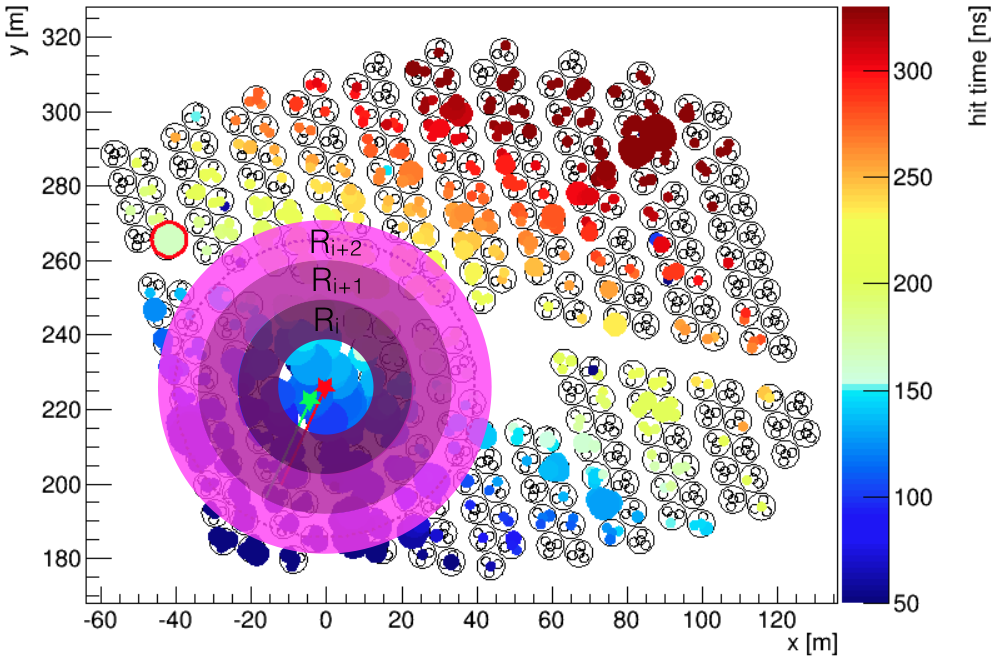}
\caption{Diagram showing calculation of PINCness variable in a simulated 47 TeV proton shower
         which consists of averaging PMT charge measurements within 5 m annuli ($R_i,R_{i+1},R_{i+2}$)
         centered on the reconstructed core location. Annuli are not drawn to scale.
         This method differs from the compactness parameter in Section \ref{sec:compactness} in that
         it only tests rotational symmetry about the shower axis, not differences in the radial distribution
         between gamma-ray and hadronic showers. The PMT measurement outlined in a RED circle at x = -20 m, y = 265 m
         marks for comparison the $Q_{max}(R>40\textrm{m})$ used in the compactness calculation of this shower.}
\label{fig:pincnessdiag}
\end{center}
\end{figure}

PINCness is a gamma-hadron separation variable that describes rotational asymmetry in the distribution of shower energy
about the shower axis. It is calculated from the reduced $\chi^2$ of all PMT measurements averaged in 5 meter annuli
according to
\begin{equation}
\textrm{PINCness} = \frac{1}{N} \sum_{i=0}^{N_{R}} \Bigg( \sum_{n=0}^{N_i} \frac{(Q_n - Q_i)^2}{\sigma_i^2} \Bigg)
\end{equation}
where $i$ denotes the 5 meter annulus with radius $R_i$ measured from the reconstructed core location as shown in Figure \ref{fig:pincnessdiag}. 
$N$ is the total number of PMTs used in the reconstruction and $N_R$ is the number annuli needed to contain the PMT positioned
furthest from the reconstructed core. The remaining parameters all pertain to measurements within the $i^{th}$ annulus where
$N_i$ is the number of contained PMTs, $Q_n$ is an individual PMT charge measurement, $Q_i$ is the average charge, and $\sigma_i$ is the uncertainty associated with $Q_i$.
As in the case of the timing corrections from Section \ref{sec:anglefit}, $\sigma_i$ is determined as a function of $Q_i$ directly from large gamma-ray showers reconstructing 
within 0.25$^\circ$ of the Crab Nebula.

\begin{figure}[ht!]
\begin{center}
\includegraphics[width=3.5in]{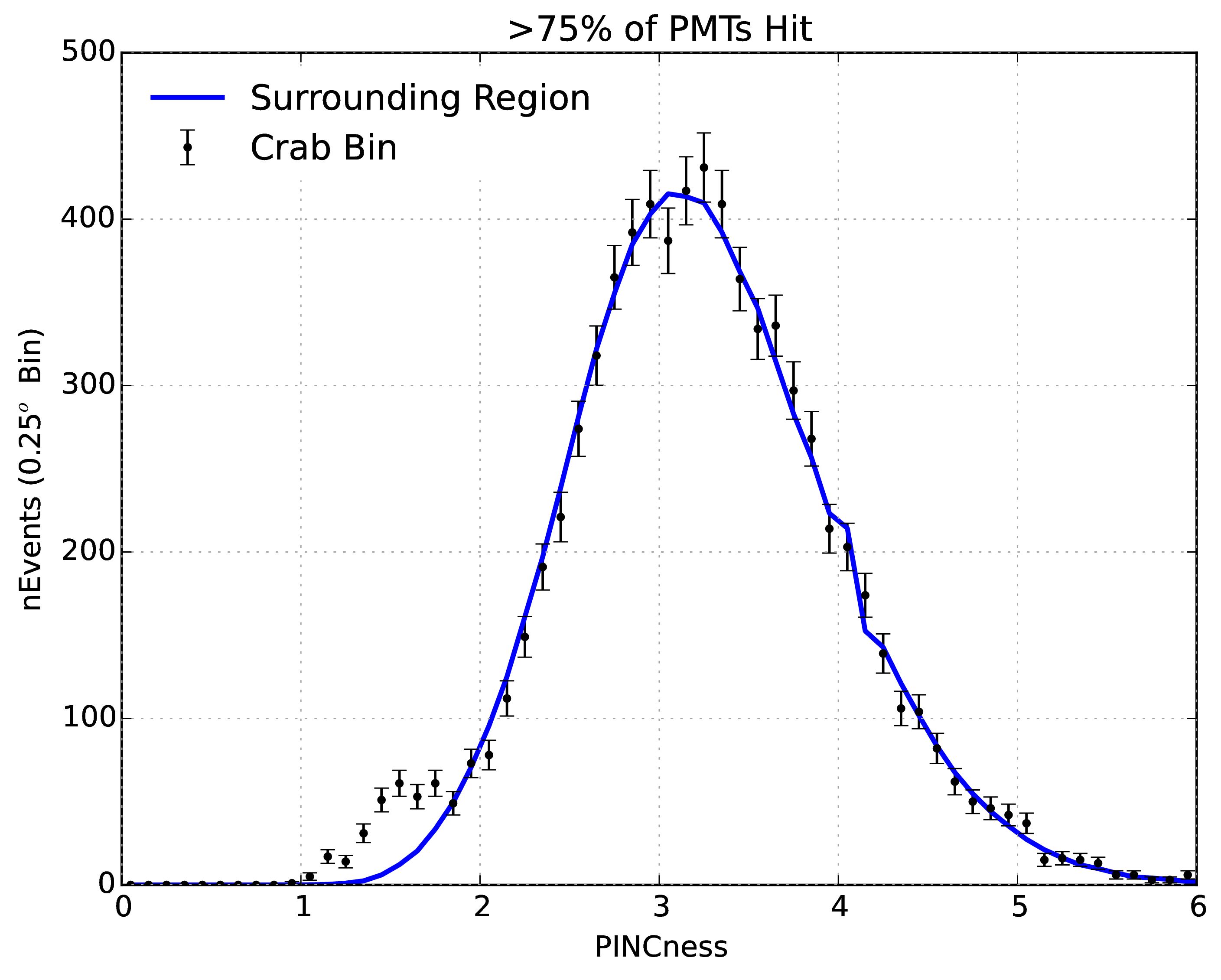}
\caption{PINCness distribution for events containing measurements in $>$75\%
         of the HAWC PMTs. The blue curve represents a selection of cosmic-ray showers
         arriving in a 1$^\circ$-3$^\circ$ annulus surrounding the location of the Crab Nebula
         where there are no known high energy photon sources. The data points represent a selection
         of showers arriving within 0.25$^\circ$ degrees of the Crab Nebula. The bulk of this
         distribution is comprised of cosmic-ray air showers that match the blue curve. The deviation
         of data points above the blue curve between PINCness values of 1-2 represent the population
         of high energy gamma-ray photons coming from the Crab Nebula. }
\label{fig:pincnessmoneyplot}
\end{center}
\end{figure}

The asymmetries present in the spatial distribution of shower energy
in hadronic showers means they will exhibit larger values of PINCness compared
to gamma-ray air showers with the same footprint in HAWC. This is clearly shown for
large showers arriving within 0.25$^\circ$ of the Crab Nebula in Figure \ref{fig:pincnessmoneyplot}.
However, one failing of the PINCness variable is that it requires showers to register
in $>10$\% of PMTs in order to accurately calculate $Q_i$. This means it has no discriminating
power in our analysis as the expected shower size for $\sim$100 GeV showers arriving from GRBs is around 5\%
the size of the detector. However, a PINCness selection is used in the point-source analysis of the Crab Nebula in Section \ref{sec:performance}.

\vspace{3cm}

\newpage

\newpage

\section{Crab Performance}\label{sec:performance}

The Crab Nebula is both the oldest detected TeV gamma-ray source \cite{Weekes:1989} as well as the brightest steady-state source in the TeV gamma-ray sky.
It therefore acts as the standard candle for verifying the performance of all ground-based gamma-ray telescopes.
The performance of the Pass 4 reconstruction algorithms was verified
by analyzing the gamma-ray signal from the Crab Nebula in a 211 day data set
beginning in November 2014 and ending in December 2015 using the standard
likelihood method for point-source analysis developed in HAWC \cite{Younk:2015}.
This corresponds to an average daily detection of 5.5$\sigma$ and broadly agrees with the design
sensitivity of the HAWC experiment \cite{Abeysekara:2013tza}. 

\begin{figure}[ht!]
\begin{center}
\includegraphics[width=3.5in]{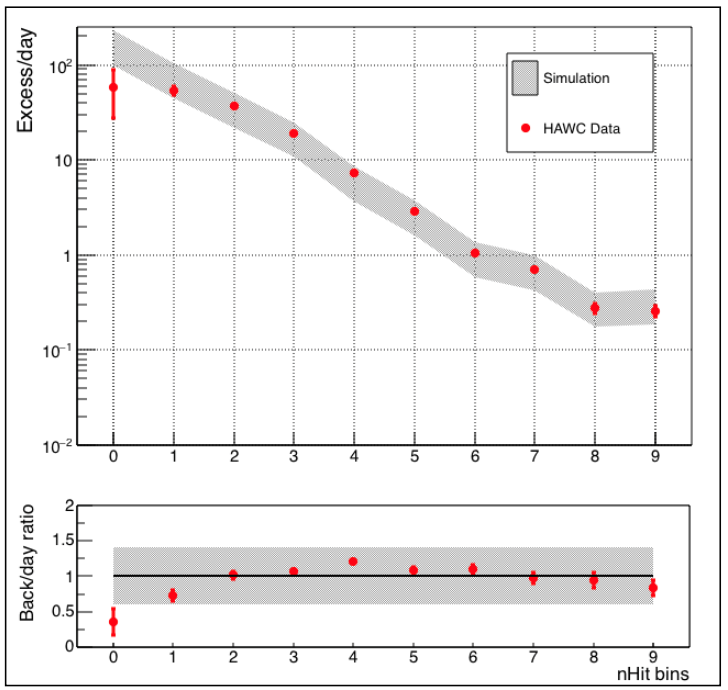}
\caption{Comparison of gamma-ray excess for the Crab Nebula in HAWC data (RED) and 
         Monte Carlo simulations of the HAWC experiment (GREY) as a function of
         point-source analysis bin. Bins 0-1 correspond to showers that trigger $\sim$5\%
         of the detector with shower sizes increases with number until reaching Bin 10 where showers saturate
         the full detector.
         The width of the simulated
         excess results from systematic studies performed by varying detector parameters.
         The lower panel shows the ratio of observed background counts compared to simulation.
         In both cases the data agree well for large nHit bins but deviate in
         Bin 0 which is where typical GRB photons should arrive.}
\label{fig:dmc}
\end{center}
\end{figure}

Figure \ref{fig:dmc} presents a comparison of the gamma-ray excess measured at the location
of the Crab Pulsar to the expectation obtained from Monte Carlo simulations of the HAWC detector
as a function of the 10 analysis bins used in the point-source likelihood analysis. Each bin represents
a selection of increasing shower sizes ranging from $\sim$5\% the size of the detector footprint for
Bin 0 up to showers that saturate the entire detector in Bin 10. These bins are used to define the angular resolution
and gamma-hadron cuts for similarly sized showers over the full sensitive energy range of the HAWC detector.
The only bins relevant to detection of $\sim$100 GeV photons from GRB sources are Bins 0 and 1. Unfortunately,
this is where the largest discrepancy exists between data and Monte Carlo. We introduce a scaling of the
simulated photon signal to account for this systematic during the optimization of our analysis in Chapter \ref{ch:sens}.

\begin{figure}[ht!]
\begin{center}
\includegraphics[width=3.5in]{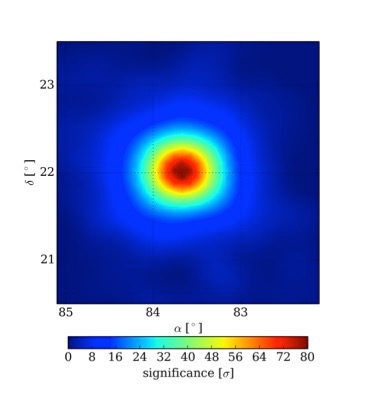}
\caption{Point source analysis of the Crab Nebula for a live time of 211 days beginning
         in November 2014 and ending in December 2015. The significance at the location of the Crab Pulsar is 80$\sigma$
         which corresponds to an average daily detection of 5.5$\sigma$. Early data utilize a a 250 tank
         configuration. The full detector came online in March, 2015.}
\end{center}
\end{figure}

\renewcommand{\thechapter}{5}

\chapter{Search Method for Short-Timescale VHE Transients}\label{ch:method}

The goal of this section is to describe in detail an overview of how we perform
our all-sky search for short-timescale VHE transients in HAWC air shower data.
The inspiration for this search is our desire to leverage the full capability
of the HAWC observatory's wide-field, continuous monitoring of the TeV sky
to discover GRB transients that occur at any time within the field-of-view,
not just during the $\sim$50\% of the time when satellites capable of providing GRB triggers are overhead. As we will
show in Section \ref{sec:senswtrials}, this allows us to have appreciable sensitivity
to detecting a GRB transient even after correcting for the trials taken to search the full overhead sky.

Our GRB search algorithm examines the $\sim$24 kHz rate of reconstructed air shower events passing
through the overhead sky in HAWC using a fixed-width sliding time window.
Inside each position of the time window, all points within 50 degrees of detector zenith are tested against the
hypothesis that the local air shower count comes from the $\sim$500 Hz rate of cosmic-ray air showers
remaining after applying gamma-hadron separation cuts. We interpret significant upward fluctuations from the
expected number of background counts as candidates for detected GRB emission.

A fixed-width window is chosen rather than attempting to fit a light curve profile in order
to maintain the computational efficiency necessary to shift the time window continuously
through the full HAWC dataset, allowing a full search of the HAWC field-of-view for all times.
Square bins are used in the spatial search for efficiency reasons as well. Overall, the search method
is able to process data at $>$2x real-time on a single CPU for timescales down to 0.01 seconds. 

The benefits of this analysis are that it eliminates the need for an overhead satellite to provide
the location and time of a GRB event, thereby increasing the search exposure compared to a externally triggered search,
and that it provides us with the ability to generate alerts to trigger other experiments for follow-ups of GRBs missed by 
the current generation of satellites.
This comes at a cost of reduced sensitivity as the trials associated with searching the field-of-view for all time requires
a higher false positive threshold. However, we will show in Section \ref{sec:senswtrials} that the sensitivity loss is only
a factor of $\sim$2 compared to the single trial case and results in roughly the same expectation for
the discovered bursts as using a triggered search. 

The following sections in this chapter describe the implementation of the search method, background calculation, and
trials correction for the three search timescales, 0.2 seconds, 1 second, and 10 seconds, used in our analysis.
Chapter \ref{ch:sens} follows with a full description of the optimization of the spatial bin size, time window
duration, and post-trials sensitivity based on Monte Carlo simulations of the HAWC detector.

\newpage

\section{Spatial Search}

The spatial search is performed in a rectangular grid of right ascension and declination using
locally smoothed 2.1$^\circ$ x 2.1$^\circ$ square bins optimized for both short and long GRB
models (See Chapter \ref{ch:sens}). This means the width of the spatial bin as measured in right ascension scales
with declination according to 2.1$^\circ$/cos($dec$) in order to account for the smaller line elements
described by right ascension on the surface of the unit sphere when not in the plane of 0$^\circ$ in declination.
The spatial bin height in declination is a constant 2.1$^\circ$ as line elements measured in declination
remain constant over the sphere.
The grid is divided using steps of 0.11$^\circ$ in right ascension and
declination to yield a total of 19 steps along each side of the square bin at a declination of 0$^\circ$. This results in $\ge$90\%
overlap between any two adjacent search bins, allowing for fine tuning on the spatial position
of air shower excesses.

\begin{figure} [ht!]
  \centering
  \includegraphics[width=4in]{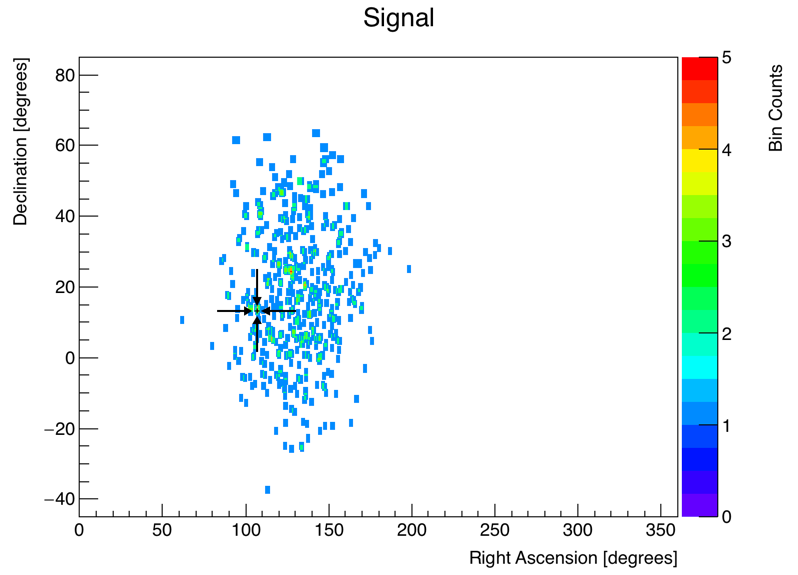}
  \caption
  {
    Recorded air shower counts in the spatial search grid for one position of a 1 second long sliding window.
    There are a total of 443 air shower events in this map.
    Detector zenith is located in the center of the count distribution at a declination of $\sim$19$^\circ$
    and a right ascension of $\sim$280$^\circ$. The low event rate far from zenith results from the attenuation
    in the larger atmospheric depth of off-axis showers. This yields in zero air shower counts at most
    points. The square shapes appearing for locations far from zenith with only a single air shower count are the 2.1$^\circ$ x 2.1$^\circ$ local smoothing applied
    to each air shower in our analysis. The cross marks the location of the most significant result found from searching this map.
  }
  \label{fig:sigsky}
\end{figure}

Figure \ref{fig:sigsky} shows the sky map produced by the spatial search for showers arriving
in one time-domain position of the 1 second long search window. Detector zenith is located in
the center of the count distribution at a declination of $\sim$19$^\circ$ and a right ascension of $\sim$280$^\circ$.
The low event rate far from zenith results from the attenuation of off-axis showers in the larger atmospheric depth.
Points outside a zenith angle of 50$^\circ$ are excluded from the spatial search as most photons
at the energies expected from a GRB signal do not have sufficient energy to reach HAWC.

\newpage

The cross in Figure \ref{fig:sigsky} marks the location of the most significant candidate
in the spatial search of this map. It contains 5 counts for a background expectation of 0.47
resulting in a pre-trials probability of 1.3$\times10^{-4}$. While this appears as a 3.7$\sigma$ result
when considering the single trial case, it corresponds to a post-trials
probability of 0.62 after accounting for spatial trials which is consistent with air shower backgrounds.
See Section \ref{sec:bgcalc} for a description of the background calculation and Section \ref{sec:trialscalc} for a 
description of the method used to calculate post-trial probabilities.

\newpage

\newpage
\section{Temporal Search}

\begin{figure} [ht!]
  \centering
  \includegraphics[width=5in]{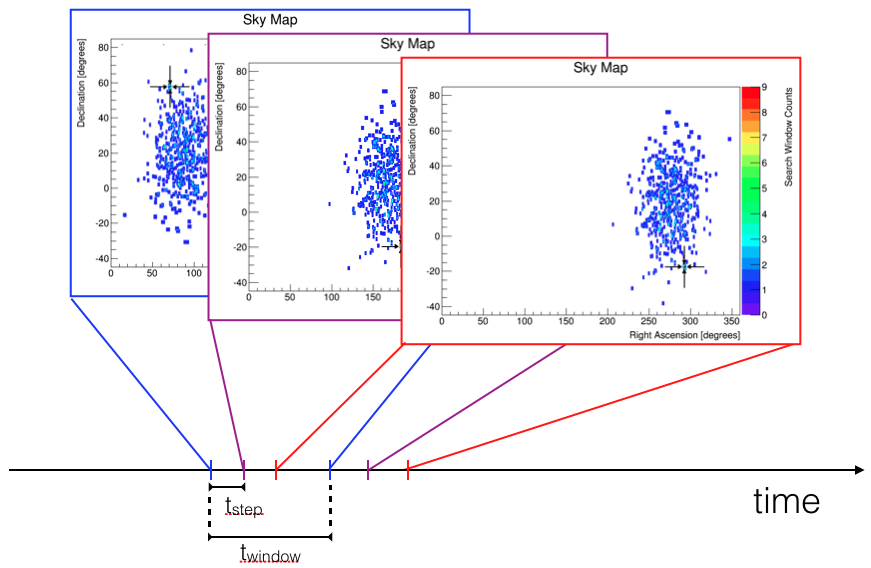}
  \caption{Diagram of the temporal search method for time window duration $t_{window}$.
           The window is advanced with $t_{step} = 0.1 \times t_{window}$ after completing
           the spatial search within that window. This results in a complete scan of
           right ascension for declinations from -31$^\circ$ to 69$^\circ$ over the
           course of one sidereal day.}
  \label{fig:timewindowdiag}
\end{figure}

Once the spatial search at one position of the time window is complete,
we advance the time window forward by 10\% the window width and repeat the search again
(Figure \ref{fig:timewindowdiag}). This yields 90\% overlap between the number of air
shower counts detected at the same position on the sky in two adjacent time windows,
allowing for fine tuning of the start time of an air shower excess. This overlap is
chosen based on Monte Carlo studies of oversampling transient signals on fixed backgrounds
for the all-sky search method used in the Milagro experiment \cite{Vlasios:2008}.

We store the most significant candidate from the spatial search at one position of the time window
and compare it to the best candidate from the next window in the same duration search after accounting for spatial trials. The
more significant post-trials candidate is chosen and stored for comparison to the following window. In this way
we search for the best candidate in a given time window over a complete scan of right ascension for declinations from -31$^\circ$ to 69$^\circ$ over the
course of one sidereal day, where the declination range is determined by the location of detector zenith
at a declination of $19^\circ$.

\newpage

\newpage

\section{Background Estimation}\label{sec:bgcalc}

The gamma-hadron separation cut applied in our analysis reduces the
number of cosmic-ray air showers contributing to the background of our search
but does not result in the complete elimination of this background.
In principle, gamma-ray showers from steady-state sources also contribute as a background
to our transient analysis but they represent a negligible contribution to the overall rate
for the timescales relevant to our search. From this it follows that the roughly 500 Hz rate of reconstructed showers
remaining after applying the gamma-hadron separation cut is entirely due to cosmic-ray air showers.
The following discussion in this section describes how we obtain accurate estimates of the cosmic-ray background
for each position on the sky as a function of time.

\begin{figure} [ht!]
  \centering
  \includegraphics[width=110mm]{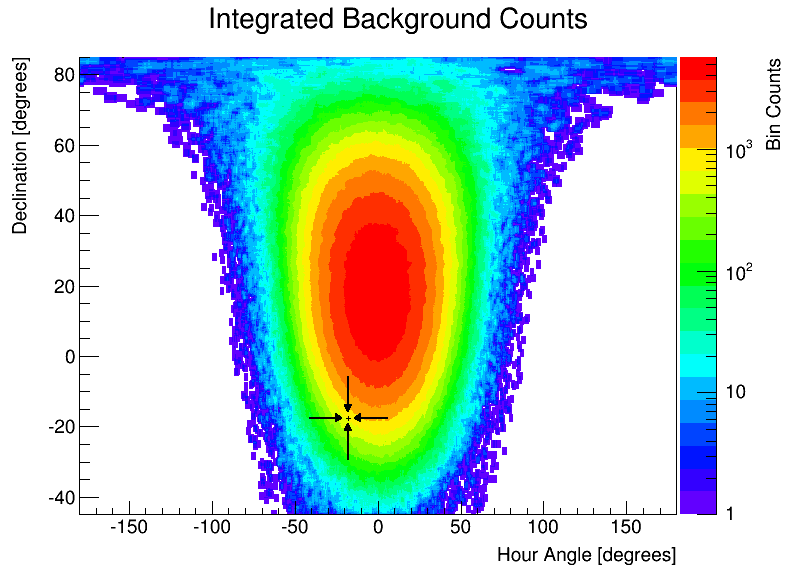}
  \caption
  {Air shower arrival distribution integrated for 1.75 hours in local detector coordinates of hour angle and declinations.
   Individual events are smoothed with the same 2.1$^\circ$ x 2.1$^\circ$ smoothing used for the spatial search.
   The statistical uncertainty in each point in this distribution goes as the square root of counts in each bin, giving
   errors of $\sim$1\% and $\sim$3\% for showers arriving directly overhead and at a zenith angle of 40$^\circ$, respectively.
  }
  \label{fig:backmap}
\end{figure}

As discussed in Chapter \ref{ch:showers}, cosmic-ray air showers arrive uniformly at the upper atmosphere. This means
the shape of the instantaneous air shower arrival distribution can be precisely measured by integrating the 
locations of reconstructed showers in local detector coordinates of hour angle and declination
on a long timescale ($\sim$1 hour) \cite{Fiorino:2015}. We use an integration time of 1.75 hours in this
analysis to obtain measurements of the arrival shape to within a statistical error of few percent (Figure \ref{fig:backmap}).
As we will see in Section \ref{sec:space}, this error does not significantly affect the probability distribution of measurements
made at individual spatial locations.

\newpage

\newpage

Normalizing Figure \ref{fig:backmap} to total number of showers recorded during the 1.75 hour integration duration yields
an acceptance map which describes the probability for an air shower count from the total rate to arrive at a given location
in the HAWC field of view. The background at a given spatial location in detector coordinates for a search window at time $t$ is then
\begin{equation}
N_{bg}(ha, dec, t) \approx \frac{N_{1.75}(ha, dec)}{1.75 \, \textrm{hr} \times 500 \, \textrm{Hz}} \times \textrm{rate}(t) \times t_{window}
\label{directint}
\end{equation}
where $N_{1.75}(ha, dec)$ is the number of showers recorded in a 2.1$^\circ$ x 2.1$^\circ$ spatial bin centered at hour angle $ha$ and declination $dec$
over the 1.75 hour background integration period,
rate$(t)$ is the instantaneous detector rate at time $t$, and $t_{window}$ is the timescale of the search window. This yields an expectation for
1 background count near zenith and 0.16 background counts at a declination of 40$^\circ$ in the 1 second timescale search.

\begin{figure} [ht!]
  \centering
  \includegraphics[width=110mm]{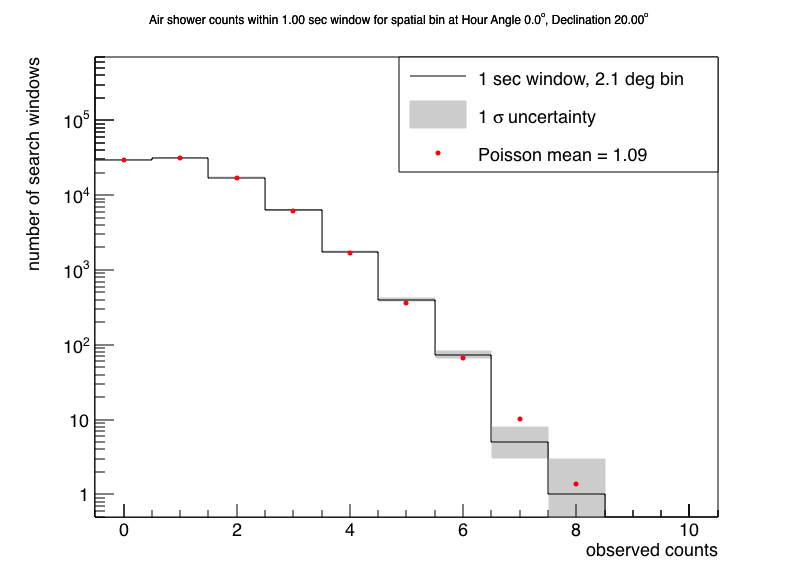}
  \caption
  {
    Histogram of observed counts in the 2.1$^\circ$ x 2.1$^\circ$ search bin at detector zenith over the course of one day for the 1 second interval search. The data follows what one expects from a Poisson distribution with the same mean as the observed data. The reported 1$\sigma$ errors are smaller than can be seen using this vertical scale for $\le$5 observed counts.
  }
  \label{fig:poisson}
\end{figure}

The observed number of counts for this expectation follow a Poisson distribution (Figure \ref{fig:poisson}).
This allows us to categorize the significance of upward fluctuations using
one-sided cumulative Poisson probabilities for finding greater than or equal to the number of observed
counts in the spatial search bin:
\begin{equation}
P(i \ge n, \mu) = \sum_i^\infty \frac{\mu^n e^{-\mu}}{n!}
\label{cumulpoisson}
\end{equation}
These are converted to significances in a standard normal distribution using the inverse of the compliment to the error function:
\begin{equation}
S(i \ge n, \mu) = \sqrt{2} \, \textrm{erfc}(2 \times P(i \ge n, \mu)) 
\end{equation}

\newpage

\newpage

For the 10 second search window duration we use the total all-sky rate within the search map
itself to estimate the instantaneous rate in Equation \ref{directint} because there are enough events to keep the
statistical uncertainty of this estimate to about $\sim$1\%. This is not true for the timescales less than 10 seconds.
In this case, we compute the instantaneous rate inside a 10 second duration centered on the location of the temporal search window
rather than inside the search window itself.

Strictly speaking, signal photons from a GRB source will contribute to both the acceptance map and the rate estimate
in our analysis leading to an artificially increased background measurement and reducing the sensitivity of the search.
This is predominantly an issue near the post-trials discovery threshold of the search where increases in the background
can transform a detection into a sub-threshold result. However, the long integration timescale used to create the acceptance
map effectively reduces signal contributions by a factor of $t_{window}$ / 1.75 hours $\ll$ 1 given the largest
time window used in our search is 10 seconds. Additionally, signal events from a single point on the sky will invariably be much
smaller than the total all sky rate summed from all points within detector zenith. The end result is that effects from signal
contamination are smaller than the statistical uncertainties in both the acceptance and rate measurements.

\newpage

We can show this for the case of the spatial bin located at zenith in the 10 second search window, which is 
the worst case scenario given the background rate of 10 counts is the highest from all three timescales used in this search. In this case,
there are 7.5$\times$10$^{12}$ effective trials taken while searching the 313 day dataset described in Chapter \ref{ch:results}.
This requires a total of 41 signal photons to yield a significant post-trials discovery on the expected cosmic-ray background
of 10 air showers, accounting for a 0.7\% contribution to the acceptance at zenith an 0.8\% contribution to the total rate.

%

\newpage

\section{Background Sample Study}

Given we expect to detect about $\sim$1 GRB per year lasting on the order of seconds,
the air showers analyzed by our search on any given day consist entirely of
cosmic-ray background events. As a result, we can verify the background
estimation obtained from Equation \ref{directint} by running our search over one randomly selected day
of HAWC data and comparing the observed probability value of $P(i \ge N, N_{bg})$ to the Poisson
probability in Equation \ref{cumulpoisson}.

We do this by binning the estimated background values logarithmically from the smallest possible
non-zero background dictated by the shortest timescale and the background integration time (0.02 s $\times$ 1 count/1.75 hr) up to the level of
the all sky rate itself on timescales of order $\sim$1 second (1 s $\times$ $\sim$500 Hz). This effectively bounds all possible values of the estimated background for all three search timescales.
We then choose a logarithmic bin spacing
which is less than 50\% of the statistical uncertainty between the largest two background values (489 and 500). This
groups backgrounds together which have similar discrete values of the Poisson probability. 

We then run each timescale separately over the same randomly selected day of data
to count the number of times we observe $i$ showers for the binned value of the estimated background
of every searched time window and spatial bin combination. From this we obtain the observed probability
\begin{equation}
P_{obs}(i \ge n, N_{bg}) = \frac{\sum_{k=i}^n N_{obs}(k,N_{bg})}{\sum_{k=0}^\infty N_{obs}(k,N_{bg})}
\label{eq:obspval}
\end{equation}
where $N_{bg}$ is now the central value of the logarithmically spaced background bin for which we observe $i$ counts.

\begin{figure}[ht!]
\begin{center}
\includegraphics[width=4.5in]{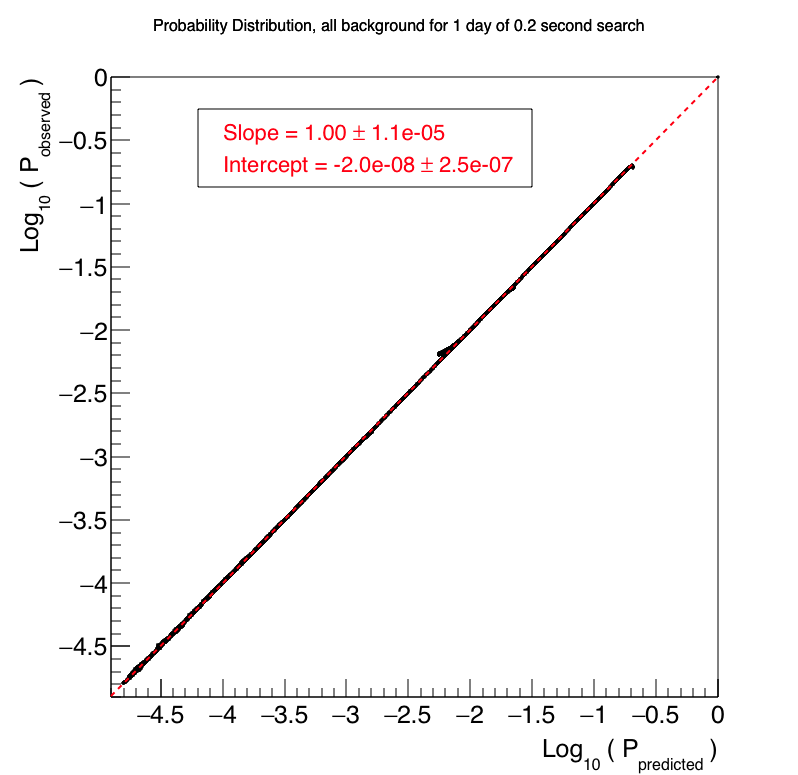}
\caption{Graph of observed probabilities (Equation \ref{eq:obspval}) versus the cumulative Poisson probability
         calculated for binned background counts determined with Equation \ref{directint}
         for all spatial bins searched with a 0.2 second long sliding window shifted over 1 full day.
         A selection is applied to require measurements with counts greater than the correlation
         scale of overlapping spatial and temporal search values ($> 19 \times 19 \times 10$).
         The distribution follows a line of slope 1, confirming that we correctly model the background.
         The small deviations from this line result from spatial bins near the $50^\circ$ extent of the spatial
         search where the background uncertainty is largest. This distribution starts at log$_{10}(P_{predicted}) = -0.74$
         because the smallest non-zero count has a cumulative Poisson probability of $P(i \ge 1, 0.2) = 0.18$ which
         occurs at zenith. }
\label{fig:pdistp2sec}
\end{center}
\end{figure}

\begin{figure}[ht!]
\begin{center}
\includegraphics[width=4.5in]{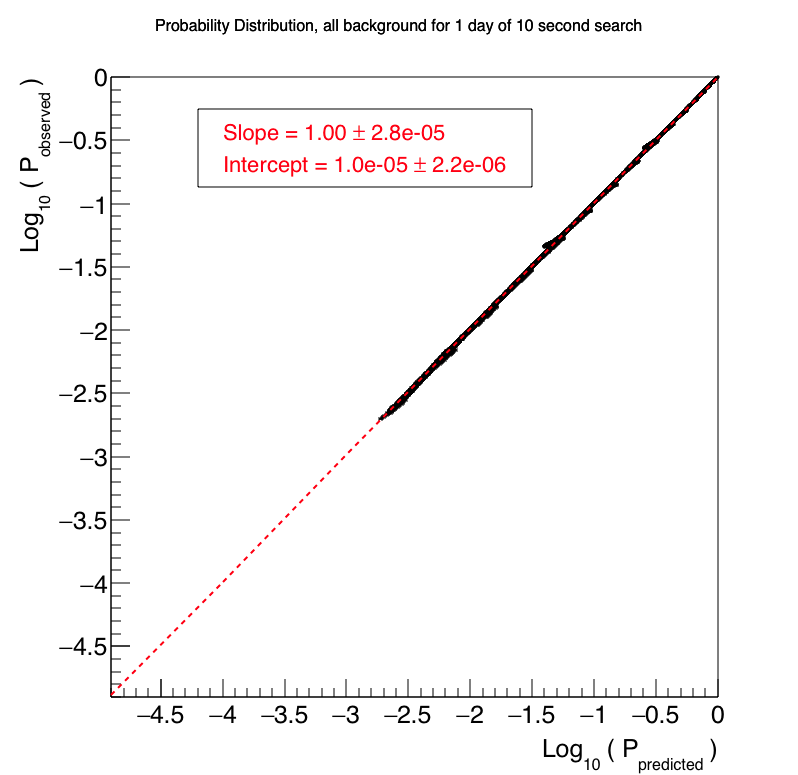}
\caption{Graph of observed probabilities (Equation \ref{eq:obspval}) versus the cumulative Poisson probability
         calculated for binned background counts determined with Equation \ref{directint}
         for all spatial bins searched with a 10 second long sliding window shifted over 1 full day.
         A selection is applied to require measurements with counts greater than the correlation
         scale of overlapping spatial and temporal search values ($> 19 \times 19 \times 10$)
         The distribution follows a line of slope 1, confirming that we correctly model the background.
         The small deviations from this line result from spatial bins near the $50^\circ$ extent of the spatial
         search where the background uncertainty is largest. This distribution ends an order of magnitude sooner
         than Figure \ref{fig:pdistp2sec} because the 10 second search timescale yields fewer spatial searches
         compared to the 1 second search given the $0.1 \times t_{window}$ step size.
         }
\label{fig:pdist10sec}
\end{center}
\end{figure}

Figures \ref{fig:pdistp2sec} - \ref{fig:pdist10sec} present graphs of the observed probabilities versus the
poisson probability calculated using $N_{bg}$. These are in good agreement with the fit $P_{obs} = P_{predicted}$
thereby confirming that the background estimation technique presented in Section \ref{sec:bgcalc} is successful
at modeling the cosmic-ray air shower background. The small deviations away from this fit are the result of locations
close to the 50$^\circ$ limit of the spatial search where the acceptance map exhibits the largest uncertainty.
These introduce only a minor effect on the overall significance of the measurement as the inverse error function
suppresses up to 20\% uncertainties on the probability to a less than 5\% error on the estimated significance.

\newpage

\newpage

\begin{figure}[ht!]
\begin{center}
\includegraphics[width=4.5in]{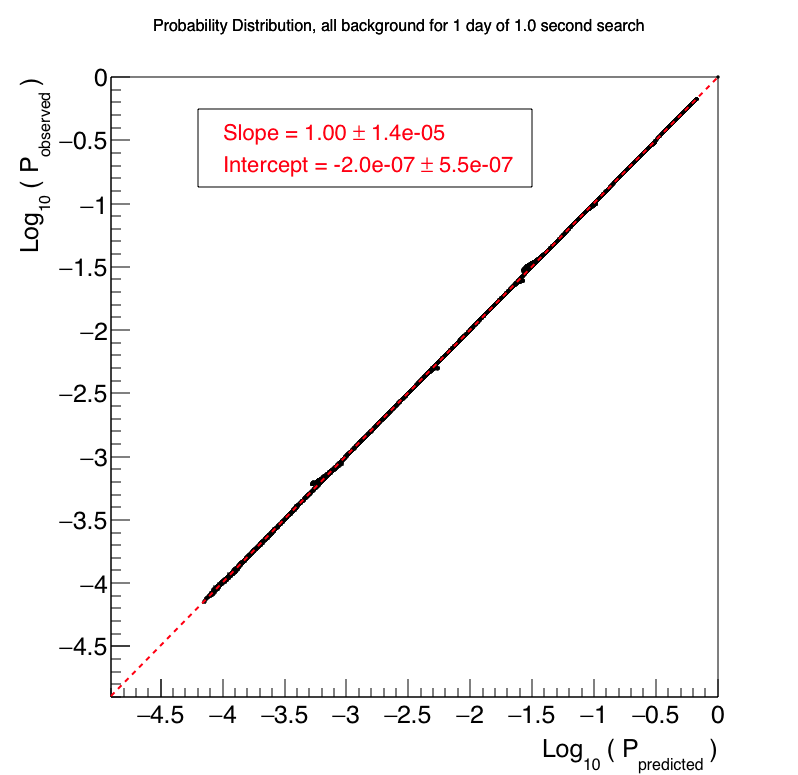}
\caption{Graph of observed probabilities (Equation \ref{eq:obspval}) versus the cumulative Poisson probability
         calculated for binned background counts determined with Equation \ref{directint}
         for all spatial bins searched with a 1 second long sliding window shifted over 1 full day.
         A selection is applied to require measurements with counts greater than the correlation
         scale of overlapping spatial and temporal search values ($> 19 \times 19 \times 10$).
         The distribution follows a line of slope 1, confirming that we correctly model the background.
         The small deviations from this line result from spatial bins near the $50^\circ$ extent of the spatial
         search where the background uncertainty is largest. This distribution starts at log$_{10}(P_{predicted}) = -0.2$
         because the smallest non-zero count has a cumulative Poisson probability of $P(i \ge 1, 1) = 0.63$ which
         occurs at zenith. }
\end{center}
\end{figure}

\newpage

\section{Trials}\label{sec:trialscalc}

While we have demonstrated that we can correctly predict the probability that an air shower count within a search bin at a given spatial location
and time is consistent with background, our search method will compare probabilities from multiple bins to select the result which
is least consistent with the steady-state background hypothesis in order to find transients. We therefore need to know the frequency that probabilities will appear as the
final result of our search to ensure we are correctly estimating the false positive rate. The following subsections will develop
our methodology for determining the post-search false positive rate directly from search data. This is possible
because we expect to discover approximately one GRB transient per year with a duration on the order of seconds. The vast majority
of data are therefore background events.

\newpage

\subsection{\v{S}id\'{a}k Correction for Independent Trials}

In the case of a simple poisson counting experiment where we expect to measure $\mu$ background counts within a fixed time window,
we can formulate the result of observing at least $n$ counts in terms of the one-sided, cumulative poisson probability
\begin{equation}
P(x \ge n, \mu) = \sum_{k = n}^\infty \frac{\mu^k e^{-\mu}}{k!}
\label{eq:cumulativepp}
\end{equation}
where this probability denotes consistency with the background expectation within a single realization of the experiment.
And if we repeat this experiment twice using independent but identical setups we will find the probability for obtaining
less than $n$ counts in both trials of the experiment is given by
\begin{equation}
P_{N = 2}(x < n, \mu) = (1 - P(x \ge n, \mu))^2
\end{equation}
Taking the compliment of $P_{N = 2}(x < n, \mu)$ then yields the probability of discovering at least $n$ counts in either of the two trials is
\begin{equation}
P_{N = 2}(x \ge n, \mu) = 1 - (1 - P(x \ge n, \mu))^2
\label{eq:twotrialcase}
\end{equation}
which is no longer equivalent to Equation \ref{eq:cumulativepp}. This probability is called the post-trials probability as it refers to the true
rate of occurrence for a result to be obtained after multiple trials.

Our argument can be extended to the case of $N$ independent trials simply by replacing the 2 in the exponent
of Equation \ref{eq:twotrialcase} with $N$ to find
\begin{equation}
P_{N}(x \ge n, \mu) = 1 - (1 - P(x \ge n, \mu))^N
\label{eq:ntrialcase}
\end{equation}
which is commonly referred to as the \v{S}id\'{a}k correction for independent trials \cite{Sidak:1967}. While this form may be difficult to interpret for large values of the single-trial probability, it can be approximated as
\begin{equation}
P_{N}(x \ge n, \mu) \approx N P(x \ge n, \mu)
\end{equation}
in the regime where $N P(x \ge n, \mu) \ll 1$. This reveals that rare background events will occur more frequently
when selecting the best result from $N$ repeated trials of the same experiment, which is expected because each trial
provides another opportunity to discover an upward fluctuation in the background. 
The probability threshold for determining the rate of false positives for a given pre-trials probability therefore needs to be set higher in multi-trial searches to yield the
same rate of occurrence as expected from the single-trial probability. This can be done by using Equation \ref{eq:ntrialcase} to transform the
pre-trial probability level to a post-trial probability before applying a detection threshold, such as a 5$\sigma$ level.

\newpage

\section{Calculating Independent Trials from Search Results}\label{sec:indeptrials}

Given a large enough set of search iterations, the number of trials can be reliably calculated
directly from the cumulative search results for an experiment with N-independent trials.
This is because the observed post-trials probability
\begin{equation}
P_{obs} = \frac{ N_{obs} }{N_{search}}
\end{equation}
for having $N_{obs}$ searches resulting in a pre-trial probability greater than $P_{pre}$
in the total number of search iterations, $N_{search}$, will be measured precisely
over some subset of $P_{pre}$. We can then use $P_{obs}$ to invert Equation \ref{eq:ntrialcase} and obtain
\begin{equation}
N = \frac{ log(1 - P_{post}) }{ log( 1 - P_{pre}) }
\label{eq:trialthat}
\end{equation}
which carries an uncertainty of
\begin{equation}
\delta N = \frac{ \delta P_{post} } { (1 - P_{post}) } \frac{1}{|log( 1 - P_{pre})|}
\label{eq:etrialthat}
\end{equation}

\begin{figure}[ht!]
\begin{center}
\subfigure[][]{\includegraphics[width=2.5in]{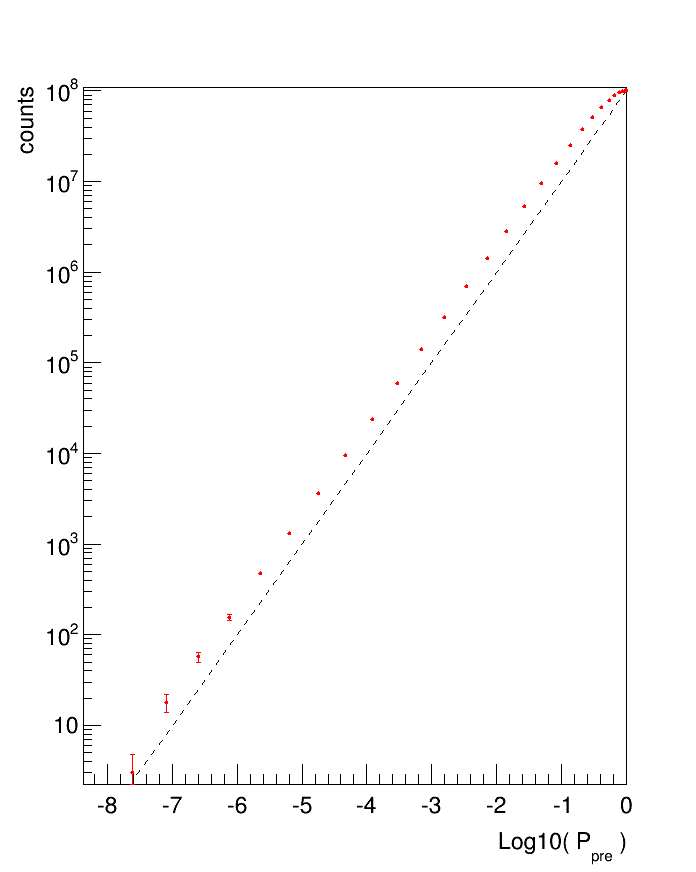}}
\quad\quad\quad\quad
\subfigure[][]{\includegraphics[width=2.5in]{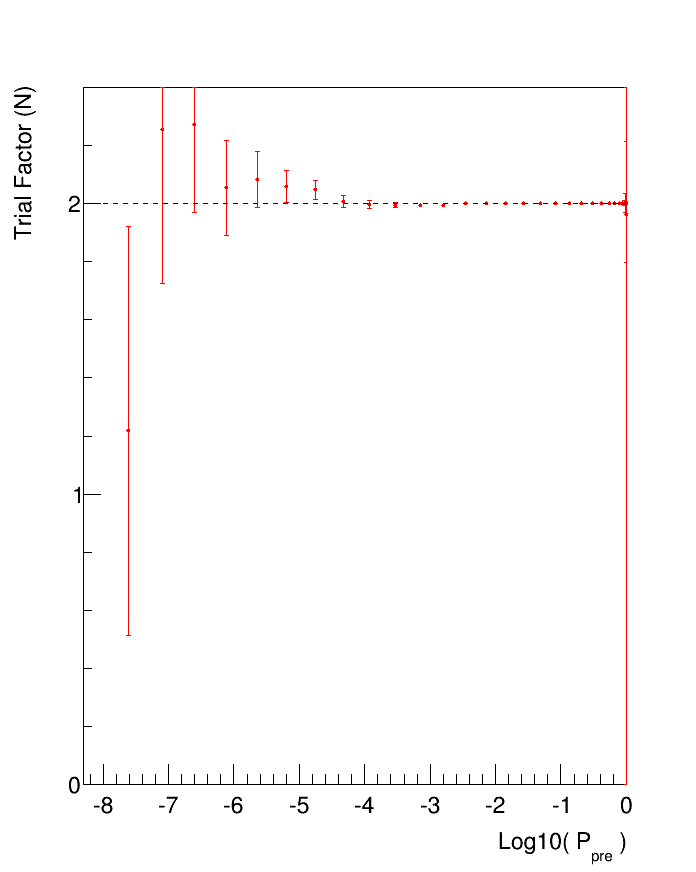}}
\caption{(a) Cumulative distribution of search results and (b) trials correction measured from these results
         as a function of pre-trials probability 
         for a simulated Poisson counting experiment with two independent bins, each with a mean expectation of 10 counts,
         after 10$^8$ iterations of the experiment.
         The two trials involved in an iteration of this experiment result in the number of search results observed at a given pre-trial probability in (a)
         being larger than expected from the value of the pre-trial probability (dashed-line), demonstrating the need to correct $P_{pre}$.
         Panel (b) shows the appropriate correction factor of 2 is precisely measured over the range 10$^{-4} < P_{pre} < 1$. }
\label{fig:indeptrials}
\end{center}
\end{figure}

This is shown in Figure \ref{fig:indeptrials} for the case of a simulated Poisson counting experiment in which
the most significant result is selected from the results of two independent bins, each with a mean expectation of 10 counts.
The left panel shows the distribution of $N_{obs}$ resulting from $N_{search} = 10^8$ graphed as a function of $P_{pre}$.
The number of trials, $N$, is computed in right panel as a function of $P_{pre}$ according to Equation \ref{eq:trialthat}
and is a precise estimate of the two trials incurred in this experiment over the range  10$^{-4} < P_{pre} < 1$. We note
that the large uncertainty present at small values of $P_{pre}$ due to fluctuations in $N_{obs}$ as well as the fact that the calculation fails at $P_{pre} = 0$
are irrelevant as the number of trials is independent of $P_{pre}$ - we need only calculate it once at a single $P_{pre}$.

\newpage

\newpage

\section{Calculating Correlated Trials from Search Results}\label{sec:corrtrials}

Although Equation \ref{eq:trialthat} is derived explicitly for the case of independent
trials, it still provides an intuitive method for mapping the pre-trial probability to
the post-trials probability in the case of correlated trials. To show this, we now
repeat the calculation of the trials factor $N$ from Section \ref{sec:indeptrials}
for a simulated Poisson counting experiment with two correlated bins, each with a mean expectation
of 10 counts where half the mean expectation of the first bin contributes to half the mean of the second bin.
This represents a case of 50\% correlation between the two bins.

\begin{figure}[ht!]
\begin{center}
\subfigure[][]{\includegraphics[width=2.5in]{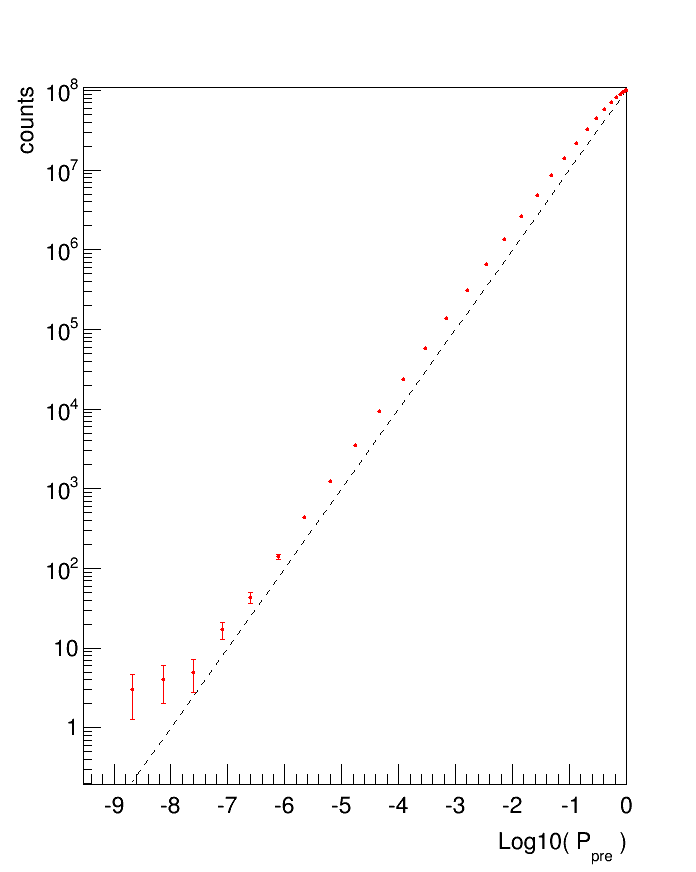}}
\quad\quad\quad\quad
\subfigure[][]{\includegraphics[width=2.5in]{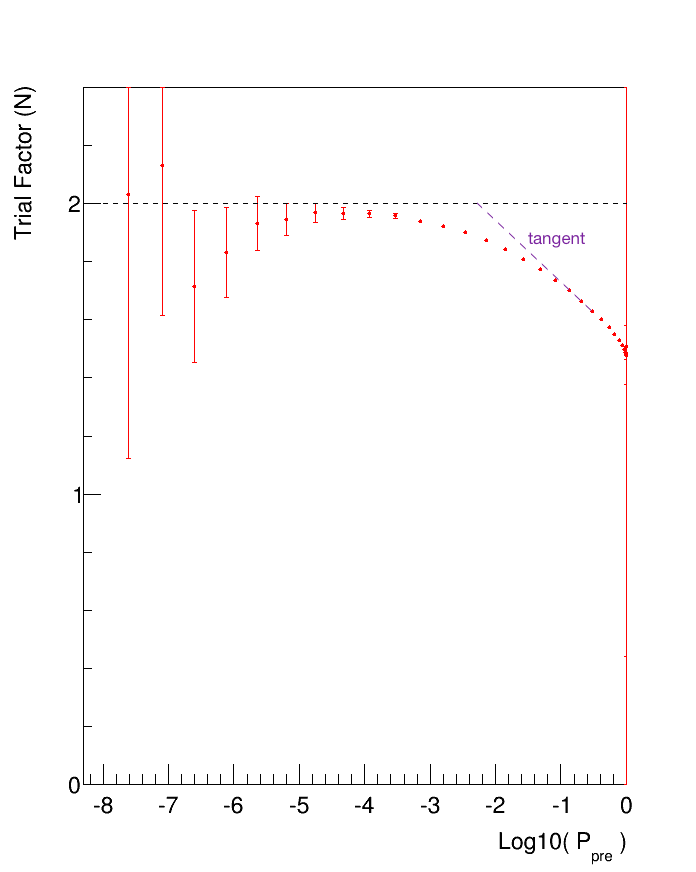}}
\caption{(a) Cumulative distribution of search results and (b) trials correction measured from these results
         as a function of pre-trials probability 
         for a simulated Poisson counting experiment with two bins correlated by 50\%, each with a mean expectation of 10 counts,
         after 10$^8$ iterations of the experiment. As in Figure \ref{fig:indeptrials} the search yields a higher number of
         observed results at a given pre-trial probability than expected from the post-trial probability but now the trial factor
         calculated according \ref{eq:trialthat} is a function of $P_{pre}$. This has an upper limit set by the total number of bins
         used in the search and a lower limit of 1 imposed by having completed at least one iteration of the search. We interpret
         it as an effective number of trials. A tangent line approximation between any two well-measured points provides
         an effective upper limit to the behavior of N($P_{pre}$).}
\label{fig:corrtrials}
\end{center}
\end{figure}

Figure \ref{fig:corrtrials} presents the trials factor calculated according to Equation \ref{eq:trialthat} from the range of observed pre-trial probabilities 
after $N_{search} = 10^8$ in the simulation of the correlated experiment.
As in Figure \ref{fig:indeptrials} the search yields a higher number of
observed results at a given pre-trial probability than expected from the post-trial probability, however, there are
now three important features to note. First, the trial factor calculated based on the search results is a function of $P_{pre}$
rather than a constant value.
We interpret this as meaning there exists an effective number of trials. It is lower at small pre-trials probabilities
because oversampling the remaining signal space cannot produce a drastically more significant result when half of the measurement
is already consistent with the background hypothesis.
Second, the trials value has an upper limit set by the total number of bins used in the search, resulting from the impossibility of obtaining more than two trials from a search of two bins.
And lastly, there exists a lower limit of one effective trial imposed by having checked the result in at least one bin.

Despite the altered interpretation of the $N$ as a function of $P_{pre}$ in the correlated trials case,
it can still be used to correct the pre-trials probabilities of the search method over the range of $P_{pre}$
where it is measured well because, by definition, it must provide the correct conversion between $P_{pre}$ and $P_{obs}$.
This is true for large $P_{pre}$. For very small $P_{pre}$ where the available data set cannot provide enough statistics
to accurately compute the number of trials we note that the upper limit of the bin number may be used
as conservative estimate of the post-trial probability as it will overcorrect the pre-trial probability to appear
as being more consistent with the null hypothesis than its true post-trial rate of occurrence. Note though that a tangent
line drawn between any two well-measured points between (-4 $<$ Log10($P_{pre}$) $<$ 0) also yields an upper limit
on the behavior of N($P_{pre}$) at decreasing values of $P_{pre}$ given that it must approach the N-trial case with decreasing probability.

\newpage

\newpage

\subsection{Spatial Trials}\label{sec:space}

\begin{figure}[ht!]
\begin{center}
\includegraphics[height=3in]{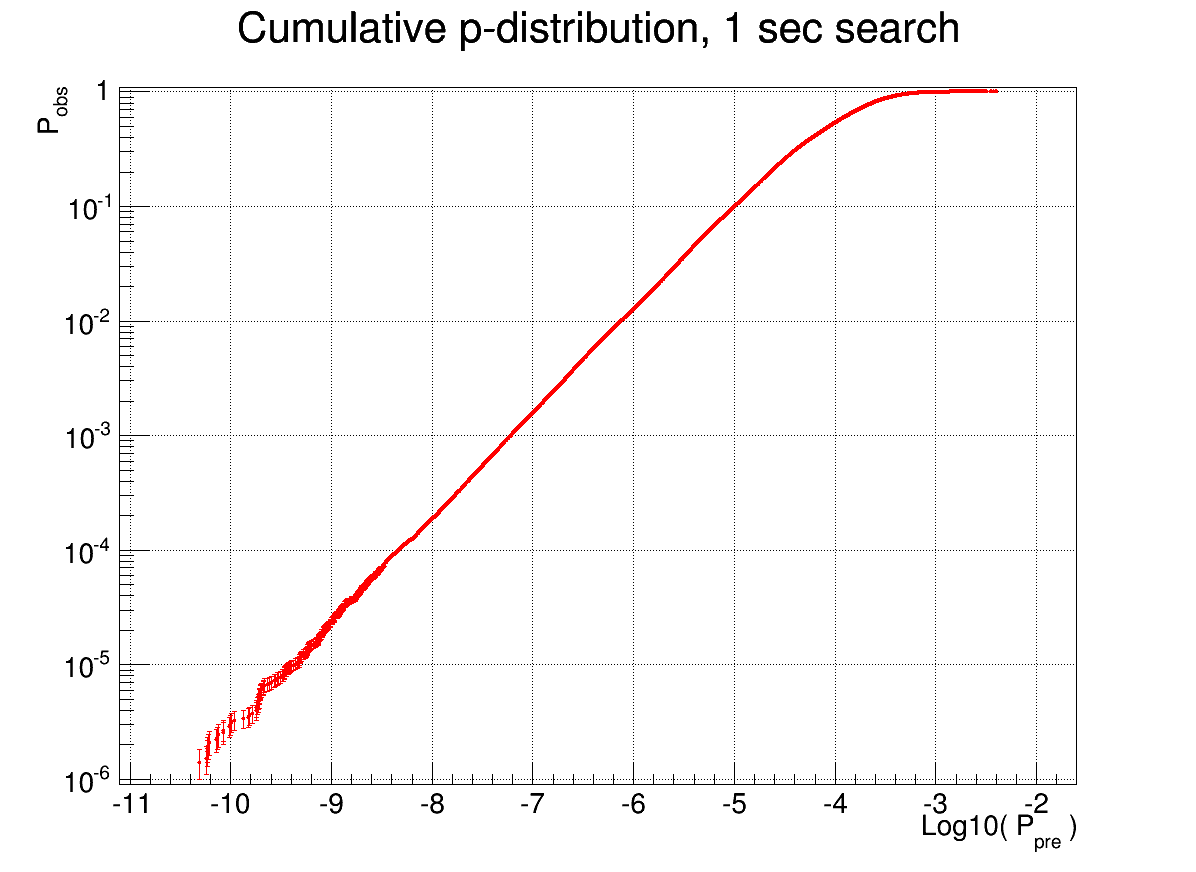}
\caption{Distribution of observed post-trial probability as a function of the pre-trial probability
         from one day of data in the 1 second time window search.}
\label{fig:1seccumdistp}
\end{center}
\end{figure}

In this section we apply the methodology developed in the previous section for calculating correlated trials from search
data to the results from the spatial portion of our all-sky search by running it through 10$^{6}$ search iterations on randomly chosen HAWC data to determine the effective
trials taken during every scan of the HAWC field-of-view. Each scan checks 7.1$\times10^5$ highly correlated (90\% overlap) points
within 50$^\circ$ of detector zenith. To do this, we create a cumulative count distribution of the best pre-trial probability from every time window
analyzed during the course of the day and normalize it to the total number of time windows to calculate the observed post-trails
probability (Figure \ref{fig:1seccumdistp}). The observed post-trials probability is then used to calculate the effective trials $N(P_{pre})$
according to Equation \ref{eq:trialthat}. 

\begin{figure}[ht!]
\begin{center}
\subfigure[][]{\includegraphics[width=2.5in]{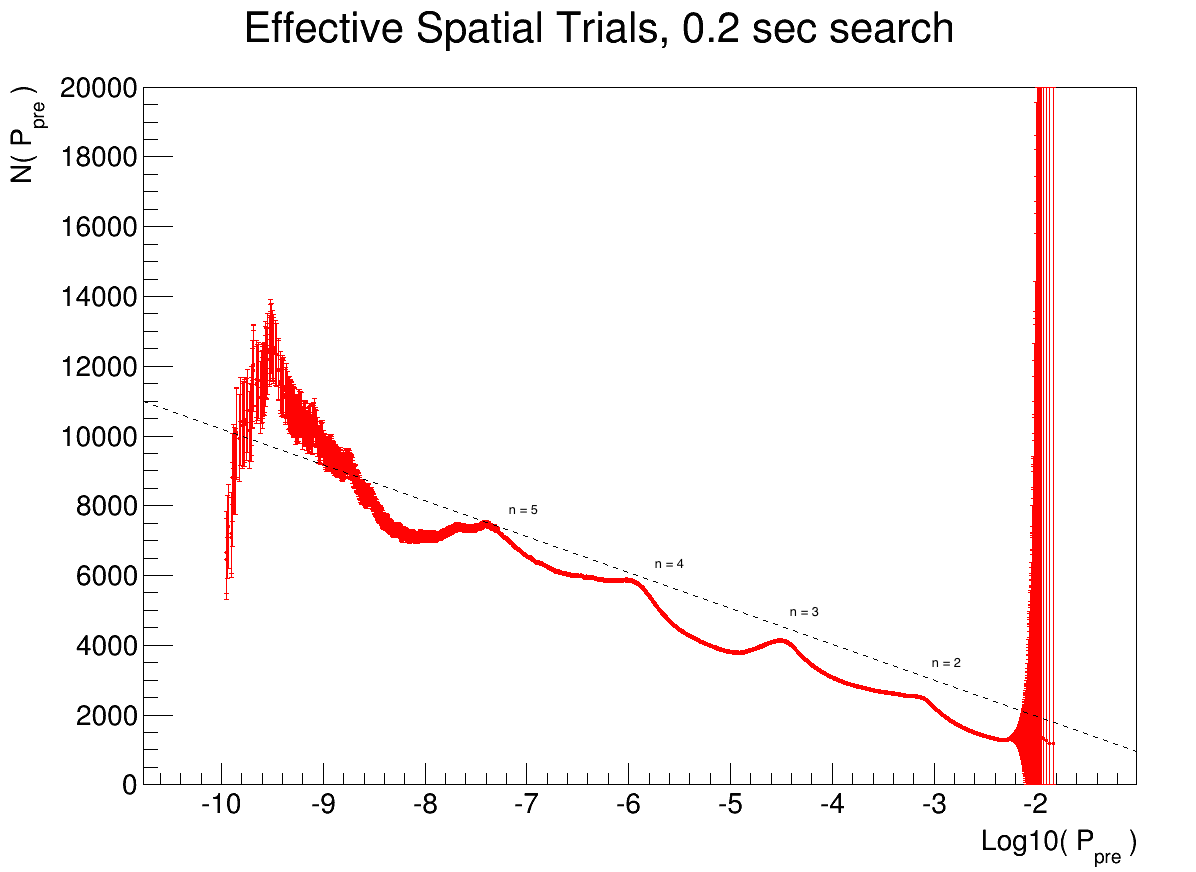}}
\quad\quad\quad\quad
\subfigure[][]{\includegraphics[width=2.5in]{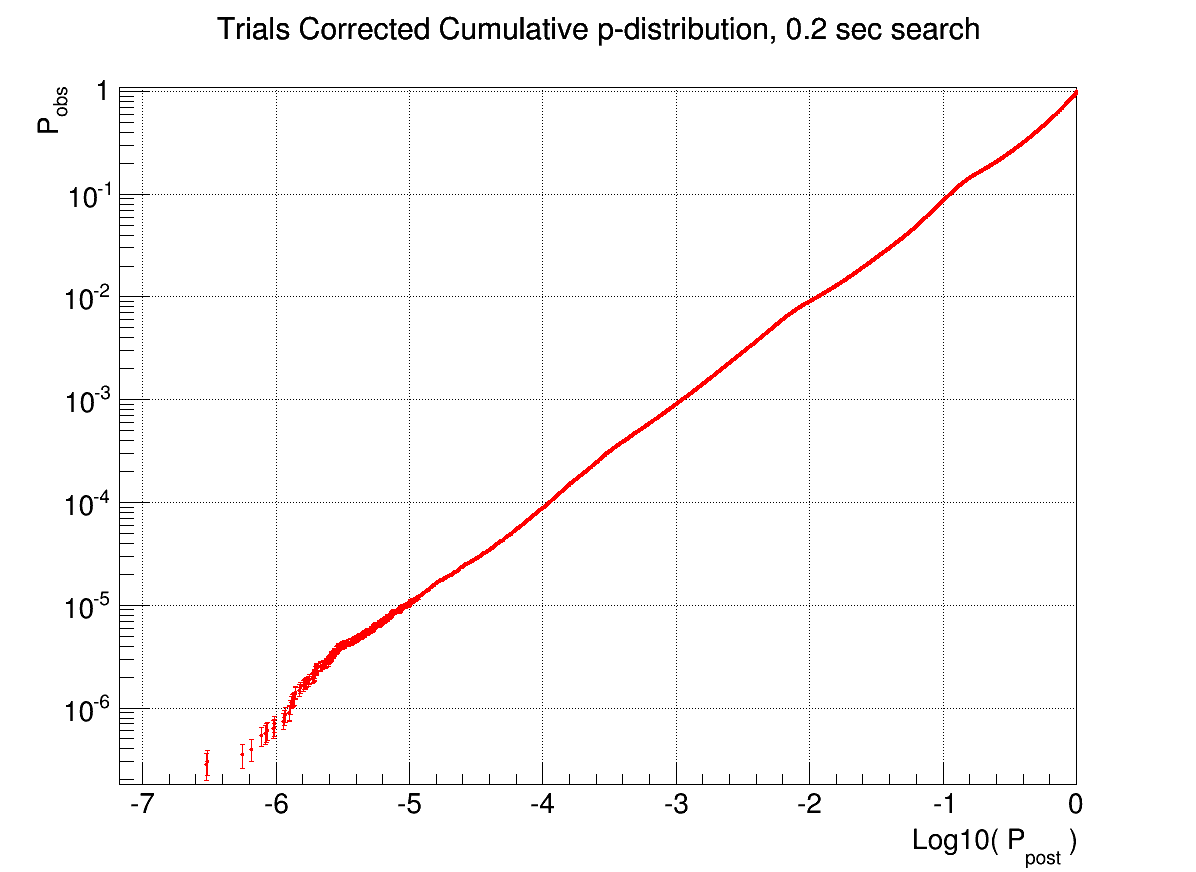}}
\caption{0.2 second duration all-sky search (a) effective trials and (b) trials corrected probability distribution for the linear fit to effective trials
          shown in (a). The wave-like shape in (a) corresponds to discrete steps in the observed number of counts at a given sky location.
          It is most apparent in the 0.2 second search because this search has the fewest expected counts in each bin. 
          The strong feature between -10 $<$ Log10($P_{pre}$) $<$ -9 results from correlations between the overlapping time windows of adjacent
          search iterations which are not modeled by our errors derived from $\sqrt{N}$ uncertainty of the observed number of counts. These
          fluctuations average out for -9 $<$ Log10($P_{pre}$) as the correlation scale of 10, set by the window step size of 0.1$\times t_{window}$,
          is much smaller than the number of observed counts contributing to the data point.
          A linear fit is applied to the effective trials for (-7 $<$ Log10($P_{pre}$) $<$ -2) and then shifted above
          the Poisson count features to yield an upper limit to the effective trials.
          The resulting trials-corrected probability distribution is in good
          agreement with the observed probability of events in our data sample.}
\label{fig:p2secspacecorr}
\end{center}
\end{figure}

\begin{figure}[ht!]
\begin{center}
\subfigure[][]{\includegraphics[width=2.5in]{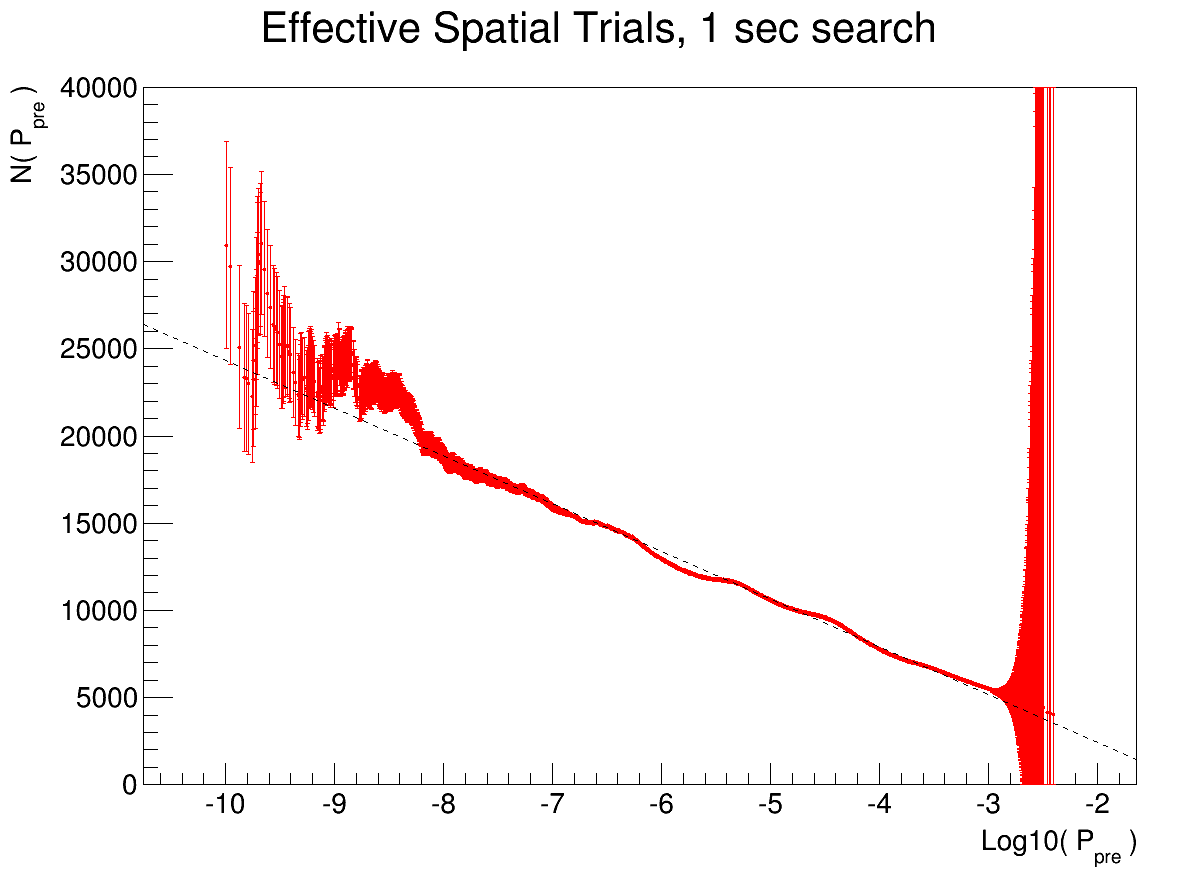}}
\quad\quad\quad\quad
\subfigure[][]{\includegraphics[width=2.5in]{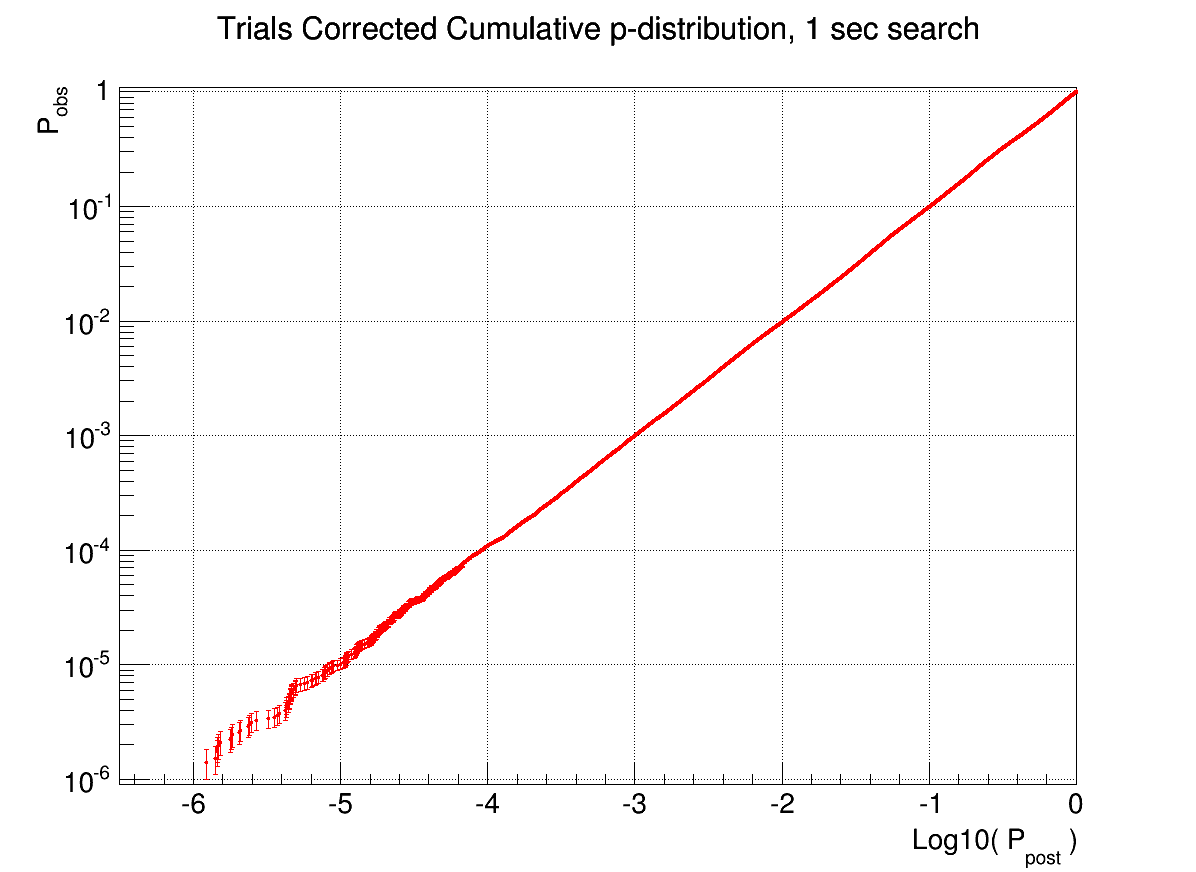}}
\caption{1 second duration all-sky search (a) effective trials and (b) trials corrected probability distribution for the linear fit to effective trials
          shown in (a).
          The feature between -10 $<$ Log10($P_{pre}$) $<$ -9 results from correlations between the overlapping time windows of adjacent
          search iterations which are not modeled by our errors derived from $\sqrt{N}$ uncertainty of the observed number of counts. These
          fluctuations average out for -9 $<$ Log10($P_{pre}$) as the correlation scale of 10, set by the window step size of 0.1$\times t_{window}$,
          is much smaller than the number of observed counts contributing to the data point.
          A linear fit is applied to the effective trials for (-7 $<$ Log10($P_{pre}$) $<$ -2).
          The resulting trials-corrected probability distribution is in good
          agreement with the observed probability of events in our data sample.}
\label{fig:1secspacecorr}
\end{center}
\end{figure}

\begin{figure}[ht!]
\begin{center}
\subfigure[][]{\includegraphics[width=2.5in]{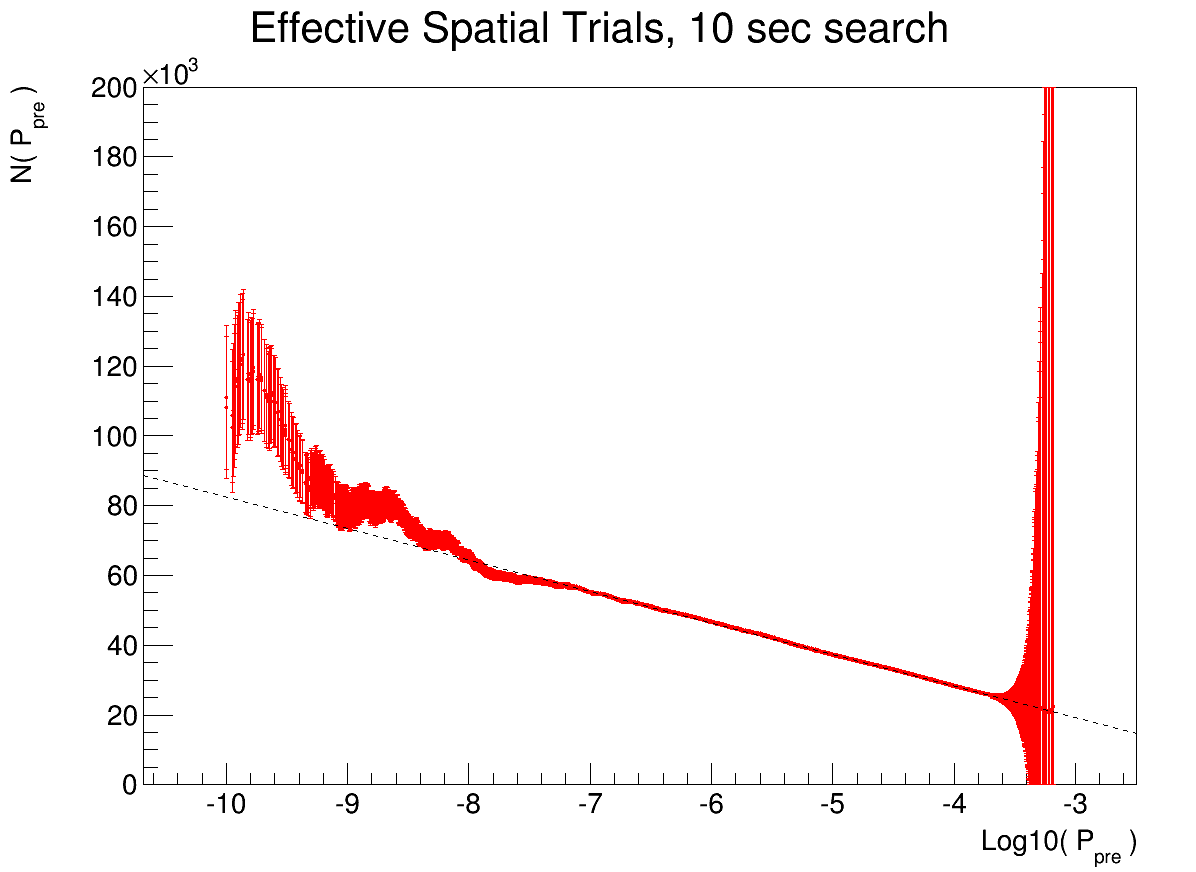}}
\quad\quad\quad\quad
\subfigure[][]{\includegraphics[width=2.5in]{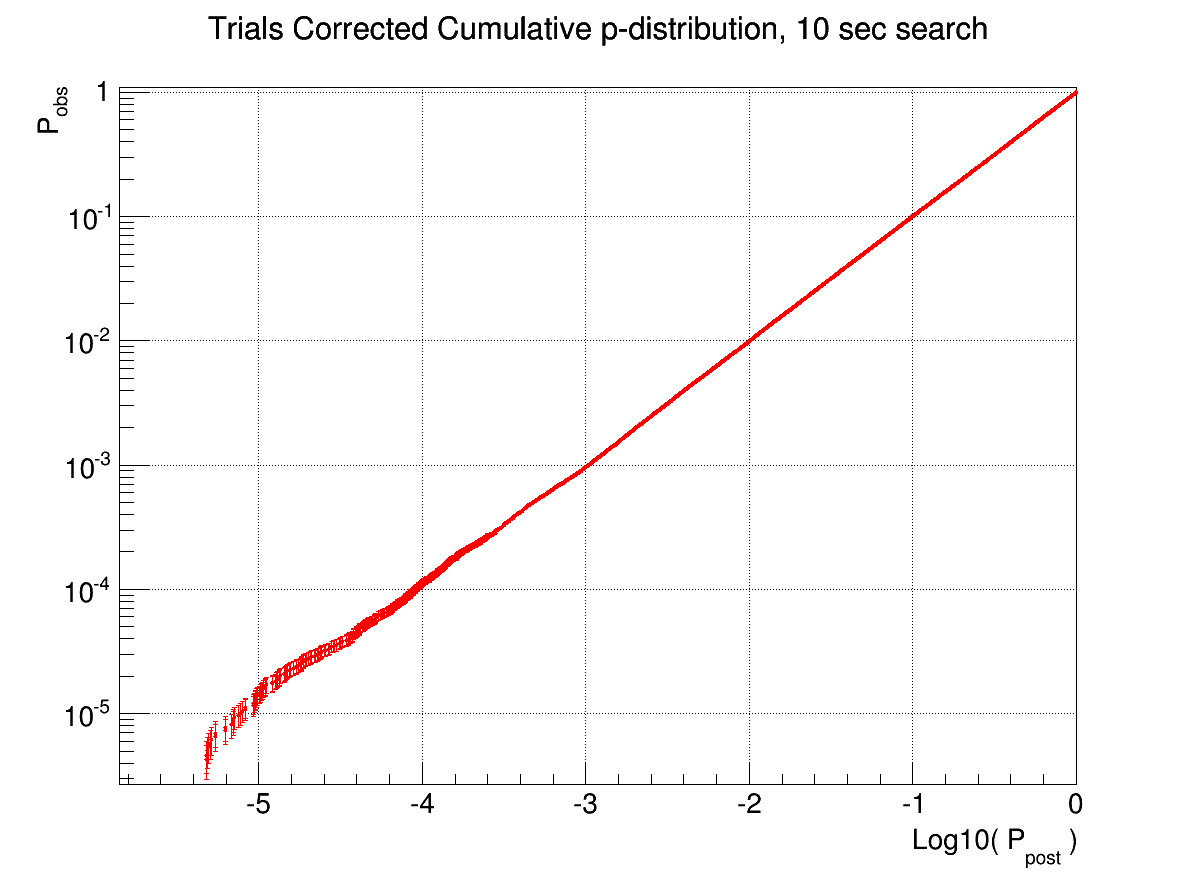}}
\caption{10 second duration all-sky search (a) effective trials and (b) trials corrected probability distribution for the linear fit to effective trials
          shown in (a).
          The feature between -10 $<$ Log10($P_{pre}$) $<$ -9 results from correlations between the overlapping time windows of adjacent
          search iterations which are not modeled by our errors derived from $\sqrt{N}$ uncertainty of the observed number of counts. These
          fluctuations average out for -9 $<$ Log10($P_{pre}$) as the correlation scale of 10, set by the window step size of 0.1$\times t_{window}$,
          is much smaller than the number of observed counts contributing to the data point.
          A linear fit is applied to the effective trials for (-7 $<$ Log10($P_{pre}$) $<$ -2).
          The resulting trials-corrected probability distribution is in good
          agreement with the observed probability of events in our data sample.}
\label{fig:10secspacecorr}
\end{center}
\end{figure}

This is done separately for the 0.2 second (Figure \ref{fig:p2secspacecorr}), 1 second (Figure \ref{fig:1secspacecorr}),
and 10 second (Figure \ref{fig:10secspacecorr}) timescales as the effective trials within the spatial search depends on the number
of empty points in the sky, which scales linearly with window duration
for the 500 Hz all sky rate. Searching consistently empty portions of sky does not yield additional trials because the pre-trial
value of unity for zero observed counts is never selected over non-zero observations. We apply linear fits in the region where
the uncertainty on the calculated value of the effective trials is low (-7 $<$ Log10($P_{pre}$) $<$ -2) in order to produce
upper limits on the evolution of N($P_{pre}$) as discussed in Section \ref{sec:corrtrials}. These fits are presented in Table \ref{tab:spacefits} and describe the N($P_{pre}$)
particularly well in the many-trial regime of the spatial search. The resulting trials-corrected probability distribution is in
good agreement with the observed probabilities (Figures \ref{fig:p2secspacecorr}-\ref{fig:10secspacecorr}).

\renewcommand{\arraystretch}{0.5}
\begin{table}[ht!]
\begin{center}
\begin{tabular}{|c|c|c|} \hline
  Window Duration & Slope & Intercept \\ \hline
  0.2 & -1.02$\times10^3$ & -9.39$\times10^1$ \\
  1.0 & -2.74$\times10^3$ & -3.04$\times10^3$ \\
  10.0 & -9.03$\times10^3$ & -7.81$\times10^3$ \\ \hline
\end{tabular}
\caption{Linear approximation to effective spatial trials correction as a function of Log10($P_{pre}$).}
\label{tab:spacefits}
\end{center}
\end{table}

\newpage

\newpage

\newpage

\newpage

\subsection{Temporal Trials}\label{sec:temporal}

In this section we apply our method for calculating correlated trials
to the temporal search method. This is done by first applying the effective spatial trial corrections
outlined in Section \ref{sec:space} to the results of the spatial search in each time window because these results act as seeds
to the time window search. We then run the time window search over 100 consecutive time windows and store the best result.
We repeat the process for approximately 1 month of HAWC data to build up enough statistics to measure the effective number of
temporal trials. We represent the effective number of temporal trials in terms of the fraction of total trials taken
as this value scales linearly with the total trials taken for the time period covered by a given sliding time window search.
Linear fits are applied to the resulting measurements of N($P_{pre}$), just as in Section \ref{sec:space}.
They are summarized in Table \ref{tab:timefits} in terms of the fraction of time windows that were searched.

\renewcommand{\arraystretch}{0.5}
\begin{table}[ht!]
\begin{center}
\begin{tabular}{|c|c|c|} \hline
  Window Duration & Slope & Intercept \\ \hline
  0.2 & -6.96$\times10^{-2}$ & 2.01$\times10^{-1}$ \\
  1.0 & -5.27$\times10^{-2}$ & 3.31$\times10^{-1}$ \\
  10.0 & -4.82$\times10^{-2}$ & 5.06$\times10^{-1}$ \\ \hline
\end{tabular}
\caption{Linear approximation to effective temporal trials correction as a function of Log10($P_{pre}$).
         Note that these values are scaled by the total number of time windows searched so they
         need to be multipled by $\Delta t_{search} / t_{step}$.}
\label{tab:timefits}
\end{center}
\end{table}

\newpage

\begin{figure}[ht!]
\begin{center}
\includegraphics[height=3in]{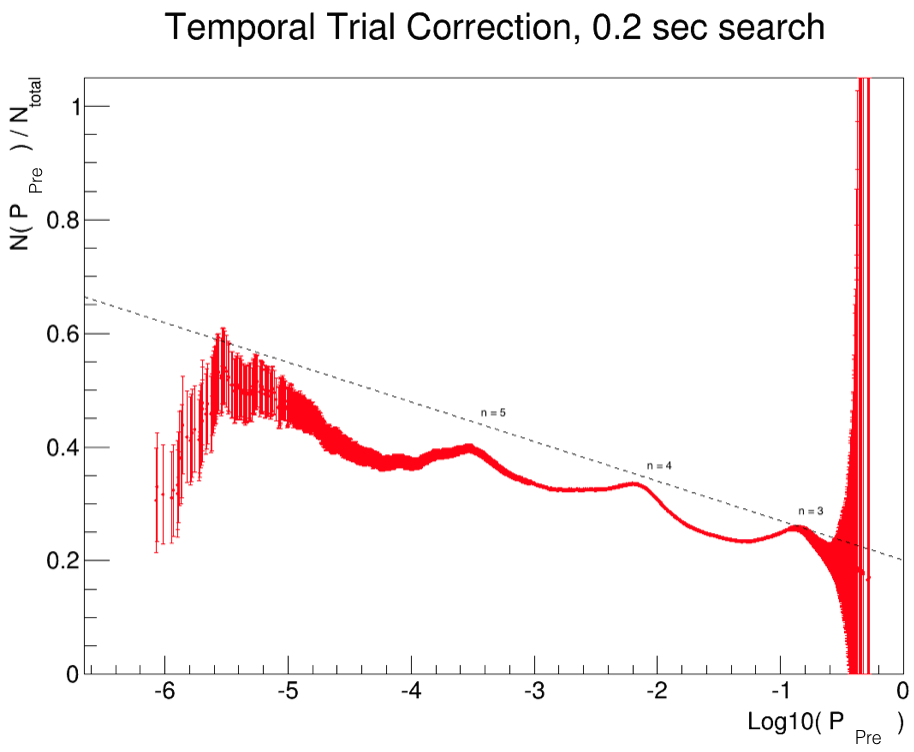}
\caption{Effective temporal trials taken in groups of 100 consecutive time windows.
          The feature between -6 $<$ Log10($P_{pre}$) $<$ -5 results from correlations between the overlapping time windows of adjacent
          search iterations which are not modeled by our errors derived from $\sqrt{N}$ uncertainty of the observed number of counts. These
          fluctuations average out for -5 $<$ Log10($P_{pre}$) as the correlation scale of 10, set by the window step size of 0.1$\times t_{window}$,
          is much smaller than the number of observed counts contributing to the data point.
          A linear fit is applied to the effective trials for (-3 $<$ Log10($P_{pre}$) $<$ -1).
          The resulting trials-corrected probability distribution is in good
          agreement with the observed probability of events in our data sample.}
\label{fig:efftemptrials}
\end{center}
\end{figure}

\newpage

\begin{figure}[ht!]
\begin{center}
\includegraphics[height=3in]{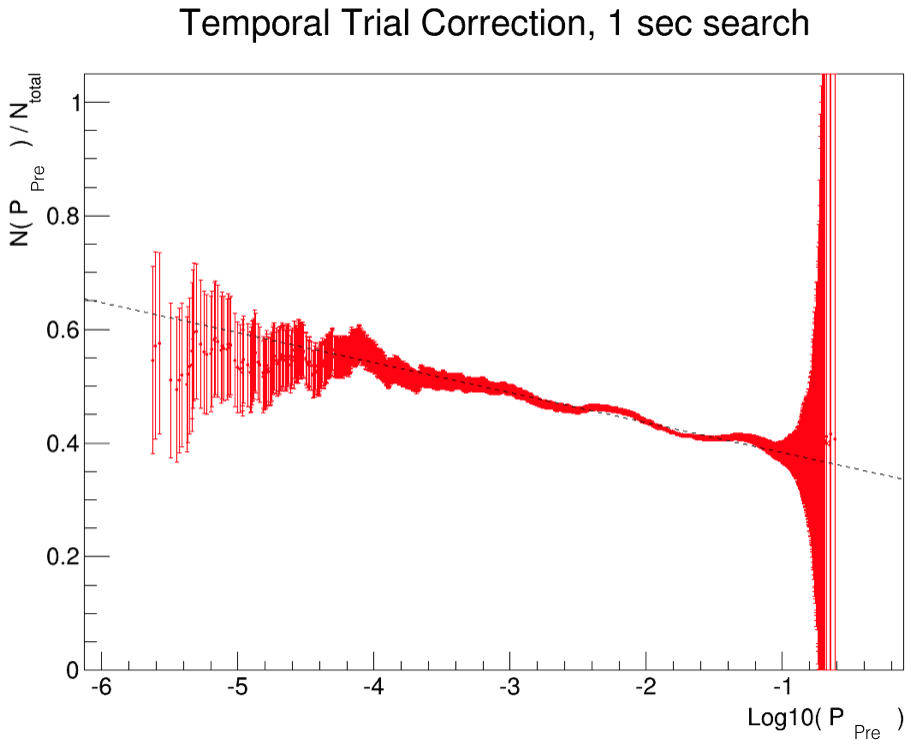}
\caption{Effective temporal trials taken in groups of 100 consecutive time windows.
          The feature between -6 $<$ Log10($P_{pre}$) $<$ -5 results from correlations between the overlapping time windows of adjacent
          search iterations which are not modeled by our errors derived from $\sqrt{N}$ uncertainty of the observed number of counts. These
          fluctuations average out for -5 $<$ Log10($P_{pre}$) as the correlation scale of 10, set by the window step size of 0.1$\times t_{window}$,
          is much smaller than the number of observed counts contributing to the data point.
          A linear fit is applied to the effective trials for (-3 $<$ Log10($P_{pre}$) $<$ -1).
          The resulting trials-corrected probability distribution is in good
          agreement with the observed probability of events in our data sample.}
\end{center}
\end{figure}

\newpage

\begin{figure}[ht!]
\begin{center}
\includegraphics[height=3in]{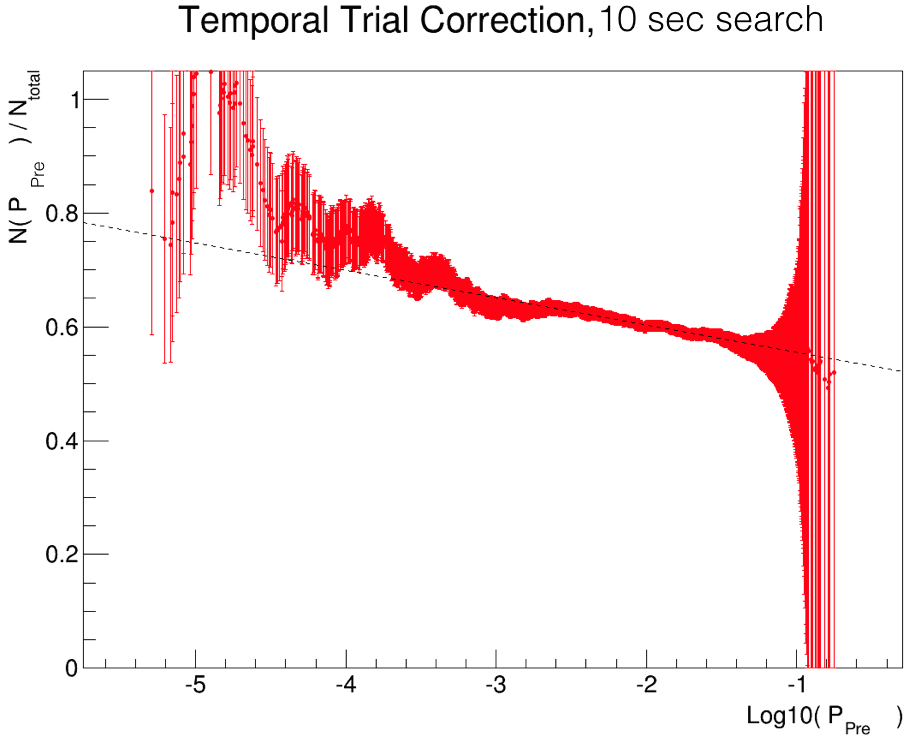}
\caption{Effective temporal trials taken in groups of 100 consecutive time windows.
          The feature between -6 $<$ Log10($P_{pre}$) $<$ -5 results from correlations between the overlapping time windows of adjacent
          search iterations which are not modeled by our errors derived from $\sqrt{N}$ uncertainty of the observed number of counts. These
          fluctuations average out for -5 $<$ Log10($P_{pre}$) as the correlation scale of 10, set by the window step size of 0.1$\times t_{window}$,
          is much smaller than the number of observed counts contributing to the data point.
          A linear fit is applied to the effective trials for (-3 $<$ Log10($P_{pre}$) $<$ -1).
          The resulting trials-corrected probability distribution is in good
          agreement with the observed probability of events in our data sample.}
\end{center}
\end{figure}

\newpage

\subsection{Sensitivity with Trials}\label{sec:senswtrials}

Given our description of how to perform the trials correction in our search,
we will now show its effect on sensitivity by accounting for trials in the case of a 5$\sigma$ 
detection in the 1 second long sliding time window performed with one year of HAWC data.
We do so by first counting every trial as independent to yield an upper limit on the number of trials.
This amounts to the total 7.1$\times10^5$ spatial bins in the spatial search multiplied by
the 3.2e$\times10^8$ temporal trials taken over one year for the 0.1 second step size used in the 1 second
search. A post-trial probability of $5\sigma$ corresponds to a pre-trials probability of $P_{pre} = 1.28\times10^{-12}$ in this case.

We expect approximately 1 background count for the spatial bin located at zenith in our spatial search.
This would yield a 5$\sigma$ detection for a signal level of 9 counts if we were to take only a single trial.
Accounting for the the larger pre-trials probability required to exceed 5$\sigma$ post-trials increases the signal requirement to 21 counts.
The sensitivity of our search then is roughly $\sim$2x worse than the single trial case. Keep in mind though that satellites provide
triggers inside the HAWC field-of-view around 25\% of the time. Combining this with the fact that the GRB fluence falls as a -3/2 power
law means that the all-sky sensitivity is still roughly comparable to the sensitivity of a triggered search in HAWC.

\newpage

Finally, accounting for correlated trials as in Sections \ref{sec:space} and \ref{sec:temporal} yields
5.1$\times10^4$ spatial trials and 3.2$\times10^8$ temporal trials where the largest
reduction in trials comes from the spatial search. This is about an order of magnitude
smaller than the independent trials case. It results in a requirement of 20 signal counts to
produce a 5$\sigma$ post-trials detection which is improved over the independent trials case.
This implies that obtaining an exact calculation of the effective trials factor isn't strictly
necessary so our simple linear fits are a reasonable approach.

\renewcommand{\thechapter}{6}
\chapter{Sensitivity}\label{ch:sens}

The overall sensitivity of the analysis method described in Chapter \ref{ch:method}
depends on a number of design choices. First, the spatial bin used to assess whether
air showers arriving from a point on the sky are consistent with a point source transient
must be large enough to include the majority of signal events while not being so large as
to contain an overwhelming number of background air showers. Similarly, the time windows
used in our search must be tuned to the characteristic timescales of GRB emission
in order to again ensure we retain a high fraction of signal events while excluding as
many background events as possible. And, finally, the choice of a cut based on gamma-hadron separation
variables must be optimized to provide the best discriminating power between background and signal showers.

All of these choices are made by modeling characteristic GRB signals as they
would appear in the HAWC detector using Monte Carlo simulations. For our studies
of the optimal spatial bin size, minimum shower size, and compactness cuts presented
in Sections \ref{sec:optimalspatialbin} - \ref{sec:finalcuts}
we employ two models for GRB emission, a short GRB model and a long GRB model, to
provide the input VHE gamma-ray photon signal to simulations of the HAWC detector.
The short GRB model consists of a 1 second long GRB with an E$^{-1.6}$ power law spectrum and the long GRB model consists of a 10 second
long GRB with an E$^{-2.0}$ power law spectrum. This roughly matches the global behavior high energy observations
of GRBs made by the Fermi LAT experiment \cite{Taboada:2013uza}. 

The studies of characteristic time structure presented in Section \ref{sec:duropt}
forgo the two global models of short and long GRBs in favor of studying individual light curves
from a set of 50 GRBs with high energy detections in the Femi LAT instrument.
This is because individual light curves, as discussed in Chapter \ref{ch:intro}, display
large variability which is not reflected in the choice of two timescales alone. However,
we find a set of just three timescales, 0.2, 1, and 10 seconds, provide a high
efficiency for detecting individual light curves when modeled in HAWC and align well with our
two global models of short and long GRBs. 

Section \ref{sec:sensflu} culminates with the resulting sensitivity to GRB fluence
in the 100 MeV - 10 GeV band as a function of source redshift corresponding to 5$\sigma$ detections
at the 50\% level. Our discussion here is informed by the known systematic error on the 
low energy excess of gamma-rays coming from the Crab Nebula shown in Chapter \ref{ch:Reco}
which reduces the overall sensitivity of our search compared to the design expectations
for the HAWC detector. Nevertheless, we still find that HAWC would detect the extraordinary
bursts of GRB 090510 and GRB 130427A if it were to occur today at favorable zenith angles in the HAWC field-of-view.

\newpage

We use the most recent version of the HAWC detector simulation.
This simulation models cosmic-ray air shower propagation through the
atmosphere with CORSIKA \cite{CORSIKA:1998} followed by a Geant4 \cite{GEANT4:2016} model of the detector's response
to the secondary air shower particles arriving within the detector plane.
The hadronic air shower background in this simulation is normalized to the measured CREAM spectrum \cite{Ahn:2010}.




\newpage
\section{Optimal Spatial Bin}\label{sec:optimalspatialbin}

The sensitivity of any analysis in HAWC depends highly on the choice of spatial bin size
used to assess whether the air showers coming from a specific point on the sky 
are consistent with a cosmic-ray background as opposed to a point-like gamma-ray source.
Choosing too large of a spatial bin will reduce sensitivity as it includes a large
number of background events, which scale linearly with bin area for typical bin sizes.
Additionally, choosing too small of a spatial bin also reduces sensitivity as it excludes much of
the desired signal.

\begin{figure}[ht!]
\begin{center}
\includegraphics[height=1.75in]{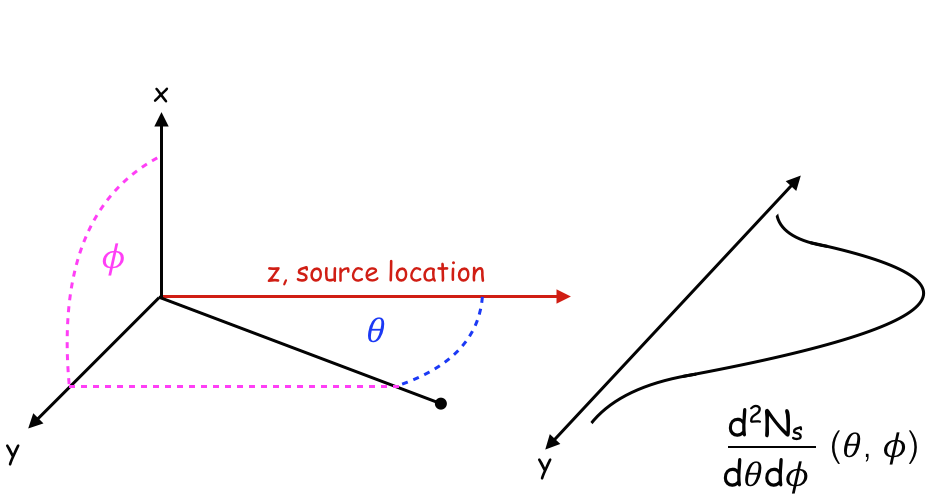}
\caption{Geometry of a simple experiment with a gaussian PSF. The source location is along the z-axis at $(\theta = 0, \phi = 0)$. A projection of $\frac{d^2N_s}{d\theta d \phi}(\theta, \phi)$
        is shown in the $z$-$y$ plane under the small angle approximation where $y \approx \theta$ assuming $\sigma \sim 1^o$. }
\label{fig:psf_geom}
\end{center}
\end{figure}

One can illustrate this by considering the simple case of an experiment with a gaussian
point spread function (PSF) for gamma-ray photons and a uniform cosmic-ray background.
Under this setup, the differential number of photons at an angular position of $\theta$ and $\phi$ defined from the location of
a gamma-ray point source is
\begin{equation}
\frac{ d^2N_s }{ d\theta d\phi } (\theta, \phi) = \frac{n_s}{ 2 \pi \sigma^2 } e^{ - \theta^2 / 2 \sigma^2 }
\label{eq:Ns_gauss_psf}
\end{equation}
where $\sigma$ is the PSF of the experiment and $n_s$ is the total number of source
photons recording during the live time of the experiment (See Figure \ref{fig:psf_geom}). Integrating this equation to obtain the total photons falling
inside a bin centered on the source with an angular extent of $\theta_{bin}$ gives
\begin{equation}
N_s(\theta_{bin})  = \int_0^{\theta_{bin}} \int_0^{2\pi} \frac{n_s}{ 2 \pi \sigma^2 } e^{ - \theta^2 / 2 \sigma^2 } \theta \, d\theta \, d\phi = n_s (1 - e^{-\theta_{bin}^2 / 2 \sigma^2})
\end{equation}
The same bin also yields the following expression for the number of uniform background events contained within $\theta_{bin}$
\begin{equation}
N_{bg}(\theta_{bin}) = \int_0^{\theta_{bin}} \int_0^{2\pi} n_{bg} \, \theta \, d\theta \, d \phi = n_{bg} \, \pi \, \theta_{bin}^2
\label{eq:Nbg_gauss_psf}
\end{equation}
where $n_{bg}$ is the density of background events recorded per steradian during the live time of the experiment.

\begin{figure}[ht!]
\begin{center}
\includegraphics[height=3in]{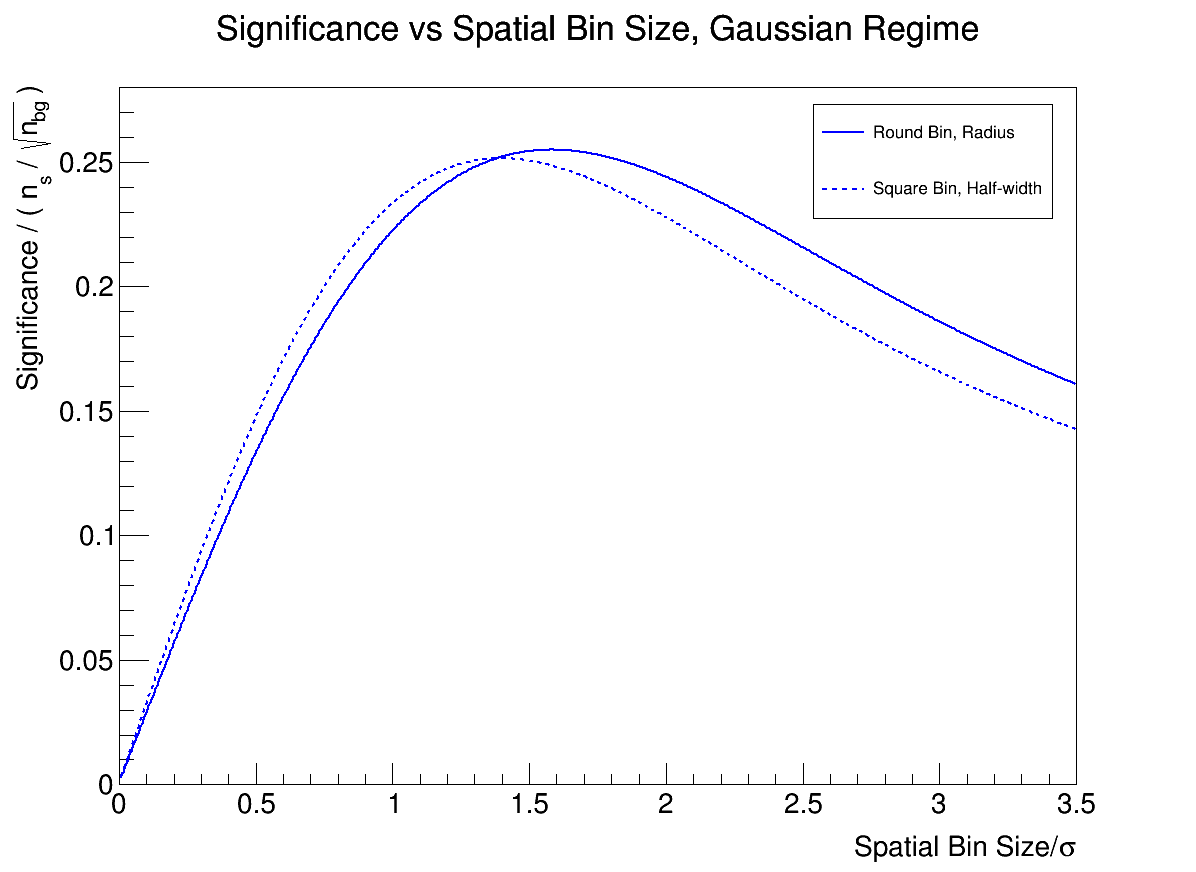}
\caption{Source significance versus spatial bin size for a Gaussian PSF with a standard deviation of $\sigma$ in the regime where $N_s$ and $N_{bg}$ follow Gaussian distributions and are large enough that $N \gg \sqrt{N}$.}
\label{fig:opt_gaus_bin}
\end{center}
\end{figure}

Assuming $N_s(\theta_{bin})$ and $N_{bg}(\theta_{bin})$ follow a gaussian distribution and that the statistical fluctuations in the measured number of events are small ($N \gg \sqrt{N}$), the significance of our gamma-ray source for one choice of $\theta_{bin}$ is
\begin{equation}
S(\theta_{bin}) = \frac{ N_s(\theta_{bin}) }{ \sqrt{  N_{bg}(\theta_{bin}) } } = \frac{n_s}{\sqrt{n_{bg} \pi}} \frac{ (1 - e^{-\theta_{bin}^2 / 2 \sigma^2}) }{ \theta_{bin} }
\label{eq:gaus_sig}
\end{equation}
Plotting this result reveals how the source significance quickly approaches zero for small bin sizes as we exclude most of the signal photons (Figure \ref{fig:opt_gaus_bin}).
The significance also drops off at very large bin sizes as we include a large number of background events. The maximum value near $\sim 1.5\sigma$ corresponds to the most sensitive
choice of $\theta_{bin}$ because a higher value of significance for a fixed signal $n_s$ indicates the analysis requires fewer total signal photons to reach a 5$\sigma$ discovery threshold.

\newpage

One can find the exact maximum value of $\theta_{bin}$ in Equation \ref{eq:gaus_sig} by
setting $dS / d\theta_{bin} = 0$ and solving for $\theta_{bin}$. Doing so results in the following expression
\begin{equation}
  \frac{ n_s } { \sqrt{ n_{bg} \, \pi } } \,
  \Bigg( \frac{1}{ \theta_{bin}^2 } \Bigg) \,
  \Bigg[ \Bigg(\frac{\theta_{bin}^2}{ \sigma^2 } + 1 \Bigg) e^{ - \theta_{bin}^2 / 2 \sigma^2 } - 1 \Bigg]
  = 0
\end{equation}
which we solve numerically to find a single solution
\begin{equation}
\theta_{bin, \, optimal} = 1.585 \, \sigma
\end{equation}
on the domain $\theta_{bin} \in [0, \pi/2]$. This solution corresponds to a containment radius of roughly 70\% of the total source photons.

\subsection{Square Bin Optimization}\label{sec:optimalsquarebin}

While a round spatial bin defined by a radius $\theta_{bin}$ is convenient for analytic integration, there is a large computational advantage
to using square spatial bins when searching a wide field-of-view in an experiment with a small PSF. This is because the choice of a rectilinear coordinate
system eliminates the need to invoke the square root function when smoothing the field-of-view with the optimal spatial bin size. We shall therefore revisit the
model of an experiment with a simple Gaussian PSF using a locally rectilinear coordinate system near the gamma-ray source.

In this case, the equation for $N_s(\theta_{bin})$ becomes
\begin{equation}
N_s(\theta_{bin})  = \int_{-\theta_{bin}}^{\theta_{bin}} \int_{-\theta_{bin}}^{\theta_{bin}} \frac{n_s}{ 2 \pi \sigma^2 } e^{ - (\theta_x^2 + \theta_y^2) / 2 \sigma^2 } \, d\theta_x \, d\theta_y
\label{eq:signal_in_square}
\end{equation}
where $\theta = \sqrt{\theta_x^2 + \theta_y^2}$ in Figure \ref{fig:psf_geom} under the small angle approximation and $\theta_{bin}$ now describes the half-width for one side of the square spatial bin.
Similarly,
\begin{equation}
N_{bg}(\theta_{bin})  = \int_{-\theta_{bin}}^{\theta_{bin}} \int_{-\theta_{bin}}^{\theta_{bin}} n_{bg} \, d\theta_x \, d\theta_y = 4 \, n_{bg} \theta_{bin}^2
\end{equation}
and again we can calculate the source significance for a specific choice of $\theta_{bin}$ using $S(\theta_{bin}) = N_s(\theta_{bin}) / \sqrt{N_{bg}(\theta_{bin})}$, albeit numerically because there is no longer an
analytic form to the result of Equation \ref{eq:signal_in_square}. 

The location of maximum significance is also found numerically by searching for a local maximum in the graph of $S(\theta_{bin})$ for the square bin case in Figure \ref{fig:opt_gaus_bin}. Again, there is a single maximum occurring at
\begin{equation}
\theta_{bin, \, optimal} = 1.40 \, \sigma
\end{equation}
Considering that $\theta_{bin, \, optimal}$ is now the half-width of a square bin, this corresponds to roughly the same area as the optimal round bin.
Additionally, it is important to note that the value of maximum significance obtained from the optimal square bin analysis is only $\sim 1$\% less than
the value obtained from the optimal round bin analysis, so the large performance gain obtained by using a computationally efficient bin type incurs just a
negligible reduction in overall sensitivity.

\subsection{Optimization in Poisson Regime}\label{sec:poissonopt}

When both the signal and expected background are small, Poisson fluctuations are large and we can no longer
use $N/\sqrt{N}$ as a good estimate of the sensitivity of our optimal bin. Instead we need to account for
fluctuations in both the signal and background. This is true for a GRB analysis
because the timescales for prompt emission, even in the case of long GRBs, are short enough that the background counts will be in the poisson regime.

To estimate sensitivity in the Poisson regime we use a simple Monte Carlo
simulation to randomly throw both signal and background counts according to a Poisson distribution, accounting for the efficiency 
of retaining gamma-ray events inside our cuts with a binomial probability. We then calculate the
average number of detections obtained at different pre-cut signal normalizations to find the
signal level that results in 5$\sigma$ detections $>50\%$ of the time.
This is done for 1000 realizations of the expected signal and background to keep the uncertainty in the average number of detections 
at the 50\% level below a 5\% precent. 

\begin{figure}[ht!]
\begin{center}
\includegraphics[height=3in]{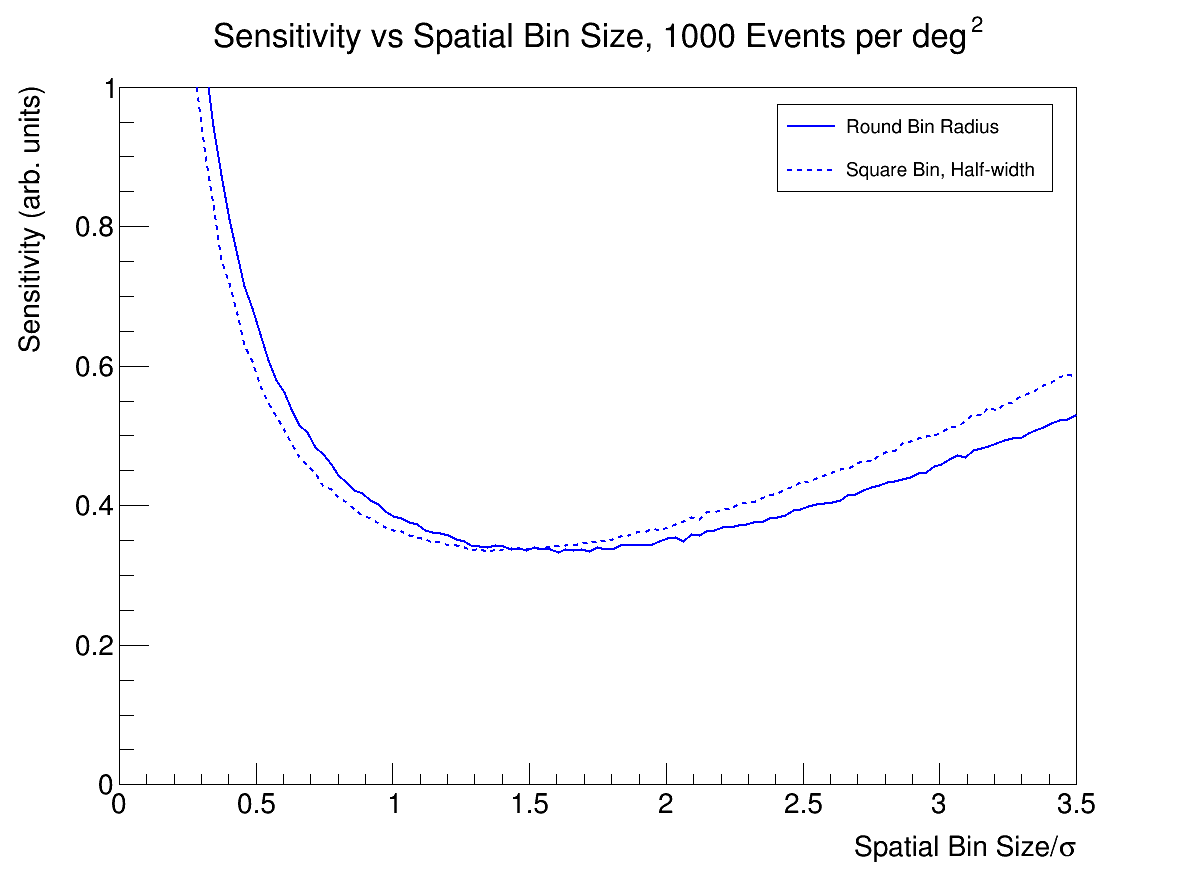}
\caption{Sensitivity versus spatial bin size for Poisson optimization of the spatial bin in the Gaussian PSF example using both round (solid curve) and square (dashed curve)
         spatial bins on a background of 1000 events per deg$^2$. This represents the limiting case in which we recover the optimal bin size
         values of 1.585$\sigma$ in the round bin and 1.40$\sigma$ in the square bin from the Gaussian regime to within 1\% of their true value.
         These are denoted by the local minimum of the sensitivity curve, which is reported in arbitrary units but generally corresponds to the number of
         signal photons needed to create an average detection at the 5$\sigma$ level.}
\label{fig:poissonopt_check}
\end{center}
\end{figure}

\begin{figure}[ht!]
\begin{center}
\includegraphics[height=3in]{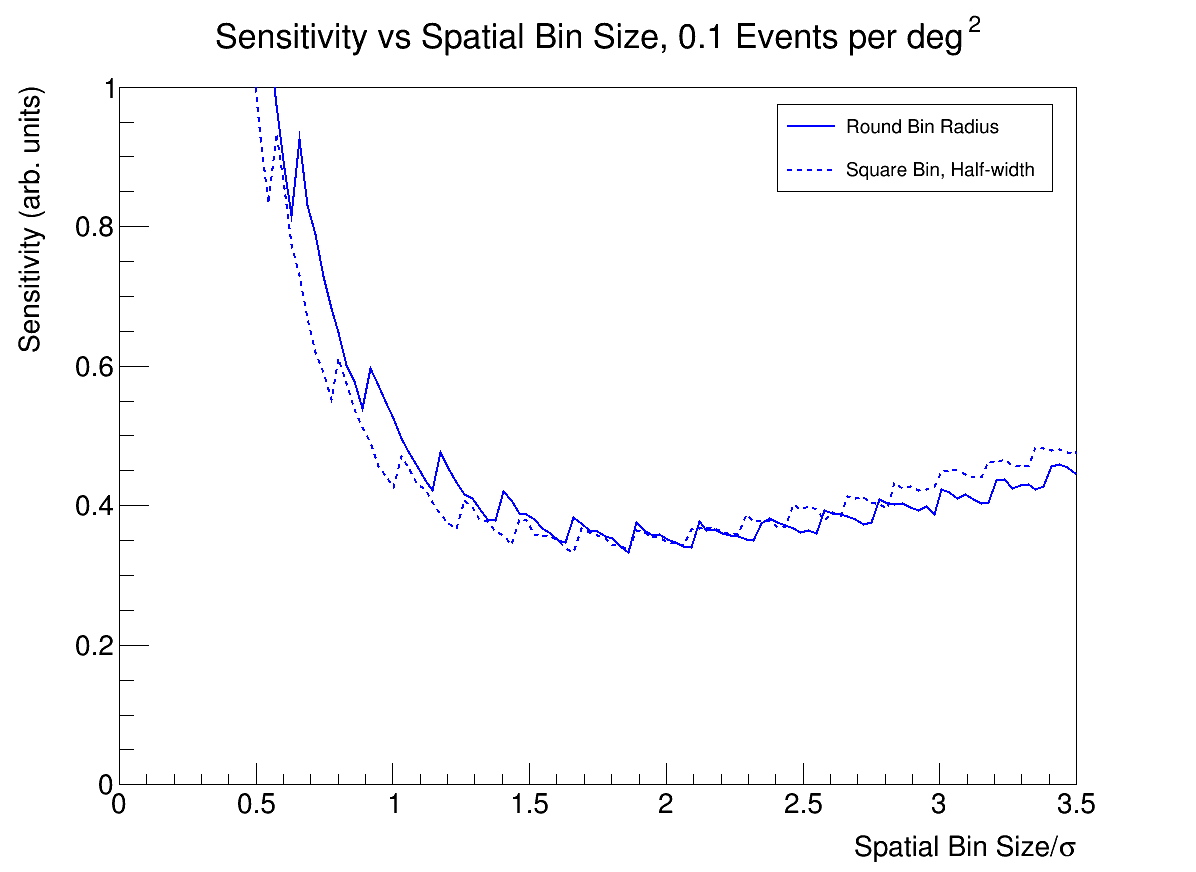}
\caption{Sensitivity versus spatial bin size for Poisson optimization of the spatial bin in the Gaussian PSF example using both round (solid curve) and square (dashed curve)
         spatial bins on a background of 0.1 events per deg$^2$.
         The sawtooth nature of both curves results from the discreteness of the Poisson distribution,
         which requires an integer number of signal counts to cross the detection threshold and causes
         the curve to sharply rise upward when the background contained within the spatial bin is large enough
         to need an additional signal photon to obtain a detection.
         The optimal bin size is 1.86$\sigma$ for the round bin and 1.65$\sigma$ for the round bin and
         are denoted by the local minima in the sensitivity curve. This maintains
         the equal area association between the two bin types. Sensitivity is reported in arbitrary units but generally corresponds to the number of
         signal photons needed to create an average detection at the 5$\sigma$ level.}
\label{fig:poissonoptrad}
\end{center}
\end{figure}

\newpage
The results of this simulation are shown in Figures \ref{fig:poissonopt_check} and \ref{fig:poissonoptrad} for the two types of spatial bins in the Gaussian
PSF example from the previous section. In the case of Figures \ref{fig:poissonopt_check} we set the background level to 1000 events per square degree to
recover the optimal bin sizes expected from the Gaussian signal optimization to within 1\%. This confirms our choice of using 1000 realizations
of the counting experiment is enough to accurately describe the optimal bin size to within negligible error. Figure \ref{fig:poissonoptrad} presents
the Poisson analysis for a background rate of 0.1 events per square degree. This is the typical event rate associated with a spatial bin in the 1 second
time window search from our analysis. In this case the optimal round bin size is 1.86$\sigma$ and the optimal square bin size is 1.65$\sigma$, again
denoting the equal area relationship between the two optimal bins. Both
are larger than the corresponding values from the Gaussian signal optimization because the reduction of sensitivity introduced from including fractionally
more background events is suppressed somewhat in the Poisson regime. This intuitively makes sense as a counting experiment in which there were no
known backgrounds would favor no spatial bin cut at all as there is no penalty for expanding the bin size, only losses in sensitivity from not
containing all of the signal events.

One striking note about Figure \ref{fig:poissonoptrad} is the sawtooth nature of the sensitivity curve, which results from the discreteness
of the Poisson distribution requiring an integral number of counts to cross the detection threshold. This technically leads to over-tuning of
the bin size for use in data because the overall background rate will fluctuate with the density of the atmosphere, thereby shifting
the locations of sawtooth minima in Figure \ref{fig:poissonoptrad} to different bin sizes. In practice though the broadness of this distribution
compared to the Gaussian optimization regime results in sensitivity losses only on the order of 10\% for not obtaining the exact local minimum at a given background rate.
This is much less than the systematic uncertainty measured on the Crab Nebula excess in small footprint showers discussed in Chapter \ref{ch:Reco}.
Furthermore, this feature actually aids the optimization of our spatial bin size over the range of
searched detector zenith values as it allows a single bin to provide appreciable sensitivity despite the worsening intrinsic PSF of the detector at increasing
zenith due to shower attenuation in the larger atmospheric slant depth.

\newpage

\newpage

\newpage
\section{Optimal nHit Cut}\label{sec:nhitcut}

In this section we discuss the determination of the optimal shower size cut used in our analysis.
This cut is implemented as a requirement of having greater than a minimum number of PMTs, nHit, participate in an shower trigger
event processed by our search method. It is important for a number of reasons. First, the $E^{-2.7}$ spectrum of the hadronic air shower
background discussed in Chapter \ref{ch:showers} increases sharply with smaller shower sizes, which correspond to lower energy primaries, and
favors introducing a higher nHit cut. Second, the quality of the reconstructed angle of the shower primary degrades for smaller, low
energy showers due to the smaller amount of energy available for ground-level measurements in HAWC although this is offset somewhat
by the larger number of low energy photons in GRB spectra.

We use a single shower size cut rather than the 10 separate analysis
bins used in the point-source analysis of the Crab described in Chapter \ref{ch:Reco} because
the typical photon energies expected from GRB signals in HAWC represent a small fraction of
HAWC's sensitive energy range and all have similar footprints within the detector. The steeply falling
spectrum of hadronic showers means we do not need to include a cut on the maximum shower size as the 
background is dominated by low energy showers. Not including a maximum shower size cut also has the
added benefit of allowing for potentially extraordinary sensitivity to rare GRBs occurring within $z \approx 0.1$
where lower EBL attenuation supports appreciable transmission of TeV photons because the 
effective area of HAWC scales roughly as $E^2$ above 100 GeV \cite{Abeysekara:2013tza}.

A key aspect of determining the nHit cut is the fact that its optimization is 
convolved with optimizations of spatial bin size and gamma-hadron separation cuts.
This is because the detected shower size in HAWC determines both the angular resolution and
compactness (See Chapter \ref{ch:Reco}) associated with a given reconstructed shower.
We therefore perform an iterative approach where we first apply a nHit cut followed 
by optimization of the spatial square bin used in our search according to the method outlined in Section \ref{sec:poissonopt}
and we finish with optimization of a compactness cut. Compactness is optimized in the
same manner as the spatial bin cut, namely by Monte Carlo simulations of the signal level
needed to pass a 5$\sigma$ detection threshold in the Poisson regime after accounting for the cut efficiency. 
We do not apply a PINCness cut because photon signals from expected GRB emission are to small
to compute this variable to an uncertainty that provides appreciable discriminating power.

\newpage

\subsection{Differential Sensitivity} \label{sec:diffsens}

We begin by applying our optimization procedure to the two case examples, a short GRB and a long GRB, for GRB emission
shown in Table \ref{tab:grbmodels}.
Both GRBs are simulated at a redshift $z=0.5$ using the 2012 WMAP Fiducial EBL model \cite{GILMORE2012}. We chose this EBL model
because it is tuned to describe attenuation at the high redshifts where typical GRBs occur.
The EBL model effectively imposes a spectral cutoff that roughly corresponds to an exponential cutoff at 300 GeV.
We do not apply an intrinsic cutoff at this stage.

\newpage

\renewcommand{\arraystretch}{0.5}
\begin{table}[ht!]
\begin{center}
\begin{tabular}{|c|c|c|} \hline
  GRB Model & Index & Duration [sec] \\ \hline
   Short & -1.6 & 1 \\
   Long & -2.0 & 10 \\ \hline
\end{tabular}
\end{center}
\caption{Simulated GRB Models for determining optimal nHit, square bin size, and compactness cut.
         Flux is assumed to be constant over the full burst duration. The spectral indices are motived by fits
         to observed high energy power laws for GRB emission measured in the Fermi LAT \cite{Ackermann:2013a}. The 1 and 10 second times
         are chose because they approximately represent the timescales of the short and long GRB T90 distributions shown in Chapter \ref{ch:intro}.}
\label{tab:grbmodels}
\end{table}
\renewcommand{\arraystretch}{1.0}

\begin{figure}[ht!]
\begin{center}
\includegraphics[height=4in]{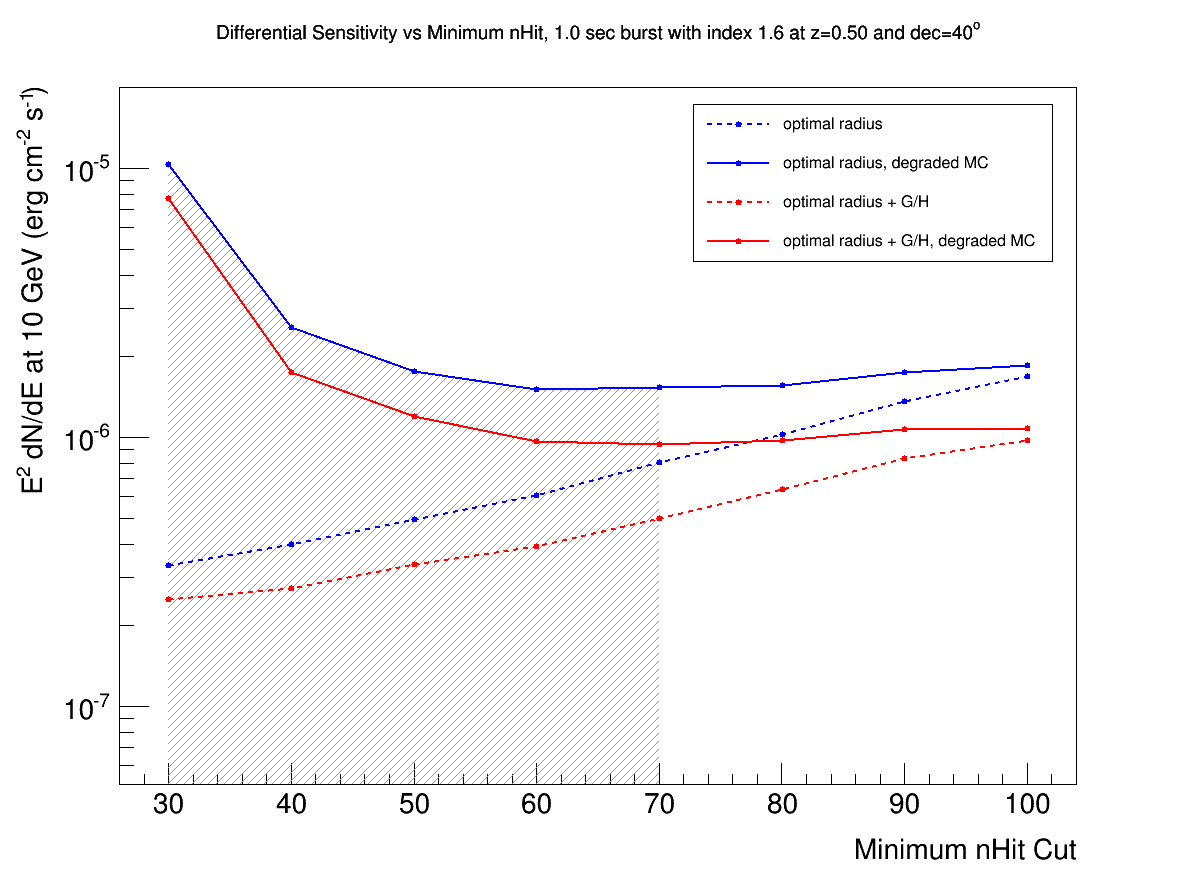}
\caption{Simulated differential sensitivity versus minimum nHit cut
         for our short GRB model (Table \ref{tab:grbmodels})
         after optimizing the spatial bin size and compactness cut individually for each nHit cut.
         The optimal nHit choice is nHit = 70 after accounting for the measured systematics
         that degrade sensitivity in real data (solid curves) and the region where compactness
         does not provide significant discrimination power in data (hatched region).} 
\label{fig:shortdiffsens}
\end{center}
\end{figure}
\begin{figure}[ht!]
\begin{center}
\includegraphics[height=4in]{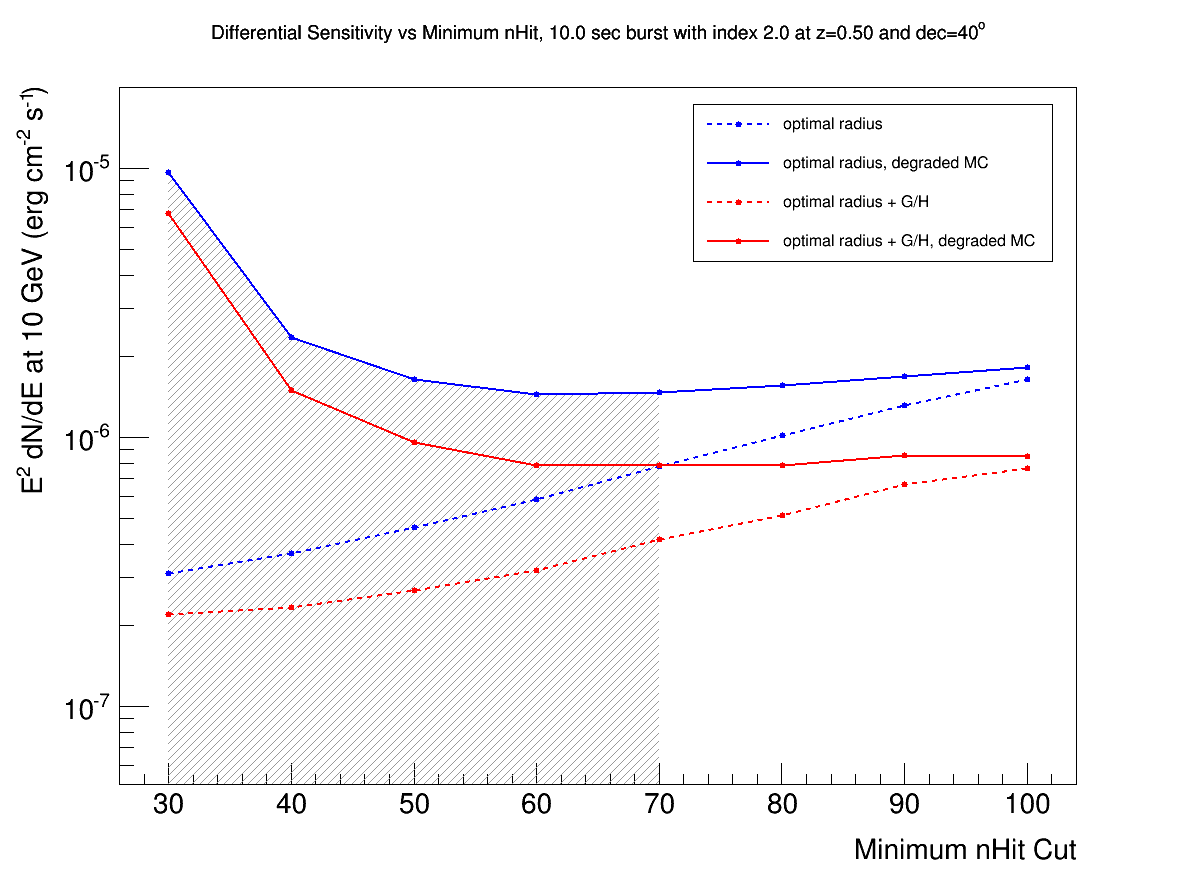}
\caption{Simulated differential sensitivity versus minimum nHit cut
         for our long GRB model (Table \ref{tab:grbmodels})
         after optimizing the spatial bin size and compactness cut individually for each nHit cut.
         The optimal nHit choice is nHit = 70 after accounting for the measured systematics
         that degrade sensitivity in real data (solid curves) and the region where compactness
         does not provide significant discrimination power in data (hatched region).} 
\label{fig:longdiffsens}
\end{center}
\end{figure}

Figures \ref{fig:shortdiffsens} and \ref{fig:longdiffsens}
show the differential sensitivity of these two bursts, defined as the normalization needed
to produce an average detection of 5$\sigma$ in a simulated set of 1000 realizations of the
expected signal and background counts within the GRB duration in the Poisson regime, as a function of the minimum
nHit cut for a detected zenith angle of 20$^\circ$. The differential sensitivity is reported in
units of the flux normalization at 10 GeV because this is directly comparable to normalizations
measured for known GRBs detected by Fermi LAT.
A major feature of these plots is the fact that we
degrade the number of signal photons reported directly from the Monte Carlo to account for the 
systematic error in the measured excess in the point-source analysis of the the Crab Nebula
signal shown in Chapter \ref{ch:Reco}. This significantly reduces the sensitivity of bins
defined by small values of nHit. Additionally, the hatched area represents where compactness
has shown no separation power in air shower data despite the Monte Carlo prediction for
it to have a small effect. The solid red curve indicating the sensitivity corrected for the
detector systematic cannot be used in this region. This leads to an optimal nHit cut of 70
in both the long and short GRB models. One interesting note is that the differential sensitivity
is the same in both models because the longer duration of the 10 second burst roughly accounts
for the spectral difference between the long and short GRB models at fixed flux normalization.
As will be seen in Section \ref{sec:sensflu}, this is not the case for fixed fluence.

\newpage

\newpage

\newpage

\subsection{Final Cuts}\label{sec:finalcuts}

Given that our search is self-triggered, we do not know the zenith angle, redshift, and intrinsic cutoff
of a burst prior to detecting it. We therefore
repeat the differential sensitivity calculation from Section \ref{sec:diffsens} for
the range of zenith angles, redshifts, and intrinsic cutoffs
shown in Table \ref{tab:paramspace} to obtain a search bin that is sensitive the range of burst parameters producing a detectable
number of photons greater than the $\sim$100 GeV shower threshold in HAWC.
We choose to use a maximum zenith angle of $\sim50^\circ$ because it corresponds to the slant depth of the atmosphere
where potential GRB signals are highly attenuated prior to reaching HAWC. The same is true for the choice of $z=1$
where the attenuation is due to EBL cutoff rather than the density of the atmosphere.
The range of intrinsic cutoffs spans the space between the $\sim$500 GeV cutoff dictated by EBL attenuation for $z=1$
and the $\sim$100 GeV shower threshold in HAWC. We find that the square bin size and compactness
cut in Table \ref{tab:fincut} yield a mean differential sensitivity that is only 15\% less sensitive
than individually tuning these values for each burst model defined by a unique combination
of zenith angle, redshift, and intrinsic and intrinsic cutoff. The median is well represented by the mean in this case.

\newpage

\renewcommand{\arraystretch}{0.5}
\begin{table}[ht!]
\begin{center}
\begin{tabular}{|c|c|c|} \hline
  Parameter & Simulated Range \\ \hline
   Zenith Angle & 1-51$^\circ$ (steps of 10$^\circ$) \\
   Redshift & 0.25-1.00 (steps of 0.25)) \\
   Intrinsic Cutoff & 150,250,500 GeV\\ \hline
\end{tabular}
\end{center}
\caption{Range of zenith angle, redshift, and intrinsic cutoff parameter
         space used to determine the 2.1$^\circ$ x  2.1$^\circ$ square bin
         and compactness $\mathcal{C} \, > 10$ cut used for the nHit cut of 70
         applied in our all-sky search method.}
\label{tab:paramspace}
\end{table}
\renewcommand{\arraystretch}{1.0}

\renewcommand{\arraystretch}{0.5}
\begin{table}[ht!]
\begin{center}
\begin{tabular}{|c|c|} \hline
  Parameter & Cut Value \\ \hline
   nHit & $\ge$70 \\
   Bin Size & 2.1$^\circ$ x 2.1$^\circ$  \\
   Compacteness ($\mathcal{C}$) & $>$10\\ \hline
\end{tabular}
\caption{Final cut values used in our search method.}
\label{tab:fincut}
\end{center}
\end{table}
\renewcommand{\arraystretch}{1.0}

\newpage

\begin{figure}[ht!]
\begin{center}
\includegraphics[height=3in]{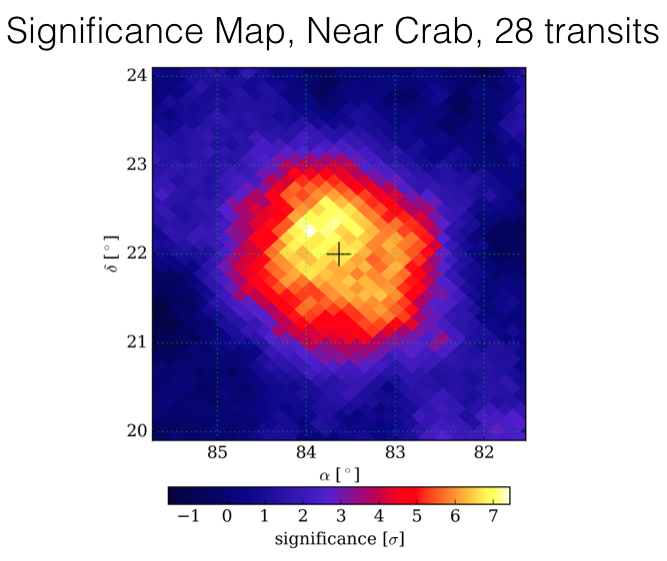}
\caption{Map of the Crab Nebula in a point-source
         analysis using the cut values defined in Table \ref{tab:fincut}.}
\label{fig:nearcrab}
\end{center}
\end{figure}

Simulation predicts that the cuts in Table \ref{tab:fincut}
retain $\sim$75\% of the original gamma-ray signal in our modeled set of GRBs.
In order to demonstrate that these cuts successfully retain gamma-ray signals
in data as well we apply them
in a standard-point source analysis of the Crab Nebula
over 1 month of HAWC data (Figure \ref{fig:nearcrab}). The Crab Nebula is clearly detected above 5$\sigma$ and confirms that our
cuts do provide sensitivity to gamma-ray air showers in HAWC. 

\vspace{3cm}

\newpage

Table \ref{tab:backpass} shows the background passing rates for ten minutes of
data taken on Feb 28, 2016 for the cuts defined in Table \ref{tab:fincut}. The 23.6 kHz
rate for all events represents the total air shower trigger rate set by the trigger criterion
of 28 PMT waveforms arriving within 150 ns discussed in Chapter \ref{ch:hawcobs}. This is reduced
to 6.6 kHz by applying the minimum nHit cut of 70 PMTs participating in the reconstructed shower.
It results in a rate of 7.7 Hz for background air showers arriving in the 2.1$^\circ \times 2.1^\circ$
spatial bin located at detector zenith. Applying the compactness cut further reduces the rate in the bin at zenith
to 0.9 Hz. Although much smaller than the total air shower rate, the final rate of 0.9 Hz obtained after applying
all cuts is still significant enough to reduce the fluence sensitivity of our analysis to long GRBs (Section \ref{sec:sensflu}).

We chose to assess the background rate in a bin located at detector 
zenith because it represents the largest background rate observed in our analysis.
This is because the atmospheric slant depth seen by air showers increases as a function of zenith angle, thereby reducing both the rate
of background and signal showers reaching ground level for increasing zenith angle. The background rate at 45$^\circ$, for example, is
reduced by a factor of 10 compared to the rate at zenith.

\renewcommand{\arraystretch}{0.5}
\begin{table}[ht!]
\begin{center}
\begin{tabular}{|c|c|}\hline
Selection & Passing Rate \\
\hline
All Events & 23.6 kHz\\
nHit $\ge$ 70 & 6.6 kHz \\
2.1$^\circ$ Bin at Zenith & 7.7 Hz \\
Gamma-Hadron Cut & 0.9 Hz \\ \hline
\end{tabular}
\caption{Background passing rates for data taken on Feb 28, 2016
         for the cuts defined in Table \ref{tab:fincut}.}
\label{tab:backpass}
\end{center}
\end{table}

\newpage

\subsection{Sensitivity to Fluence}\label{sec:sensflu}

\begin{figure}[ht!]
\begin{center}
\includegraphics[height=3in]{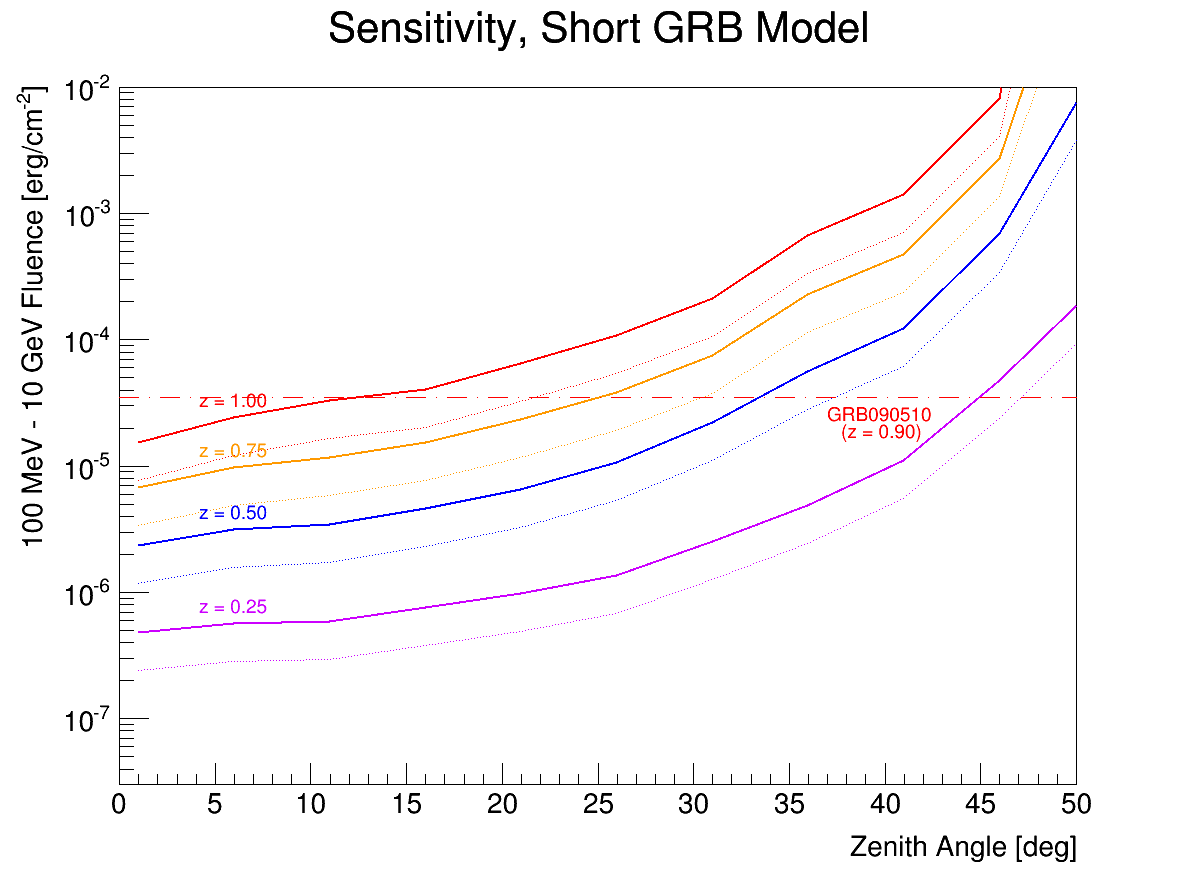}
\caption{Short GRB sensitivity in terms of the fluence required
         to obtain a 5$\sigma$ detection in 50\% of
         bursts at each redshift for a given zenith angle.
         The dashed line marks the measured fluence of GRB 090510
         which had a redshift of $z = 0.90$ \cite{Ackermann:2013a}.
         GRB 090510 would be detectable in both our all-sky search method
         and a triggered search method if it occurred at a favorable zenith angle.}
\label{fig:flushort}
\end{center}
\end{figure}

\begin{figure}[ht!]
\begin{center}
\includegraphics[height=3in]{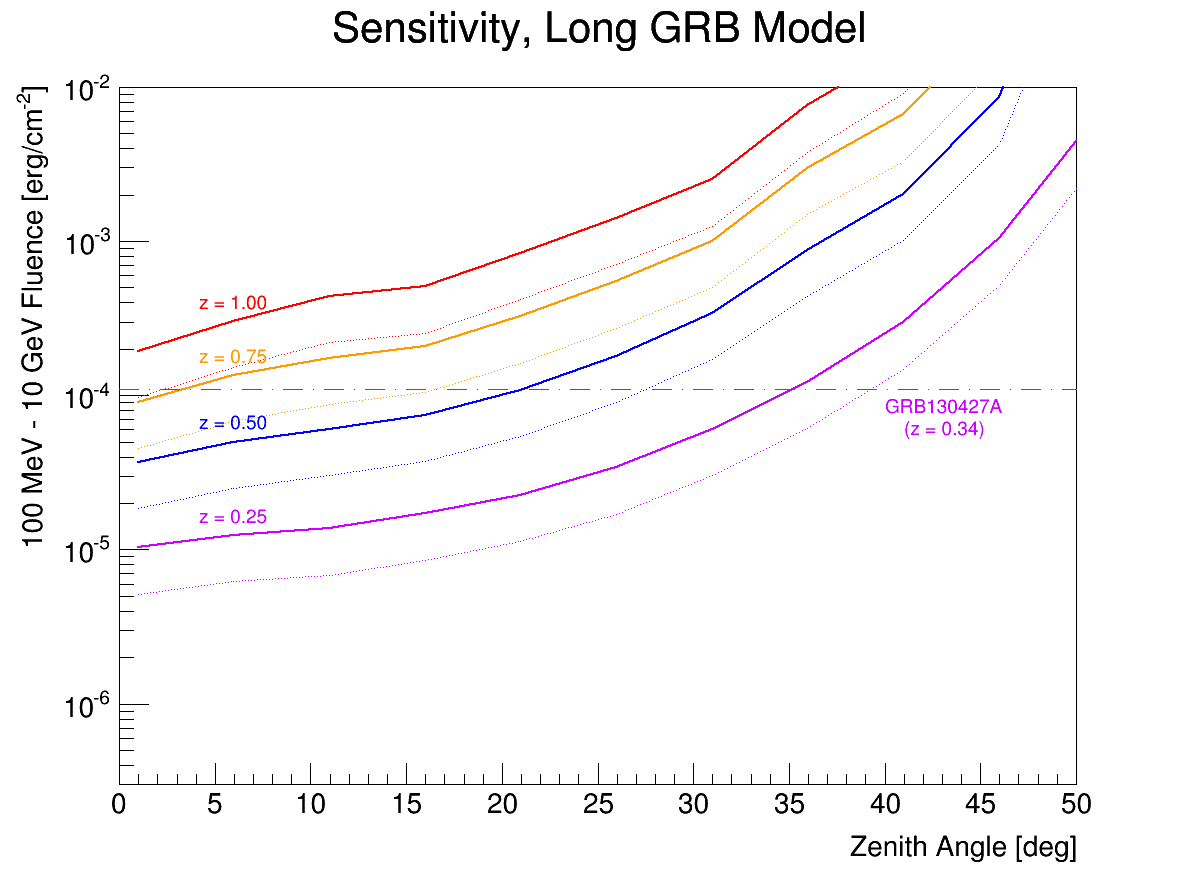}
\caption{Long GRB sensitivity in terms of the fluence required
         to obtain a 5$\sigma$ detection in 50\% of
         bursts at each redshift for a given zenith angle.
         The dashed line marks the measured fluence of GRB 130427A
         which had a redshift of $z = 0.34$.
         GRB 130427A would be easily detectable in both our all-sky search method
         and a triggered search method. The fluence value for GRB 130427A is technically
         for $>$100 MeV \cite{Zhu:2013GCN} rather than being restricting by an upper bound of 10 GeV but it
         still provides a representative estimate of the fluence in the 100 MeV - 10 GeV
         band.}
\label{fig:flulong}
\end{center}
\end{figure}

While we have shown sensitivity to the Crab Nebula
with our cuts we would also like to show that our cuts are sensitive
enough to detect high energy GRBs seen by the Fermi LAT. We therefore
generate curves in Figures \ref{fig:flushort} and \ref{fig:flulong} for the fluence range of 100 MeV - 10 GeV typically
reported in very bright GRBs by the Fermi LAT. The fine-dashed curves colored according to redshift
represent an average burst detection of 5$\sigma$ for a single trial analysis
using our optimized cuts from Section \ref{sec:diffsens}. The solid curves account for the
$\sim$2x sensitivity loss of our search compared to the single trial case after
accounting for trials as described in Chapter \ref{ch:method}. Overall
these figures show that HAWC is more sensitive to the short GRB burst model.
This is because fluence, unlike the differential flux discussed early,
is an integral over the duration of emission. At fixed fluence then, the long GRB model is less
sensitive because it contains 10x more background than the short GRB model for the same integral of signal photons.
This is compounded by the high energy index of typical long GRBs which is softer than for short GRBs.

The fluences of two seminal bursts, GRB 090510 and GRB 130427A, are also shown in Figures \ref{fig:flushort} and \ref{fig:flulong}
with dot-dashed lines. We find we can detect the short burst GRB 090510 out to a zenith angle of 10$^\circ$
even with HAWC's currently degraded sensitivity and the relatively large redshift of this burst for HAWC.
We also find that GRB 130427A is easily detectable out to a zenith angle of $\sim$25$^\circ$ given
that HAWC's sensitivity to its redshift of $z=0.34$ will be similar to the curve shown for $z=0.25$.
Additionally, we note that the single trial curves,
shown with dashed lines for each simulated redshift, correspond to HAWC's sensitivity
to these same bursts in the triggered search analysis that runs in parallel to our method.
In this case, the use of a single trial lowers the overall sensitivity enough to view
GRB 090510 and GRB 130427A out to zenith angles of about 20$^\circ$ and 40$^\circ$, respectively.
These would be easily detectable if they were to trigger in a satellite coincident with the HAWC field of view.

\newpage

\newpage

\section{Optimal Search Duration}\label{sec:duropt}

As we showed in the previous section, the sensitivity of our search is strongly related to the
duration needed to encompass gamma-ray burst emission. This is because the hadronic air shower background in HAWC
scales linearly with the width of the time window, effectively burying the signal inside fluctuations in the background counts.
We must therefore carefully choose the durations over which we perform our search for GRB emission.

We begin this process by noting the temporally extended emission measured for bright bursts in Fermi LAT quickly decays
as $\sim t^{-1.5}$ after the end of the low energy T90 measured in Fermi GBM \cite{Ghisellini:2010}.
This decay is so rapid that any time window integrating over this shape in HAWC is collecting
more background events without significantly increasing the number of signal photons.
This combined with the occurrence of peak GeV flux inside low energy T90 \cite{Beloborodov:2014} convinces
us we should be looking for VHE emission associated with the prompt light curve of the GRB.

We therefore wish to analyze a set of characteristic light curves for high energy emission
during the prompt phase of the GRB to tune the width of our sliding time window. Such a data set is available from the Fermi LAT collaboration in the 
form of LAT Low Energy (LLE) light curves \cite{Ferrara:2015} for 50 LAT bursts. These data have high enough statistics to provide well mapped time structure, are easy to analyze compared
to a full analysis of LAT transient events, and are readily available from a public database of LAT GRBs \cite{LAT:2016}.
Furthermore, they exhibit much of the behavior found in a more detailed LAT analysis with a higher energy threshold, such as a delayed start time
compared to GBM T90.

\begin{figure}[ht!]
\begin{center}
\includegraphics[height=3in]{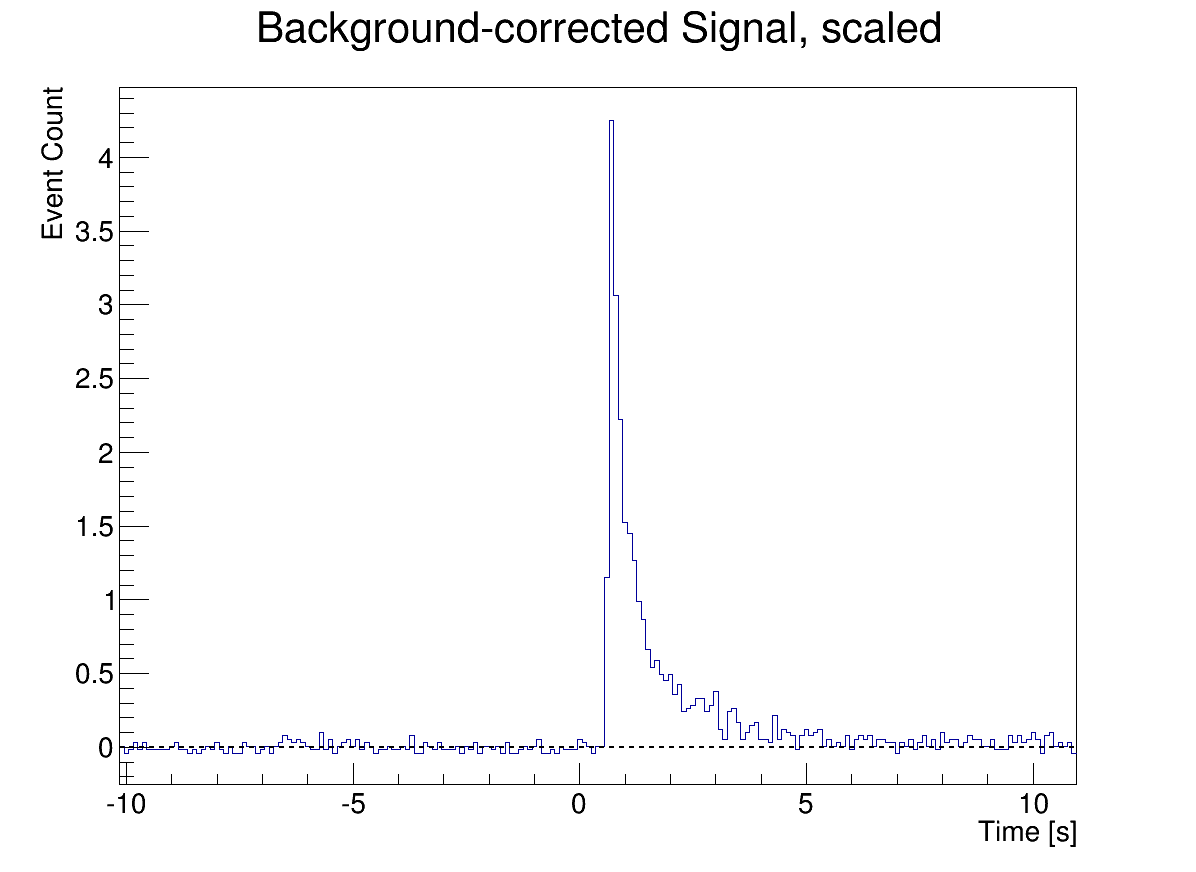}
\caption{Background corrected LLE light curve for GRB 090510.}
\label{fig:bgcorrlc}
\end{center}
\end{figure}

\begin{figure}[ht!]
\begin{center}
\includegraphics[height=3in]{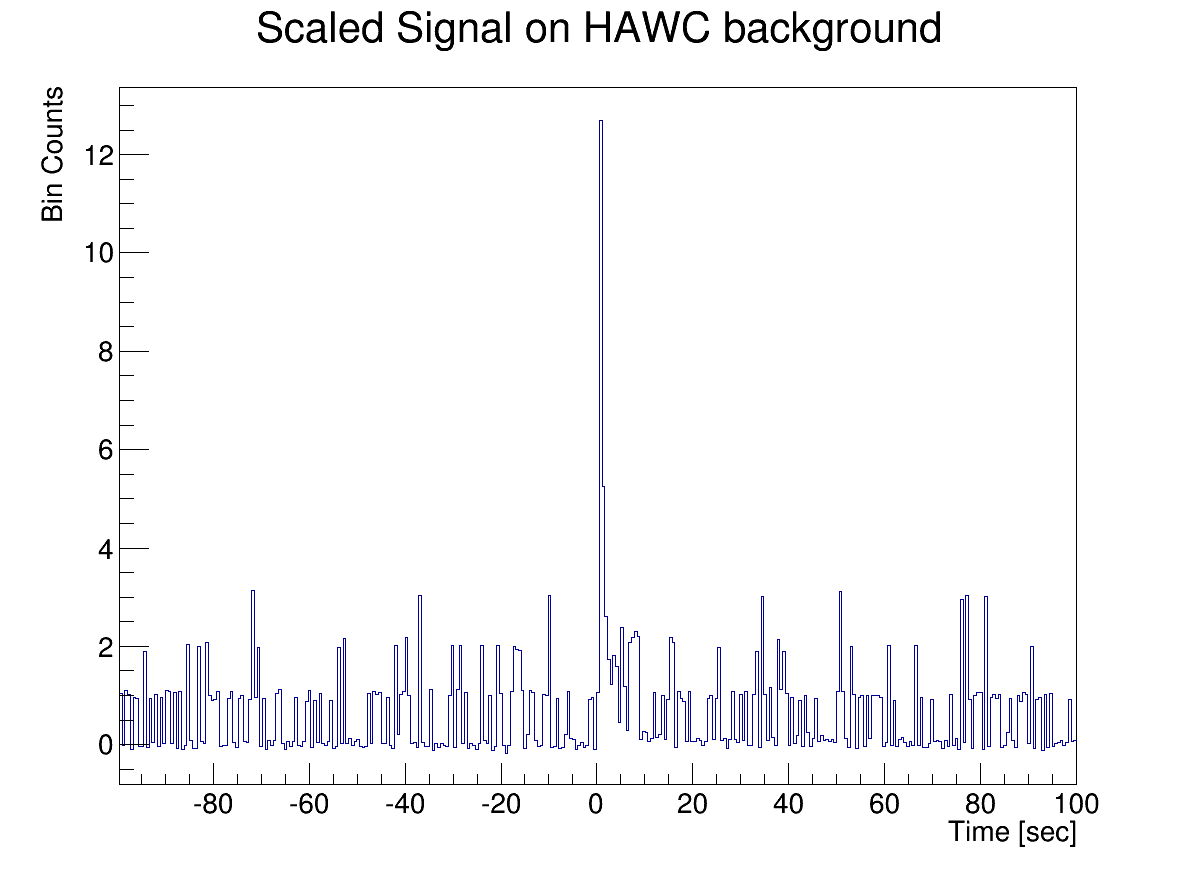}
\caption{LLE light curve for GRB 090510 injected on HAWC background at zenith
         for the 2.1$^\circ$ x 2.1$^\circ$ spatial bin used in the all-sky search.}
\label{fig:lcinhawc}
\end{center}
\end{figure}

To analyze each light curve in the context of our time window analysis, we perform a background correction
on the original LLE light curve and place it on top of a randomly thrown background that matches the rate
of air shower events expected in HAWC at zenith for the 2.1$^\circ$ x 2.1$^\circ$ spatial bin size of the all-sky search (Figures \ref{fig:bgcorrlc} and \ref{fig:lcinhawc}).
This effectively models the light curve as it would appear on top of the hadronic air shower background in our search.
We use a linear fit to events that are 50 seconds outside the reported low and high energy T90s for each burst
to perform the background correction. This results in good agreement with an average of zero counts before and after the
extent of the light curve.

\vspace{1cm}

\newpage
\subsection{Spanning All Durations}\label{sec:spandur}

Initially, we search the set of 50 LLE light curves (46 long, 4 short GRBs) over a range of time windows that span typical burst durations (0.1 - 100) seconds.
to test the effect of saturating timescale space. 
This range covers both the total duration as well as the duration of substructure inside most GRB light curves.
Our goal is to understand if taking many trials reduces the sensitivity of the search or if the additional ability
gained from fine tuning on each light curve yields more post-trials discoveries.

We model the temporal part of our all-sky search by sliding different time windows over the time range defined by the background fit of the original LLE light curve
to find the largest counts excess. 
We calculate the probability of this excess using a Poisson distribution  with the true mean
used to randomly generate the HAWC background. We correct for trials taken in the search with the total number of search windows. This
represents an upper limit due to the correlations introduced by our sliding window (See Chapter \ref{ch:method}).

For comparison, we also assess a single window that exactly matches the T90 reported by Fermi GBM.
This represents the best case scenario for a triggered search as a low energy T90 is usually reported by most
satellite triggers. However, it does not necessarily indicate an optimal search window as the start time of
LLE light curves typically occur after the start of low energy T90 \cite{Beloborodov:2014}.

Tables \ref{tab:spanlong} and \ref{tab:spanshort} presents the number of detected bursts from both the single, T90 window and the search method spanning (0.1 - 100) seconds
for fixed signal levels inside the low energy T90. The search method typically recovers many more near threshold detections than simply looking
  in T90 alone. This is because the search method is able to locate the start time and duration that maximize the number of signal
  events over background in each light curve. The fact that the search method efficiently detects bursts at
  nearly the same level as the 
  single trial in the 7$\sigma$ signal scaling shows that the extra trials taken by spanning the full timescale space of prompt
  emission, which are extraneous in this case because there is a guaranteed discovery, do not significantly reduce the search's ability
  to discover transient events. We therefore conclude that taking many trials when you expect to see a near threshold detection,
  as is the case if a burst like GRB 130427A goes off inside the HAWC field-of-view, actually enhances the overall sensitivity of the 
  experiment.

\newpage

\renewcommand{\arraystretch}{0.5}
\begin{table}[ht!]
\begin{center}
\begin{tabular}{|c|c|c|c|} \hline
   & \multicolumn{3}{|c|}{Injected Signal Level} \\
  Timescale [sec] & 4$\sigma$ & 5 $\sigma$ & 7 $\sigma$ \\ \hline \hline
   GBM T90 & 6/46 & 23/46 & 42/46 \\ \hline \hline
   100 & 5/46 & 9/46 & 27/46 \\
   56 & 8/46 & 17/46 & 32/46 \\
   31 & 11/46 & 21/46 & 32/46 \\
   17 & 14/46 & 23/46 & 37/46 \\
   10 & 14/46 & 21/46 & 38/46 \\
   5 & 14/46 & 18/46 & 37/46 \\
   3 & 13/46 & 17/46 & 35/46 \\
   2 & 13/46 & 19/46 & 28/46 \\
   1 & 15/46 & 18/46 & 24/46 \\
   0.5 & 13/46 & 16/46 & 21/46 \\
   0.3 & 13/46 & 15/46 & 16/46 \\
   0.2 & 10/46 & 13/46 & 15/46 \\
   0.1 & 9/46 & 11/46 & 14/46 \\ \hline \hline
   All Searches & 19/46 & 30/46 & 45/46 \\ \hline
\end{tabular}
\end{center}
\caption{Number of $>5\sigma$ light curve detections found inside T90 versus sliding time window searches
         for a fixed signal level injected into T90 for all 46 long GRBs modeled from LLE light curve data.
         The range of search time windows spans (0.1 - 100) seconds in logarithmically spaced intervals.
         The search window discoveries are trials-corrected for the number of start time positions searched
         by each window as well as the total number of windows. The final row labeled all searches represents
         the total number of unique light curves discovered from the set of search durations after trials correction.}
\label{tab:spanlong}
\end{table}
\renewcommand{\arraystretch}{1.0}

\newpage

\renewcommand{\arraystretch}{0.5}
\begin{table}[ht!]
\begin{center}
\begin{tabular}{|c|c|c|c|} \hline
   & \multicolumn{3}{|c|}{Injected Signal Level} \\
  Timescale [sec] & 4$\sigma$ & 5 $\sigma$ & 7 $\sigma$ \\ \hline \hline
   GBM T90 & 0/4 & 3/4 & 4/4 \\ \hline \hline
   100 & 0/4 & 0/4 & 1/4 \\
   56 & 0/4 & 0/4 & 0/4 \\
   31 & 0/4 & 0/4 & 0/4 \\
   17 & 0/4 & 0/4 & 0/4 \\
   10 & 0/4 & 0/4 & 1/4 \\
   5 & 0/4 & 0/4 & 2/4 \\
   3 & 0/4 & 1/4 & 2/4 \\
   2 & 0/4 & 1/4 & 2/4 \\
   1 & 0/4 & 1/4 & 4/4 \\
   0.5 & 0/4 & 2/4 & 4/4 \\
   0.3 & 0/4 & 2/4 & 3/4 \\
   0.2 & 0/4 & 2/4 & 3/4 \\
   0.1 & 0/4 & 1/4 & 3/4 \\ \hline \hline
   All Searches & 0/4 & 3/4 & 4/4 \\ \hline
\end{tabular}
\end{center}
\caption{Number of $>5\sigma$ light curve detections found inside T90 versus sliding time window searches
         for a fixed signal level injected into T90 for all 4 short GRBs modeled from LLE light curve data.
         The range of search time windows spans (0.1 - 100) seconds in logarithmically spaced intervals.
         The search window discoveries are trials-corrected for the number of start time positions searched
         by each window as well as the total number of windows. The final row labeled all searches represents
         the total number of unique light curves discovered from the set of search durations after trials correction.
         Long search time windows typically do not detect short GRBs. The only exception to this is the 100 second.}
\label{tab:spanshort}
\end{table}
\renewcommand{\arraystretch}{1.0}

\newpage

\subsection{Using Three Durations}\label{threedur}

We now explore the question of whether we can retain the same efficiency for detecting
near-threshold transients using fewer time windows than Section \ref{sec:spandur}. We motivate this
by noting that many of the time windows used in Tables \ref{tab:threelong} and \ref{tab:threeshort} discover similar numbers of
bursts with appreciably increasing the number of total discovered bursts. This indicates
there is significant overlap between the sets of bursts discovered in each window.

To do this we begin by noting that the 1 second and 10 second windows are separated
by an order of magnitude and recover the largest number of discoveries within the
set of 46 long GRBs. This is intuitive in that typical long GRBs have light curve pulse
widths on the order of 1 second and total durations on the order of 10 seconds.
We then note the 0.2 second window represents roughly the pulse width size expected in short GRBs
and yields the largest number of distinct burst discoveries when used with the 1 second window.
We therefore repeat the analysis in Section \ref{sec:spandur} with the 0.2 second, 1 second, and 10 second timescales
to see how it compares against spanning the full timescale space. Overall it yields roughly the same
total number of burst detections, revealing that taking more trials involves diminishing returns.
We therefore choose to use these three timescales because it achieves effectively
the same sensitivity while also reducing the computing core hours needed to complete our search.

\renewcommand{\arraystretch}{0.5}
\begin{table}[ht!]
\begin{center}
\begin{tabular}{|c|c|c|c|} \hline
   & \multicolumn{3}{|c|}{Injected Signal Level} \\
  Timescale [sec] & 4$\sigma$ & 5 $\sigma$ & 7 $\sigma$ \\ \hline \hline
   GBM T90 & 6/46 & 23/46 & 42/46 \\ \hline \hline
   10 & 15/46 & 22/46 & 39/46 \\
   1 & 15/46 & 18/46 & 24/46 \\
   0.2 & 10/46 & 13/46 & 15/46 \\ \hline \hline
   All Searches & 18/46 & 26/46 & 42/46 \\ \hline
\end{tabular}
\end{center}
\caption{Number of $>5\sigma$ light curve detections found inside T90 versus sliding time window searches
         for a fixed signal level injected into T90 for all 46 long GRBs modeled from LLE light curve data.
         The search window discoveries are trials-corrected for the number of start time positions searched
         by each window as well as the total number of windows. The final row labeled all searches represents
         the total number of unique light curves discovered from the set of search durations after trials correction.}
\label{tab:threelong}
\end{table}
\renewcommand{\arraystretch}{1.0}

\renewcommand{\arraystretch}{0.5}
\begin{table}[ht!]
\begin{center}
\begin{tabular}{|c|c|c|c|} \hline
   & \multicolumn{3}{|c|}{Injected Signal Level} \\
  Timescale [sec] & 4$\sigma$ & 5 $\sigma$ & 7 $\sigma$ \\ \hline \hline
   GBM T90 & 0/4 & 3/4 & 4/4 \\ \hline \hline
   10 & 0/4 & 0/4 & 1/4 \\
   1 & 0/4 & 1/4 & 4/4 \\
   0.2 & 1/4 & 2/4 & 3/4 \\ \hline \hline
   All Searches & 1/4 & 3/4 & 4/4 \\ \hline
\end{tabular}
\end{center}
\caption{Number of $>5\sigma$ light curve detections found inside T90 versus sliding time window searches
         for a fixed signal level injected into T90 for all 4 short GRBs modeled from LLE light curve data.
         The search window discoveries are trials-corrected for the number of start time positions searched
         by each window as well as the total number of windows. The final row labeled all searches represents
         the total number of unique light curves discovered from the set of search durations after trials correction.}
\label{tab:threeshort}
\end{table}
\renewcommand{\arraystretch}{1.0}



\newpage

\renewcommand{\thechapter}{7}

\chapter{Results}\label{ch:results}

This chapter presents the results from the all-sky, self-triggered
search described in Chapter \ref{ch:method} for the three timescales, 0.2 seconds, 1 second, and 10 seconds,
and sensitivity optimizations outlined in Chapter \ref{ch:sens}. These results are based
on the latest available off-site data reconstructed using the Pass 4 algorithms described in Chapter \ref{ch:Reco}.
These data consist of approximately one year of data with the full HAWC detector and are described in detail
in Section \ref{sec:dataset}.
Section \ref{sec:bestcand} presents the best candidates for transient VHE emission found in
each timescale. They are all consistent with cosmic-ray air shower backgrounds after accounting
for trials. Section \ref{sec:limits} uses this null detection in conjunction with the search sensitivity
outlined in Chapter \ref{ch:sens} to place upper limits on the rate per year of GRBs with high-energy emission.

\newpage

\section{Data Set}\label{sec:dataset}

The start of our data set is marked by the inauguration of the HAWC observatory
on March 19, 2015. This date represents the first day of stable operation of the full detector.
The end date of our data set is March 1, 2016 and corresponds to the most recent Pass 4 reconstructed data available
off-site at the time of this analysis. We choose to use a shower reconstruction produced off-site because it is
the most sensitive reconstruction to date. 

We apply a data quality selection requiring that the detector remain in continuous operation
for at least the 1.75 hour duration needed to build the acceptance map used for the background
calculation in our search method. This cut excludes 17.8 days of reconstructed data but still yields
a total searched time of 295.9 days that covers 85\% of the total live time.
This is far greater than can be achieved by any IACT and represents an enormous amount
of sky coverage compared to the field-of-view of an IACT given the 50$^\circ$ zenith cut used¬
in our analysis.

\begin{figure}[ht!]
\begin{center}
\includegraphics[height=3in]{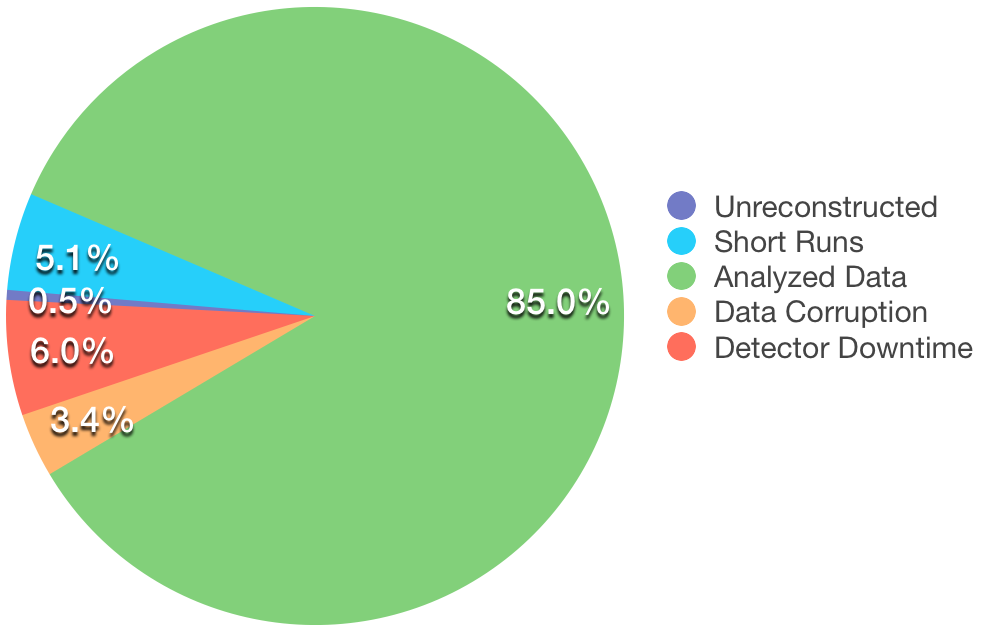}
\caption{Chart showing the distribution of the total 348.3 day live time
         of the data used in our search.
         The vast majority of live time (85\%)
         was successfully analyzed for GRB transients. 6\% of live time was lost 
         due to down time for detector maintenance, 3.4\% was lost to due data corruption
         introduced by failure of a disk array during a power outage, and 5.1\% was excluded
         from the search due to our stability requirement of at least 1.75 hours of continuous detector operation.
         0.5\% of data were accidentally left unreconstructed during creation of the Pass 4 data set due to a database error
         and are currently undergoing reconstruction.}
\label{fig:datapie}
\end{center}
\end{figure}

The total live time of our data set is 348.2 days during which there were 
roughly 21 days of downtime for detector maintenance. Part of this time 
involved recovery from an exceptional power outage that corrupted 11.7 days of
data recorded prior to the shutdown. Additionally, 1.8 days were excluded from the
creation of the reconstructed data due to a database error. Figure \ref{fig:datapie} shows 
a breakdown of the percentage of live time occupied by each type of data loss as well
as the amount of data analyzed by our search.

\newpage

\newpage

\section{Candidate Events}\label{sec:bestcand}

Table \ref{tab:besties} presents the best candidates for VHE gamma-ray transients
from each of the three time searches. The pre-trials probabilities of these candidates
represent significant fluctuations in the background for the single trial case but are
not significant after accounting for trials.  We obtain the temporal and spatial trials
shown for each search by applying the methods for trial calculation described in Chapter \ref{ch:method}.
We multiply these values together to obtain the total searched trials within a given window.
We then use the total number of trials in each window to compute the post-trial probability shown in Table \ref{tab:besties}.

\renewcommand{\arraystretch}{0.5}
\begin{table}[ht!]
\begin{center}
\begin{tabular}{|c|c|c|c|c|} \hline
  Duration  & Pre-Trial   & Effective      & Effective       & Post-Trial \\
  (seconds) & Probability & Spatial Trials & Temporal Trials & Probability \\ \hline
  0.20  & 3.91$\times$10$^{-14}$ & 1.37$\times$10$^4$ & 1.15$\times$10$^9$ & 0.46 \\
  1.00  & 8.97$\times$10$^{-15}$ & 3.54$\times$10$^4$ & 2.25$\times$10$^8$ & 0.07 \\
  10.00 & 2.51$\times$10$^{-13}$ & 1.06$\times$10$^5$ & 2.36$\times$10$^7$ & 0.47 \\ \hline
\end{tabular}
\caption{Best candidate obtained in each search duration. These candidates have significant
         pre-trial probabilities but are consistent with background after accounting for
         trials. Effective spatial and temporal trials are calculated according to the methods
         outlined in Chapter \ref{ch:method}.}
\label{tab:besties}
\end{center}
\end{table}
\renewcommand{\arraystretch}{1.0}

\newpage

\begin{figure}[ht!]
\begin{center}
\includegraphics[height=3in]{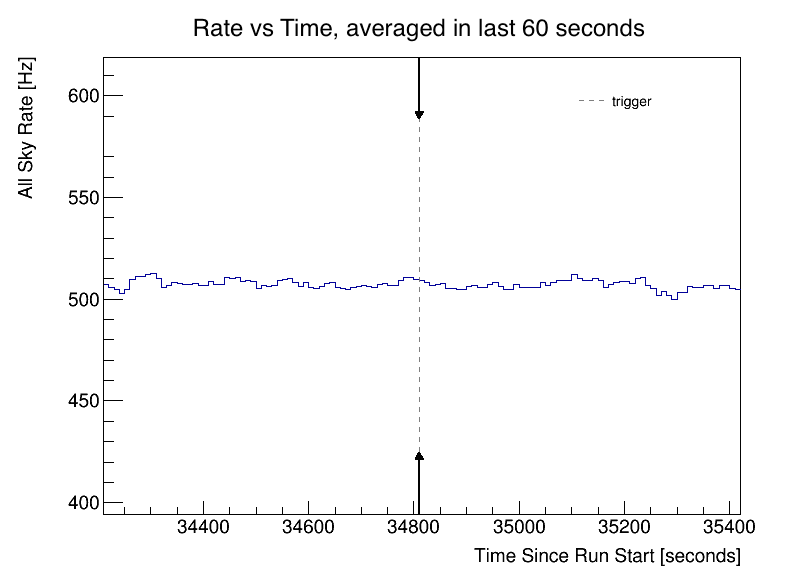}
\caption{All-sky rate near the best candidate averaged in a 60 second long sliding window
         shifted in steps of 10 seconds. The rate is stable indicating normal detector operation.
         The excess at the candidate does not result from detector instabilities.}
\label{fig:rateplot}
\end{center}
\end{figure}

\begin{figure}[ht!]
\begin{center}
\includegraphics[width=6in]{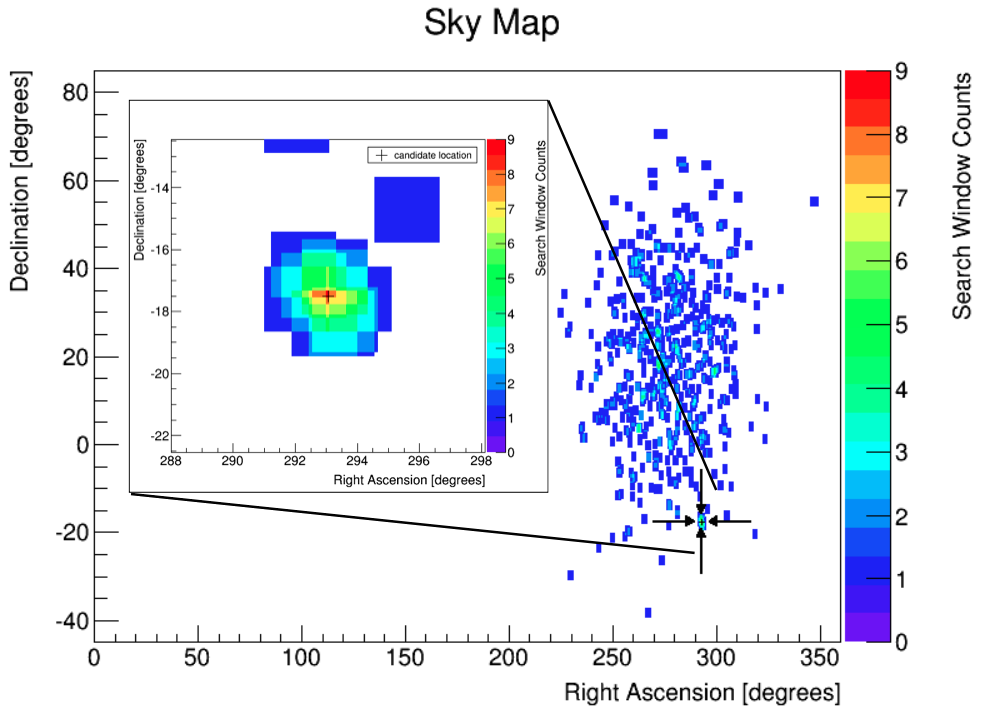}
\caption{Sky map from the 1 second window containing the best candidate event in the all-sky GRB search. The cross
         marking the location of the candidate is at a detector zenith angle of 40.5$^\circ$.}
\label{fig:myskymap}
\end{center}
\end{figure}

\begin{figure}[ht!]
\begin{center}
\includegraphics[height=3in]{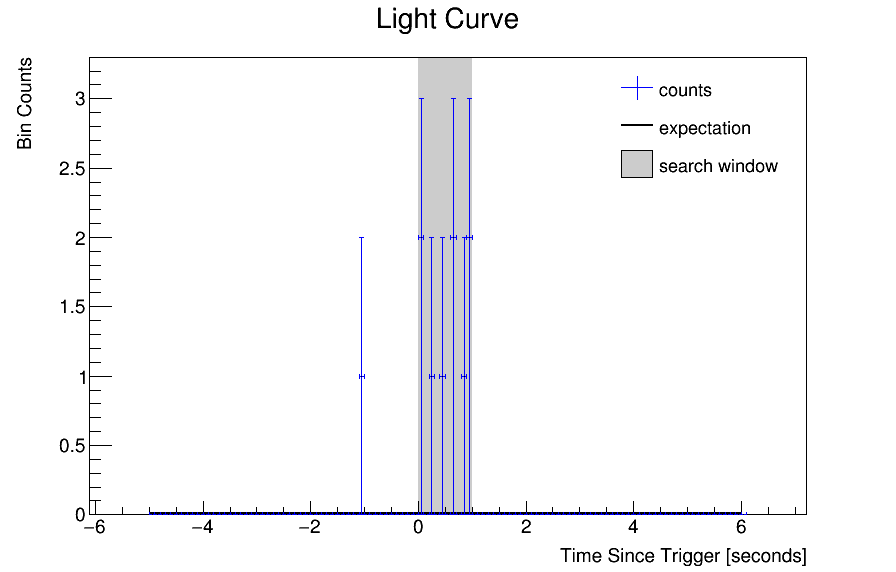}
\caption{Light curve of event counts binned in intervals of the sliding time window step size $t_{step} = 0.1 \times t_{window}$
         for the location of the best candidate. Unlike Figure \ref{coarselc}, these intervals are independent. The 1 second window
         containing the best candidate is shaded in GREY and the BLACK line marks the background expectation
         in each light curve bin. The background expecation is so low that it cannot be distinguished
         from zero in this plot.
         }
\label{fig:thatsfine}
\end{center}
\end{figure}

The overall best candidate was found inside the 1 second search window and
occurred on May 30, 2015 at the location reported in Table \ref{tab:1seccand}.
The all-sky rate shown in Figure \ref{fig:rateplot} shows the HAWC detector was
stable near the time of the candidate event.
The sky map for the window containing this candidate is shown in Figure \ref{fig:myskymap} and its light curve, binned
in intervals of the 0.1 second time step used to advance the 1 second sliding window, can be seen in Figure \ref{fig:thatsfine}.

\begin{figure}[ht!]
\begin{center}
\includegraphics[height=3in]{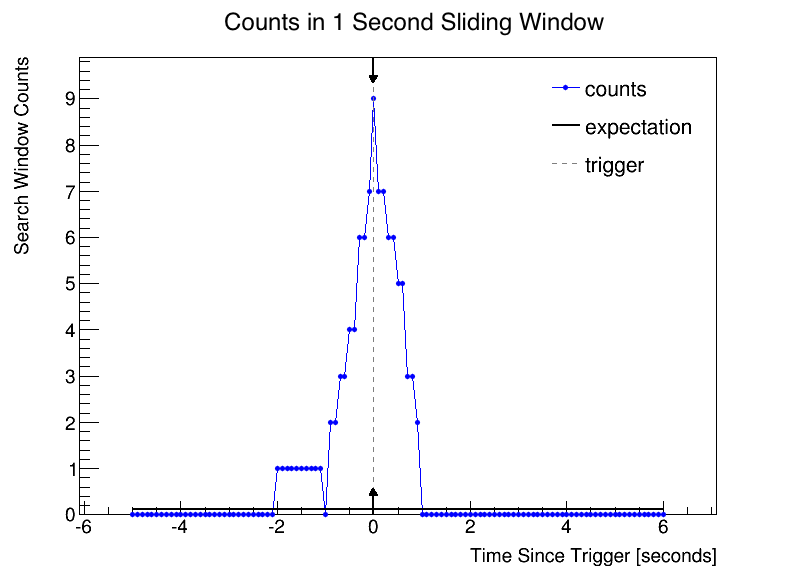}
\caption{Light curve of observed counts detected within the 1 second sliding
         window at the location of the best candidate. The ordinate represents
         the start time of the 1 second search window. Error bars are not shown
         as this figure is only intended to demonstrate the raw event counts assessed by 
         the search window. The background expectation is shown in black. Its uncertainty is
         much smaller than can be seen on the scale of maximum counts in the window.
         Adjacent points correspond to time windows that overlap by 90\%, introducing
         strong correlations between points. These correlations account for step function
         shape from -2 to -1 seconds which is the result of a single event moving through
         ten steps of the time window. The peaked nature of the light curve which increases to
         a maximum as the sliding window includes more events while
         moving forward in time is also the result of overlap between the time bins.
         The decay phase occurs after reaching the maximum as more events shift to being outside the time window.}
\label{coarselc}
\end{center}
\end{figure}

This candidate has a pre-trial probability of 8.97$\times10^{-15}$ corresponding to 9 observed counts on a
background of 0.115 (Figure \ref{coarselc}). Applying an additional factor of 3 trials to account
for choosing the best result from the total of 3 windows increases the post-trials probability from 0.07 to 0.19.
This represents the independent-trial upper limit for using 3 different windows on the same data set
but cannot be significantly different than accounting for the effective number of trials as
the correlations induced by running all three windows over the same data set must still yield a number of trials $\ge$1.

\begin{figure}[ht!]
\begin{center}
\includegraphics[width=6in]{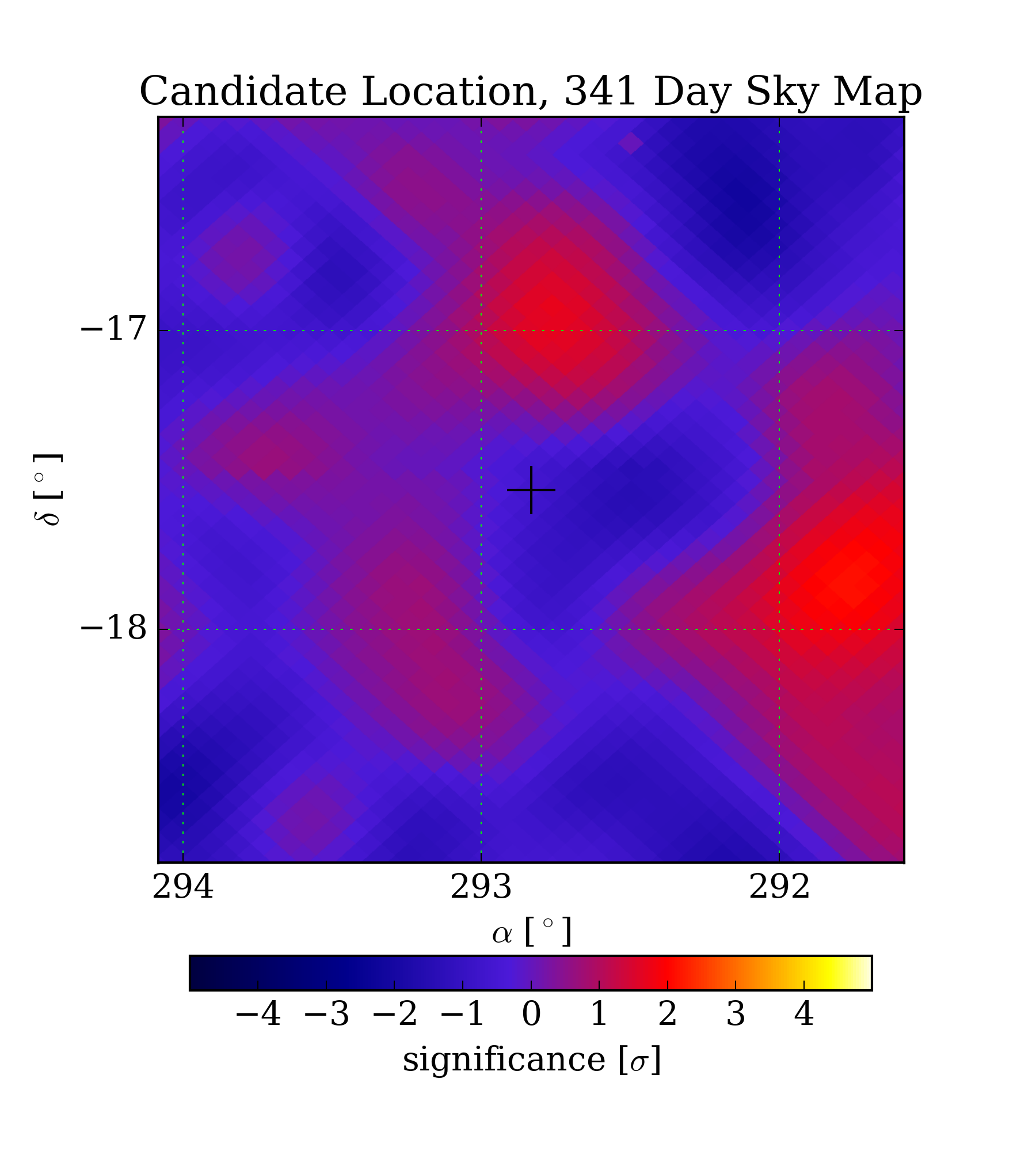}
\caption{View of candidate location in standard HAWC point source analysis with 341 days of data.
         There are no significant sources near this location. The Crab significance is $\sim$80$\sigma$ in this data set.}
\label{fig:psmap}
\end{center}
\end{figure}

This yields a post-trials probability at the 1$\sigma$ level, which is not significant though we note only 3 additional counts were needed to yield a 5$\sigma$ result.
There are no transients reported by other experiments near the location of this candidate at the time of its trigger within HAWC.
Furthermore, there are no indications of a steady-state source at the location of this candidate in the point-source
sky map produced with 341 days of HAWC data (Figure \ref{fig:psmap}). All evidence suggests this candidate is not a significant event.

\renewcommand{\arraystretch}{0.75}
\begin{table}[ht!]
\begin{center}
\begin{tabular}{|l|c|} \hline
 \multicolumn{2}{|c|}{1 Second Candidate} \\ \hline
 Date & 2015/05/30 \\
 Trigger Time & 08:20:59.67 UTC \\
 Duration & 1.0 second \\
 Obs. Counts & 9 \\
 Bkg. Counts & 0.115 \\
 Right Ascension & 292.83$^\circ$ (J2000) \\
 Declination & -17.53$^\circ$ (J2000) \\
 Zenith & 40.48$^\circ$ \\ \hline
\end{tabular}
\caption{Details of the overall best candidate from all three searches.
         This candidate was found in the 1 second sliding time window.}
\label{tab:1seccand}
\end{center}
\end{table}
\renewcommand{\arraystretch}{1.0}

\newpage


\newpage

\newpage

\newpage

\newpage

\newpage

\section{Upper Limit Calculation}\label{sec:limits}

The null-detection presented in the previous section is entirely consistent
with the expectation for observing $\sim$1 GRB per year from
the design sensitivity of the HAWC experiment. It is even more consistent 
with expectations after accounting for the currently reduced sensitivity
presented in Chapter \ref{ch:sens}. Yet, we have still shown that our
search method is capable of detecting extraordinary bursts like GRB 090510 and GRB 130427A.
We can therefore place upper limits on the rate of these rare bursts using the time period
of our data set and test
how they compare to their rate of detection in the Fermi LAT, the only experiment
which currently detects GRBs at high energies. 

We begin by noting that our null-detection yields a 90\% CL upper limit
of 2.3 GRBs occurring within our data set given the Poisson probability $P(i=0,\mu=2.3) = 0.1$. In principle
this corresponds to the integral number of GRBs producing VHE emission at
redshifts relevant to HAWC and a population study accounting for the measured redshift
distributions of both long and short GRBs is needed to fully interpret the limit
from of 2.3 GRBs occurring over the live time covered by our search. This study is currently underway but not yet finished.

In the absence of a full population study, we can still place constraints on the number
of GRBs occurring at specific redshifts by noting that the number of GRBs at any given
redshift must be less than the integral number of bursts. Limits obtained in this way
are much less constraining than a study of the integral number of bursts and do not
indicate that HAWC will not detect VHE emission from a GRB. However, they still provide
insight into the parameter space of bursts currently accessible to our search. 

\begin{figure}[ht!]
\begin{center}
\includegraphics[height=3in]{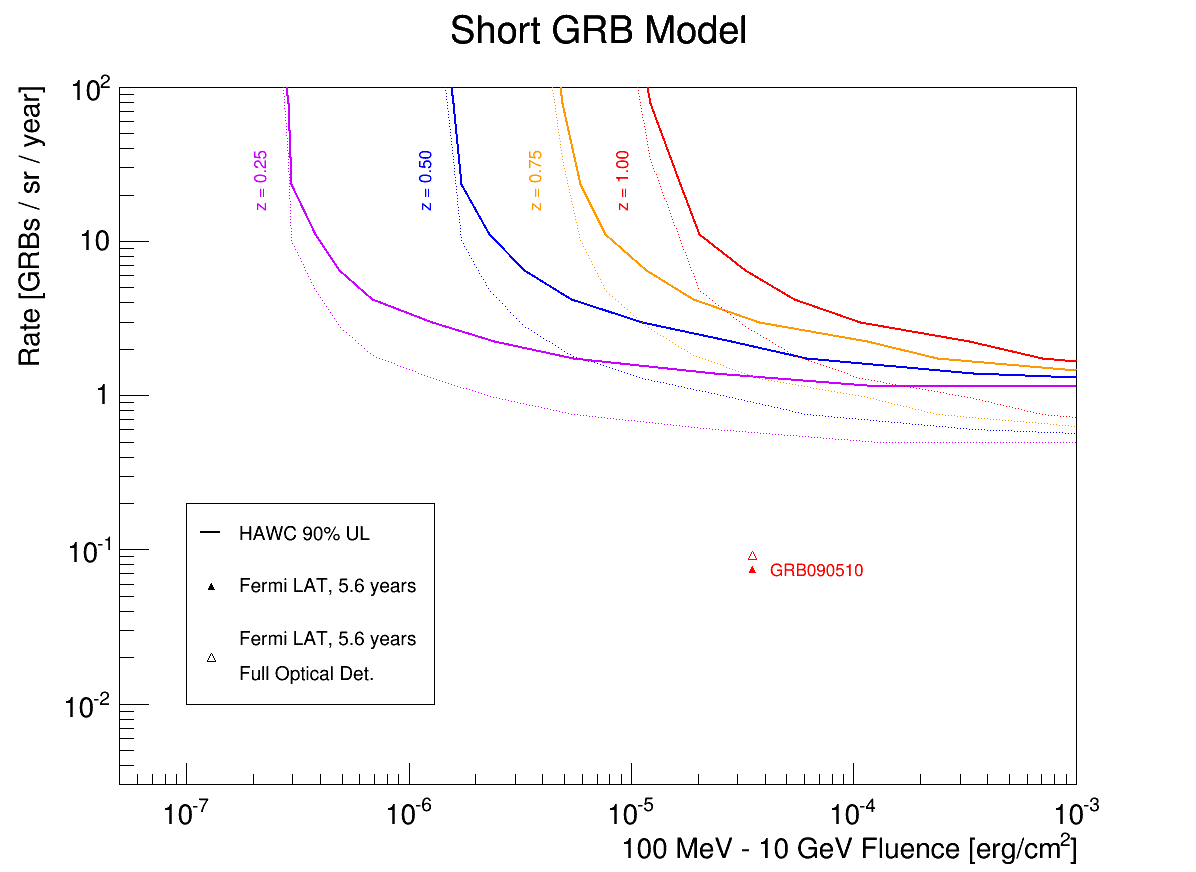}
\caption{HAWC upper limit on the rate of GRBs per steradian per year for simulated short GRBs coming from four different redshifts
         in the range where EBL is low enough to achieve appreciable detection at VHE photon energies in HAWC.
         Solid curves mark the upper limit obtained from applying the 90\% CL upper limit of 2.3 GRBs 
         at each redshift over the sensitivity of our search described in Chapter \ref{ch:sens}.
         The dashed curves mark the rate of GRBs if HAWC were to detect a single GRB.
         The solid triangle indicates the single short GRB detected by the LAT over its exposure during 
         the first 5.6 years of operations with measured redshift in the volume of space viewable to HAWC
         and reported fluence
         in the 100 MeV - 10 GeV energy band. The open triangle predicts the total potential rate of GRBs based on the single Fermi LAT detection after accounting
         for the lack of optical detections determining redshift.}
\label{fig:grbrate}
\end{center}
\end{figure}

\begin{figure}[ht!]
\begin{center}
\includegraphics[height=3in]{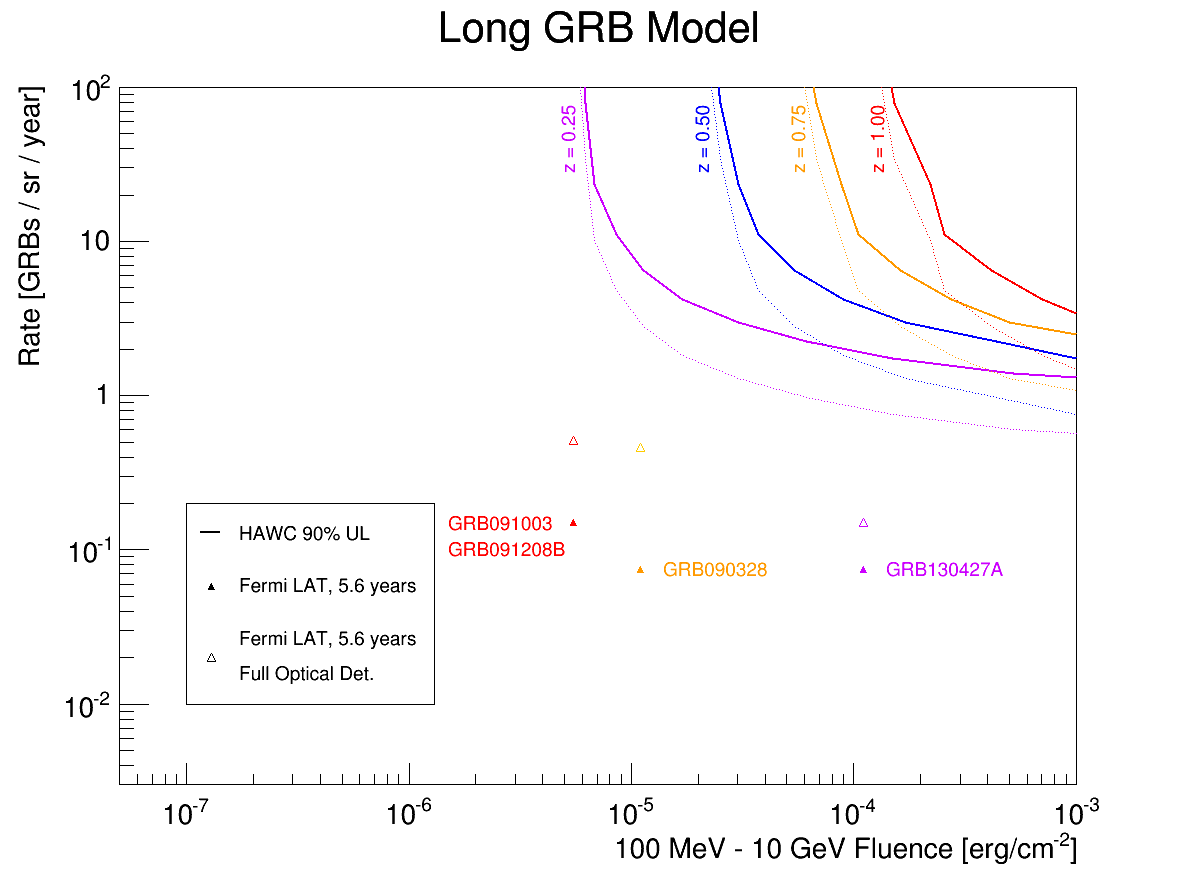}
\caption{HAWC upper limit on the rate of GRBs per steradian per year for modeled long GRBs coming from four different redshifts
         in the range where EBL is low enough to achieve appreciable detection at VHE photon energies in HAWC.
         Solid triangles indicate GRBs detected by the LAT with measured redshift and reported fluence
         in the 100 MeV - 10 GeV energy band.}
\label{fig:grbratelong}
\end{center}
\end{figure}

Figures \ref{fig:grbrate} and \ref{fig:grbratelong} present the upper limits for the rate of GRBs 
per steradian year occurring at modeled redshifts ranging from $z=0.25-1$ as a function
of the sensitivity of our search to fluence in the 100 MeV - 10 GeV
energy band. These are created by applying the upper limit on the integral number of bursts in our data
to each redshift and accounting for the portion of the HAWC field-of-view sensitive to a given fluence level
for this redshift. As we will see below, these rates are not very constraining compared to measurements of 
high energy bursts by Fermi LAT but they do provide insight into the exposure of our data set
as a function of fluence. High fluency bursts have the largest exposure because our analysis 
can detect them out to high zenith angles (Chapter \ref{ch:sens}) and therefore provide
the best limits on the rate of GRBs at a given redshift. Low fluency bursts are
only visible from directly overhead, resulting in a very low exposure and much higher limit.

For comparison, we also show the rate of detected bursts in the Fermi LAT
with measured redshifts over the same range described above and with reported values of fluence in 
the 100 MeV - 10 GeV energy band. Nearly all of these bursts come from the 
First Fermi LAT Burst Catalog \cite{Ackermann:2013a} which represents the most complete set of measurements
for Fermi LAT detected bursts. The one exception is the extraordinary burst GRB 130427A
which is very well studied due to its extremely high fluence. We choose to use the fluence
of reported during the first 163 seconds of this burst because it is most
representative of the prompt signal where our search would make a measurement.
Table \ref{tab:latburstlist} presents the full list of Fermi LAT detected bursts that
pass our criteria.

We calculate the rate of bursts per steradian year in the Fermi LAT data set by 
coarsely binning the bursts in redshift bins with $z = 0.25$
for both the short and long GRB populations. We then account
for the 2.4 steradian LAT field-of-view \cite{Ackermann:2013a}
and the 5.6 years between the launch of the Fermi satellite
and the end of 2013, the year containing GRB 130427A. The rates obtained this way
are much lower than the current upper limit from HAWC even if we
attempt to account for the selection bias in our sample by using the
distribution of known GRB redshifts to account for missing redshift
measurements in other LAT detected bursts in the 5.6 period containing our
sample. This is largely because the HAWC limits we present are
drawn from the original limit on the integral of bursts of bursts in our data
set and have little power to constrain individual bursts.

Nevertheless, Figure \ref{fig:grbrate} demonstrates a very important point
about the exposure of our search, namely that it is appreciable near 10$^{-6}$ erg/cm$^2$.
This is significant because typical LAT detected bursts typically have fluences on the order of 10$^{-5}$ erg/cm$^2$ \cite{Ackermann:2013a},
which is a result of the threshold required for triggering the LAT detector. Our search in HAWC therefore probes a largely unexplored parameter space
in fluence that could yield discoveries of a population of GRBs that the Fermi LAT has difficulty detecting. 











\newpage

\renewcommand{\arraystretch}{0.5}
\begin{table}[ht!]
\begin{center}
\begin{tabular}{|c|c|c|c|} \hline
  GRB  & Redshift & Type &Fluence \\
  & & & (100 MeV - 10 GeV) \\ \hline
  130427A & 0.34 & L & 1.1$\times10^{-4}$ erg/cm$^2$/s\\
  090510 & 0.90 & S & 3.5$\times10^{-5}$ erg/cm$^2$/s\\
  090328 & 0.74 & L & 1.1$\times10^{-5}$ erg/cm$^2$/s\\
  091003 & 0.90 & L & 0.6$\times10^{-5}$ erg/cm$^2$/s\\
  091208B & 1.06 & L & $<$0.5$\times10^{-5}$ erg/cm$^2$/s\\ \hline
\end{tabular}
\caption{LAT GRBs with measured redshifts less than 1.12
         and reported fluence in the 100 MeV - 10 GeV band.
         (S) indicates a short GRB and (L) indicates a long GRB. Bursts are sorted according to decreasing fluence.
         Data for GRB130427A come from References \cite{Maselli:2013uza} \cite{FLUENCE:2013}.
         Data for the remain bursts are from the First LAT Burst Catalog \cite{Ackermann:2013a}}.
\label{tab:latburstlist}
\end{center}
\end{table}
\renewcommand{\arraystretch}{1.0}

\newpage

\newpage

\newpage

\newpage

\section{Conclusion}

The analysis that we have developed here therefore shows a very promising future.
While we have not yet detected a burst, we know we are sensitive to bursts like GRB 090510
and GRB 130427A and would be able to detect a similar burst if it occurred today. 
And in the case of short GRBs we also have sensitivity to a fluence range where current satellites have difficulty
observing high energy emission, not because high energy photons do not exist in this range but rather because the $\sim$1 m$^2$
effective area of the Fermi LAT limits observations.
The effective area of HAWC, which is about 100$\times$ the size of the Fermi LAT,
provides our analysis with the sensitivity to probe this population of relatively unstudied
high energy GRBs.
It may therefore be only a few years before we obtain a detection of a burst.

Additionally, we expect the sensitivity of our method will improve over time
as the HAWC observatory is a young experiment. Already there have been great
strides towards matching the original design sensitivity with the latest Pass 4 analysis
and we expect more improvements to come as the HAWC collaboration plans to understand
the current systematic associated with reconstruction of small air showers.
Furthermore, recent changes have unified the on-site reconstruction performed in real-time
at HAWC with the off-site reconstruction methods of the Pass 4 data set.
This allows us to now run our analysis in real-time with the same sensitivity presented in this work.

\newpage

The ability to run this analysis in real-time and alert the astrophysical
community to a positive detection is key to obtaining the redshift measurements necessary to interpret
the result our search. These measurements, combined with a detected spectral
cutoff from the number of observed signal counts within our search, would provide an estimate of the bulk Lorentz factor in
the emission region of high energy photons and thereby yield great insights into
the environment where high energy emission occurs.
And with the VERITAS experiment currently observing the same TeV sky as HAWC,
our search's ability to provide real-time triggering on VHE emission from a GRB offers the tantalizing prospect
of the first IACT follow-up of a burst as well.

\appendix
\titleformat{\chapter}
      {\normalfont\large}{Appendix \thechapter:}{1em}{}
\renewcommand{\thechapter}{A}
\renewcommand{\chaptername}{Appendix}

\chapter{Calculation of $\Gamma_{min}$ in the one-zone model}\label{appendixC}

In the one-zone emission model, we assume both the low energy photons in the keV - MeV range
and the high energy photons in the GeV range are made in the same environment moving with bulk Lorentz factor $\Gamma$
during the prompt emission phase of the gamma-ray burst. It is therefore possible
for the highest energy photons to collide with lower energy photons immediately after their production and create
electron-positron pairs. This process results in appreciable attenuation of the highest energy photons. The impact of this attenuation
on the population of low energy photons is negligible because GRB spectra contain far more low energy events than there are high energy photons.

The pair production cross section is
\begin{equation}
\sigma_{\lambda \lambda}(y) = \sigma_T \, g(y), \,\,\,\,\, g(y) = \frac{3}{16} (1 - y^2) \Bigg[ (3 - y^4) ln \frac{1+y}{1-y} - 2 y (2 - y^2) \Bigg]
\end{equation}
where $\sigma_T = 6.65\times10^{-25}$ cm$^2$ is the the Thomson cross section and
\begin{equation}
y^2 \equiv 1 - \frac{ 2 m_e^2 c^4 }{ E_0' \, E' \, (1-cos\theta')}
\end{equation}
with $\theta'$ being the collision angle, $E_0'$ being the incident photon energy, and $E'$ being the target photon energy in the co-moving frame. \cite{Gould:1966}. 

Noting that the form of $y$ yields larger cross sections for higher energy target photons, we choose to only consider photons above energy $E_c$ in the Band fit
which yields the following expression for fluence measured at Earth
\begin{equation}
F(E) = T_{90} \, A \, ( \frac{E_c}{100 keV} )^\alpha \, e^{\beta - \alpha} \, ( \frac{E}{E_c} )^\beta = F(E_c) \, ( \frac{E}{E_c} )^\beta, \,\,\,\,\,\, E \ge E_c
\end{equation}
The photon number at energy $E$ is then
\begin{equation}
N(E) = \frac{ 4 \pi d_L(z)^2 }{(1+z)^2} \, F(E) \, dE
\label{eq:Nthatthing}
\end{equation}
where $z$ is the redshift and $d_L(z)$ is the luminosity distance of the source. This expression can be integrated to find the total photon number is
\begin{equation}
N_{tot} = \frac{ 4 \pi d_L(z)^2 }{(1+z)^2} \int_{E_c}^{\infty} \, F(E) \, dE= \frac{ 2 \pi d_L(z)^2 F(E_c) } { (1+z)^2 (-\beta - 1) } E_c
\end{equation}
Additionally, the conservation of photon number results in the relation
\begin{equation}
N'(E') = 4 \pi R^2 \, W' \, n'_\lambda(E') dE' = N(E)
\end{equation}
where $n'_\lambda(E')$ is the photon number density, $R$ is the emission region radius, and $W'$ is the emission region width in the co-moving frame. Substituting for the full expression of N(E) from Equation 
\ref{eq:Nthatthing} then gives
\begin{equation}
n_\lambda'(E') = \Bigg( \frac{d_L(z)}{R} \Bigg)^2 \frac{ \Gamma \, F(E_c) } {(1+z)^3 W'} \Bigg( \frac{E'}{E_c'} \Bigg)^\beta \,\,\,\, [\textrm{photons/cm$^3$/keV}]
\end{equation}
where primed energies satisfy the relation $E' = (1+z)E/\Gamma$. This results in the following expression for opacity due to pair production in the rest frame of emission
\begin{equation}
\tau_{\lambda \lambda}(E_0') = \int \Omega \int_{E_c'}^\infty dE' \, \frac{n_\lambda'(E')}{4 \pi} \, \sigma_{\lambda \lambda}(E_0',E', \theta') \, (1 - cos\theta') \, W'
\end{equation}
Defining
\begin{equation}
G(\beta) \equiv \frac{4}{(1-\beta)} \int_0^1 dy (1 - y^2)^{-\beta-2}g(y)y
\end{equation}
and converting all energies back to the observer frame yields
\begin{equation}
\tau_{\gamma\gamma}(E) = \sigma_T \, \Bigg( \frac{d_L(z)}{c \Delta t} \Bigg)^2 \, E_c F(E_c) \, (1 + z)^{-2(\beta + 1)} \, \Gamma^{2(\beta-1)} \, \Bigg( \frac{E E_c}{m_e^2 c^4} \Bigg)^{-\beta-1} \, G(\beta)
\end{equation}
where $\Delta t$ is the measured variability time of prompt emission and $W' = c \Delta t$. Setting the opacity equal to 1 and solving
for $\Gamma$ then gives the minimum bulk Lorentz factor required to detect a photon of energy $E_0$ during the prompt emission phase:
\begin{equation}
\Gamma > \Gamma_{min} =  \Bigg[ \sigma_T \Big( \frac{d_L(z)}{c \delta t} \Big)^{\frac{1}{2(1-\beta)}} (1 +z)^{ \frac{\beta + 1}{\beta - 1} } \Big( \frac{E_0 E_c}{m_e^2 c^4} \Big)^{\frac{beta+1}{2(\beta-1)}} \Bigg]
\end{equation}

\noindent Note: This calculation is reproduced from the supplementary material associated with \cite{Fermi:2008} with additional steps for clarity.

\renewcommand{\thechapter}{B}
\renewcommand{\chaptername}{Appendix}

\chapter{Full Solution to Transmission Line Equation}\label{appendixB}

\noindent Applying Kirchhoff's voltage law to the outermost
elements in Figure \ref{fig:coax_circuit} gives
\begin{equation}
v(x,t) = (r \Delta x) \,\, i(x,t) + (l \Delta x) \,\, \frac{d}{dt} i(x,t) + v(x+\Delta x,t)
\label{eq:voltage}
\end{equation}
Rearranging and taking the limit $\Delta x \to 0$ results in the differential equation
\begin{equation}
- \frac{dv}{dx}(x,t) = r \,\, i(x,t) + l \,\, \frac{di}{dt} (x,t) \label{eq:redvolt}
\end{equation}
Additionally, noting that the output voltage, $v(x+\Delta x,t)$, is applied across both the
capacitance and conductance yields the following expressions for currents $i_C$ and $i_G$
\begin{equation}
i_{C} = (c \Delta x) \, \frac{d}{dt} \, v(x + \Delta x,t) \qquad i_{G} = (g \Delta x) \, v(x + \Delta x,t)
\end{equation}
which can be used with Kirchhoff's current law to show
\begin{equation}
i(x,t) = (g \Delta x) \,\, v(x+\Delta x,t) + (c \Delta x) \,\, \frac{d}{dt} v(x+\Delta x,t) + i(x+\Delta x,t)
\label{eq:current}
\end{equation}
Again, rearranging and taking the limit $\Delta x \to 0$ results in a differential equation
\begin{equation}
- \frac{di}{dx}(x,t) = g \,\, v(x,t) + c \,\, \frac{dv}{dt} (x,t) \label{eq:redcurr}
\end{equation}
Since $x$ and $t$ are independent variables, we can differentiate Equation \ref{eq:redvolt} with respect to
$x$ and swap the order of integration for the last term on the right
\begin{equation}
-\frac{d^2v}{dx^2}(x,t) = r \,\, \frac{di}{dx}(x,t) + l \,\, \frac{d}{dt} \frac{di}{dx} (x,t)
\end{equation}
Using Equation \ref{eq:redcurr} to substitute for $\frac{di}{dx}(x,t)$ then gives
\begin{eqnarray}
-\frac{d^2v}{dx^2}(x,t) &= - r \,\, \bigg( g \,\, v(x,t) + c \,\, \frac{dv}{dt} (x,t) \bigg) - l \,\, \frac{d}{dt} \bigg( g \,\, v(x,t) + c \,\, \frac{dv}{dt} (x,t) \bigg) \\ [0.5em]
 \frac{d^2v}{dx^2}(x,t) &= rg \,\, v(x,t) + (rc + lg) \,\, \frac{dv}{dt}(x,t) + lc \,\, \frac{d^2v}{dt^2}(x,t) \hspace{2.5cm}
\end{eqnarray}
Taking $g = 0$, as is true in most insulators, simplifies this result to
\begin{equation}
\frac{d^2v}{dx^2}(x,t) = rc \,\, \frac{dv}{dt} (x,t)  + lc \,\, \frac{d^2v}{dt^2} (x,t)
\label{wave_eq_app}
\end{equation}
Looking for wave solutions of the form
\begin{equation}
v(x,t) = V e^{i(\omega t - kx)}
\end{equation}
results in the following relation
\begin{equation}
-k^2 \, V e^{i(\omega t - kx)} = (rc \, i\omega - lc \, \omega^2) \, V e^{i(\omega t - kx)}
\end{equation}
which must hold for all $x$ and $t$. The non-trivial solution for which this is true is
\begin{equation}
k = \sqrt{ lc \, \omega^2 - rc \, i\omega }
\label{eq:k_with_r}
\end{equation}

At high frequencies the skin effect allows us to treat the series resistance per unit length as occurring
inside a skin depth of $\delta$, and results in the expression for $r$ seen in Table \ref{tab:coaxcab_hf}.
The same effect also introduces an inductance which we treat by re-writing $r$ as a complex impedance
\begin{equation}
r \rightarrow z = K \sqrt{ i \omega }
\label{complex_z_per_dx}
\end{equation}
where $K \equiv \frac{1}{2 \pi a} \sqrt{ \frac{\mu_c}{\sigma_c} }$ after accounting for $a \ll b$ as is
true in typical coaxial cables \cite{Wigington:1957}. Combining this result with Equation \ref{eq:k_with_r}
then yields the following expression for $k$
\begin{equation}
k = \sqrt{ lc \, \omega^2 + \frac{K c \, \omega^{3/2}}{\sqrt{2}} - i \frac{ K c \, \omega^{3/2}}{\sqrt{2}} }
\label{eq:k_w_K}
\end{equation}
Applying Euler's formula $x + iy = A e^{i\theta}$ to the argument of the square root function in Equation \ref{eq:k_w_K} then gives the following relations
\begin{eqnarray}
A &= \sqrt{x^2 + y^2} \\
cos(\theta) &= \frac{x}{ \sqrt{x^2 + y^2} } \\
sin(\theta) &= \frac{y}{ \sqrt{x^2 + y^2} }
\end{eqnarray}
where $x \equiv lc \, \omega^2 + \frac{K c \, \omega^{3/2}}{\sqrt{2}}$ and $y \equiv -\frac{ K c \, \omega^{3/2}}{\sqrt{2}}$ which results in
\begin{eqnarray}
Re(k) &= \sqrt{ \frac{\sqrt{x^2 + y^2} + x}{2} }\\
Im(k) &= \frac{y} {|y|} \sqrt{ \frac{\sqrt{x^2 + y^2} - x}{2} }
\end{eqnarray}
Note, though, that we can re-write the equation for $k$ as
\begin{equation}
k = \sqrt{lc} \, \omega \, \sqrt{ 1 + (1 - i) \frac{K}{l \, \sqrt{2\omega}} }
\end{equation}
and in the high frequency limit where $\frac{K^2}{2 \omega l^2} \ll 1$ this becomes
\begin{equation}
k \approx \Bigg(\sqrt{lc} \, \omega + \frac{K}{2 Z_0} \sqrt{ \frac{\omega}{2} }\Bigg) - \, i \frac{K}{2 Z_0} \sqrt{ \frac{\omega}{2}}
\end{equation}
where $Z_0 \equiv \sqrt{l / c}$ is the intrinsic impedance of the cable. The expressions for the real and imaginary
parts of $k$ in the high frequency limit are therefore reduced to
\begin{eqnarray}
Re(k) &= \sqrt{lc} \, \omega + \frac{K}{2 Z_0} \sqrt{ \frac{\omega}{2} }\\
Im(k) &= -\frac{K}{2 Z_0} \sqrt{ \frac{\omega}{2}} \,\,\,\,\,\,\,\,\,\,\,\,\,\,\,\,\,\,
\end{eqnarray}
resulting in a frequency dependent wave velocity
\begin{equation}
v = \frac{\omega}{Re(k)} = \frac{1}{\sqrt{lc} + K / 2Z_0 \, \sqrt{2\omega}}
\end{equation}

\renewcommand{\thechapter}{C}
\renewcommand{\chaptername}{Appendix}

\chapter{Full Solution to Analog Input Circuit}\label{appendixD}

\begin{figure}[ht!]
\begin{center}
\includegraphics[width=2.5in]{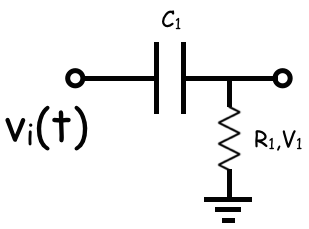}
\end{center}
\caption{Input circuit from simplified analog circuit diagram in Figure \ref{fig:analog_circuit}}
\label{eq:analog_input_circuit}
\end{figure}

\noindent Applying Kirchhoff's voltage loop law to Figure \ref{eq:analog_input_circuit} gives
\begin{equation}
v_i(t) = \frac{Q(t)}{C_1} + R_1 \, I(t)
\end{equation}
which results in 
\begin{equation}
\frac{1}{R_1} \, \frac{dv_i}{dt}(t) \, e^{t/R_1C_1} = \Big( \frac{I(t)}{R_1C_1} +  \frac{dI}{dt}(t) \Big) \, e^{t/R_1C_1} = \frac{d}{dt} \Big( I(t) \, e^{t/R_1C_1} \Big)
\end{equation}
after differentiating both sides with respect to $t$, dividing by $R_1$, and multiplying by $e^{t/R_1C_1}$. Integrating both sides of this equation with respect to $t$ yields the following
expression for $I(t)$
\begin{equation}
I(t) = e^{-t/R_1C_1} \int_0^t dt' \, \frac{1}{R_1} \, \frac{dv_i}{dt'}(t') \, e^{t'/R_1C_1}
\end{equation}
assuming the initial condition $v_i(t=0) = 0$ which forces $I(t=0) = 0$. The voltage $V_1$ is then
\begin{equation}
V_1(t) = R_1 \, I(t) = e^{-t/R_1C_1} \int_0^t dt' \, \frac{dv_i}{dt'}(t') \, e^{t'/R_1C_1}
\end{equation}
according to Ohm's law.

\renewcommand{\thechapter}{D}
\renewcommand{\chaptername}{Appendix}

\chapter{Full Solution to Analog Load Circuit}\label{appendixE}

\begin{figure}[ht!]
\begin{center}
\includegraphics[width=3.5in]{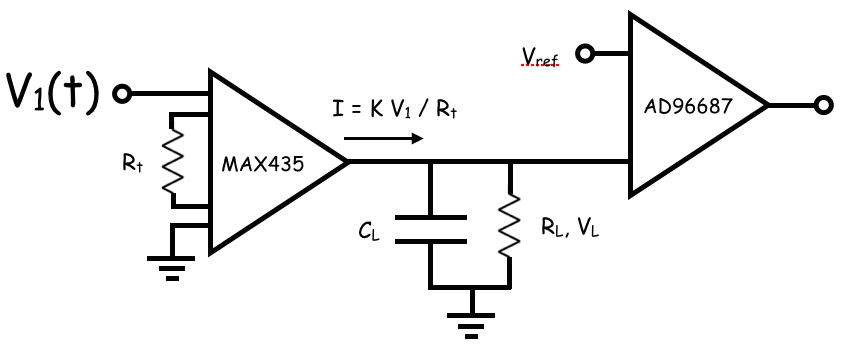}
\end{center}
\caption{Load circuit from simplified analog circuit diagram in Figure \ref{fig:analog_circuit}}
\label{eq:analog_load_circuit}
\end{figure}

\noindent The MAX435 transconductance amplifier creates a current
\begin{equation}
I(t) = \frac{K}{R_t} V_1(t)
\label{eq:trans_gain}
\end{equation}
for input voltage $V_1(t)$. This current then travels entirely across $R_L$ and $C_L$ because the AD96687 ultra-fast comparator chip has a large input impedance.
Since $R_L$ and $C_L$ are in parallel, there is an identical voltage drop across both elements
\begin{equation}
V_L = R_L \, I_R = \frac{Q_C}{C_L}
\end{equation}
and the current through each element can be written in terms of $V_L(t)$
\begin{eqnarray}
I_R &= V_L / R_L \,\,\,\,\,\,\,\,\,\,\,\,\,\,\,\,\,\,\,\,\,\,\,\,\,\,\,\,\,\, \\
I_C &= dQ_C /dt = C_L \, dV_L/dt
\end{eqnarray}
Kirchhoff's current law dictates
\begin{equation}
I(t) = I_R + I_C = \frac{V_L(t)}{R_L} + C_L \frac{dV_L}{dt}(t)
\end{equation}
Multiplying both sides by $e^{t/R_LC_L}$ and dividing by $C_L$ gives
\begin{equation}
\frac{1}{C_L} \, I(t) \, e^{t/R_LC_L} = \frac{d}{dt} \Big( V_L(t) \,  e^{t/R_LC_L} \Big)
\end{equation}
which can be integrated with respect to $t$ find
\begin{equation}
V_L(t) = \frac{K}{C_LR_t} \, e^{-t/R_LC_L} \int_0^t dt' \, V_1(t') \, e^{t'/R_LC_L}
\end{equation}
where we have used Equation \ref{eq:trans_gain} to express $I(t)$ in terms of $V_1(t)$ and we assume
$V_L(t)$ is equal to 0 at $t=0$.

\renewcommand{\thechapter}{E}
\renewcommand{\chaptername}{Appendix}

\chapter{HAWC PMT Base Design}\label{appendixbase}

\begin{figure}[ht!]
\begin{center}
\includegraphics[width=6in]{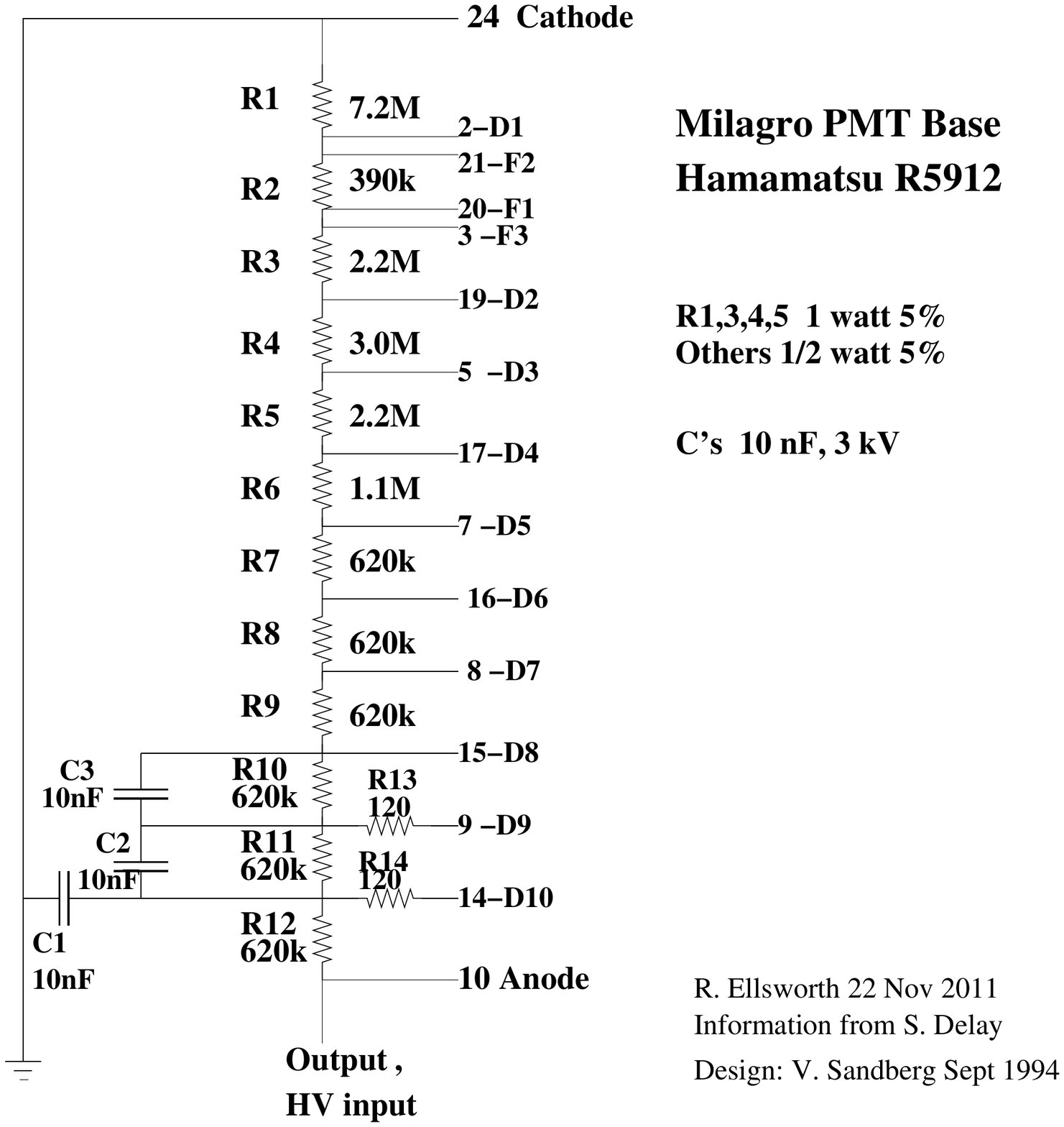}
\end{center}
\caption{HAWC PMT Base Design}
\label{fig:pmtbase}
\end{figure}

Dynode voltages in the HAWC PMTs are set by a resistor chain which acts as a voltage divider (Figure \ref{fig:pmtbase}).
The cathode is set to ground and the anode is held at positive high voltage. The voltages differences across each resistor
for an operating voltage of 1500V are shown in Table \ref{tab:resvolts}.

\renewcommand{\arraystretch}{0.5}
\begin{table}[ht!]
\begin{center}
\begin{tabular}{|c|c|c|} \hline
  Resistor & Resistance [M$\Omega$] & $\Delta$V [Volts] \\ \hline
  R1 & 7.20 & 545 \\
  R2 & 0.39 & 29.5 \\
  R3 & 2.20 & 167 \\
  R4 & 3.00 & 227 \\
  R5 & 2.20 & 167 \\
  R6 & 1.10 & 83.3 \\
  R7 & 0.62 & 47.0 \\
  R8 & 0.62 & 47.0 \\
  R9 & 0.62 & 47.0 \\
  R10 & 0.62 & 47.0 \\
  R11 & 0.62 & 47.0 \\
  R12 & 0.62 & 47.0 \\ \hline
\end{tabular}
\end{center}
\caption{Voltage differences across each resistor in Figure \ref{fig:pmtbase} for an operating voltage of 1500V}
\label{tab:resvolts}
\end{table}
\renewcommand{\arraystretch}{1.0}

\renewcommand{\baselinestretch}{1}
\small\normalsize
\newpage
\bibliographystyle{unsrt}
\bibliography{Bibliography}
\end{document}